\documentclass[10pt, final, journal, letterpaper, oneside, twocolumn]{IEEEtran}
\usepackage{graphicx}
\usepackage{amsmath}
\usepackage{amssymb}
\usepackage{mathrsfs}
\usepackage{stfloats}
\usepackage{dsfont}
\usepackage{setspace}
\usepackage{array}
\usepackage{bbm}
\usepackage{color, soul, framed}
\usepackage[centerlast,small]{caption}
\begin{document}
\title{Smart Radio Environments Empowered by Reconfigurable Intelligent Surfaces:\\ How it Works, State of Research, and Road Ahead}
\author{Marco~Di~Renzo,~\IEEEmembership{Fellow,~IEEE},
Alessio~Zappone,~\IEEEmembership{Senior~Member,~IEEE},
Merouane~Debbah,~\IEEEmembership{Fellow,~IEEE},
Mohamed-Slim~Alouini,~\IEEEmembership{Fellow,~IEEE},
Chau~Yuen,~\IEEEmembership{Senior~Member,~IEEE},
Julien~de~Rosny, and
Sergei~Tretyakov,~\IEEEmembership{Fellow,~IEEE}
\thanks{Manuscript received April 20, 2020.} 
\thanks{M. Di Renzo is with Universit\'e Paris-Saclay, CNRS and CentraleSup\'elec, Laboratoire des Signaux et Syst\`emes,  Gif-sur-Yvette, France. (e-mail: marco.direnzo@centralesupelec.fr). The research activity of M. Di Renzo was supported by the European Commission through the H2020 ARIADNE project under grant 871464.} 
\thanks{A. Zappone is with University of Cassino and Lazio Meridionale, Cassino, Italy. (e-mail: alessio.zappone@unicas.it).}
\thanks{M. Debbah is with Huawei France R\&D, Boulogne-Billancourt, France. (e-mail: merouane.debbah@huawei.com).}
\thanks{M.-S. Alouini is with King Abdullah University of Science and Technology (KAUST), Kingdom of Saudi Arabia. (e-mail: slim.alouini@kaust.edu.sa).}
\thanks{C. Yuen is with Singapore University of Technology and Design (SUTD), Singapore. (e-mail: yuenchau@sutd.edu.sg).}
\thanks{J. de Rosny is with Paris Sciences \& Lettres, CNRS, Institut Langevin, Paris, France. (e-mail: julien.derosny@espci.psl.eu).}
\thanks{S. Tretyakov is with Aalto University, Helsinki, Finland. (e-mail: sergei.tretyakov@aalto.fi).}
}
%
%
%\markboth{Journal on Selected Areas in Communications} {M. Di Renzo et al.: Smart Radio Environments Empowered by Reconfigurable Intelligent Surfaces: How it Works, State of Research, and Road Ahead}
%
%
%
%
\maketitle
\begin{abstract}
What is a reconfigurable intelligent surface? What is a smart radio environment? What is a metasurface? How do metasurfaces work and how to model them? How to reconcile the mathematical theories of communication and electromagnetism? What are the most suitable uses and applications of reconfigurable intelligent surfaces in wireless networks? What are the most promising  smart radio environments for wireless applications? What is the current state of research? What are the most important and challenging research issues to tackle? 

These are a few of the many questions that we investigate in this short opus, which has the threefold objective of introducing the emerging research field of smart radio environments empowered by reconfigurable intelligent surfaces, putting forth the need of reconciling and reuniting C. E. Shannon's mathematical theory of communication with G. Green's and J. C. Maxwell's mathematical theories of electromagnetism, and reporting pragmatic guidelines and recipes for employing appropriate physics-based models of metasurfaces in wireless communications.
\end{abstract}
\section{Introduction} \label{Introduction}
\textbf{Increasing data traffic}. Wireless connectivity is regarded as a fundamental need for our society. Between 2020 and 2030, it is forecast that the data traffic of the global Internet protocol (IP) will increase by 55\% each year, eventually reaching 5,016 exabytes \cite{ITU_2015}, with data rates scaling up to 1 Tb/s \cite{Letaief_AI}. Besides supporting very high data rates, future wireless networks are expected to offer several other heterogeneous services, which include sensing, localization, low-latency and ultra-reliable communications. Fifth-generation (5G) networks are, however, not designed to meet these requirements. As the demands and needs become more stringent, in fact, fundamental limitations arise, which are ultimately imposed by the inherent nature of wireless operation. 

\textbf{Current network design assumptions}. The first five generations of wireless networks have been designed by adhering to the postulates that the wireless environment between communicating devices (i) is fixed by nature, (ii) cannot be modified, (iii) can be only compensated through the design of sophisticated transmission and reception schemes. After five generations of wireless networks, however, the improvements that can be expected by operating only on the end-points of the wireless environment may not be sufficient to fulfill the challenging requirements of future wireless networks. The sixth generation (6G) of communication networks is, on the other hand, envisioned to require a new architectural platform that performs joint communication, sensing, localization, and computing, while ensuring ultra-high throughput, ultra-low latency, and ultra-high reliability, which need to be flexibly customized in real-time. 

\textbf{An emerging paradigm: Programming the environment}. Major performance gains can be expected by breaking free from the postulate that regards the wireless environment as an uncontrollable element. For example, a typical base station transmits radio waves of the order of magnitude of Watts while a user equipment detects signals of the order of magnitude of $\mu$Watts. The rest of the power is, in general, wasted in different ways through the environment by, e.g., generating interference to other network elements or creating security threats, since the propagation of radio waves through the wireless channel cannot be controlled and customized after they are emitted from the transmitters and before they are received by the receivers. An intriguing question was recently brought to the attention of the wireless community: \textit{Can this status quo be fundamentally overcome}?

\begin{figure}[!t]
\begin{centering}
\includegraphics[width=\columnwidth]{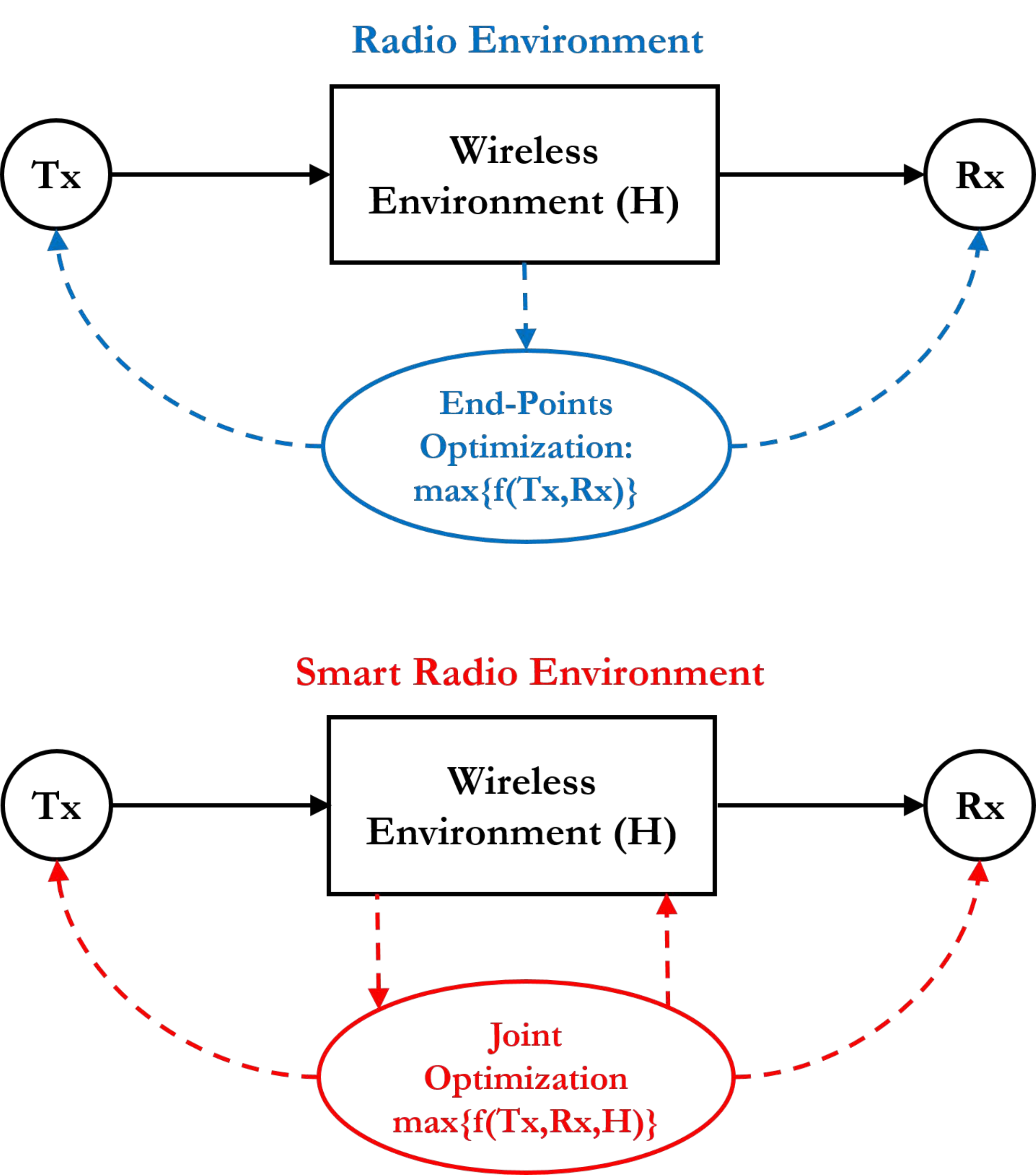}
\caption{Radio environments vs. smart radio environments.}
\label{Fig_1}
\end{centering} 
\end{figure}
\textbf{The road to smart radio environments}. At the time of writing, no precise answer to this question can be given. A plethora of research activities have, however, recently flourished in an attempt of tackling and putting this question in the context of the most promising technologies that were developed during the last decades and that are envisioned to constitute the backbone of 5G networks. The current long-term vision for overcoming the limitations of 5G networks consists of turning the wireless environment into an optimization variable, which, jointly with the transmitters and receivers, can be controlled and programmed rather than just adapted to. This approach is widely referred to as smart radio environment (SRE) or, more recently, intelligent radio environment (IRE), or ``Wireless 2.0'' in order to emphasize the conceptual and fundamental difference with the designs and optimization criteria adopted in current and past generations of wireless networks. Conceptually, the vision of SREs is depicted in Fig. \ref{Fig_1}.

\textbf{Structure of the paper}. The objective of the present paper is to provide the readers with a comprehensive and critical overview of the fundamental technology enablers, the main operating principles and envisioned potential applications, the current state of research, and the open research challenges of the emerging concept of SREs. To this end, the present paper is organized in six sections.
\begin{itemize}
\item In Section II, the concept of reconfigurable intelligent surface (RIS), as the technology enabler to realize the vision of SREs, is introduced.
\item In Section III, the concept of SREs is introduced in more detail by reporting major examples of application and use cases.
\item In Section IV, a communication-theoretic perspective to RIS-empowered SREs is given, with a focus on analytical and computational methods for modeling RISs and their interactions with the radio waves.
\item In Section V, a comprehensive survey of the current state of research is given.
\item In Section VI, the major open research issues that need to be tackled to realize the vision of SREs are discussed.
\item In Section VII, final conclusions are provided.
\end{itemize}

\section{Reconfigurable Intelligent Surfaces} \label{RISs}
\textbf{General definition}. The key enabler to realize the vision of SREs, by making the wireless environment programmable and controllable, is the so-called RIS. Broadly speaking, an RIS can be thought of as an inexpensive adaptive (\textit{smart}) thin composite material sheet, which, similar to a \textit{wallpaper}, covers parts of walls, buildings, ceilings, etc., and is capable of modifying the radio waves impinging upon it in ways that can be programmed and controlled by using external stimuli. A prominent property of RISs is, therefore, the capability of being (re-)configurable after their deployment in a wireless environment. 

\textbf{General operation}. Based on this general definition, the operation of an RIS can, in general, be split into two phases that are executed periodically based on the coherence time of the environment.
\begin{itemize}
\item \textbf{\textit{Control and programming phase}}. During this phase, the necessary environmental information for configuring the operation of the RIS is estimated, and it is configured for subsequent operation.
\item \textbf{\textit{Normal operation phase}}. During this phase, the RIS is configured already and assists the transmission of other devices throughout the network.
\end{itemize}

In further text, we elaborate on different implementations of this general working operation, which include centralized, distributed, and hybrid network architectures, and encompass the control/programming and normal operation phases.

\subsection{Two Practical Examples of Reconfigurable Intelligent Surfaces} \label{Examples_RISs}
\begin{figure}[!t]
\begin{centering}
\includegraphics[width=\columnwidth]{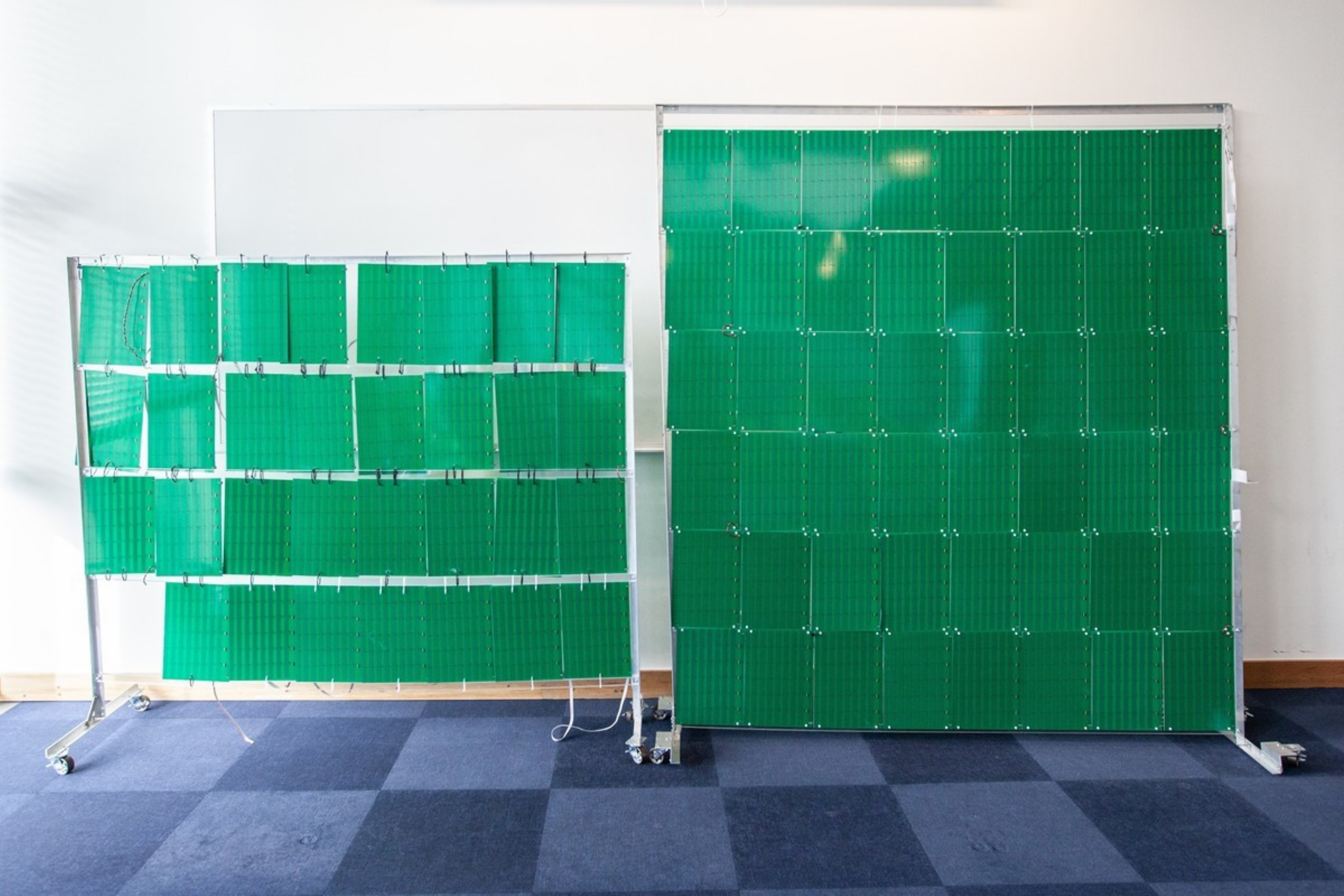}
\caption{MIT's RFocus prototype (photo: Jason Dorfman, CSAIL).}
\label{Fig_2}
\end{centering} 
\end{figure}

\begin{figure}[!t]
\begin{centering}
\includegraphics[width=\columnwidth]{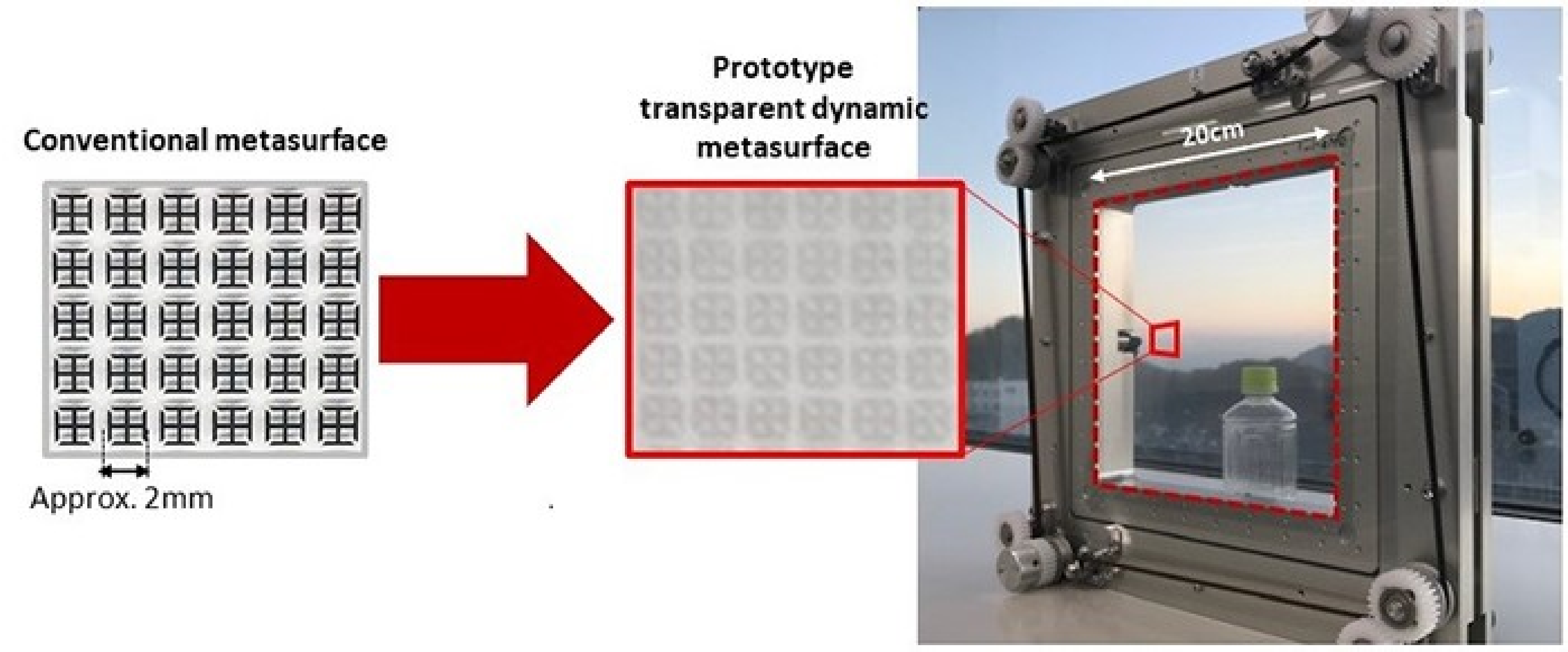}
\caption{NTT DOCOMO's prototype (photo: NTT DOCOMO).}
\label{Fig_3}
\end{centering} 
\end{figure}

\textbf{Two examples of RISs}. Although the current state of research may be far from realizing RISs according to the just mentioned definition, several researchers are working towards the realization of smart surfaces that behave, conceptually, as a programmable thin \textit{wallpaper} and as a programmable thin \textit{glass}, which are capable of manipulating the radio waves as desired. Two recent examples of these research activities are illustrated in Figs. \ref{Fig_2} and \ref{Fig_3}. 

\textbf{MIT's RFocus prototype}. In Fig. \ref{Fig_2}, the RFocus prototype, recently designed by researchers of the Massachusetts Institute of Technology (MIT), USA, is depicted \cite{RFocus}. The RFocus prototype is made of 3,720 inexpensive antennas arranged on a six square meter surface. At scale, each antenna element is expected to have a cost of the order of a few cents or less. The structure operates in a nearly-passive mode, since the surface itself does not emit new radio waves, but it can be adaptively configured by means of low power electronic circuits in order to beamform and to focus the impinging radio waves towards specified direction and locations, respectively. 

\textbf{NTT DOCOMO's prototype}. In Fig. \ref{Fig_3}, a prototype of smart glass, recently designed by researchers of NTT DOCOMO, Japan, is depicted \cite{Docomo_Glass}. The manufactured smart glass is an artificially engineered thin layer (i.e., a metasurface) that comprises a large number of sub-wavelength unit elements placed in a periodic arrangement on a two-dimensional surface covered with a glass substrate. By moving the glass substrate slightly, it is possible to dynamically control the response of the impinging radio waves in three modes: (i) full penetration of the incident radio waves; (ii) partial reflection of the incident radio waves; and (iii) full reflection of all radio waves. The smart glass is highly transparent and, hence, is suitable for unobtrusive use. For example, it can manipulate the radio waves in accordance with the specific installation environment, particularly in locations that are not suited for installing base stations, such as in built-up areas or in indoor areas where the reception of signals needs to be blocked selectively, (e.g., high-security areas). In addition, the transparent substrate does not interfere aesthetically or physically with the surrounding environment or with the line-of-sight of people, thus making the structure suitable for use within buildings and on vehicles or billboards.

\subsection{Nearly-Passive Reconfigurable Intelligent Surfaces} \label{NearlyPassive_RISs}
\textbf{Different kinds of RISs}. Different kinds of RISs are currently under research and design. These include smart surfaces that are or are not capable of amplifying and performing signal processing operations on the impinging radio waves (active vs. passive surfaces), as well as surfaces whose functions cannot or can be modified after being manufactured and deployed (static vs. dynamic/reconfigurable surfaces). A detailed classification of these options is provided in further text. For ease of writing, however, we feel important to mention that in the present paper we refer, unless otherwise stated, to RISs that can be broadly classified as nearly-passive and dynamic. 

\textbf{Definition of nearly-passive RISs}. We define an RIS as nearly-passive and dynamic (or simply reconfigurable) if the following three conditions are fulfilled simultaneously.
\begin{enumerate}
\item No power amplification is used after configuration (during the normal operation phase).
\item Minimal digital signal processing capabilities are needed only to configure the surface (during the control and programming phase).
\item Minimal power is used only to configure the surface (during the control and programming phase).
\end{enumerate}

Based on this definition, the next sub-section reports the conceptual architecture of a nearly-passive RIS. Then, the subsequent sub-section presents a broader classification of RISs for which the above-mentioned three conditions may not be fulfilled simultaneously. The relevance and broad interest in nearly-passive RISs is elaborated at the end of this section.

\begin{figure}[!t]
\begin{centering}
\includegraphics[width=\columnwidth]{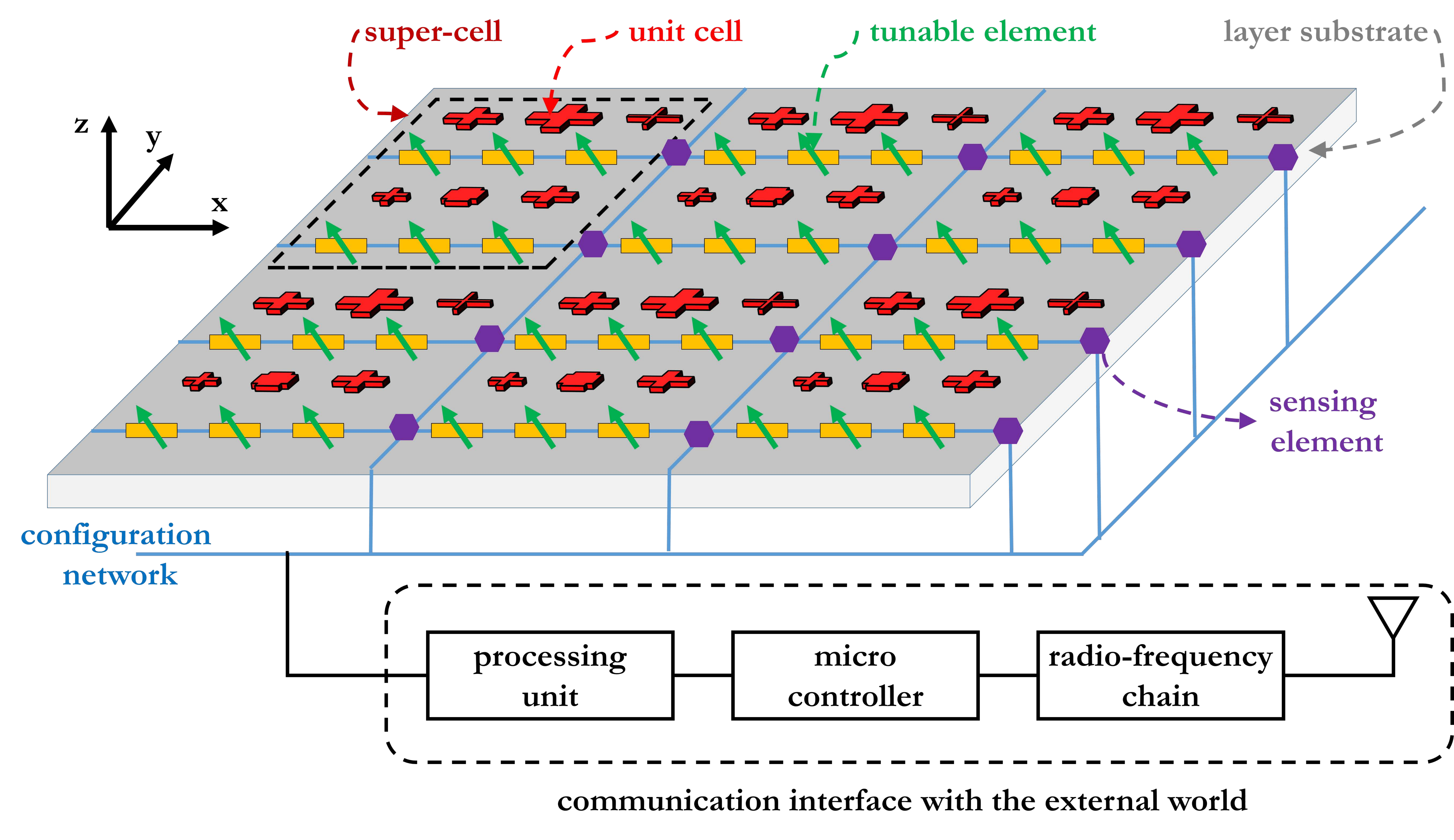}
\caption{Conceptual structure of a reconfiguration intelligent surface.}
\label{Fig_4}
\end{centering} 
\end{figure}
\subsection{Conceptual Structure of Reconfigurable Intelligent Surfaces} \label{RIS_Structure}
\textbf{Reference structure}. The two examples in Figs. \ref{Fig_2} and \ref{Fig_3} indicate that an RIS can be realized by employing different technologies and that it can be designed based on conceptually different approaches. Regardless of the specific design methods and engineering implementations, a conceptual schematic representation of an RIS, which can be employed for analyzing and synthesizing it, is reported in Fig. \ref{Fig_4}. 

\textbf{Two-dimensional structure}. The RIS in Fig. \ref{Fig_4} is modeled as a two-dimensional structure of man-made material, whose transverse size is much larger than its thickness. Usually, the transverse size of an RIS is much larger than the wavelength of the radio waves (e.g., a few tens or a few hundreds times larger than the wavelength depending on the functions to realize), and its thickness is much smaller than the wavelength of the radio waves. For this reason, an RIS is often referred to as a \textit{zero-thickness sheet of electromagnetic material}. The two-dimensional structure in Fig. \ref{Fig_4} makes RISs easier to design and to deploy, less lossy, and less expensive to realize, as compared with their three-dimensional counterpart whose thickness is not negligible.

\textbf{Composite material layers of unit cells}. The RIS in Fig. \ref{Fig_4} is constituted by composite material layers that are made of metallic or dielectric patches printed on a grounded dielectric substrate. Each patch can be modeled as a passive scattering element and it is often referred to as \textit{unit cell} or scattering particle. The microscopic design of each unit cell determines the macroscopic response of the RIS to the impinging radio waves. This includes the material with which the unit cells are made of, the size of the unit cells, and the inter-distance among the unit cells. As detailed in further text, the size and inter-distance of the unit cells can be either of the order of the wavelength (usually half the wavelength) or can be smaller than the wavelength (usually 5-10 times smaller than the wavelength). We anticipate that an RIS can be either locally passive or locally active, even if it is globally nearly-passive, i.e., the sum of the reflected and transmitted powers is equal to the incident power. If the response of an RIS is locally active in some parts of the metasurface structure and is locally passive in some other parts of the metasurface structure, this implies that some surface waves may exist and that they may travel along the metasurface structure so as to transfer energy from the (virtual) passive regions to the (virtual) active regions. In other terms, the metasurface may transform the incident electromagnetic (EM) fields into desired EM fields through appropriate, locally distributed, balanced absorptions and gains. This design ensures that no active elements are employed during the normal operation phase and that the metasurface structure is globally passive. An example of a metasurface structure that realizes perfect anomalous reflection based on this design principle is analyzed in Section IV.

\textbf{Configuration network}. The dynamic operation of the RIS in Fig. \ref{Fig_4}, i.e., its reconfigurability, is ensured by low power tunable electronic circuits, e.g., positive intrinsic negative (PIN) diodes or varactors. By appropriately configuring the on/off state of the PIN diodes or the bias voltage of the varactors, one can control and program the (macroscopic) transformations that are applied to the impinging radio waves. Let us consider, for example, that the RIS is made of two PIN diodes and that the unit cells are designed to simply rotate the phase of the impinging radio wave of 0, 90, 180, and 270 degrees. Then, the rotation phase can be controlled by coding it into the four possible states of the two PIN diodes, which need only two bits for being configured. In general terms, the specific design of the unit cells (their size, their inter-distance, the material they are made of, etc.), and the inter-cell communication network, which define the microscopic behavior of the RIS, determine the functions that the RIS can apply, at the macroscopic level, to the impinging radio waves.

\textbf{Communication with the external world}. In order to be controlled and programmed remotely, Fig. \ref{Fig_4} highlights that RISs need to be equipped with at least one gateway (with transmit and receive capabilities), which constitutes the interface of the RIS with the external world. In addition,  a micro-controller and a wireless or wired (on-chip) inter-cell communication network, which enables the transfer of information from the gateway throughout the surface, are needed. As mentioned, nearly-passive RISs have minimal power requirements and signal processing capabilities. These are needed, only during the control and programming phase, for: (i) operating the low power tunable electronic circuits that ensure the reconfigurability of the RIS; and (ii) communicating with the external world, e.g., to receive the control and configuration signals.

\textbf{On-board sensing capabilities}. The nearly-passive RIS depicted in Fig. \ref{Fig_4} may or may not be equipped with low power sensing elements, whose role is to help estimating the channel (or, more in general, the environmental) state information that is necessary for optimizing the operation of the RIS based on some key performance indicators, e.g., the desired signal-to-noise ratio at a given location. Equipping RISs with low power sensors increases the cost and the power consumption of the entire surface. Dispensing RISs with low power sensors makes, on the other hand, more challenging the estimation of the necessary environmental state information, since the RISs cannot sense and learn the environment on their own. The vast majority of current research activities rely on the assumption that nearly-passive RISs are not equipped with sensing elements, and, therefore, different algorithms and protocols are under analysis for efficiently estimating the channel state information that is needed for optimizing their operation. These research activities are discussed in Section V.

\subsection{Surfaces vs. Smart Surfaces} \label{Surfaces_vs_SmartSurfaces}
Having defined the conceptual structure of a nearly-passive RIS, one may wonder what the difference between a conventional surface (e.g., a wall) and a smart surface (e.g., a smart wall) like the one sketched in Fig. \ref{Fig_4} is. 

\textbf{Conventional surfaces}. In general terms, a radio wave that impinges upon a conventional wall induces some surface currents that determine the radio waves that are scattered off. The surface currents are determined by the boundary conditions at the interface of the wall, which depend on the permittivity and permeability of the material that the wall is made of, its thickness, and the wavelength of the radio waves. 

\textbf{Smart surfaces}. When the same radio wave impinges upon a smart wall, the distribution of the surface currents is, in general, different, and is determined by the characteristics of the unit cells (their size, their inter-distance, the material they are made of, etc.) and the status of the electronic circuits that constitute the configuration network. The different surface currents result in different radio waves that are scattered off. In general, an RIS can be approximately modeled by specific boundary conditions at the interface of a smart wall, which define discontinuities of the electric and magnetic fields in the close vicinity of the surface. For this reason, an RIS is often referred to as an electromagnetic discontinuity in space. This concept is further elaborated and discussed in Section IV.

\textbf{A simple example: Specular vs. anomalous reflection}. A typical example to understand this difference is the relation between specular reflection and anomalous reflection. When a radio wave impinges upon a uniform surface, the angle of incidence and the angle of reflection (with respect to the normal of the surface) are the same. This is dictated by the boundary conditions at the surface and the corresponding surface currents that are induced by the impinging radio waves. When a radio wave impinges upon a smart surface, on the other hand, the unit cells and the configuration network can be designed to make the angle of incidence and the angle of reflection different. This is obtained because the design of the unit cells is realized in a way that the induced surface currents generate radio waves that are reflected, predominantly, in a specified direction that may be different from the direction of incidence (with respect to the normal of the surface).

\subsection{Metamaterials-Based Reconfigurable Intelligent Surfaces} \label{Metasurfaces}
The architecture of the smart surface illustrated in Fig. \ref{Fig_4} is general enough for accommodating different practical implementations of nearly-passive RISs. At the time of writing, the vast majority of researchers in wireless communications have focused their attention on two main practical implementations.
\begin{itemize}
\item \textbf{\textit{Smart surfaces made of discrete tiny antenna elements}}. This implementation is exemplified in the hardware prototype illustrated in Fig. \ref{Fig_2}. In this case, the unit cells depicted in Fig. \ref{Fig_4} can be regarded as tiny antenna elements whose size and inter-distance are usually equal to half of the wavelength of the radio waves. Conceptually, each unit cell is often considered as a reflecting element that modifies the phase of the impinging radio wave. Since the unit cell are spaced by half of the wavelength, the mutual coupling among them is usually ignored and each unit cell is designed independently of the others.
\item \textbf{\textit{Smart metasurfaces}}. This implementation is exemplified in the hardware prototype illustrated in Fig. \ref{Fig_3}. In this case, the unit cells depicted in Fig. \ref{Fig_4}, which can be full or slotted patches, straight or curves strips, etc., are arranged in non-uniform repeating patterns. The repeating pattern of the unit cells is referred to as \textit{super-cell}, and it determines the period of the structure. The unit cells that constitute the super-cell have, usually, different geometric shapes and sizes, and are jointly optimized in order to realize the specified functions. This implies that the mutual coupling among the unit cells of a super-cell cannot be ignored and need to be carefully engineered. These two concepts are further elaborated in Section IV. If the unit cells of a smart metasurface are made of a reconfigurable material, the tunable elements depicted in Fig. \ref{Fig_4} may not be needed anymore, since the reconfigurability of the surface is ensured through the material of the unit cells itself. The metasurface structure needs, however, to be equipped with a configuration network in order to ensure its control and programmability.
\end{itemize}

\textbf{Key properties of metasurfaces}. The prefix \textit{meta} is a Greek word whose meaning is, among others, ``beyond''. In the context of metamaterials and metasurfaces, it refers to a three-dimensional and a two-dimensional structure that exhibits some kind of exotic properties that natural materials and surfaces, respectively, do not usually posses. By definition, a metasurface has the following properties \cite{Book_SurfaceElectromagnetics}: (i) it is \textit{electrically thin}, i.e., its thickness is considerably smaller than the wavelength; (ii) it is \textit{electrically large}, i.e., its transverse size is relatively large as compared with the wavelength; (iii) it is \textit{homogenizable}, i.e., the distance between adjacent unit cells is much smaller than the wavelength; and (iv) it is a \textit{sub-wavelength structure}, i.e., the size of each unit cell is much smaller than the wavelength. 

\textbf{Impact of the sub-wavelength thickness}. The deeply sub-wavelength thickness of the surface ensures that the propagation or resonance effects in the direction perpendicular to the surface can be safely ignored in the process of synthesis and analysis of the surface. This implies that the EM field on the transmission side of the surface (e.g., $z=0^+$) depends only on the EM field on the incidence and reflection side (e.g., $z=0^-$) of the surface, and that the surface can be effectively modeled as a sheet of induced surface electric and magnetic currents. In other words, the effects of the EM fields within the substrate of sub-wavelength thickness can be averaged out and ignored. This specific property allows one to define a metasurface as a local entity, as a zero-thickness sheet, or as a sheet discontinuity. \textit{It is worth mentioning that the term local entity needs not to be interpreted as the absence of spatial coupling among the unit cells, which, on the other hand, cannot be ignored due to the sub-wavelength inter-distance among the unit cells of the metasurface}. 

\textbf{Impact of the sub-wavelength inter-distance}. The sub-wavelength inter-distance among the unit cells make the metasurface equivalent to a sub-wavelength particle lattice that can be locally homogeneized, and, therefore, can be described through \textit{continuous} mathematical tensor functions that are, in general, simpler to handle as compared with the actual physical structure of the metasurface. Analytical models for the metasurfaces are detailed in Section IV, where the difference between microscopic and macroscopic modeling is discussed, along with the analytical convenience, especially for wireless applications, of macroscopic models.

\subsection{Nomenclature and Classification} \label{Nomenclature_and_Classification}
In the present paper, we have adopted the term RIS in order to refer to any kind of smart surfaces, as those depicted in Figs. \ref{Fig_2} and \ref{Fig_3}, which have the capability of being reconfigurable after their deployment in the network. In Fig. \ref{Fig_4}, in addition, we have sketched the conceptual architecture of nearly-passive RISs, which can be representative of different practical implementations of RISs. 

\textbf{RISs and friends}. In the literature, however, several other terms and acronyms are often employed to refer to smart surfaces. The most widely used are briefly discussed as follows.
\begin{itemize}
\item \textbf{\textit{Large intelligent surfaces (LISs)}}. The term LIS is referred to surfaces that are viewed as the next step beyond massive multiple-input-multiple-output (MIMO) technology. LISs are typically defined as active surfaces whose individual antenna elements are equipped with dedicated radio frequency (RF) chains, power amplifiers, and signal processing capabilities. Conceptually, their architecture is similar to that shown in Fig. \ref{Fig_4}. However, each unit element may have a complete RF chain and an independent baseband unit. 
\item \textbf{\textit{Intelligent reflecting surfaces (IRSs)}}. The term IRS is typically referred to surfaces that operate as reflectors and that are made of individually tunable unit elements whose phase response can be individually adjusted and optimized for beamsteering, focusing, and other similar functions. Usually, the unit elements are assumed not to be capable of amplifying the impinging radio waves, so that only their phase response can be modified (not their amplitude response).
\item \textbf{\textit{Digitally controllable scatterers (DCSs)}}. The term DCS is the most similar to RIS, and it is typically employed to emphasize the possibility of controlling, in a digital manner, the behavior of objects and devices coated or made of smart surfaces. In this case, the emphasis is put on the individual elements of the smart surface that are viewed as local scatterers. DCSs are often made of passive elements that cannot amplify the received signals. If made of passive elements, the operation of DCSs is based on the mutual coupling among the elements, and, hence, there exists a high spatial correlation among the unit cells depicted in Fig. \ref{Fig_4}.
\item \textbf{\textit{Software controllable surfaces (SCSs)}}. The term SCS is typically employed when the emphasis is to be given to the capability of the smart surfaces of being controlled and optimized by using software-defined networking technologies. The term SCS is often employed when the unit elements of the smart surface are equipped with a nano-communication network for enabling the communication among the unit cells. The smart surface is often equipped with low power sensors for environmental monitoring. The joint functionality of sensing and communication provides the smart surface with the capability of performing simple local operations, thus making it more autonomous. This may, however, affect the complexity and the power consumption of the entire smart surface.
\end{itemize}

\textbf{Formal classification of metasurfaces}. A comprehensive classification of electromagnetic surfaces is available in \cite[Fig. 1.10]{Book_SurfaceElectromagnetics}. As far as smart surfaces with general (passive or nearly-passive) functions are concerned, which is the main interest of the present paper, three main definitions can be found in \cite[Fig. 1.10]{Book_SurfaceElectromagnetics}.
\begin{itemize}
\item \textit{\textbf{Huygens's surfaces}}. These are defined as smart surfaces that manipulate the wavefront of incident radio waves in ways that a secondary wavefront with the desired features can be obtained.
\item \textit{\textbf{Metasurfaces}}. These are defined as any artificial smart surfaces with unconventional features.
\item \textit{\textbf{Reconfigurable surfaces and programmable metasurfaces}}. These are defined as smart surfaces that are equipped with active control devices, such as PIN diodes, micro electro mechanical systems (MEMS) switches, and varactors, which are integrated into the smart surface in order to provide real-time control of the surface properties.
\end{itemize}

Based on the classification in \cite[Fig. 1.10]{Book_SurfaceElectromagnetics} and the names usually adopted by researchers working in wireless communications and networks, we evince that the acronym RIS can be considered to be the most suitable definition for the conceptual architecture of smart surface depicted in Fig. \ref{Fig_4}. This justifies and corroborate the adoption of the term RIS in the present paper.

\subsection{Macroscopic Functions of Reconfigurable Intelligent Surfaces} \label{MacroscopicFunctions}
\textbf{Microscopic and macroscopic views}. Based on the conceptual architecture depicted in Fig. \ref{Fig_4}, it is apparent that the functions that an RIS is capable of applying, at the macroscopic level, to the impinging radio waves can be synthesized by appropriately optimizing its operation at the microscopic level, i.e., by appropriately designing the unit cells and the configuration network made of low power electronic circuits. Simple examples on how to optimize the unit cells of a phase-gradient RIS in order to synthesize specified macroscopic functions are reported in Section IV.

\textbf{Classification of macroscopic functions}. In general terms, the functions realized by RISs for application to wireless networks can be classified into two main categories.
\begin{itemize}
\item \textbf{\textit{EM-based design of RISs}}. These functions correspond to elementary transformations of the radio waves that can be directly specified at the EM level. Under this design paradigm, an RIS is optimized in order to realize some fundamental manipulations of the radio waves, which may be employed in wireless networks. According to this design paradigm, communication engineers view RISs as black boxes in which some parameters (knobs) can be optimized for improving the network performance. 
\item \textbf{\textit{Communication-based design of RISs}}. The functions realized by RISs under a communication-based design paradigm may not necessarily correspond to elementary EM-based manipulations of the radio waves. For a communication engineer, a fundamental question may, for example, be: \textit{What is the optimal design or use of an RIS in order to maximize the channel capacity}? According to this design paradigm, the functions of RISs are not specified a priori, but they are the result of an optimization problem and, therefore, may be different depending on the performance metric of interest. For example, RISs may be employed for realizing advanced modulation and coding schemes by directly operating at the EM level.
\end{itemize}

Some representative examples of these two design approaches are given in the following two sub-sections.

\begin{figure}[!t]
\begin{centering}
\includegraphics[width=\columnwidth]{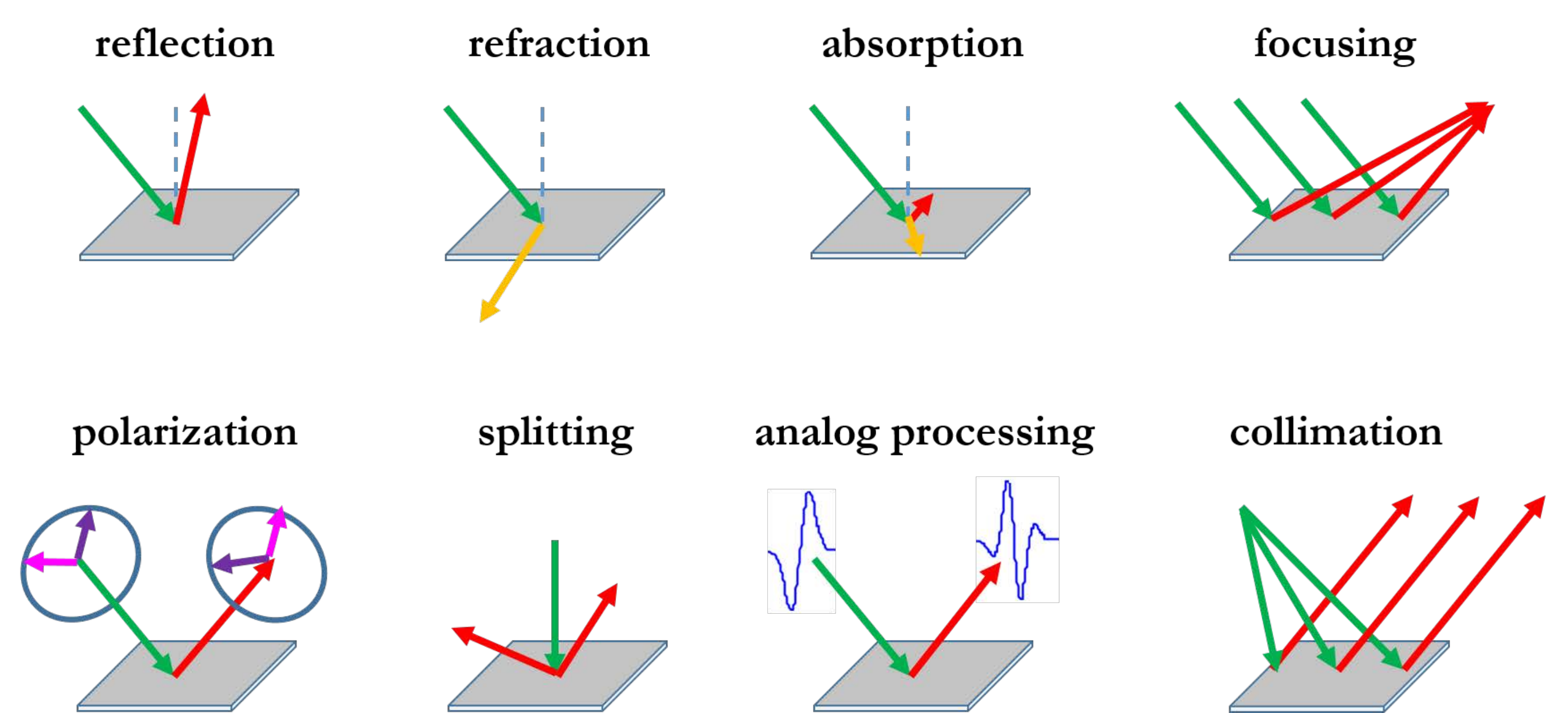}
\caption{Electromagnetic-based elementary functions.}
\label{Fig_5}
\end{centering} 
\end{figure}
\subsubsection{EM-Based Design of RISs} In Fig. \ref{Fig_5}, we report some examples of EM-based elementary functions that may be applied by RISs and that may have useful applications in wireless communications. These elementary functions are briefly discussed as follows.
\begin{itemize}
\item \textbf{\textit{Reflection}}. This function consists of reflecting an impinging radio wave towards a specified direction that may not necessarily coincide with the direction of incidence. 
\item \textbf{\textit{Transmission/refraction}}. This function consists of refracting an impinging radio wave towards a specified direction that may not necessarily coincide with the direction of incidence. 
\item \textbf{\textit{Absorption}}. This function consists of designing a smart surface that nulls, for a given incident radio wave, the corresponding radio waves that are reflected and refracted.
\item \textbf{\textit{Focusing/beamforming}}. This function consists of focusing (i.e., concentrating the energy of) an impinging radio wave towards a specified location.
\item \textbf{\textit{Polarization}}. This function consists of modifying the polarization of an incident radio wave (e.g., the impinging radio wave is transverse electric polarized and the reflected radio wave is transverse magnetic polarized).
\item \textbf{\textit{Collimation}}. This function is the complementary of focusing.
\item \textbf{\textit{Splitting}}. This function consists of creating multiple reflected or refracted radio waves for a given incident radio wave.
\item \textbf{\textit{Analog processing}}. This function consists of realizing mathematical operations directly at the EM level. For example, the radio wave refracted by a smart surface may be the first-order derivative or the integral of the impinging radio wave.
\end{itemize}
\begin{figure}[!t]
\begin{centering}
\includegraphics[width=\columnwidth]{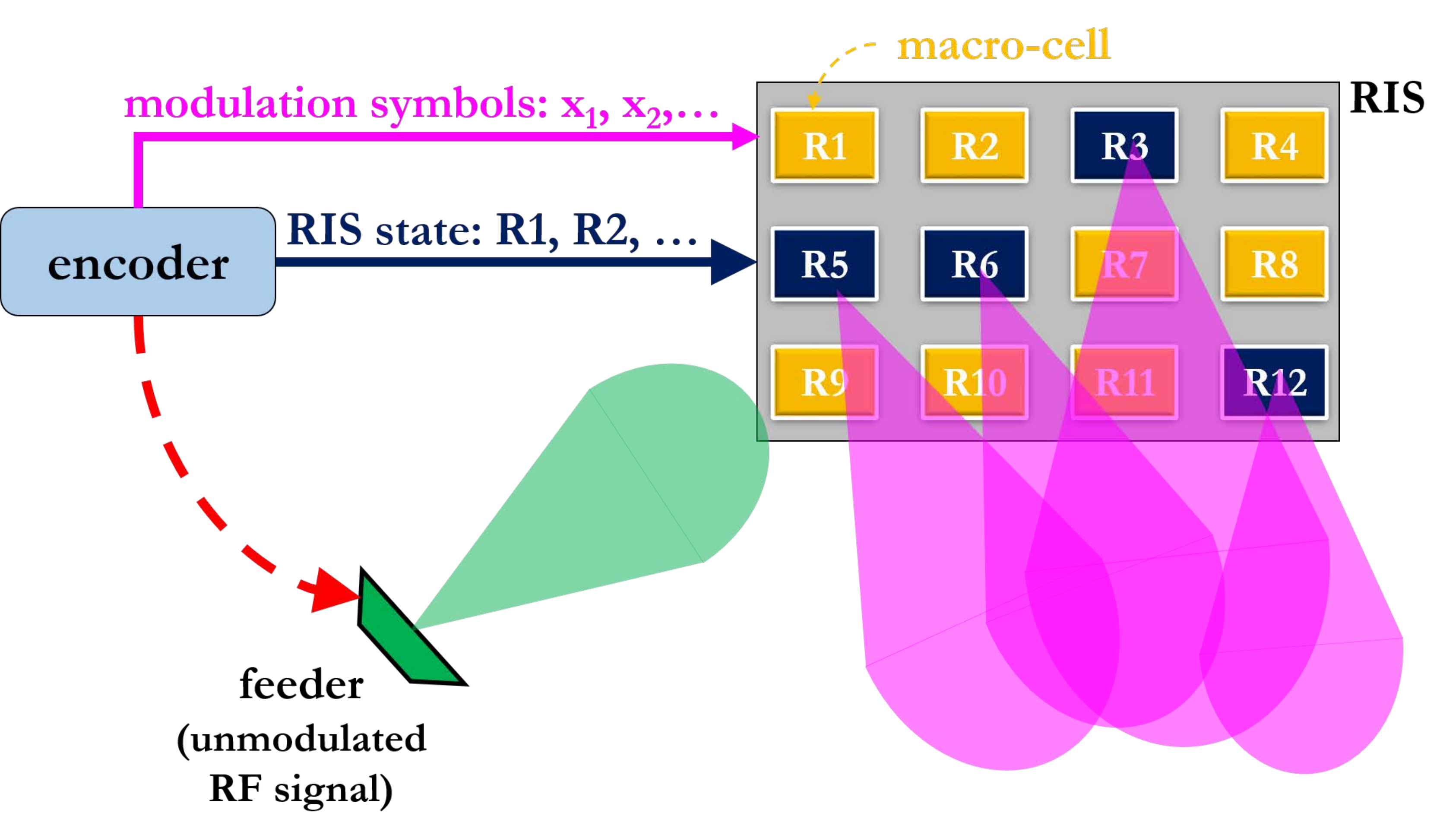}
\caption{Communication-based (modulation and encoding) design of reconfigurable intelligent surfaces.}
\label{Fig_6}
\end{centering} 
\end{figure}
\subsubsection{Communication-Based Design of RISs} In addition to functions that directly pertain the manipulation of the impinging radio waves at the EM level, an RIS can be employed for operations and applications that capitalize on EM-based manipulations of the radio waves but go beyond them. For example, a conceptual block diagram that depicts the application of RISs for implementing different types of transmitter designs is illustrated in Fig. \ref{Fig_6}. In this case, an RIS is viewed as an integral part of a transmitter. Other communication-based designs, applications, and examples beyond transmit-related operations (i.e., modulation and encoding) are described in further text. 

\textbf{Metasurface-based transmitters}. Figure \ref{Fig_6} shows an RIS that is realized based on the conceptual architecture reported in Fig. \ref{Fig_4}. From a communication-design perspective, a number of super-cells (it can be even a single super-cell depending on the function to be realized and the technology employed) are grouped together to form a macro-cell. A macro-cell can be thought of as the atomic element to realize RIS-based transmitters. As a concrete example, one may think of a macro-cell as a physical structure that allows one to mimic a phase shift keying (PSK) modulation. If an 8-PSK modulation is of interest, the macro-cell needs to be capable of realizing eight distinct phase shifts that correspond to those of a PSK modulation employed by conventional modulators. In practice, this is realized by jointly designing the unit cells and the super-cells that constitute the macro-cell, as elaborated in Section IV. The metasurface structure is illuminated by a feeder that is located in close proximity of the RIS. The feeder emits only an un-modulated signal. The modulation is, in fact, realized uniquely by the RIS through appropriate reflections of the signal emitted by the feeder. To modulate data, the RIS is controlled by an encoder, which outputs two data streams that are employed for configuring the RIS. The first data stream is employed to set the reflection coefficient (R1, R2, ...) of each macro-cell. Continuing with our example, each reflection coefficient corresponds to one phase shift of an 8-PSK modulation. It may, however, be an arbitrary reflection coefficient that is chosen based on any design criteria. The second data stream corresponds to conventional modulation symbols. The two data streams are employed to simultaneously control the macro-cells that are activated for transmission at a given time instance (the macro-cells in blue color in Fig. \ref{Fig_6}), and the modulated signals that they emit (the beams in purple color in Fig. \ref{Fig_6}). The transmitter architecture sketched in Fig. \ref{Fig_6} is general enough to realize multiple practical implementations. Three examples are described as follows.
\begin{itemize}
\item \textbf{\textit{RIS-based modulation}}. Let us assume that the data stream that corresponds to the state of the RIS controls only whether a macro-cell is either activated (ON state) or not activated (OFF state). This implies that the reflection coefficient of each macro-cell is only one or zero, respectively. Let us consider, in particular, that only a single macro-cell is activated at any transmission instance. Let us assume, in addition, that the data stream of the modulation symbols contains a single PSK symbol. Then, the RIS-based transmitter scheme in Fig. \ref{Fig_6} can be employed to realize a metasurface-based version of spatial modulation and index modulation, in which the transmitted data is encoded into the activated macro-cell and the PSK-modulated symbol \cite{MDR_SM_COMMAG}, \cite{MDR_SM_PIEEE}, \cite{MDR_SM_CST2015}, \cite{MDR_SM_CST2016}, \cite{MDR_SM_MILCOM}, \cite{MDR_IM_Access}, \cite{MDR_SM_JSAC}. This specific implementation of spatial modulation is attractive because of the large number of macro-cells that may be deployed on an RIS, and, therefore, the large number of bits that can be modulated onto the ON-OFF states of the macro-cells. For example, the RIS prototype in Fig. \ref{Fig_2} is made of 3,720 inexpensive individually tunable antennas and the RIS prototype in \cite{Wankai_PathLoss} is made of 10,000 unit cells. This implies that tens of bits per channel use may be modulated by using this approach, while using a single feeder. Ultimately, however, the achievable rate depends on the speed at which the macro-cells can be configured.
\item \textbf{\textit{RIS-based multi-stream transmitter}}. Let us assume that the data stream that corresponds to the state of the RIS controls the reflection coefficient of each macro-cell so as to mimic a PSK modulation. Let us assume, in addition, that the data stream of the modulation symbols is not enabled, i.e., the RIS is not fed with any bits through this control signal. Then, the RIS-based transmitter scheme in Fig. \ref{Fig_6} can be employed to realize a metasurface-based version of multi-antenna spatial multiplexing, in which the number of data streams that are simultaneously transmitted depend on the number of macro-cells that are activated at any transmission instance. If all the macro-cells are activated simultaneously, the RIS-based transmitter scheme in Fig. \ref{Fig_6} is capable of emitting twelve data streams simultaneously. In general, the larger the number of data streams is, the higher the complexity of the control and configuration network of the RIS is. Examples of existing prototypes for RIS-based multi-stream transmitters can be found in \cite{Wankai_ElectronLetterEditor}, \cite{Wankai_ElectronLetter}, \cite{Wankai_ChinaCommun}, \cite{Wankai_WCM}, \cite{Wankai_JSAC}. This specific implementation of spatial multiplexing is attractive because multiple data streams are transmitted simultaneously, while employing a single feeder, i.e., a single power amplifier and a single RF chain.
\item \textbf{\textit{RIS-based encoding}}. Let us assume that the data stream that corresponds to the state of the RIS controls the reflection coefficient of each macro-cell so as to mimic a discrete set of values. For generality, let us consider that all the macro-cells are activated at the same time. Let us assume, in addition, that the data stream of the modulation symbols contains a single symbol that belongs to a given constellation diagram. Then, the RIS-based transmitter scheme in Fig. \ref{Fig_6} can be employed to realize a transmitter that jointly encodes data onto the modulation symbol $x$ and the set of reflection coefficients R1, R2, ... of the macro-cells. This implementation has recently been studied, from an information-theoretic standpoint, in \cite{MDR_ISIT2020}. Therein, it is proved that jointly encoding information on the modulation symbol and the configuration of an RIS, as a function of the channel state information, provides a better channel capacity as compared with the baseline scheme in which the configuration of the RIS is not exploited for data modulation. This result is important because it proves that maximizing the received power is not necessarily optimal from an information-theoretic point of view.
\end{itemize}

Several other transmitter designs may be realized, based on these three examples, which capitalize on the possibility of shaping the radio waves emitted by an RIS through a simple RF feeder and an encoder that controls the configuration network sketched  in Fig. \ref{Fig_4}.

\subsection{Distinctive Peculiarities of Reconfigurable Intelligent Surfaces} \label{DistinctivePeculiarities}
\textbf{RISs vs. competing technologies}. As mentioned in previous text, some instances of RISs with active unit elements (usually referred to as LISs) are often viewed as the next step beyond massive MIMO. Similarly, nearly-passive RIS that are implemented by using surfaces made of discrete tiny antenna elements with inter-distances equal to half of the wavelength are often compared to multi-antenna relays with no power amplification capabilities or to reflectarrays that are illuminated by a feeder that is not located in close proximity of the surface. Comparative studies on the differences and similarities between RISs, massive MIMO, relays, and other technologies are reported and discussed in Section V. 

\textbf{The main reasons that make RISs different}. In this sub-section, we are interested in elaborating on some fundamental aspects that, in our opinion, make nearly-passive RISs different from currently available technologies, and that, at the same time, offer opportunities for innovative solutions that have never been in the mainstream of wireless communications before. More specifically, the following aspects are considered to be distinctive and peculiar features of nearly-passive RISs.
\begin{itemize}
\item \textbf{\textit{Unique design constraints}}. The nearly-passive nature of RISs introduces challenging design constraints, e.g., the impossibility of performing channel estimation directly on the smart surface and on the way that the radio waves can be manipulated by the smart surface. 
\item \textbf{\textit{Communication without new waves}}. The nearly-passive nature of RISs offers unique opportunities for redefining the notion of communication, in which information can be exchanged without producing new EM signals but by recycling existing radio waves. This may be highly beneficial for reducing the EM pollution and for decreasing the level of EM exposure of human beings, which is usually increased by deploying additional network infrastructure and by using more spectrum. This may be extremely relevant for the successful deployment of wireless technologies in EM-sensitive environments (e.g., in hospitals).
\item \textbf{\textit{Sustainable wireless by design}}. The use of innovative eco-friendly metamaterials in order to realize and manufacture nearly-passive RISs opens the possibility for building future wireless networks that are sustainable by design. The materials that are employed to realize the smart surfaces can be chosen to be cost-effective, to have a light environmental footprint, and to be highly recyclable. This, in turn, may impose some design constraints on the functions that the RISs may be able to realize.
\begin{figure}[!t]
\begin{centering}
\includegraphics[width=\columnwidth]{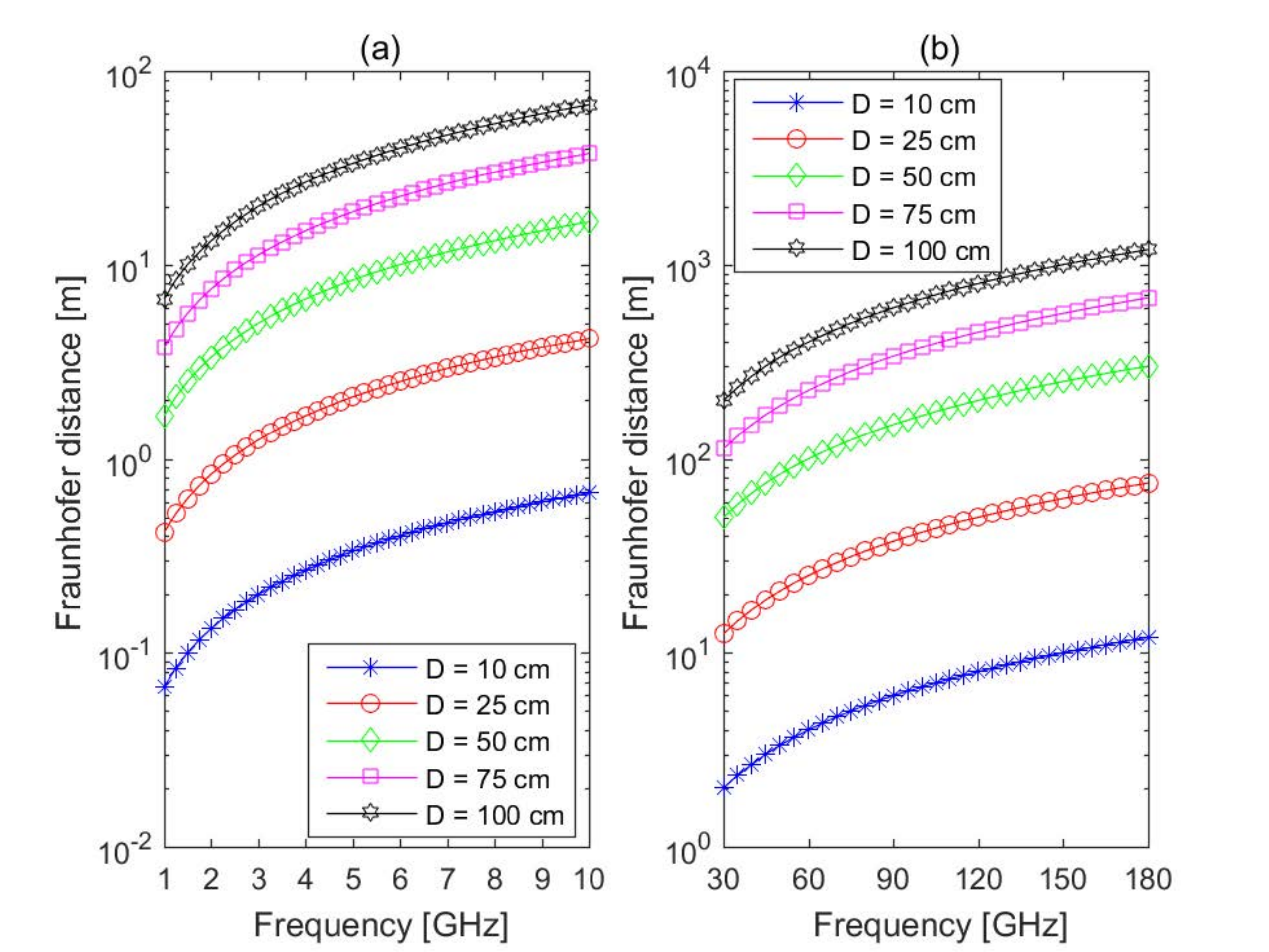}
\caption{Fraunhofer's distance at sub-6 GHz (a) and millimeter-wave (b) frequencies ($D$ is fixed).}
\label{Fig_7}
\end{centering} 
\end{figure}
\begin{figure}[!t]
\begin{centering}
\includegraphics[width=\columnwidth]{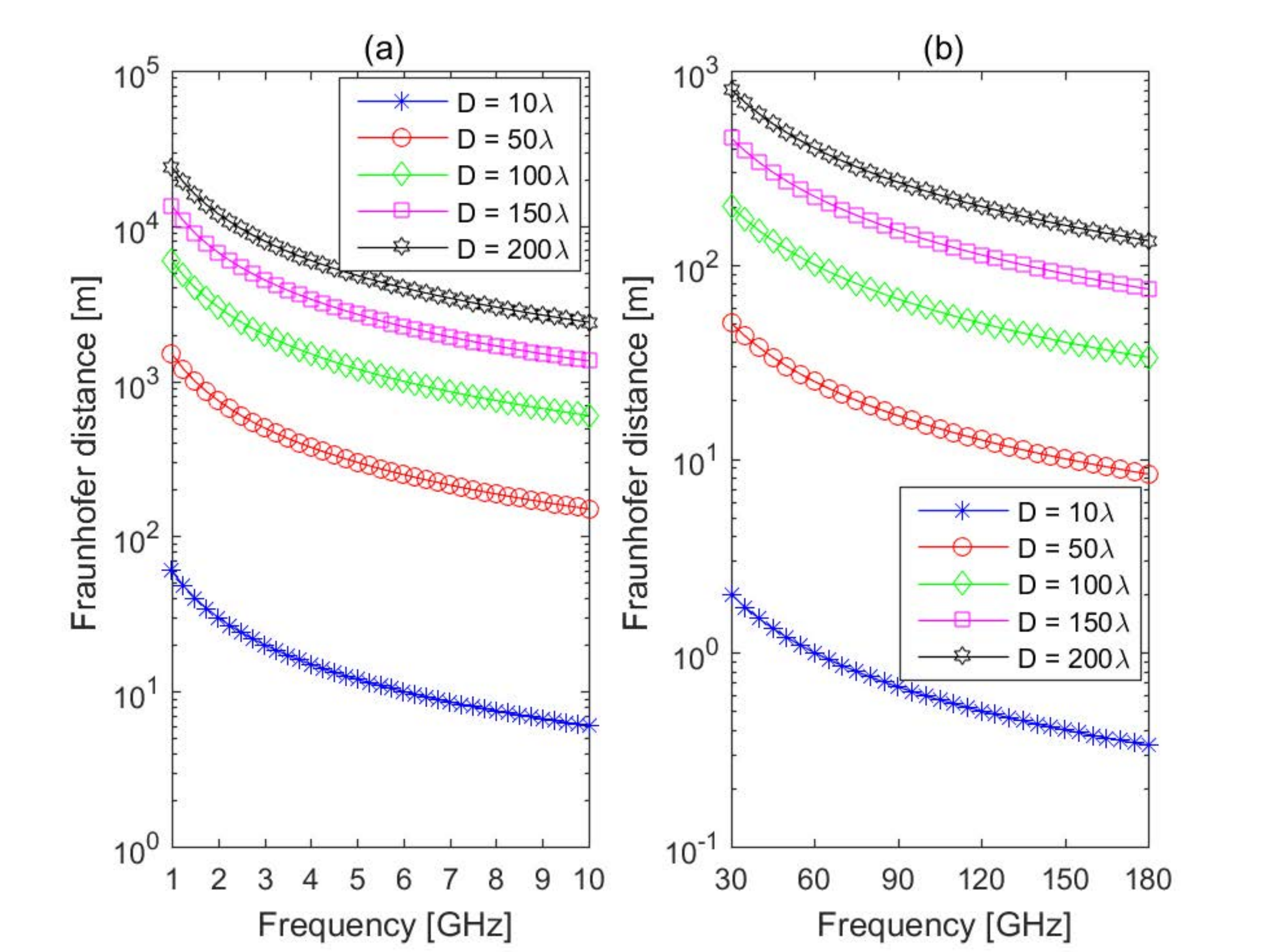}
\caption{Fraunhofer's distance at sub-6 GHz (a) and millimeter-wave (b) frequencies ($D$ depends on $\lambda$).}
\label{Fig_8}
\end{centering} 
\end{figure}
\item \textbf{\textit{Beyond the far-field regime}}. The RISs may be made of geometrically large surfaces of the order of a few square meters. Figure \ref{Fig_2} reports an example of smart surface whose size is of the order of six square meters. A recent prototype of metasurface reported in \cite{Wankai_PathLoss} has a size of one square meter and operates at 10.5 GHz. If one assumes that the far-field of the smart surface can be defined in a similar manner as for conventional antenna arrays, the far-field propagation regime for such an RIS may start at tens of meters far away from the smart surface. This implies that such a smart surface may operate in the near-field in some relevant scenarios, e.g., in indoor environments. The use of geometrically large RISs opens, therefore, the possibility of building new wireless networks that operate in the near-field regime, which is not a conventional design assumption in wireless communications. As an illustrative example, Figs. \ref{Fig_7} and \ref{Fig_8} reports the Fraunhofer distance that is usually employed for identifying the limit between the radiative near-field and the far-field, i.e., $d_F = 2D^2/\lambda$, where $D$ is the largest dimension of the RIS under analysis and $\lambda$ is the wavelength of the radio wave. It is apparent that, depending on the setup, the radiative near-field operating regime may be sufficiently large for typical indoor and mobile outdoor communication systems.
\item \textbf{\textit{Dense deployment of scatterers}}. If RISs are made of smart metasurfaces, they comprise a large number of sub-wavelength unit cells. Such envisioned sub-wavelength dense deployments of tiny sub-wavelength scattering elements is not commonly employed in wireless communications, where the mutual coupling among the radiating elements is often avoided by design, i.e., by ensuring that the scattering elements are sufficient distant from each other. This paves the way for developing new signal and propagation models that may affect the ultimate performance limits of wireless networks, and for introducing new design paradigms according to which communication systems and networks are engineered to be mutual-coupling-aware and mutual-coupling-robust.
\item \textbf{\textit{High focusing capabilities in the radiative near-field}}. As illustrated in Figs. \ref{Fig_7} and \ref{Fig_8}, RISs may have a sufficiently large size and the operating wavelength may be sufficiently small that the radiative near-field region may not be ignored in typical wireless applications. This operating regime is not usual in wireless communications, and may pave the way for realizing near-field focused RISs that highly concentrate the EM power in small spot regions \cite{NearFieldFocused_Antennas}. This high focusing capability may have multiple applications, e.g., (i) for enabling interference-free communication in areas with high densities of devices; (ii) for making possible the precise radio localization of users and mapping of environments; and (iii) for recharging the batteries of low power devices via wireless power transfer methods.
\end{itemize}

\subsection{Beyond Planar Reconfigurable Intelligent Surfaces: Conformal Smart Surfaces} \label{ConformalSurfaces}
\textbf{RISs are not necessarily planar structures}. We close this section by mentioning that the present paper is focused on RISs that are made of smart surfaces that are planar. This assumption originates from the ease of deployment of such surfaces on, e.g., the internal walls of indoor environments, the external facades of buildings, and the glasses of windows. The same principles apply, however, to free-form conformal bi-dimensional smart surfaces, which are not necessarily planar. These structures may have several applications in wireless communications, e.g., for coating objects that are not planar and for shaping the radio waves in ways that planar smart surfaces many not be capable of.

\section{Smart Radio Environments} \label{SREs}
\textbf{Designing wireless networks today: The environment is fixed by nature}. Current methods to design wireless networks usually rely on the optimization of the so-called end-points of communication links, e.g., transmitters and receivers. Over the last decades, therefore, many advanced techniques have been proposed for improving the performance of wireless networks, which encompass advanced modulation/encoding schemes and protocols based on using, e.g., multiple antennas at the transmitters, powerful transmission and retransmission protocols, and robust demodulation and decoding methods at the receivers. The wireless environment has, on the other hand, been conventionally modeled as an exogenous entity that cannot be controlled but can only be adapted to. According to this design paradigm, communication engineers usually design the transmitters, the receivers, and the transmission protocols based on the specific properties of the wireless channels and in order to achieve the desired performance. For example, transmitters equipped with multiple radiating elements may be configured differently as a function of the specific characteristics of the wireless channel where they operate, in order to achieve the desired trade-off in terms of spatial multiplexing, spatial diversity, and beamforming gains. 

\textbf{SREs: The environment is generated by nature but is programmable by design}. The overarching paradigm that characterizes the design of current wireless networks consists, therefore, of pre-processing the signals at the transmitters and/or post-processing the signals at the receivers, in order to compensate the effect of the wireless channel and/or in order to capitalize on specific features and characteristics of the wireless channel. RISs provide wireless researchers with more opportunities for designing and optimizing wireless networks, which are built upon a different role played by the wireless environment. RISs are, in fact, capable of shaping the radio waves that impinge upon them, after the radio waves are emitted by the transmitters and before they are observed by the receivers, in ways that the wireless environment can be customized, in principle as one desires, in order to fulfill specific system requirements. The wireless environment is, therefore, not treated as a random uncontrollable entity, but as part of the network design parameters that are subject to optimization for supporting diverse performance metrics and quality of service requirements.

\begin{figure}[!t]
\begin{centering}
\includegraphics[width=\columnwidth]{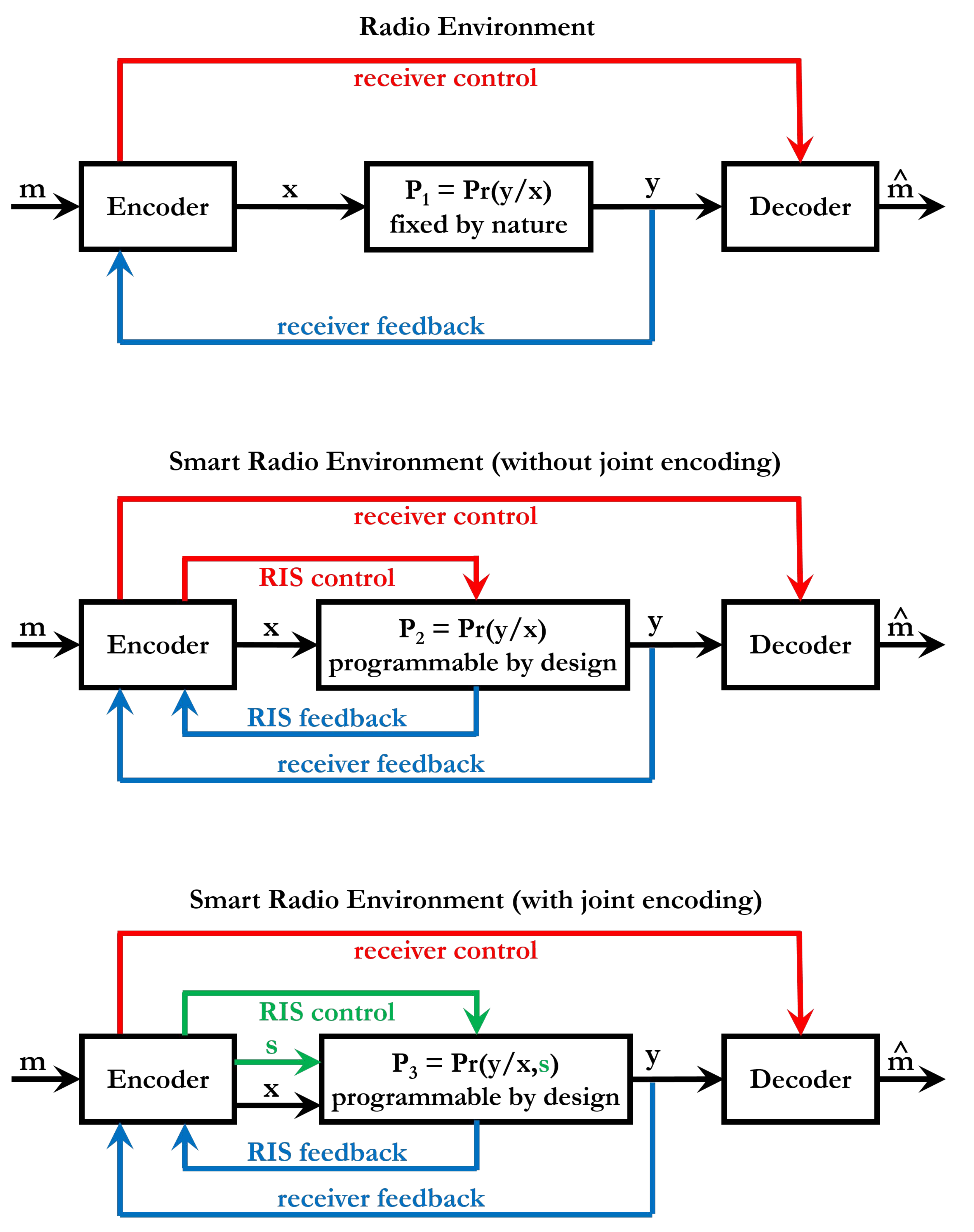}
\caption{Communication-theoretic models for radio environments and smart radio environments (with and without joint encoding).}
\label{Fig_9}
\end{centering} 
\end{figure}
\textbf{Revisiting communication-theoretic models}. The concept of SREs introduces, therefore, a new communication-theoretic view of wireless systems and offers new opportunities for optimization. In Fig. \ref{Fig_9}, the conceptual block diagram of a conventional point-to-point communication system and the corresponding block diagram under a SRE-based framework are illustrated. Under the conventional communication-theoretic framework, the system is modeled through transition probabilities that are not considered to be optimization variables. Under the SRE-based communication-theoretic framework, on the other hand, the system is modeled through transition probabilities that can be customized, thanks to the deployment of RISs throughout the environment and thanks to the possibility of controlling and programming the functions that RISs apply to the impinging radio waves. Therefore, the system model itself becomes an optimization variable, which can be jointly optimized with the transmitter and the receiver: Rather than optimizing the input signal for a given system model, one can now jointly optimize the input signal and the system.

\textbf{SREs with and without joint encoding and modulation at the transmitter and at the RIS}. In Fig. \ref{Fig_9}, in particular, three system models are illustrated. The first communication-theoretic model is referred to a conventional radio environment in which, via a feedback channel that provides the encoder with channel state information, the transmitter and receiver are jointly optimized, e.g., by designing appropriate transmit and receive channel-aware vectors. The second communication-theoretic model is referred to an RIS-empowered SRE in which the feedback channels from the receiver and the environment (e.g., the RIS) are exploited for optimizing the setup (i.e., configuration, state, or action) of the RIS besides the transmit and receive channel-aware vectors. In this case, therefore, the transition probabilities that describe the wireless environment can be customized by appropriately optimizing the state of the RIS. These first two communication-theoretic models are analyzed and compared in \cite{MDR_OverheadAware}. The third communication-theoretic model is referred to an RIS-empowered SRE in which the state of the RIS is employed for customizing the wireless environment while at the same time encoding information jointly with the transmitter. In this setup, in particular, the transition probabilities of the wireless environment depend on both the state of the RIS, which affects the wireless channel, and the data $s$ encoded by the transmitter on the state of the RIS. This setup is analyzed in \cite{MDR_ISIT2020}, where it is proved that performing joint transmitter-RIS encoding yields, in general, a better channel capacity.

\textbf{Communication model with state-dependent channels}. From an information-theoretic standpoint, broadly speaking, the conventional block diagram in Fig. \ref{Fig_9} is described by the conditional probability law of the channel output given the channel input. For example, a binary symmetric channel is a model for describing the communication of binary data in which the noise may cause random bit-flips with a fixed probability. In the context of SREs, the conditional probability law of the channel output given the channel input can be customized by using RISs. The wireless environment can be programmed to evolve through multiple states (or configurations) that depend on how the RISs shape the impinging radio waves. The possibility of controlling the possible states of operation of the wireless environment jointly with the operation of the transmitter and the receiver offers opportunities for enhancing the overall communication performance. The resulting communication-theoretic model well fits, therefore, the transmission of information through state-dependent wireless environments (or simply channels) that are generated by nature but that are controlled and affected by the communication system. The block diagram in Fig. \ref{Fig_9} illustrates the case study in which the transmitter takes actions (i.e., it configures the operation of the RISs distributed throughout the environment) that affect the formation of the states of the environment. In general, the specific state of the wireless environment can be controlled by the transmitter, the receiver, or by an external controller that oversees the operation of portions or the entire network. In general, in fact, the operating state of the wireless environment depends on the configuration of all the RISs distributed throughout it, which can be jointly optimized for achieving superior performance. The block diagram of SREs well fits, therefore, a communication model with state-dependent channels, whose states are directly controlled by the communication system rather than being generated and being dependent only by nature \cite{TIT_StateDependentChannels}.

\textbf{Overcoming fundamental limitations of wireless networks design}. The possibility of creating wireless systems with state-dependent channels, which are generated by nature but are controlled by communication designers, opens new opportunities for overcoming some fundamental limitations in designing current wireless networks. Four of them are the following.
\begin{itemize}
\item \textbf{\textit{The ultimate performance limits of wireless networks may not have been reached yet}}. Recent research works have proved that by jointly optimizing the transmitter, the receiver, and the environment, the channel capacity of a point-to-point wireless communication system can be further improved \cite{MDR_ISIT2020}. In particular, the channel capacity can be increased by capitalizing on RISs as a means for encoding and modulating additional information besides the transmitters.
\item \textbf{\textit{Customizing and controlling the wireless environment may open new opportunities for network optimization}}. In some application scenarios, the transmitters and receivers may not be equipped with multiple antenna-elements because of the challenges that their practical realization entail. For example: (i) devices such as sensors and handhelds are usually small in size, and multiple antennas cannot be accommodated; (ii) connecting each antenna to an independent RF chain and transmit/receive circuitry increases the cost and power consumption; and (iii) multiple-antenna structures are often bulky and are not easy to deploy even at the base stations. Recent results have shown that these challenges can be overcome by capitalizing on RISs and by moving the operations that are typically executed by multiple-antenna transmitters directly to the radio environment \cite{RFocus}. An example is the RFocus prototype illustrated in Fig. \ref{Fig_2}.
\item \textbf{\textit{The radio waves may be used more efficiently}}. When reflected or refracted by an object, for example, the energy of the impinging radio waves is scattered towards unwanted directions, thus reducing the efficiency of utilization of the emitted power. The possibly large size of RISs and their fine ability of controlling the radio wave thanks to their sub-wavelength structure offer the opportunity for realizing smart surfaces that are capable of increasing, in the radiative near-field of the surface, the EM power density in spot regions of very limited size \cite{NearFieldFocused_Antennas}. Therefore, very high focusing capabilities may be obtained by focusing the energy only where it is needed and by avoiding to create interference in unwanted locations.
\item \textbf{\textit{The spatial capacity density may be increased}}. Recent results in \cite{Dardari_DegreesFreedom} have proved that RISs offer opportunities for increasing the available degrees of freedom of wireless communications, by creating large numbers of orthogonal communication links per square meter. These communication channels, which are orthogonal at the EM level, can be employed to simplify the wireless access in multiple-access systems, by realizing communication functionalities directly at the EM level, i.e., the ``layer-0'' of the protocol stack. Based on these findings, RISs may be employed for achieving spatial multiplexing gains in wireless environments in which conventional multiple-antenna schemes cannot, e.g., in environments without a sufficiently rich multipath scattering. These results can find applications in scenarios with high densities of devices that can be simultaneously served over the EM-orthogonal channels, thus offering, in principle, very high capacity densities.
\end{itemize}

\textbf{SREs: Centralized vs. distributed}. RIS-empowered SREs offer, therefore, several opportunities for improving the performance of wireless networks and, possibly, for further moving ahead the fundamental limits of wireless communications. RISs can turn wireless networks into SREs in multiple ways. Broadly speaking, \textit{one could think of wireless networks in which users and devices interact with the environment anytime that they are in close vicinity of a smart surface}, so as to either enhance the communication performance or to reduce the utilization of resources. SREs can be realized in a centralized, distributed, or hybrid manner.
\begin{itemize}
\item \textbf{\textit{Centralized implementation}}. In this case, it is assumed that a central controller exists and that it coordinates the activity of a network neighborhood. In a centralized implementation, RISs can be realized as simple as possible, since they are not required to be equipped with significant on-board, sensing, signal processing, and communication capabilities. RISs could be realized based on nearly-passive implementations, which need to be only able to receive the configuration signals and to set the configuration network depicted in Fig. \ref{Fig_4} accordingly. In a centralized implementation, however, the central controller needs to be able to gather the channel state information without relying on any on-board sensing and signal processing capabilities at the RISs. Therefore, appropriate protocols to obtain this information are needed, and the associated channel estimation overhead needs to be taken into account for the optimal design and operation of RISs \cite{MDR_OverheadAware}.
\item \textbf{\textit{Distributed implementation}}. In this case, a central controller is not necessarily needed. RISs need, however, to have the required functionalities for autonomously identifying the optimal function to apply based on the environmental state information and on the network topology, e.g., the locations of users and access points. It may be difficult, as a consequence, to employ nearly-passive RISs in a distributed implementation, since they need to be able to sense the channel, to extract the necessary channel state information, and to communicate this information throughout the structure of the smart surface. This may be realized by employing low power sensors that are possibly equipped with energy harvesting capabilities and nano-communication protocols for enabling the self-configuration of the unit cells. The higher implementation complexity of the RISs is the price to pay for dispensing SREs from the need of a central controller and the associated signaling and channel estimation overhead.
\item \textbf{\textit{Hybrid implementation}}. This can be viewed as an intermediate network realization of SREs, where RISs with different capabilities may be deployed in the network in order to strike a balance between (i) the higher implementation complexity and power consumption of RISs in distributed implementations, and (ii) the higher channel estimation and signaling overhead that are necessary to feed the central controller and to report the environmental information to other network elements.
\end{itemize}

\textbf{Potential applications and scenarios}. In the next two sub-sections, we report some applications of RISs and some scenarios for SREs. For ease of representation, the illustrations are given without explicitly referring to a centralized, distributed, or hybrid implementation. In principle, the three approaches can be realized with their advantages and limitations.

\begin{figure}[!t]
\begin{centering}
\includegraphics[width=\columnwidth]{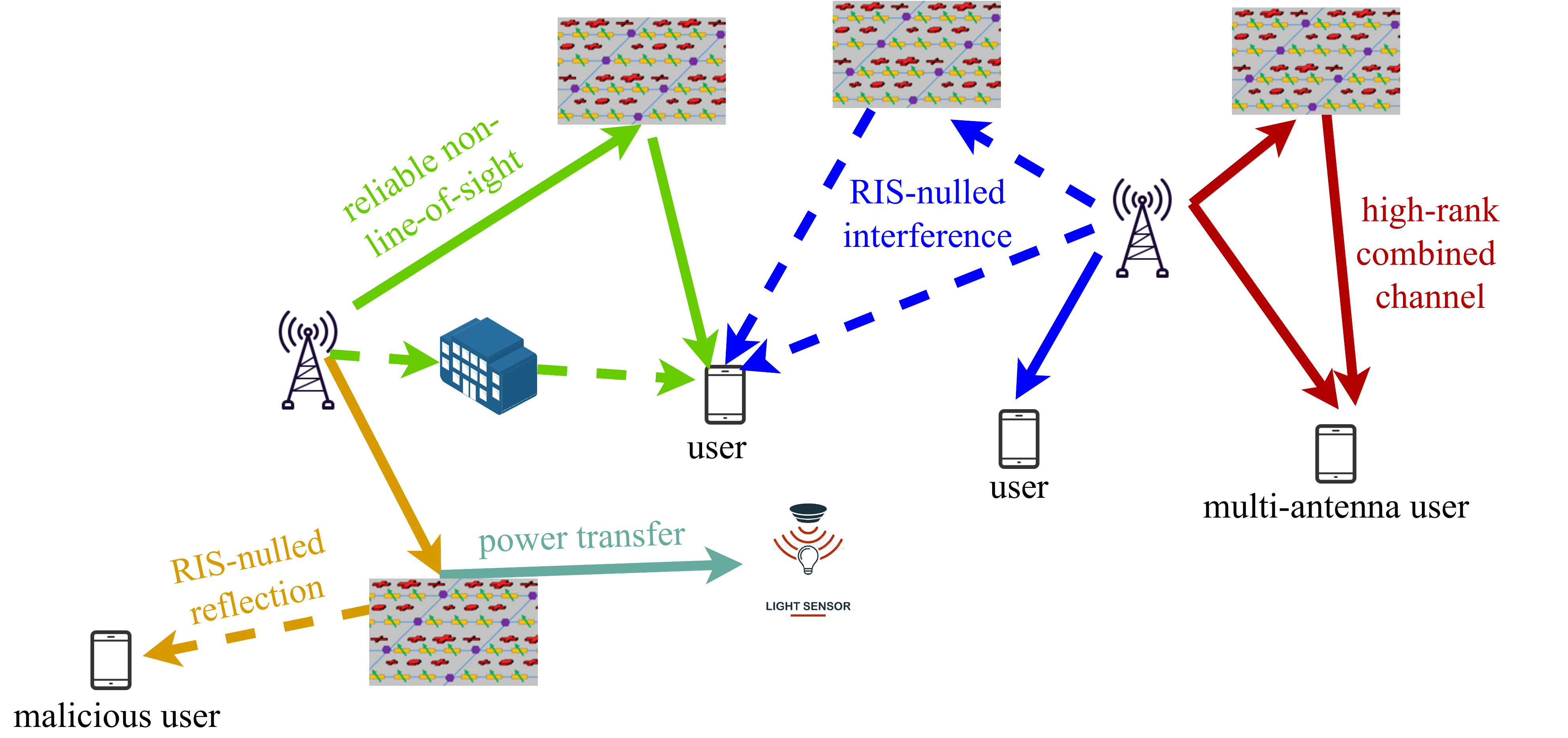}
\caption{Potential applications of reconfigurable intelligent surfaces in smart radio environments.}
\label{Fig_10}
\end{centering} 
\end{figure}
\subsection{Potential Applications of Reconfigurable Intelligent Surfaces in Smart Radio Environments} \label{RISs_Applications}
In this section, we briefly describe some potential applications of RISs. Some of them are illustrated in Fig. \ref{Fig_10} as examples.
\begin{itemize}
\item \textbf{\textit{Coverage enhancement}}. An RIS can be configured in order to create configurable non-line-of-sight links in dead-zone (or low coverage) areas in which line-of-sight communication is not possible or it is not sufficient.
\item \textbf{\textit{Interference suppression}}. An RIS can be configured to steer signals towards specified directions or locations not only for enhancing the signal quality, but for suppressing unwanted signals that may interfere with other communication systems.
\item \textbf{\textit{Security enhancement}}. This application is similar to interference suppression, with the difference that an RIS can be configured to worsen the signal detected by eavesdroppers by either creating destructive interference or by altering the reflection of signals towards locations not occupied by unauthorized users.
\item \textbf{\textit{Channel rank enhancement}}. The spatial multiplexing gain that can be achieve in multiple-antenna systems depends on how well conditioned the channel matrix is. An RIS can be appropriately configured in order shape the wireless environment in a way that the channel matrix has a high rank and a condition number close to one, so as to increase the channel capacity.
\item \textbf{\textit{Focusing enhancement}}. An RIS of large geometric size (in further text referred to as electrically large RIS) can operate in the radiative near-field at transmission distances up to a few tens of meters, as illustrated in Figs. \ref{Fig_7} and \ref{Fig_8}. Therefore, the scattered radio waves can be focused towards spatial spots of narrow size, so as to capillary serve dense deployments of users without creating mutual interference.
\item \textbf{\textit{Radio localization enhancement}}. The high focusing capabilities of RISs of large geometric size can be capitalized for finely estimating the location of mobile terminals and devices, so as to support high-precision ranging, radio localization, and mapping applications. 
\item \textbf{\textit{Information and power transfer}}. The high focusing capabilities of RISs of large geometric size can be exploited for concentrating the energy towards tiny and energy-autonomous sensor nodes, so that the radio waves can be employed, simultaneously, to recharge the sensors and to transmit information.
\item \textbf{\textit{Ambient backscattering}}. Consider a low power sensor node that is embedded into a smart surface for environmental monitoring. Any time that a radio wave impinges upon the smart surface, the RIS may be configured to modulate/encode the data sensed by the low power sensor into the scattered signal, e.g., by transmitting the sensed data through the time-domain scattered waveform (see Fig. \ref{Fig_5}). This enables low power sensors to piggyback information into ambient radio waves without creating new radio signals and, de facto, by recycling existing radio waves for communication \cite{ORANGE_BackScattering_1}, \cite{ORANGE_BackScattering_2}.
\end{itemize}

Other applications, such as RIS-based modulation, RIS-based multi-stream transmission, and RIS-based encoding are depicted in Fig. \ref{Fig_6} and are described in Section \ref{MacroscopicFunctions}.

\begin{figure*}[!t]
\begin{centering}
\includegraphics[width=2\columnwidth]{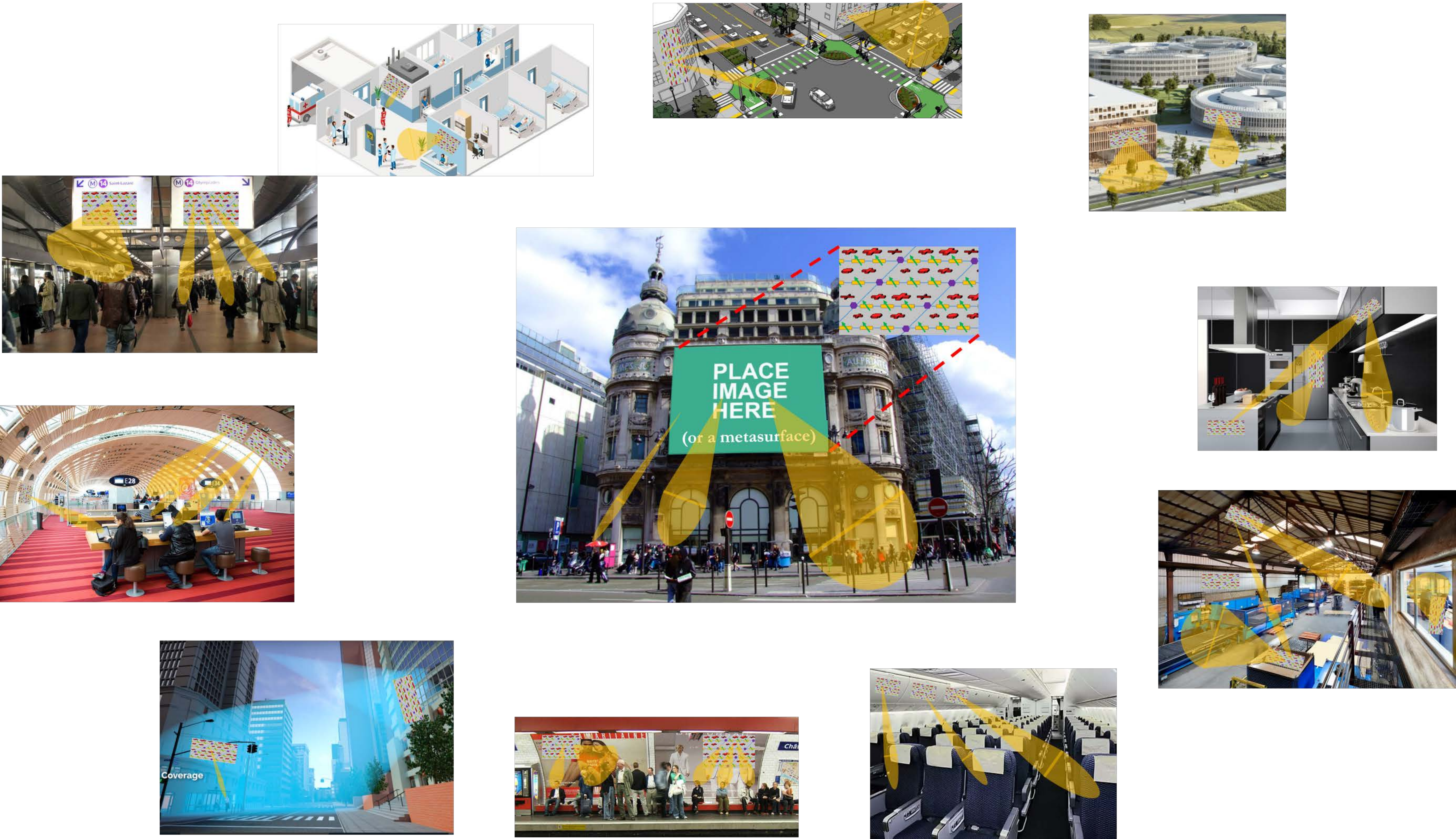}
\caption{Potential scenarios of smart radio environments: A world of nearly-passive reconfigurable intelligent surfaces.}
\label{Fig_11}
\end{centering} 
\end{figure*}
\subsection{Potential Scenarios of Smart Radio Environments} \label{SREs_Scenarios}
Several potential scenarios can benefit from the concept of SREs. Some promising case studies are briefly discussed as follows, and a sub-set of them is illustrated in Fig. \ref{Fig_11}.
\begin{itemize}
\item \textbf{\textit{Smart cities}}. In cities, the facades of large buildings may be coated with RISs of large geometric size. This  offers opportunities for, e.g., enhancing the coverage, increasing the spectral efficiency, and reducing the exposure to the EM radiation in outdoor environments, since the deployment of RISs may reduce the amount of network infrastructure (e.g., based stations) to deploy.
\item \textbf{\textit{Smart homes}}. In homes, the interior walls may be coated with RISs of different sizes for enhancing the local connectivity of several kinds of handlers (mobile phones, tablets, etc.) and other devices that rely on wireless connectivity for operation.
\item \textbf{\textit{Smart buildings}}. In buildings, large windows may be made of special glasses that can selectively enable indoor-to-outdoor and outdoor-to-indoor connectivity. This may be suitable for enhancing the connectivity at high transmission frequencies.
\item \textbf{\textit{Smart factories}}. In factories, the presence of large metallic objects usually result in harsh wireless propagation environments. RISs constitute a suitable approach for turning strong reflections of radio waves into a benefit for enhancing the coverage and the transmission rate.  
\item \textbf{\textit{Smart hospitals}}. Hospitals are EM-sensitive environments, where the intensity of the radio waves needs to be kept at a low level. RISs can be employed in order to enhance the local coverage without the need of increasing the transmitted power.
\item \textbf{\textit{Smart university campuses}}. In university campuses, large buildings, offices, classrooms, etc. can be coated with RISs that can offer the desired high-speed connectivity without the need of installing several access points.
\item \textbf{\textit{Smart undergrounds}}. In undergrounds, the coverage of wireless signals is usually negatively affected by the complex topology of the radio environment. RISs can be deployed in, e.g., ceilings, billboards, and walls, so as to provide the necessary connectivity to a large number of users simultaneously.
\item \textbf{\textit{Smart train stations}}. In train stations, several users may be waiting on the platforms before the arrival of trains. RISs can be deployed for illuminating areas of the platforms in order to enhance the received signals of individual users or clusters of users.
\item \textbf{\textit{Smart airports}}. In airports, large numbers of users may make take different directions when disembarking from airplanes. RISs can be employed for steering different beams towards different hallways, so as to enhance the quality of the received signals. Also, RISs may be used to enhance the signals in high-speed download areas, e.g., Internet areas in close proximity of the gates. 
\item \textbf{\textit{Smart billboards}}. Billboards may constitute a promising approach for deploying RISs of different geometric size in a simple and effective manner, both in indoor and outdoor environments.
\item \textbf{\textit{Smart glasses}}. In urban areas, large portions of the surface of buildings are made of glass. The example in Fig. \ref{Fig_3} shows that smart glasses realized with metasurfaces can be built and can effectively control the propagation of radio waves. These smart glasses can be employed for enhancing the coverage in, e.g., dead-zone areas.
\item \textbf{\textit{Smart clothing}}. Clothing can be realized with metamaterials and several embedded smart sensors in order to create wearable body networks for monitoring the health of people. 
\item \textbf{\textit{Smart cars}}. The glasses and the roof of cars may be coated with RISs that can provide reliable communications within the car itself, as well as can serve as moving nearly-passive relays for enhancing vehicle-to-vehicle and vehicle-to-infrastructure communications.
\item \textbf{\textit{Smart trains}}. The interior of trains may be coated with RISs in order to provide a better signal coverage and to reduce the levels of EM radiation and the exposure of passengers to EM fields.
\item \textbf{\textit{Smart airplanes}}. The overhead bins inside airplanes may be coated with RISs that may provide high-speed Internet to passengers while reducing their EM field exposure and/or decreasing the power consumption.
\end{itemize}

\textbf{Aesthetically-friendly environments}. In our everyday life, in conclusion, RISs may offer countless opportunities to realize SREs. One of the advantages of metasurfaces is that they can be built, e.g., by using the smart glass in Fig. \ref{Fig_3}, so as not to interfere aesthetically or physically with the surrounding environment or people line-of-sight, making them ideal for use within buildings and on vehicles or billboards. RISs can be deployed every time and in locations that are not suited for the installation of base stations, such as built-up areas or in indoor areas in which the reception of signals needs to be blocked selectively (e.g., in high-security areas).

\subsection{On the Role of Machine Learning in Smart Radio Environments} \label{SREs_ML}
\textbf{Awareness of the surrounding environment}. In order to realize a ``truly smart'' SRE, it is not sufficient to be able to customize the radio environment as desired. It is necessary, in addition, to optimally configure the most appropriate operation of spatially distributed RISs, in order to provide the users with the best communication performance and quality of experience. RISs necessitate, therefore, full awareness of the complex and non-stationary surrounding environments in which they are deployed. This requires efficient and on-demand network intelligence to cope with complex deployment planning, real-time programmability for optimization, and dynamic control for service provisioning. Machine learning and artificial intelligence (AI) may constitute efficient approaches for leveraging the potential benefits of RIS-empowered SREs, especially because the computational complexity of deploying, programming, and controlling SREs rises significantly with the increase of the network-to-infrastructure and user-to-network interactions.

\textbf{AI is the answer}. Therefore, machine learning computational methods may be suitable enablers for optimizing and operating SREs. AI-enabled machines, in particular, can be designed to perform ``intelligent'' tasks without being programmed to accomplish any single (repetitive) task, but by adapting themselves to different environments. To this end, AI provides methods for designing networks that autonomously interact with the environment, in ways that humans consider intelligent, including the characteristics of human cognitive abilities, i.e., planning, perceiving, reasoning, learning, and problem solving. AI defines general frameworks for knowledge manipulation (building new knowledge and exploiting already gained knowledge) through perception, reasoning (specifying what is to be done, but not how) and acting. SREs may be designed by leveraging and capitalizing on AI-based reasoning, acting, planning, and learning. Further information on the application of machine learning and AI to SREs can be found in \cite{MDR_AI_VTM}, \cite{MDR_AM_TCOM}, \cite{MDR_AI_RIS}.

\section{RISs in Wireless Networks -- A Macroscopic Homogenized Communication-Theoretic Approach} \label{CommTheory}
\textbf{Communication-theoretic modeling and analysis}. This section introduces analytical methods and methodologies for modeling RISs that are made of metasurfaces, and for computing the EM field scattered by metasurfaces. To this end, this section is split in two main parts.
\begin{itemize}
\item \textit{\textbf{Modeling metasurfaces}}. In the first sub-section, we move from a physics-based \textit{microscopic} description of a metasurface and introduce a \textit{macroscopic} representation for it, which is shown to be suitable for application in wireless communications. In particular, a metasurface is represented by using continuous inhomogeneous functions that allow one to describe the signal transformations applied by a metasurface directly on the EM fields. The use of continuous functions is allowed, even though the (conceptual) physical structure of the metasurface in Fig. \ref{Fig_4} is made of discrete unit cells (scatterers), because a metasurface can be homogenized. Therefore, its EM properties can be completely described through macroscopic parameters, in analogy with volumetric materials that can be completely described through their effective permittivity and permeability parameters. The homogenized macroscopic representation of a metasurface based on continuous inhomogeneous functions is shown to be useful for the \textit{synthesis} (i.e., to design a metasurface based on specified/desired signal transformations) and for the \textit{analysis} (i.e., to compute the reflected and refracted EM fields in close proximity of a metasurface, for a given incident EM field and the physical structure of the metasurface) of RISs.

\item \textit{\textbf{Modeling radio waves}}. In the second sub-section, we consider the interaction between the radio waves emitted by a source and a metasurface. We introduce an analytical approach, based on the \textit{theory of diffraction} and the \textit{Huygens-Fresnel principle}, which allows us to compute the EM field at any point of a given volume that contains the metasurface. In particular, we show that the EM field at any point of the volume can be formulated in terms of the EM field in close proximity of the metasurface (i.e., on its surface), as specified in the first sub-section.
\end{itemize}

\textbf{Modeling radio waves in the presence of metasurfaces}. It is apparent, therefore, that the next two sub-sections are intertwined, and are both needed for designing, analyzing, and optimizing RISs in wireless networks. In particular, the first sub-section introduces the analytical tools for modeling the discontinuity of the EM field at the incidence-reflection side and refraction side of a metasurface. The second sub-section, on the other hand, departs from the latter surface EM fields and introduces the analytical tools for calculating the EM field at any point of a volume, which, for example, allows one to unveil how the EM fields scale as a function of the transmission distance, the size of the metasurface, and the physical structure of the metasurface.

\subsection{Surface Electromagnetics} \label{SEM}
Before introducing the details of the concept of homogenized modeling for metasurfaces, we feel important to briefly acquaint the readership of the present paper with the research discipline of surface electromagnetics (SEM), which is the enabling tool for modeling, analyzing, and synthesizing metasurfaces. The content of this sub-section is, in particular, based on the textbook \cite{Book_SurfaceElectromagnetics}, to which interested readers are referred for further information.

\textbf{Definition of SEM}. Electromagnetics is a fundamental discipline of sciences that describes the temporal and spatial behavior of the electric and magnetic fields. Broadly speaking, electromagnetics can be defined as the theory of EM fields and waves. SEM is a sub-discipline of electromagnetics. From the temporal point of view, electromagnetics is usually classified into different categories according to the oscillation frequency of the EM fields, such as direct current (DC), RF, microwaves, terahertz (THz), optics, X-rays, and beyond. The definition of SEM can be traced back to the classification of electromagnetics from the spatial point of view. Four regimes can be identified in the space domain, and each of them is usually modeled through \textit{effective parameters}, which are viewed as an adequate simplification of Maxwell's equations under the corresponding spatial regime \cite[Chapter 1]{Book_SurfaceElectromagnetics}. 
\begin{itemize}
\item \textit{\textbf{Zero-dimensional EM phenomena}}. This regime occurs when the spatial variations of a device or an EM phenomenon are much smaller than the wavelength of the radio waves in all three spatial dimensions. \textit{Circuit theory} is considered to be an accurate and efficient approach for modeling zero-dimensional EM phenomena. The effective parameters that are usually employed in circuit theory are the resistor, the inductor, and the capacitor.
\item \textit{\textbf{One-dimensional EM phenomena}}. This regime occurs when the longitudinal dimension and the transverse dimensions of a device or an EM phenomenon are comparable to and much smaller than, respectively, the wavelength of the radio waves. In this regime, circuit theory is no longer valid and it is replaced by \textit{transmission line theory}. The effective parameters that are usually employed in transmission line theory are the characteristic impedance and the propagation constant.
\item \textit{\textbf{Two-dimensional EM phenomena}}. This regime occurs when the longitudinal dimension and the transverse dimensions of a device or an EM phenomenon are much smaller than and comparable to, respectively, the wavelength of the radio waves. This is the regime of interest of this paper and it will be comprehensively elaborated in further text through its corresponding effective (surface-averaged homogenized) parameters.
\item \textit{\textbf{Three-dimensional EM phenomena}}. This is the most general regime in which the variations of the EM fields are comparable to the wavelength of the radio waves in all three spatial dimensions. Maxwell's equations are usually employed to analyze this regime. The effective parameters that are employed in this regime are the electric permittivity and magnetic permeability of the volumetric material.
\end{itemize}
\begin{figure}[!t]
\begin{centering}
\includegraphics[width=\columnwidth]{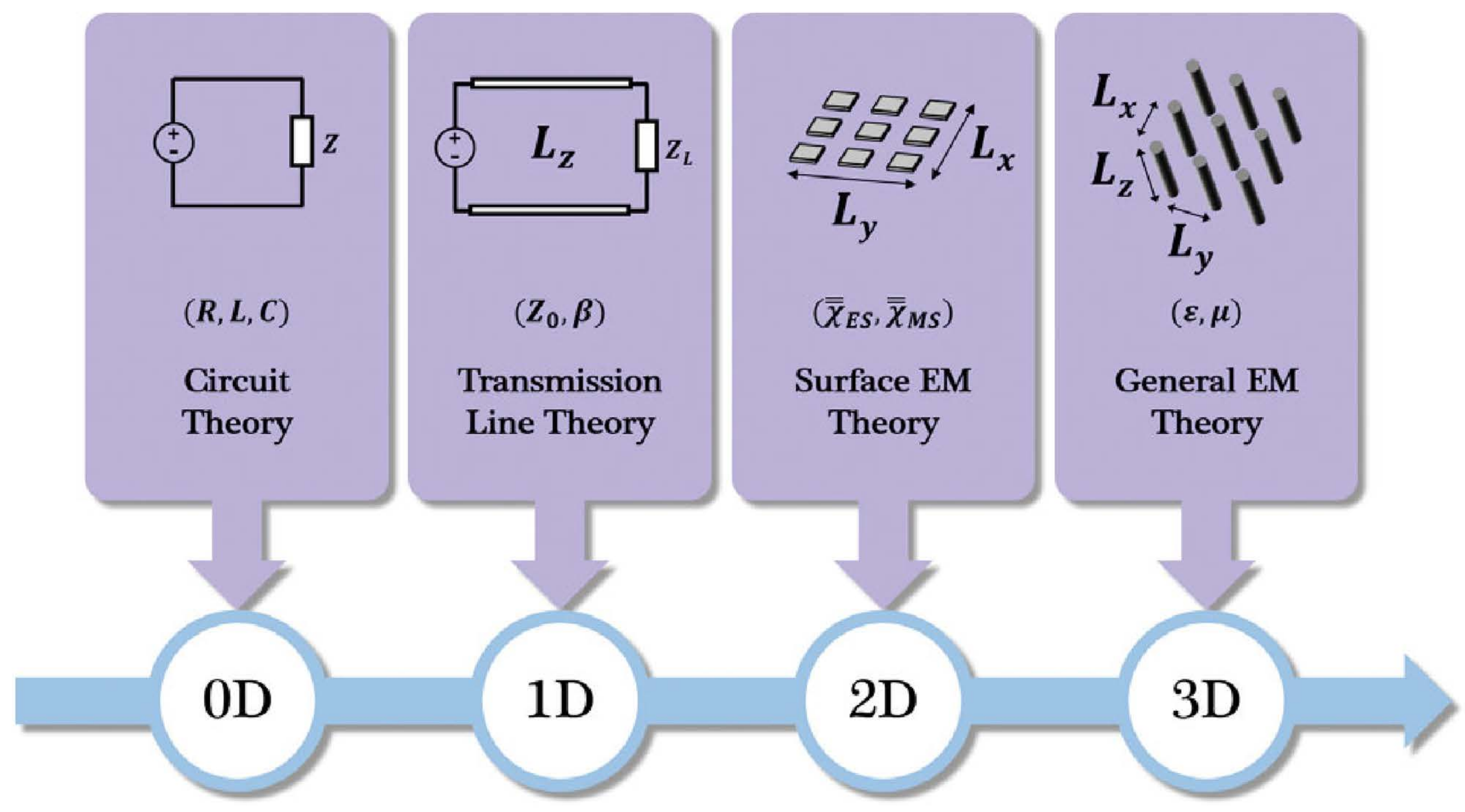}
\caption{Classification of electromagnetics in the spatial domain (reproduced from \cite{Book_SurfaceElectromagnetics}).}
\label{Fig_Book_1}
\end{centering} 
\end{figure}
A graphical comparison among these four operating spatial regimes is sketched in Fig. \ref{Fig_Book_1}, which is reproduced from \cite{Book_SurfaceElectromagnetics}. Broadly speaking, effective parameters allow one to (approximately) model natural and artificial structures and EM phenomena as a whole, instead of modeling their many constituent elements individually. Therefore, they are a convenient tool for studying complex EM phenomena and structures from the macroscopic point of view.

\textbf{From uniform to quasi-periodic metasurfaces}. In the context of SEM research and development, it is important to distinguish three milestones that characterize the evolution of the spatial variations along the transverse dimensions of a metasurface.
\begin{itemize}
\item \textit{\textbf{Uniform metasurfaces}}. By definition, (natural) surfaces are uniform surfaces. They are characterized by variations of the properties of the medium, which surrounds the surface, along the longitudinal direction. The properties of the surface does not change, on the other hand, along the tangential directions, i.e., on the surface itself ($z=0$ in Fig. \ref{Fig_4}).
\item \textit{\textbf{Periodic metasurfaces}}. In contrast to uniform surfaces, periodic surfaces exhibit spatial variations along the tangential directions. Concretely, this implies that the unit cells in Fig. \ref{Fig_4} are arranged in a periodic lattice and each unit cell has the same geometry and size.
\item \textit{\textbf{Quasi-periodic metasurfaces}}. Similar to periodic surfaces, quasi-periodic surfaces exhibit spatial variations along the tangential directions. In contrast to periodic surfaces, the unit cells are still arranged in a periodic lattice, but each unit cell differs from the others in terms of, e.g., geometrical variations, shape, size, orientation angle, etc. This is the configuration reported in Fig. \ref{Fig_4}.
\end{itemize}
\begin{figure}[!t]
\begin{centering}
\includegraphics[width=\columnwidth]{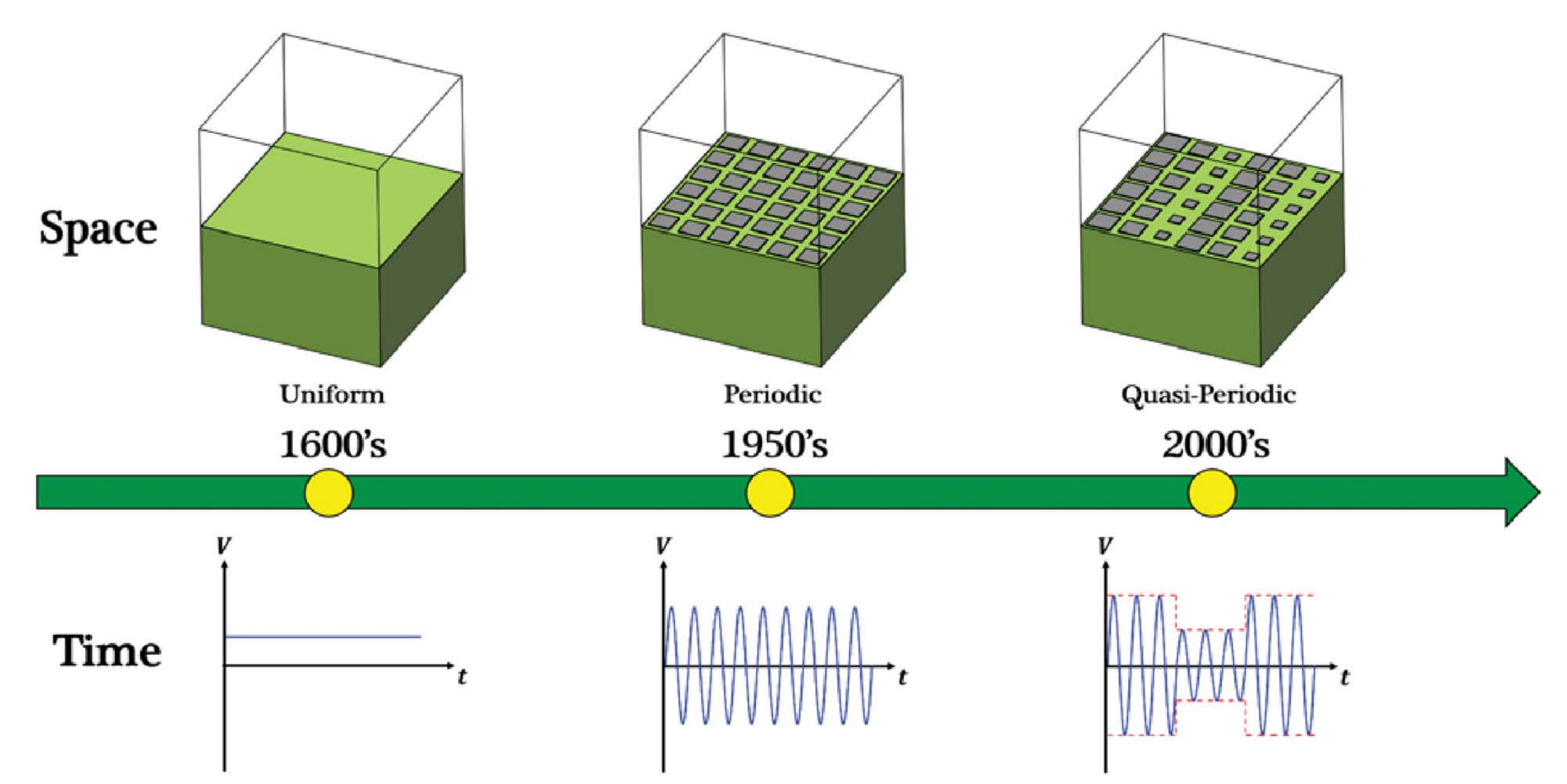}
\caption{The road to quasi-periodic metasurfaces (reproduced from \cite{Book_SurfaceElectromagnetics}).}
\label{Fig_Book_2}
\end{centering} 
\end{figure}
\textbf{Similarity with modulated signals}. The evolution from the uniform, periodic, and quasi-periodic feature of the transverse dimensions of a metasurface can be considered to be equivalent to the evolution of the analysis of signals in circuits, i.e., from DC signals, to alternate current (AC) signals, and eventually to modulated signals. A graphical comparison among the uniform, periodic, and quasi-periodic design structures is reported in Fig. \ref{Fig_Book_2}, which is reproduced from \cite{Book_SurfaceElectromagnetics}. The quasi-periodic structure of the smart surface in Fig. \ref{Fig_4} is, therefore, apparent and completely justified.

\textbf{Naive interpretation of quasi-periodic metasurfaces}. The core concept behind the development and widespread utilization of quasi-periodic metasurfaces lies in their more versatile, unprecedented, and exotic capabilities of manipulating EM waves. Naively, the operating principle of quasi-periodic metasurfaces can be thought of as a two-step manipulation of the radio waves: (i) first, each unit cell manipulates the incident radio waves locally and individually; and (ii) next, the spatial distribution of the obtained wave manipulations collectively alters the wavefront of the incident radio waves, leading to extraordinary
EM responses and transformations of EM fields.

\subsection{Theoretical Foundation of Surface Electromagnetics} \label{SEM_Theory}
\textbf{Volumetric (bulk) vs. surface effective parameters}. The behavior and the properties of metamaterials, which are volumetric engineered structures, are determined by the electric and magnetic properties of its constituent scatterers. The traditional and most convenient approach for modeling metamaterials is the \textit{effective medium theory} (EMT). EMT pertains to the theoretical and analytical models and methods for describing the macroscopic properties of composite materials. At the constituent level, composite materials can be viewed as a micro-inhomogeneous medium. The precise calculation and analysis of the many elements that constitute a composite material is a nearly impossible task. EMT provides one with a set of effective parameters that describe, approximately, a composite material as a whole. These parameters are, e.g., the effective electric permittivity and the effective magnetic permeability of the composite material, which are obtained by averaging (over a small volume) the response of the multiple constituents that compose the material. More precisely, the notions of electric permittivity and magnetic permeability result from the volumetric averaging of microscopic electric and magnetic currents over volumes that are small compared to the wavelength of the radio waves. A similar bulk-parameters representation is, in general, not appropriate for metasurfaces \cite[Chapter 3]{Book_SurfaceElectromagnetics}. In this case, surface-averaged effective parameters are a more physically sound and accurate choice, as compared with volumetric-averaged effective parameters that account for an arbitrary non-zero thickness parameter in order to model the sub-wavelength thickness of metasurfaces. If a metasurface is, in particular, modeled by using an effective permittivity and and effective permittivity, then these two parameters  have to be infinite. This is because the reflection coefficient from a material slab tends to zero when the its thickness tends to zero. Therefore, a model based on effective permittivities and permeabilities implies the need of having a surface of finite (non-zero) thickness. In detail, surface-averaged effective parameters result from the surface averaging of microscopic currents over a surface area of the order of the wavelength. This implies that the conventional notions of electric permittivity and magnetic permeability lose their meaning because there is no volume over which the EM fields and their corresponding induced currents can be averaged out. Surface averaging methods lead to the introduction of effective surface parameters, which include surface susceptibility functions and sheet impedances. These parameters are formally introduced in further text along with the concept of homogenized models for metasurfaces. We anticipate that surface-averaged effective parameters constitute an essential notion for developing macroscopic models and representations of metasurfaces.

\begin{figure}[!t]
\begin{centering}
\includegraphics[width=\columnwidth]{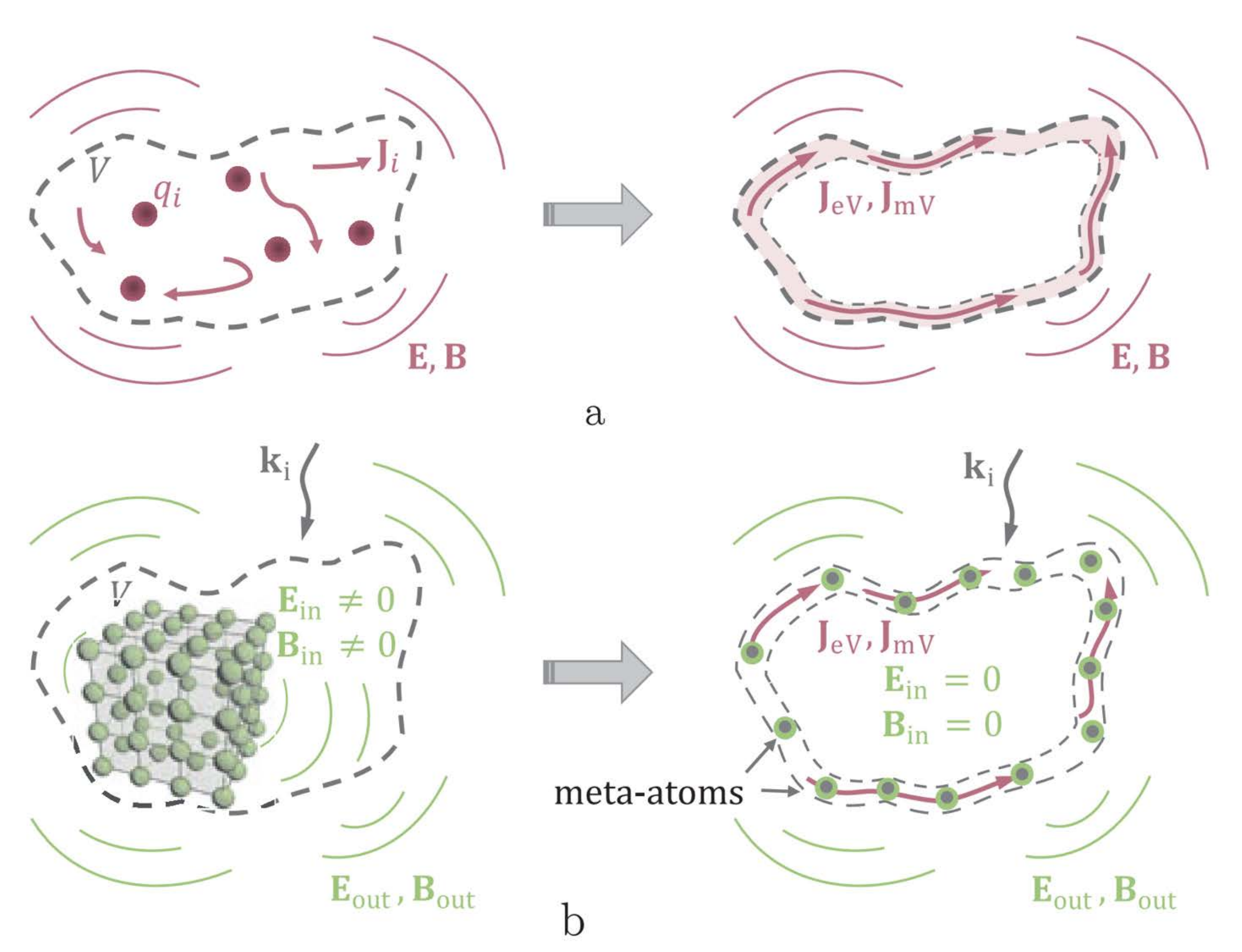}
\caption{Illustration of the surface equivalent theorem applied to the scattering of volumetric electromagnetic sources (a) and to the concept of metamaterials (b) (reproduced from \cite{Book_SurfaceElectromagnetics}).}
\label{Fig_Book_3}
\end{centering} 
\end{figure}
\textbf{The surface equivalence theorem}. The surface equivalence theorem can be regarded as the physics-based foundation of SEM \cite{Book_SurfaceElectromagnetics}. The basic idea of the surface equivalence theorem originates from the \textit{Huygens's principle}, which states that ``each point on a primary wavefront can be considered to be a new source of a secondary spherical wave, and that a secondary wavefront can be constructed as the envelope of these secondary spherical waves'', and from the \textit{uniqueness theorem}, which, in the context of this paper, can be stated as ``if the tangential electric and magnetic fields are completely known over a closed surface, the fields in the source-free region can be determined'' \cite[Chapter 1]{Book_SurfaceElectromagnetics}. The surface equivalence theorem is illustrated in Fig. \ref{Fig_Book_3}, which is reproduced from \cite[Chapter 2]{Book_SurfaceElectromagnetics}. Let us consider a volume $V$ that is filled with arbitrary sources of EM radiation, e.g., some charges $q_i$ and some currents $\mathbf{J}_i$. These sources create an electric field, $\mathbf{E}$, and a magnetic induction field, $\mathbf{B}$, outside the volume $V$. According to the Huygens principle, the system of scatterers can be replaced by an arbitrarily thin layer of specific electric currents, $\mathbf{J}_{\rm{eV}}$, and magnetic currents, $\mathbf{J}_{\rm{mV}}$, that encloses the volume $V$. The thickness of the layer can be electrically small but non-zero since magnetic currents can be generated only via loops of electric currents with finite thickness. The equivalent (surface) currents $\mathbf{J}_{\rm{eV}}$ and $\mathbf{J}_{\rm{mV}}$ scatter EM fields only outward of the volume $V$, and these EM fields are the same as those created by the original system of sources. Such currents, which scatter EM fields only in one side, are referred to as Huygens's surfaces or Huygens's sources. This concept can be applied to metamaterials. Let us consider an arbitrary volumetric metamaterial sample that is excited by an arbitrary external EM wave whose wavevector is $\mathbf{k}_i$. The external wave induces some charges $q_i$ and currents $\mathbf{J}_i$ in the inclusions (or unit cells) of the sample, which irradiate secondary EM fields $\mathbf{E}_{\rm{out}}$ and $\mathbf{B}_{\rm{out}}$ into the space outside the volume $V$ that encloses the metamaterial sample. According to the Huygens principle, one can replace the bulky metamaterial sample with induced polarization charges and currents by equivalent surface currents $\mathbf{J}_{\rm{eV}}$ and $\mathbf{J}_{\rm{mV}}$ that scatter the same EM fields $\mathbf{E}_{\rm{out}}$ and $\mathbf{B}_{\rm{out}}$ outside the volume $V$. By knowing these equivalent currents, one can determine appropriate topologies of unit cells (meta-atoms in the figure), which are placed along the surface of the volume $V$ and generate, if illuminated by an external EM wave with wavevector $\mathbf{k}_i$, the same currents $\mathbf{J}_{\rm{eV}}$ and $\mathbf{J}_{\rm{mV}}$ as those of the original volumetric setup. The resulting arrangement of unit cells is the metasurface structure that yields the same EM response as that of the original metamaterial sample. Broadly speaking, therefore, the surface equivalence theorem states that volumetric metamaterials
can be replaced by electrically thin and, in general, curved metasurfaces, i.e., structures whose surface is engineered to provide field transformation capabilities. The surface equivalence theorem is, therefore, the core tenet behind the development and evolution from metamaterials to metasurfaces.

\begin{figure}[!t]
\begin{centering}
\includegraphics[width=\columnwidth]{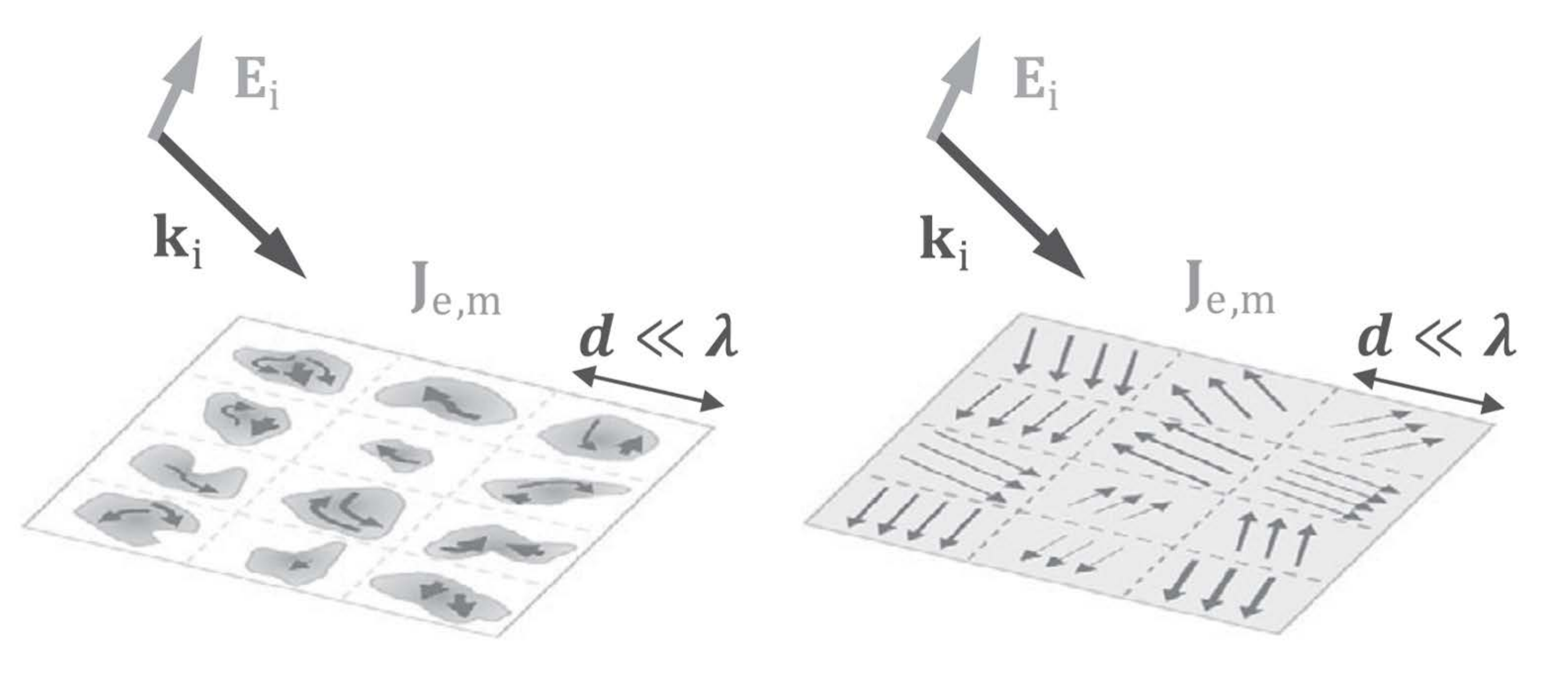}
\caption{Illustration of homogenization: A metasurface consisting of arbitrary sub-wavelength unit cells is modeled as a continuous sheet of surface-averaged (over the unit cell) currents (reproduced from \cite{Book_SurfaceElectromagnetics}).}
\label{Fig_Book_4}
\end{centering} 
\end{figure}
\textbf{Homogenized modeling of metasurfaces}. As anticipated in previous text, a metasurface can be homogenized owing to the sub-wavelength inter-distance among its constituent sub-wavelength unit cells. Concretely, this implies that the EM properties of a metasurface can be completely described by using macroscopic effective (surface) parameters, similar to the macroscopic effective (volumetric) parameters that describe three-dimensional materials, e.g., the effective permittivity and permeability. Knowledge of the macroscopic effective (surface) parameters of a metasurface allows one to formulate its EM response to arbitrary (in terms of wavefront, incidence angle, and polarization) impinging EM waves. This implies that macroscopic models for metasurfaces, which are described in the next sub-sections, do not assume that the impinging EM fields are plane wave. It is necessary, however, that the incident EM fields do not change significantly over the scale of one unit cell. This condition is usually fulfilled in metasurface structures whose unit cells have sub-wavelength inter-distances and sub-wavelength sizes. It is worth mentioning that, on the other hand, it is not possible to introduce effective macroscopic parameters for non-homogenizable structures, even if such structures are electrically thin. Constructing a homogenization model corresponds (i) to determine the effective parameters that can appropriately describe the response of a metasurface to incident EM fields and (ii) to identify such effective parameters from the physical and EM properties of the unit cells. The last step can be performed by using experimental data or simulations of the metasurface response as well. Homogenization is a fundamental prerequisite for understanding, modeling, and designing metasurfaces. By using effective parameters, the complexity of the problem is remarkably reduced, since modeling the collective response of many individually small unit cells is a complex brute-force numerical optimization problem. In (volumetric) metamaterials, as mentioned, homogenized models are obtained through effective parameters that are averaged over small volumes that contain many unit cells. Due to the sub-wavelength thickness of metasurfaces, homogenized models are obtained through effective parameters that are averaged over small surface areas (whose dimensions are of the order of one wavelength) that contain several unit cells. If the impinging EM fields are plane waves, surface averaging over the area of one unit cell is usually sufficient. The concept of homogenized equivalent of a metasurface is sketched in Fig. \ref{Fig_Book_4}, which is reproduced from \cite[Chapter 2]{Book_SurfaceElectromagnetics}. The figure depicts a metasurface that consists of an array of sub-wavelength unit cells. Since the inter-distance between adjective unit cells is sufficiently small compared to the wavelength of the impinging EM fields, the incident radio waves can be assumed to be uniform or homogeneous along the size of a single unit cell. The impinging EM fields induce electric and magnetic currents in each unit cell. In homogenized models, the position-dependent induced electric and magnetic currents in each unit cell are replaced by surfaced-averaged (hence macroscopic) effective parameters. By using homogenization, therefore, a metasurface is modeled as a planar thin sheet of electric and magnetic surface-averaged current densities or, as described in further text, polarization surface densities. This implies that the modeling and analysis of the scattering from a metasurface that is made of finite-size (or discrete) unit cells is turned into the modeling and analysis of the scattering from an equivalent continuous sheet of electric and magnetic currents.

\subsection{On Modeling Metasurfaces: A Macroscopic Homogenized Approach} \label{MacroscopicApproach} 
Starting from the introductory material on SEM, this sub-section is focused on reporting tractable analytical models for metasurfaces. Special emphasis is, in particular, put on the theory that leads from a microscopic to a macroscopic representation of metasurfaces through surface-averaged homogenized effective parameters.

\textbf{From a microscopic to a macroscopic representation}. We consider an RIS made of a metasurface according to the definition given in Section II, i.e., the smart surface is electrically thin, electrically large, homogenizable, and has a sub-wavelength structure. Under these assumptions, we begin our development by introducing a microscopic description of a metasurface that is based on and accounts for the spatial coupling between adjacent unit cells, which is due to their sub-wavelength inter-distance. The obtained physics-based model may not, however, be sufficiently flexible for designing and analyzing complex metasurfaces. Subsequently, for this reason, we introduce a macroscopic description of a metasurface, which is based on continuous tensor functions, that yields a more suitable representation of a metasurface, especially for application to wireless networks. The macroscopic approach, however, still allows us to obtain a direct connection with the material parameters, i.e., the surface susceptibility of the smart surface. Based on the introduced macroscopic description of a metasurface, finally, we discuss its synthesis and analysis.

\textbf{Reconciling communication and electromagnetism}. It is worth mentioning that the approach reported in this sub-section is widely employed in the fields of electromagnetism and metamaterials, and our treatise is mainly based on the theories available in \cite{Holloway_Oct2003}, \cite{Holloway_Nov2005}, \cite{Caloz_GSTCs2015}, \cite{Caloz_Synthesis2018}, \cite{Caloz_Computational2018}. These methods are, however, not necessarily known to wireless researchers. The main objective of this sub-section is to fill this gap of knowledge in a simple but rigorous manner, and with the aid of some examples and case studies.

\subsubsection{EM-Based Model of a Metasurface} Our departing point is the EM-based model of a metasurface reported in \cite{Holloway_Oct2003}, \cite{Holloway_Nov2005}. 

\begin{figure}[!t]
\begin{centering}
\includegraphics[width=\columnwidth]{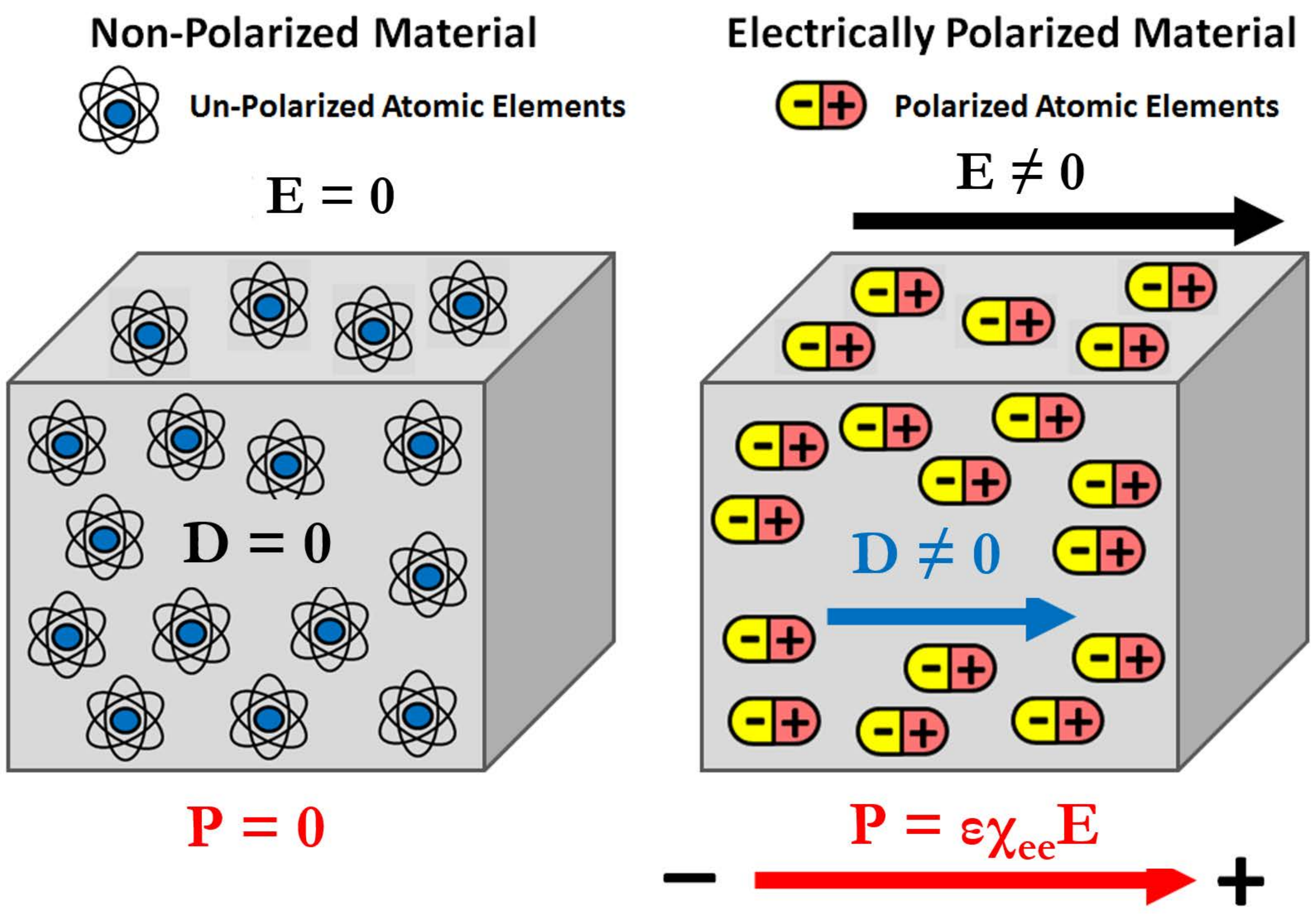}
\caption{Example of induced electric dipole moment and electric polarization density ($\mathbf{P}$) in volumetric materials.}
\label{Fig_12}
\end{centering} 
\end{figure}

\textbf{A metasurface as a surface distribution of electrically small resonant scatterers}. In the most general sense, a metasurface (also called a metafilm in \cite{Holloway_Oct2003}, \cite{Holloway_Nov2005}) is viewed as a surface distribution of electrically small resonant scatterers, and it is characterized by electric and magnetic surface polarization densities. In simple terms, the electrically small scatterers are the unit cells in Fig. \ref{Fig_4}. In volumetric, three-dimensional materials, the electric polarization density is a vector field that expresses the density of permanent or induced electric dipole moments in a dielectric material. When a dielectric is placed in an external electric field, its molecules gain electric dipole moment and the dielectric is said to be polarized. The electric dipole moment induced per unit volume of the dielectric material is referred to as the electric polarization of the dielectric. The electric polarization density describes how a material responds to an applied electric field and how the material changes the electric field. More precisely, an external electric field that is applied to a dielectric material causes a displacement of bound charged elements, which are elements bound to molecules and, hence, are not free to move around the material. Positive charged elements are displaced in the direction of the electric field and negative charged elements are displaced in the opposite direction of the field. The molecules may remain neutral in charge, but an electric dipole moment is formed. The electric polarization density corresponds to the induced dipole moment per unit volume of the material. For ease of understanding, this concept is sketched in Fig. \ref{Fig_12}. Similarly, the magnetic polarization density describes how a material responds to an applied magnetic field. To model metasurfaces as sheets of electrically negligible thickness, the concept of electric and magnetic surface polarization densities is used. The electric (magnetic) surface polarization density is the electric (magnetic) dipole moment per unit area of the surface, while the usual volumetric electric (magnetic) polarization density is the electric (magnetic) dipole moment per unit volume.

\textbf{A metasurface as an array of polarizable unit cells}. A metasurface can be broadly defined as an array of polarizable unit cells that induce discontinuities of the electric and magnetic fields at the two sides of the surface, i.e., at $z=0^+$ and $z=0^-$ in Fig. \ref{Fig_4}. The objective of a macroscopic description of a metasurface consists of formulating and expressing the discontinuities of the electric and magnetic fields as a function of surface-averaged electric and magnetic surface polarization densities. Since a metasurface is homogenizable, i.e., the distance between adjacent unit cells is much smaller than the wavelength of the radio waves, the approach consists of replacing the discrete distribution of unit cells with a continuous distribution, which, in turn, results in a continuous function of electric and magnetic surface polarization densities. It is worth nothing that the unit cells may be of arbitrary shape and are not infinitely thin. Their thickness is only required to be small in comparison with the wavelength of the radio waves. 

\begin{figure}[!t]
\begin{centering}
\includegraphics[width=\columnwidth]{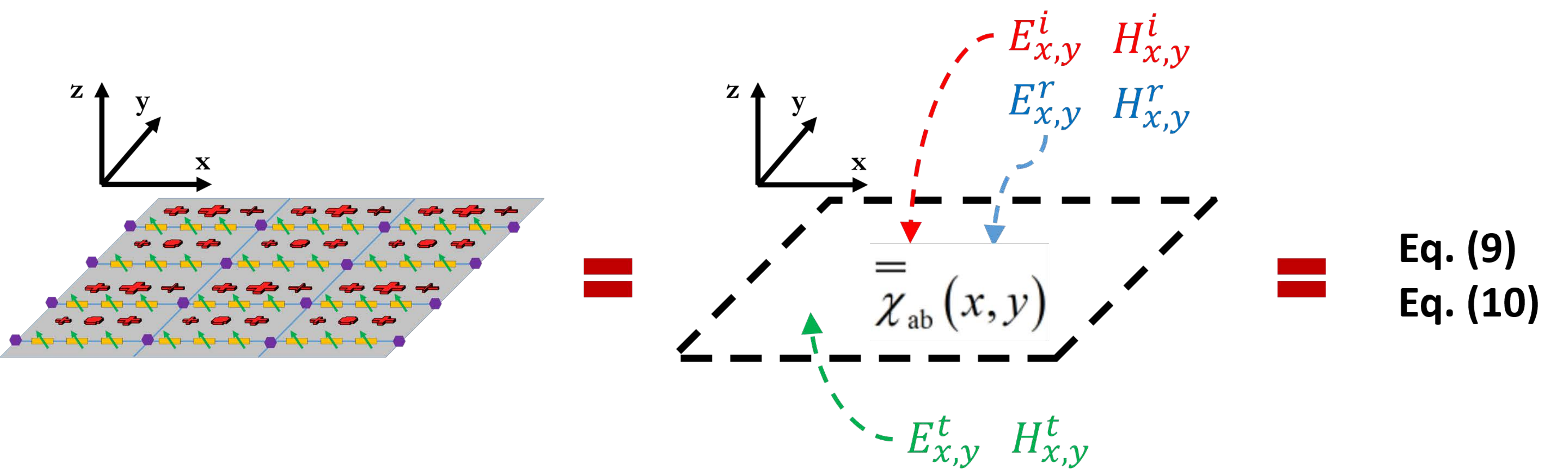}
\caption{Equivalent analytical representation of a metasurface: A zero-thickness material sheet whose EM properties are modeled through surface susceptibility functions that introduce EM discontinuities (or ``jumps'') quantified by the generalized sheet transition conditions.}
\label{Fig_13}
\end{centering} 
\end{figure}
\textbf{Towards a macroscopic description}. A macroscopic description of a metasurface consists of replacing the actual spatial distribution of the unit cells on the smart surface with analytical relations between the electric and magnetic fields at the two sides of the metasurface (at $z=0^+$ and $z=0^-$ in Fig. \ref{Fig_4}). Conceptually, this equivalence is illustrated in Fig. \ref{Fig_13}. The size, the shape, and the physical characteristics of the unit cells are directly incorporated into these analytical relations through the homogenized (continuous) electric and magnetic surface polarization densities. The analytical relations that express the electric and magnetic fields at the two sides of a metasurface as functions of the homogenized (continuous) electric and magnetic surface polarization densities are referred to as \textit{generalized sheet transition conditions (GSTCs)}. It is worth anticipating that the EM fields and the effective parameters that appear in a macroscopic formulation of the GSTCs are surface-averaged quantities, i.e., they are averaged over small areas of the order of a wavelength.

\textbf{Generalized sheet transition conditions}. As far as the GSTCs are concerned, three remarks are worth mentioning.
\begin{itemize} 
\item \textit{\textbf{The tangential components of the EM fields are sufficient}}. The electric and magnetic fields that appear in the GSTCs involve only their tangential (or transverse) components across the surface. In other words, the cartesian or longitudinal component of the EM fields along the $z$-axis in Fig. \ref{Fig_4} does not explicitly appear. This is because they can be uniquely determined by the transverse components by virtue of the uniqueness theorem \cite{Caloz_GSTCs2015}.
\item \textit{\textbf{The EM fields have variations on scales larger than the wavelength}}. The electric and magnetic fields that appear in the GSTCs do not exhibit variations on a length scale that is comparable with the dimension of the unit cells and with their inter-distance. They only exhibit variations on a length scale that is larger than the wavelength of the radio waves. In the present paper, however, the corresponding model is still referred to as microscopic because it requires an appropriate definition of the spatial coupling between adjacent unit cells \cite{Holloway_Oct2003}. We use the term macroscopic modeling to identify similar analytical relations in which this coupling is implicitly taken into account by using surface-averaged susceptibility tensor functions \cite{Caloz_GSTCs2015}. This concept is better elaborated in further text.
\item \textit{\textbf{Generalization of conventional interface conditions}}. The GSTCs, which best characterize a metasurface as an EM discontinuity, constitute a generalization of conventional interface conditions for the EM fields at the interface of two media. For example, it is known that the tangential components of the electric field are continuous at the interface of two media. This is not true if the interface is constituted by a zero-thickness metasurface that acts as an EM discontinuity. The relations between the tangential components of the electric and magnetic fields are, in fact, specified by the GSTCs. In general, the tangential components of the electric and magnetic fields are different at the two sides of a metasurface, i.e., they are discontinuous. It is usual jargon to say that the GSTCs formulate the ``jumps'' (i.e., discontinuities) of the electric and magnetic fields at the two sides of a metasurface.
\end{itemize}
\begin{table}[!t]
\centering
\caption{Recurrent notation ($k = 2\pi / \lambda = \omega \sqrt{\mu \varepsilon}$).}
\newcommand{\tabincell}[2]{\begin{tabular}{@{}#1@{}}#2\end{tabular}}
 \begin{tabular}{l||l}
\hspace{0.15cm} Symbol & \hspace{2.50cm} Definition \\ \hline \hline
${()}^*$ & Complex conjugate \\
${\mathop{\rm Re}\nolimits} \left\{  \cdot  \right\}$ & Real part \\
${()}^T$ & Matrix transpose \\
${()}^H$ & Hermitian conjugate \\
${()}^\dag$ & Pseudo inverse \\
$\cdot$ & Scalar product \\
$\times$ & Vector product \\
$\widehat {()}$ & Unit vector \\
$\lambda$ & Wavelength \\  
$\omega$ & Angular frequency \\  
$\varepsilon$ & Permittivity in vacuum \\
$\mu$ & Permeability in vacuum \\
$\eta = \sqrt{\mu / \varepsilon}$ & Impedance in vacuum \\
$k$ & Wavenumber \\
$\textbf{P}(x,y)$ & Electric surface polarization density \\ 
$\textbf{M}(x,y)$ & Magnetic surface polarization density \\ 
${{\bf{P}}_{_\parallel }}\left( {x,y} \right)$ & Longitudinal component of $\textbf{P}(x,y)$\\
${{\bf{M}}_{_\parallel }}\left( {x,y} \right)$ & Longitudinal component of $\textbf{M}(x,y)$\\
$\textbf{E}(x,y,z)$ & Electric field \\
$\textbf{H}(x,y,z)$ & Magnetic field \\
$\textbf{D}(x,y,z)$ & Electric displacement field \\
${E}_{x,y} (x,y)$ & Tangential components of the electric field \\
${H}_{x,y} (x,y)$ & Tangential components of the magnetic field \\
$i$, $r$, $t$ & Incident, reflected, transmitted field \\
$\theta_i$, $\theta_r$, $\theta_t$ & Angle of incidence, reflection, transmission \\
${\overline{\overline \chi } _{{\rm{ab}}}}\left( {x,y} \right)$ & Surface susceptibility dyadics \\ 
$R(x,y)$ & Surface reflection coefficient \\ \hline \hline
\end{tabular}
\label{Table_Notation}
\end{table}
\subsubsection{Microscopic Description of a Metasurface} \textbf{From unit cells to mathematics}. After introducing the general definition of GSTCs and the general meaning of macroscopic homogenized description of a metasurface, we are ready to formulate these two concepts in analytical terms. To this end, we depart from the physical structure of a metasurface and its microscopic representation. The corresponding macroscopic homogenized description is derived in the next sub-section. 

\textbf{Metasurface structure}. We consider a two-dimensional metasurface that lies on the $xy$-plane at $z=0$. The metasurface is centered at the origin and has a finite size equal to $2L_x$ and $2L_y$ along the $x$-axis and $y$-axis, i.e., $-L_x \le x \le L_x$ and $-L_y \le y \le L_y$, respectively. Without loss of generality, we assume that the incidence-reflection side of the metasurface is $z=0^+$ and the transmission side of the metasurface is $z=0^-$. Also, we assume that the medium surrounding the metasurface is vacuum. The symbols and definition used throughout the present paper are reported in Table \ref{Table_Notation}. Our analytical formulation is mostly based on \cite{Caloz_GSTCs2015}, \cite{Caloz_Synthesis2018}, \cite{Caloz_Computational2018}. For simplicity, we consider the sufficient general case of metasurfaces whose longitudinal components of the electric ($\textbf{P}$) and magnetic ($\textbf{M}$) surfaces polarization densities are equal to zero, i.e., $P_z = 0$ and $M_z = 0$, respectively. We show in further text that this assumption allows one to design metasurfaces that realize EM-based functions that are sufficiently general for wireless applications, e.g., reflection and refraction towards arbitrary directions. The advantage of this assumption is that closed-form analytical expressions for the EM fields are obtained.

\textbf{Analytical formulation of GSTCs}. Let us consider an EM field that illuminates a metasurface. Under the assumption $P_z = 0$ and $M_z = 0$, the GSTCs, i.e., the ``jumps'' of the electric and magnetic fields at the two sides of a metasurface, can be formulated as follows:
\begin{equation} \label{Eq_GSTCs}
\begin{split}
\hat z \times \Delta {\bf{H}}\left( {x,y} \right) = j\omega {{\bf{P}}_{_\parallel }}\left( {x,y} \right)\\
\Delta {\bf{E}}\left( {x,y} \right) \times \hat z = j\omega \mu {{\bf{M}}_{_\parallel }}\left( {x,y} \right)
\end{split}
\end{equation}
\noindent where ${{\bf{P}}_{_\parallel }}\left( {x,y} \right)$ and ${{\bf{M}}_{_\parallel }}\left( {x,y} \right)$ are the longitudinal components of the surface-averaged electric and magnetic surface polarization densities, respectively, and $\Delta {\bf{E}}\left( {x,y} \right)$ and $\Delta {\bf{H}}\left( {x,y} \right)$ are the differences between the surface-averaged electric and magnetic fields at the two sides of the metasurface, whose cartesian components can be written, respectively, as follows:
\begin{equation} \label{Eq_FieldsDifference}
\begin{split}
 \Delta {\Psi _u}\left( {x,y} \right) &= {\bf{\hat u}} \cdot \left. {\Delta {\bf{\Psi }}\left( {x,y} \right)} \right|_{z = {0^ - }}^{z = {0^ + }}\\
& = \Psi _u^t\left( {x,y} \right) - \left( {\Psi _u^i\left( {x,y} \right) + \Psi _u^r\left( {x,y} \right)} \right)
\end{split}
\end{equation}
\noindent where ${\bf{\Psi }} = \left\{ {{\bf{E}},{\bf{H}}} \right\}$, $u = \left\{ {x,y,z} \right\}$, and the superscripts $i$, $r$ and $t$ denote the incident, reflected, and transmitted components of the fields, respectively. 

\textbf{Computing $\textbf{P}$ and $\textbf{M}$}. The GSTCs depend on the longitudinal components of the electric and magnetic surface polarization densities. In order to formulate and compute them, we need to investigate the working operation of a metasurface at the microscopic level and we need to introduce the concept of acting (or local) fields.

\textbf{Acting (or local) field: Excitation field at the location of an individual unit cell}. At the microscopic level, a metasurface is viewed as an array of electrically small polarizable unit cells. When an EM field illuminates a metasurface, the unit cells get polarized and, since they are small (sub-wavelength) in size, the induced polarizations of each individual unit cell can be modeled through induced electric and magnetic dipole moments (see Fig. \ref{Fig_12}). Since the unit cells are arranged in a dense array, i.e., the inter-distance between adjacent unit cells is sub-wavelength, knowing the polarization induced on a single unit cell is not sufficient to determine the response of the entire metasurface. This is because the EM field that excites a unit cell that occupies a specific location, which is referred to as the acting or local EM field, is given by the summation of the incident EM field and the, so-called, interaction EM field, which is the field created by the induced electric and magnetic dipoles of all the unit cells of the metasurface with the exception of the unit cell under analysis.

\textbf{Acting field: Modeling the mutual coupling among unit cells}. By definition, therefore, the acting electric and magnetic fields account for the contribution of all the unit cells of a metasurface with the exception of the given unit cell under consideration. This implies that they account for the mutual (spatial) coupling among the unit cells of a metasurface and that they yield a microscopic description of a metasurface, since the acting fields are referred to individual unit cells. If the acting fields are known, the electric ($\textbf{P}$) and magnetic ($\textbf{M}$) surfaces polarization densities in \eqref{Eq_GSTCs} can be formulated as follows \cite{Holloway_Oct2003}:
\begin{equation} \label{Eq_PolarizationMicro__P}
\begin{split}
{\bf{P}}\left( {x,y} \right) &= \varepsilon N\langle {\overline{\overline \alpha } _{{\rm{ee}}}}\left( {x,y} \right)\rangle {{\bf{E}}_{{\rm{act}}}}\left( {x,y} \right) \\ &+ \sqrt {\mu \varepsilon } N\langle {\overline{\overline \alpha } _{{\rm{em}}}}\left( {x,y} \right)\rangle {{\bf{H}}_{{\rm{act}}}}\left( {x,y} \right)
\end{split}
\end{equation}
\begin{equation} \label{Eq_PolarizationMicro__M}
\begin{split}
{\bf{M}}\left( {x,y} \right) & = \sqrt {{\varepsilon  \mathord{\left/ {\vphantom {\varepsilon  \mu }} \right. \kern-\nulldelimiterspace} \mu }} N\langle {\overline{\overline \alpha } _{{\rm{me}}}}\left( {x,y} \right)\rangle {{\bf{E}}_{{\rm{act}}}}\left( {x,y} \right) \\ &+ N\langle {\overline{\overline \alpha } _{{\rm{mm}}}}\left( {x,y} \right)\rangle {{\bf{H}}_{{\rm{act}}}}\left( {x,y} \right)
\end{split}
\end{equation}
\noindent where $N$ is the number of unit cells per unit area, and $\langle {\overline{\overline \alpha } _{{\rm{ab}}}}\left( {x,y} \right)\rangle$, for $a = \left\{ {\rm{e},\rm{m}} \right\}$ and $b = \left\{ {\rm{e},\rm{m}} \right\}$, are the average (electric-electric, electric-magnetic, magnetic-electric, and magnetic-magnetic) polarizability dyadics of the unit cell in which the electric and magnetic surface polarization densities are computed. In particular, the average $\langle  \cdot \rangle$ is calculated over the unit cells in the vicinity of the unit cell being considered and where the electric and magnetic surfaces polarization densities are evaluated. The polarizability dyadics depend on the physics and EM properties of the unit cells. As far as the present paper is concerned, it is sufficient to know that they can be obtained from analysis, simulations, or measurements.

\textbf{Relating local (microscopic) fields to average (macroscopic) fields}. Based on \eqref{Eq_PolarizationMicro__P} and \eqref{Eq_PolarizationMicro__M}, the electric and magnetic surface polarization densities can be formulated in terms of the acting electric (${{\bf{E}}_{{\rm{act}}}}\left( {x,y} \right)$) and magnetic (${{\bf{H}}_{{\rm{act}}}}\left( {x,y} \right)$) fields at the position of a given unit cell. By definition, the acting fields can be formulated as follows \cite{Holloway_Oct2003}: 
\begin{equation} \label{Eq_ActingField}
\begin{split}
& {{\bf{E}}_{{\rm{act}}}}\left( {x,y} \right) = {{\bf{E}}_{{\rm{av}}}}\left( {x,y} \right) - {\bf{E}}_{{\rm{scattering}}}^{{\rm{unit \; cell}}}\left( {x,y} \right)\\
& {{\bf{H}}_{{\rm{act}}}}\left( {x,y} \right) = {{\bf{H}}_{{\rm{av}}}}\left( {x,y} \right) - {\bf{H}}_{{\rm{scattering}}}^{{\rm{unit \; cell}}}\left( {x,y} \right)
\end{split}
\end{equation}
\noindent where ${{\bf{E}}_{{\rm{av}}}}\left( {x,y} \right)$ and ${{\bf{H}}_{{\rm{av}}}}\left( {x,y} \right)$ are the average electric and magnetic fields, respectively, at the two sides of the metasurface, which are defined as follows:
\begin{equation} \label{Eq_AverageField}
\begin{split}
& {E_{{\rm{av}},u}}\left( {x,y} \right) = \frac{{E_u^t\left( {x,y} \right) + \left( {E_u^i\left( {x,y} \right) + E_u^r\left( {x,y} \right)} \right)}}{2}\\
& {H_{{\rm{av}},u}}\left( {x,y} \right) = \frac{{H_u^t\left( {x,y} \right) + \left( {H_u^i\left( {x,y} \right) + H_u^r\left( {x,y} \right)} \right)}}{2}
\end{split}
\end{equation}
\noindent where $u = \left\{ {x,y,z} \right\}$, and ${\bf{E}}_{{\rm{scattering}}}^{{\rm{unit \; cell}}}\left( {x,y} \right)$ and ${\bf{H}}_{{\rm{scattering}}}^{{\rm{unit \; cell}}}\left( {x,y} \right)$ are the electric and magnetic fields scattered by the single unit cell under consideration, respectively. 

\textbf{Contribution of a single unit cell}. In practice, the contribution of a single unit cell can be calculated by expressing the unit cell as a combination of electric and magnetic dipoles that are contained within a small disk of a given radius. The electric and magnetic fields scattered by the resulting small disk are, by definition, ${\bf{E}}_{{\rm{scattering}}}^{{\rm{unit \; cell}}}\left( {x,y} \right)$ and ${\bf{H}}_{{\rm{scattering}}}^{{\rm{unit \; cell}}}\left( {x,y} \right)$. 

\textbf{Surface-averaged fields}. It is worth emphasizing that \eqref{Eq_ActingField} relates local, hence microscopic, fields to average, hence macroscopic, fields. More specifically, ${{\bf{E}}_{{\rm{av}}}}\left( {x,y} \right)$ and ${{\bf{H}}_{{\rm{av}}}}\left( {x,y} \right)$ in \eqref{Eq_AverageField} are surface-averaged fields, where the average is computed over a small surface area of the order of the wavelength (including one or more unit cells depending on the setup). In other words, the rapid variations of the EM fields over distances of the order of the typical separation between adjacent unit cells along the surface are eliminated (i.e., averaged out) in macroscopic EM fields. This is not the case for microscopic EM fields.

\textbf{The need for a macroscopic representation}. By inserting \eqref{Eq_PolarizationMicro__P} and \eqref{Eq_PolarizationMicro__M} into \eqref{Eq_GSTCs}, one can obtain the analytical relations between the transverse components of the electric and magnetic fields at the two sides of a metasurface. The issue with this approach is that the acting electric and magnetic fields need to be estimated, which depend on a specific unit cell, i.e., ${\bf{E}}_{{\rm{scattering}}}^{{\rm{unit \; cell}}}\left( {x,y} \right)$ and ${\bf{H}}_{{\rm{scattering}}}^{{\rm{unit \; cell}}}\left( {x,y} \right)$, and may not be easy to deal with. This issue is overcome next, by introducing an explicit macroscopic description of a metasurface, which depends only on the average (macroscopic) fields.

\subsubsection{Macroscopic Description of a Metasurface}
\textbf{Explicit macroscopic formulation}. In \cite{Holloway_Oct2003}, the authors succeed in calculating a closed-form expression for ${\bf{E}}_{{\rm{scattering}}}^{{\rm{unit \; cell}}}\left( {x,y} \right)$ and ${\bf{H}}_{{\rm{scattering}}}^{{\rm{unit \; cell}}}\left( {x,y} \right)$. Based on their analytical formulation, the electric and magnetic surface polarization densities in \eqref{Eq_PolarizationMicro__P} and \eqref{Eq_PolarizationMicro__M} can be formulated in terms of the average electric and magnetic fields in \eqref{Eq_AverageField}, which, in contrast with the acting fields, yield a macroscopic description of a metasurface that is viewed as an electromagnetic discontinuity in space. This is because the average fields do not exclude the fields generated by individual unit cells. 

\textbf{Surface-averaged susceptibility functions: The effective parameters of metasurfaces}. More specifically, a macroscopic description of a metasurface, which depends only on average (macroscopic) electric and magnetic fields, can be obtained by formulating the electric and magnetic surface polarization densities in terms of \textit{susceptibility tensors} rather than in terms of polarizabilities. Similar to the average fields, the susceptibility tensors are macroscopic surface-averaged parameters. They can can be viewed as the effective parameters of metasurfaces, in analogy with the electric permittivity and magnetic permeability of volumetric metamaterials, and they depend on the topology and geometry of the unit cells, the inter-distance among the unit cells, the properties of the material, and the wavelength. In particular, the following relations hold true \cite{Caloz_GSTCs2015}:
\begin{equation} \label{Eq_PolarizationMacro__P}
\begin{split}
{\bf{P}}\left( {x,y} \right) &= \varepsilon {\overline{\overline \chi } _{{\rm{ee}}}}\left( {x,y} \right){{\bf{E}}_{{\rm{av}}}}\left( {x,y} \right) \\ & + \sqrt {\mu \varepsilon } {\overline{\overline \chi } _{{\rm{em}}}}\left( {x,y} \right){{\bf{H}}_{{\rm{av}}}}\left( {x,y} \right)
\end{split}
\end{equation}
\begin{equation} \label{Eq_PolarizationMacro__M}
\begin{split}
{\bf{M}}\left( {x,y} \right) &= \sqrt {{\varepsilon  \mathord{\left/
 {\vphantom {\varepsilon  \mu }} \right.
 \kern-\nulldelimiterspace} \mu }} {\overline{\overline \chi } _{{\rm{me}}}}\left( {x,y} \right){{\bf{E}}_{{\rm{av}}}}\left( {x,y} \right) \\& + {\overline{\overline \chi } _{{\rm{mm}}}}\left( {x,y} \right){{\bf{H}}_{{\rm{av}}}}\left( {x,y} \right)
\end{split}
\end{equation}
\noindent where ${\overline{\overline \chi } _{{\rm{ab}}}}\left( {x,y} \right)$, for $a = \left\{ {\rm{e},\rm{m}} \right\}$ and $b = \left\{ {\rm{e},\rm{m}} \right\}$, are the (electric-electric, electric-magnetic, magnetic-electric, and magnetic-magnetic) surface-averaged (or simply surface) susceptibility dyadics, which are the macroscopic quantities of interest for a simple but accurate synthesis and analysis of metasurfaces.

\textbf{GSTCs: A macroscopic algebraic formulation}. With the aid of the analytical formulation of the electric and magnetic surface polarization densities as a function of only macroscopic fields and surface susceptibility dyadics, one can obtain an explicit analytical (algebraic) formulation of the GSTCs. More specifically, by inserting \eqref{Eq_PolarizationMacro__P} and \eqref{Eq_PolarizationMacro__M} into \eqref{Eq_GSTCs}, the following explicit relation can be obtained (for ease of notation, the dependency on $\left( {x,y} \right)$ for all fields and susceptibility dyadics is explicitly omitted but implied):
\begin{equation} \label{Eq_GSTCs_Explicit_H}
\begin{split}
& \left[ {\begin{array}{*{20}{c}}
{ - H_y^t + H_y^i + H_y^r}\\
{H_x^t - H_x^i - H_x^r}
\end{array}} \right]\\
& \hspace{1cm} = \frac{{j\omega \varepsilon }}{2}\left[ {\begin{array}{*{20}{c}}
{\chi _{{\rm{ee}}}^{xx}}&{\chi _{{\rm{ee}}}^{xy}}\\
{\chi _{{\rm{ee}}}^{yx}}&{\chi _{{\rm{ee}}}^{yy}}
\end{array}} \right]\left[ {\begin{array}{*{20}{c}}
{E_x^t + E_x^i + E_x^r}\\
{E_y^t + E_y^i + E_y^r}
\end{array}} \right]\\
& \hspace{1cm} + \frac{{j\omega \sqrt {\mu \varepsilon } }}{2}\left[ {\begin{array}{*{20}{c}}
{\chi _{{\rm{em}}}^{xx}}&{\chi _{{\rm{em}}}^{xy}}\\
{\chi _{{\rm{em}}}^{yx}}&{\chi _{{\rm{em}}}^{yy}}
\end{array}} \right]\left[ {\begin{array}{*{20}{c}}
{H_x^t + H_x^i + H_x^r}\\
{H_y^t + H_y^i + H_y^r}
\end{array}} \right]
\end{split}
\end{equation}
\begin{equation} \label{Eq_GSTCs_Explicit_E}
\begin{split}
& \left[ {\begin{array}{*{20}{c}}
{E_y^t - E_y^i - E_y^r}\\
{ - E_x^t + E_x^i + E_x^r}
\end{array}} \right]\\
& \hspace{1cm} = \frac{{j\omega \mu }}{2}\left[ {\begin{array}{*{20}{c}}
{\chi _{{\rm{mm}}}^{xx}}&{\chi _{{\rm{mm}}}^{xy}}\\
{\chi _{{\rm{mm}}}^{yx}}&{\chi _{{\rm{mm}}}^{yy}}
\end{array}} \right]\left[ {\begin{array}{*{20}{c}}
{H_x^t + H_x^i + H_x^r}\\
{H_y^t + H_y^i + H_y^r}
\end{array}} \right]\\
& \hspace{1cm} + \frac{{j\omega \sqrt {\mu \varepsilon } }}{2}\left[ {\begin{array}{*{20}{c}}
{\chi _{{\rm{me}}}^{xx}}&{\chi _{{\rm{me}}}^{xy}}\\
{\chi _{{\rm{me}}}^{yx}}&{\chi _{{\rm{me}}}^{yy}}
\end{array}} \right]\left[ {\begin{array}{*{20}{c}}
{E_x^t + E_x^i + E_x^r}\\
{E_y^t + E_y^i + E_y^r}
\end{array}} \right]
\end{split}
\end{equation}
\noindent where, for $a = \left\{ {\rm{e},\rm{m}} \right\}$ and $b = \left\{ {\rm{e},\rm{m}} \right\}$, we have introduced the notation:
\begin{equation} \label{Eq_SusceptibilityMatrix}
{\overline{\overline \chi } _{{\rm{ab}}}}\left( {x,y} \right) = \left[ {\begin{array}{*{20}{c}}
{\chi _{{\rm{ab}}}^{xx}\left( {x,y} \right)}&{\chi _{{\rm{ab}}}^{xy}\left( {x,y} \right)}\\
{\chi _{{\rm{ab}}}^{yx}\left( {x,y} \right)}&{\chi _{{\rm{ab}}}^{yy}\left( {x,y} \right)}
\end{array}} \right]
\end{equation}

\textbf{GSTCs: A mathematical tool for analysis and synthesis}. The analytical expressions in \eqref{Eq_GSTCs_Explicit_H} and \eqref{Eq_GSTCs_Explicit_E} provide one with the necessary equations for designing metasurfaces, which are capable of applying specified transformations to the impinging radio waves, and for analyzing the performance of wireless networks in the presence of metasurfaces. These two important issues are discussed in the next two sub-sections. Before tackling them, we close this sub-section by emphasizing four important properties of the constitutive relations in \eqref{Eq_GSTCs_Explicit_H} and \eqref{Eq_GSTCs_Explicit_E}: (i) the constitutive relations are formulated only in terms of macroscopic surface-averaged EM fields and macroscopic surface-averaged effective parameters; (ii) the analysis of the surface-averaged susceptibility functions allows one to understand whether the metasurface is locally lossy, i.e., their imaginary part is always negative, or whether some local gains are needed, i.e., their imaginary part takes positive values; (iii) even though the constitutive relations offer a macroscopic description of a metasurface, they inherently account for the mutual coupling among the unit cells, since they originate from a microscopic description of a metasurface. In particular, the mutual coupling between adjacent unit cells is implicitly taken into account in the surface-averaged susceptibility matrices in \eqref{Eq_SusceptibilityMatrix}. This point will be further clarified in the next sub-section; and (iv) the constitutive relations in \eqref{Eq_GSTCs_Explicit_H} and \eqref{Eq_GSTCs_Explicit_E} can be applied to formulate the EM response of a metasurface to arbitrary (in terms of wavefront, incidence angle, and polarization) impinging EM waves. This implies that \eqref{Eq_GSTCs_Explicit_H} and \eqref{Eq_GSTCs_Explicit_E} do not assume that the impinging EM fields are plane waves. It is necessary, however, that the incident EM fields do not change significantly over the scale of one unit cell. This condition is usually fulfilled in metasurface structures whose unit cells have sub-wavelength inter-distances and sub-wavelength sizes.

\textbf{Main takeaway}. In summary, the main takeaway message is that, from an analytical point of view, a metasurface can be approximately replaced by its corresponding constitutive relations in \eqref{Eq_GSTCs_Explicit_H} and \eqref{Eq_GSTCs_Explicit_E}, which are sufficient for the analysis and design of wireless networks (see Fig. \ref{Fig_13}).

\begin{figure}[!t]
\begin{centering}
\includegraphics[width=\columnwidth]{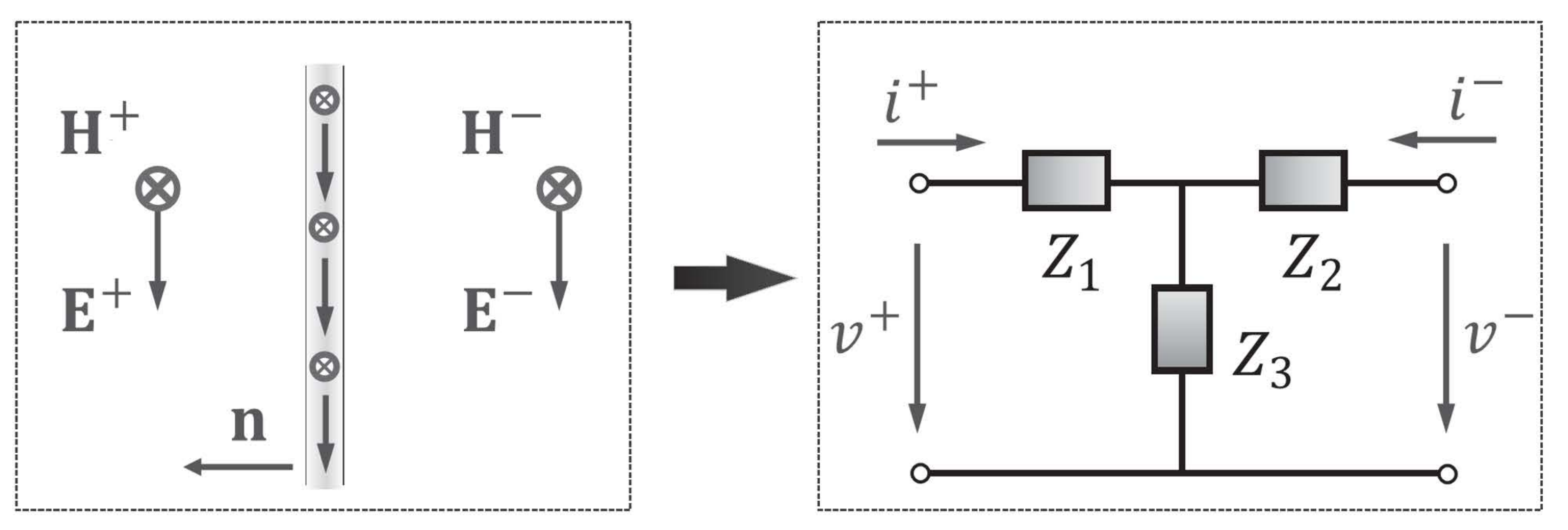}
\caption{Homogenization modeling of a metasurface based on surface-averaged impedances (reproduced from \cite{Book_SurfaceElectromagnetics}).}
\label{Fig_SurfaceImpedances}
\end{centering} 
\end{figure}
\textbf{Equivalent representation in terms of surface-averaged impedances}. An alternative homogenization model for a metasurface can be obtained by establishing an analogy between the propagation of radio waves in vacuum and the propagation of signals in transmission lines \cite[Chapter 2]{Book_SurfaceElectromagnetics}. In this analogy, more precisely, the electric and magnetic fields of a radio wave that propagates in vacuum are matched to the voltages and currents of a signal that propagates in an equivalent transmission line, respectively. The impedances that determine the equivalent transmission line constitute a macroscopic surface-averaged representation of a metasurface, which is equivalent to the representation of a metasurface in terms of surface-averaged susceptibility functions. An example of representation of a metasurface in terms of transmission line equivalent is reported in Fig. \ref{Fig_SurfaceImpedances}, which is reproduced from \cite[Chapter 2]{Book_SurfaceElectromagnetics}. The T-circuit transmission line in Fig. \ref{Fig_SurfaceImpedances} ensures, in particular, that there exist ``jumps'' (discontinuities) of both the electric and magnetic fields, i.e., the voltages and currents represented in the figure, in agreement with the GSTCs in \eqref{Eq_GSTCs}. By applying Kirchhoff's circuit laws to the equivalent transmission line representation of the metasurface, and by inserting the resulting equations in \eqref{Eq_GSTCs_Explicit_H} and \eqref{Eq_GSTCs_Explicit_E}, one can compute analytical relations between the surface-averaged susceptibility functions and the surface-averaged impedances of the transmission line equivalent in Fig. \ref{Fig_SurfaceImpedances}. An example of this computation and an explicit relation between the two representations for a metasurface can be found in \cite[Appendix]{ComputationImpedances}. The homogenization model formulated in terms of surface-averaged impedances provides one with useful engineering insights on the properties of a metasurface structure. For example, a metasurface is capacitive and inductive if the imaginary part of the impedances is negative and positive, respectively, and a metasurface is lossy and active if the real part of the impedances is positive and negative, respectively. As remarked in \cite{ComputationImpedances}, more in general, a representation of a metasurface in terms of circuital parameters, i.e., surface-averaged impedances, may be potentially more useful to engineering-oriented communities, while a representation of a metasurface in terms of surface-averaged susceptibility functions may be more widely used in physics-oriented communities. Both representations are, however, homogenized, macroscopic, and equivalent with each other. An example of application of the homogenized modeling of metasurfaces based on  surface-averaged impedances is given in the next sub-section.

\subsubsection{Synthesis of a Metasurface}
\textbf{How to build my own metasurface}? The synthesis of a metasurface corresponds to the following problem formulation: \textit{Given some specified radio wave transformations, i.e., the incident, reflected, and transmitted electric and magnetic fields are known, identify the corresponding susceptibility matrices in \eqref{Eq_SusceptibilityMatrix} that realize them, and then map the susceptibility matrices into physical structures that are made of specific arrangements of unit cells}. Therefore, this is a two-part design problem. We elaborate the details of both sub-problems in this sub-section, first in general terms and then with the aid of an example that elucidates some important details. \\

\noindent \underline{\textbf{\textit{Computation of susceptibility functions}}}. Given $E_v^{\xi}(x,y)$ and $H_v^{\xi}(x,y)$ for $v = \left\{ x,y \right\}$ and $\xi = \left\{ i,r,t \right\}$, the surface susceptibility functions in \eqref{Eq_SusceptibilityMatrix} can be obtained by solving the two systems of intertwined equations in \eqref{Eq_GSTCs_Explicit_H} and \eqref{Eq_GSTCs_Explicit_E}. 

\textbf{An undetermined system of equations}. The combined system of \eqref{Eq_GSTCs_Explicit_H} and \eqref{Eq_GSTCs_Explicit_E} is constituted by four equations and sixteen unknowns. Therefore, it is an undetermined system of equations that admits, in general, an infinite number of solutions. In practical terms, this implies that there exist many different combinations of surface susceptibility functions in \eqref{Eq_SusceptibilityMatrix} that realize the same transformations of the incident electric and magnetic fields into the reflected and transmitted electric and magnetic fields. A simple approach for overcoming the undetermined nature of the system of equations in \eqref{Eq_GSTCs_Explicit_H} and \eqref{Eq_GSTCs_Explicit_E} is to reduce to four the number of independent surface susceptibility functions in \eqref{Eq_SusceptibilityMatrix}. For example, some of them may be set equal to zero, provided that the resulting metasurface admits a physical structure that can be realized in practice. 

\textbf{A simple example}. To better understand, let us consider the case study in which the constraints ${\overline{\overline \chi } _{{\rm{em}}}}\left( {x,y} \right) = {\overline{\overline \chi } _{{\rm{me}}}}\left( {x,y} \right) = 0$ and $\chi _{{\rm{ee}}}^{xy}\left( {x,y} \right) = 0$, $\chi _{{\rm{ee}}}^{yx}\left( {x,y} \right) = 0$, $\chi _{{\rm{mm}}}^{xy}\left( {x,y} \right) = 0$, $\chi _{{\rm{mm}}}^{yx}\left( {x,y} \right) = 0$ are imposed to the surface susceptibility functions. From a physical standpoint, this implies that the metasurface structure is monoanisotropic (i.e., its physical properties exhibit variations along one direction) and uniaxial (i.e., it has a single direction of symmetry), and, therefore, it is reciprocal. Under these assumptions, the following analytical expressions for the remaining surface susceptibility functions can be obtained from \eqref{Eq_GSTCs_Explicit_H} and \eqref{Eq_GSTCs_Explicit_E}:
\begin{equation} \label{Eq_SusceptibilityExample}
\begin{split}
& \chi _{{\rm{ee}}}^{xx}\left( {x,y} \right) =  - \frac{2}{{j\omega \varepsilon }}\frac{{H_y^t\left( {x,y} \right) - H_y^i\left( {x,y} \right) - H_y^r\left( {x,y} \right)}}{{E_x^t\left( {x,y} \right) + E_x^i\left( {x,y} \right) + E_x^r\left( {x,y} \right)}}\end{split}
\end{equation}
\begin{equation} \nonumber
\begin{split}
& \chi _{{\rm{ee}}}^{yy}\left( {x,y} \right) = \frac{2}{{j\omega \varepsilon }}\frac{{H_x^t\left( {x,y} \right) - H_x^i\left( {x,y} \right) - H_x^r\left( {x,y} \right)}}{{E_y^t\left( {x,y} \right) + E_y^i\left( {x,y} \right) + E_y^r\left( {x,y} \right)}}\\
& \chi _{{\rm{mm}}}^{xx}\left( {x,y} \right) = \frac{2}{{j\omega \mu }}\frac{{E_y^t\left( {x,y} \right) - E_y^i\left( {x,y} \right) - E_y^r\left( {x,y} \right)}}{{H_x^t\left( {x,y} \right) + H_x^i\left( {x,y} \right) + H_x^r\left( {x,y} \right)}}\\
& \chi _{{\rm{mm}}}^{yy}\left( {x,y} \right) =  - \frac{2}{{j\omega \mu }}\frac{{E_x^t\left( {x,y} \right) - E_x^i\left( {x,y} \right) - E_x^r\left( {x,y} \right)}}{{H_y^t\left( {x,y} \right) + H_y^i\left( {x,y} \right) + H_y^r\left( {x,y} \right)}}
\end{split}
\end{equation}

\textbf{The case of perfect anomalous reflection}. Given the explicit expressions of the incident, reflected, and transmitted fields, \eqref{Eq_SusceptibilityExample} yields a metasurface that  can realize both anomalous reflection and anomalous transmission. If, for example, perfect reflection is of interest, i.e., $E_x^t\left( {x,y} \right) = E_y^t\left( {x,y} \right) = 0$ and $H_x^t\left( {x,y} \right) = H_y^t\left( {x,y} \right) = 0$, then one obtains $\chi _{{\rm{ee}}}^{xx}\left( {x,y} \right)\chi _{{\rm{mm}}}^{yy}\left( {x,y} \right) =  - {4 \mathord{\left/ {\vphantom {4 {{k^2}}}} \right. \kern-\nulldelimiterspace} {{k^2}}}$ and $\chi _{{\rm{ee}}}^{yy}\left( {x,y} \right)\chi _{{\rm{mm}}}^{xx}\left( {x,y} \right) =  - {4 \mathord{\left/ {\vphantom {4 {{k^2}}}} \right. \kern-\nulldelimiterspace} {{k^2}}}$, where $k = \omega \sqrt {\mu \varepsilon }$. If a perfect reflector needs to be realized, therefore, only one independent surface susceptibility function is needed. This case study is elaborated in more detail in further text. \\

\noindent \underline{\textbf{\textit{Realization of physical lattices of unit cells}}}. Once the analytical expressions of the surface susceptibility functions are obtained, e.g., the formulas in \eqref{Eq_SusceptibilityExample}, the next step is to map these functions into a physical structure of unit cells that can be realized and manufactured. This latter step is, in particular, the most difficult one to implement.

\textbf{Periodicity of a metasurface}. Before introducing the corresponding procedure, it is worth mentioning that the surface susceptibility functions that are obtained from \eqref{Eq_GSTCs_Explicit_H} and \eqref{Eq_GSTCs_Explicit_E} may or may not be periodic functions in $\left( {x,y} \right)$. If they are periodic functions, the periodicity of a metasurface coincides with the period of the surface susceptibility functions, which is usually expressed in terms of number of wavelengths. If the incident, reflected, and transmitted radio waves are plane waves, for example, the surface susceptibility functions are periodic functions in $\left( {x,y} \right)$. If more advanced functionalities are needed, e.g., focusing, aperiodical metasurfaces are usually needed, and the corresponding surface susceptibility functions are not periodic functions $\left( {x,y} \right)$. As far as periodic metasurfaces are concerned, it is worth mentioning that the period of the surface susceptibility functions is usually greater than the wavelength. The periodic feature of the surface susceptibility functions is clarified in further text with an example. In the rest of the present paper, for simplicity, we will assume that the surface susceptibility functions are periodic functions in $\left( {x,y} \right)$.

\textbf{Step-by-step procedure}. The procedure that is usually adopted for mapping the surface susceptibility functions into metasurface structures can be summarized, in general terms, as follows \cite{CalozSergei_Mar2018}.
\begin{enumerate}
\item \textbf{Identifying the super-cell}. The continuous surface susceptibility functions are discretized (or sampled) in unit cells whose size and inter-distance are sub-wavelength. Thanks to the periodic nature of the surface susceptibility functions, this operation is executed for a single period. The discretization is performed by ensuring that a sufficient number of unit cells per period is obtained in agreement with the Nyquist theorem. This step yields the discretized version of \eqref{Eq_SusceptibilityMatrix}, i.e., ${\overline{\overline \chi } _{{\rm{ab}}}}\left( {{x_p},{y_q}} \right)$, for $p=1,2,\ldots,N_p$ and $q=1,2,\ldots,N_q$, where $N_p$ and $N_q$ are the numbers of unit cells, in one period, along the $x$-axis and $y$-axis, respectively. The obtained $N_p \times N_q$ unit cells correspond to the \textit{super-cell} depicted in Fig. \ref{Fig_4}. We anticipate that the entire metasurface is obtained as the periodic repetition of the super-cell structure.
\item \textbf{Optimizing the super-cell}. Once the super-cell structure made of $N_p \times N_q$ unit cells is identified, the geometry of each unit cell of the super-cell needs to be optimized. In general, in fact, the unit cells of the super-cell may have a different geometry (size and shape). This optimization is executed by using a full-wave simulator that allows one to solve Maxwell's equations numerically. In general, this implies that the $N_p \times N_q$ unit cells of the super-cell are \textit{jointly optimized} in this step. To perform this optimization by using a full-wave simulator, first a generic structure and geometry for the unit cell are selected. This choice takes into account the fact that the structure and geometry of the unit cells need to be appropriate in order to synthesize the entire phase and amplitude range of the discretized surface susceptibility functions. In \cite{CalozSergei_Mar2018}, for example, the authors employ a dog-bone-shaped geometry for the generic unit cell, which has four optimization parameters ($G$, $L$, $S$, $W$), as depicted in Fig. \ref{Fig_14}. In \cite{CalozSergei_Mar2018}, the super-cell is constituted by $N_p \times N_q$ unit cells that are all dog-bone-shaped. Each unit cell of the super-cell is, however, characterized by a different quadruplet of parameters ($G$, $L$, $S$, $W$).
\begin{figure}[!t]
\begin{centering}
\includegraphics[width=\columnwidth]{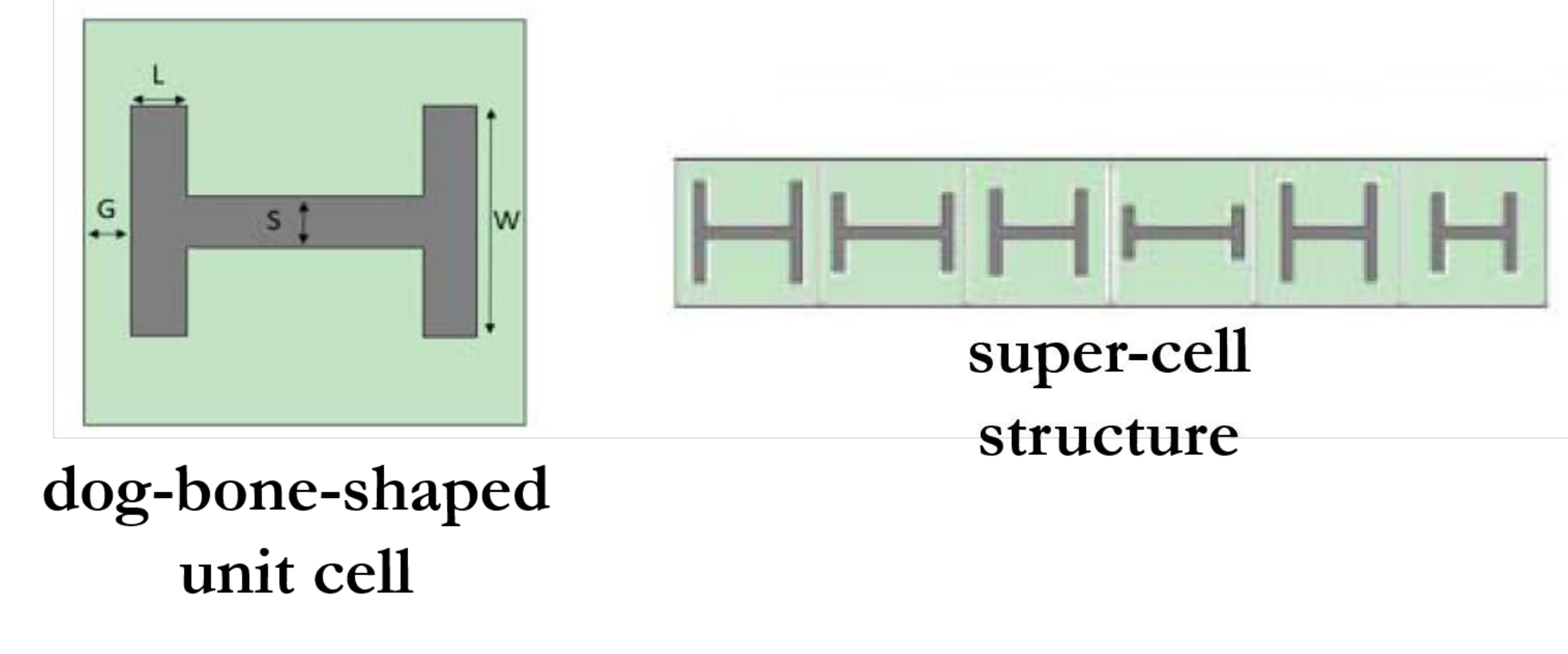}
\caption{Dog-bone-shaped unit cell and super-cell structure (example and illustration are reproduced from \cite{CalozSergei_Mar2018}).}
\label{Fig_14}
\end{centering} 
\end{figure}
\item \textbf{Metasurfaces: A periodic repetition of super-cells}. Finally, as anticipated, the metasurface is obtained as the periodic repetition of the optimized super-cell structure. This is allowed thanks to the periodicity of the surface susceptibility functions, provided that the super-cell is appropriately defined by taking the exact period into account.
\end{enumerate}

\textbf{Rationale behind the super-cell structure}. As far as the optimization of the structure and geometry of the super-cell (the above-mentioned second step) is concerned, three remarks are dutiful.
\begin{itemize}
\item \textit{\textbf{Periodic boundary conditions}}. The optimization of the geometry and structure of the entire super-cell is executed by enforcing \textit{periodic boundary conditions} in a full-wave simulator. This is a plausible choice thanks to the periodicity of the surface susceptibility functions, whose period is, by design, entirely covered by the super-cell. More specifically, the super-cell constitutes the so-called representative surface (or, in general, volume) element of the periodic structure that determines a metasurface. Therefore, it is sufficient, under periodic boundary conditions, to account for the behavior of the entire metasurface. It is important to mention, however, that this approximation is appropriate if the size of the metasurface is large as compared with the wavelength of the radio waves and if the area of the surface contains many periods. With this approximation, in particular, edge effects are ignored.
\item \textit{\textbf{Mutual (spatial) coupling}}. The joint optimization of the $N_p \times N_q$ unit cells of the super-cell and the application of the periodic boundary conditions when solving Maxwell's equations with a full-wave simulator allow one to effectively account for the mutual coupling among the unit cells. The mutual coupling is, in fact, needed in order to synthesize, in practice, the desired discretized surface susceptibility functions. The resulting structure of the metasurface is locally aperiodic, since it is made of different adjacent unit cells, but the the super-cell structure enforces the required periodicity across the entire metasurface.
\item \textit{\textbf{Efficient optimization}}. The joint optimization of the unit cells of the super-cell may not be computationally affordable. For example, a super-cell with only ten unit cells, each of which has four parameters to be optimized, necessitates the joint optimization of forty variables. To circumvent this issue and to efficiently design metasurfaces of practical use, iterative and often sub-optimal methods may be employed. For example, one may first optimize the geometry and size of each unit cell of the super-cell individually, by enforcing periodic boundary conditions to each unit cell. Since a single unit cell does not cover an entire period of the surface susceptibility functions, the resulting design is an approximation. The obtained sub-optimal design may, however, be subsequently used as the starting point for jointly optimizing all the unit cells of the super-cell. For example, one may iteratively optimize one unit cell by keeping the others fixed, and then iterate the procedure until convergence (under some error tolerance). 
\end{itemize}

\textbf{The mutual coupling among the unit cells is taken into account by design}. The actual realization of a metasurface into a physical lattice of unit cells may not necessarily be the main objective of researchers working in wireless communications. It is, however, constructive to be aware of and to understand the general procedure employed by researchers working on the actual implementation of metasurfaces, in order to appreciate the fact that the mutual coupling among local neighbors of unit cells is taken into account by design through the periodic repetition of super-cells and the periodic nature of the surface susceptibility functions. \\

\noindent \underline{\textbf{\textit{Constructive Example -- Perfect anomalous reflection}}}. In order to elucidate the procedures for the synthesis and analysis of a metasurface, we consider a single example that allows us to highlight some aspects that play an important role for the analysis and optimization of RIS-empowered wireless networks. 

\textbf{Setup and assumptions}. The chosen example corresponds to the design of a metasurface that realizes perfect anomalous reflection (zero transmission). For ease of description, in addition, the incident and reflected electric and magnetic fields are assumed to be transverse electric plane waves with an arbitrary angle of incidence and an arbitrary angle of reflection.

\textbf{Transverse electric plane waves}. Under the assumption of transverse electric plane waves, the tangential electric and magnetic fields can be formulated as follows \cite{Holloway_Nov2005}:
\begin{equation} \label{Eq_FullReflection_PlaneWave}
\begin{split}
& E_x^i\left( {x,y} \right) = E_x^r\left( {x,y} \right) = 0\\
& E_y^i\left( {x,y} \right) = \sqrt {{2 \eta P_0}} \exp \left( { - jk\sin \left( {{\theta _i}} \right)x} \right)\\
& E_y^r\left( {x,y} \right) = \sqrt {{2 \eta P_0}} {A_r}\exp \left( { - jk\sin \left( {{\theta _r}} \right)x} \right)\\
& H_x^i\left( {x,y} \right) =  \left(\sqrt {{2P_0}/ {\eta}}\right)\cos \left( {{\theta _i}} \right)\exp \left( { - jk\sin \left( {{\theta _i}} \right)x} \right)\\
& H_x^r\left( {x,y} \right) = - \left(\sqrt {{2P_0}/ {\eta}} \right){A_r}\cos \left( {{\theta _r}} \right)\exp \left( { - jk\sin \left( {{\theta _r}} \right)x} \right)\\
& H_y^i\left( {x,y} \right) = H_y^r\left( {x,y} \right) = 0
\end{split}
\end{equation}
\noindent where: (i) $\eta  = \sqrt {{\mu  \mathord{\left/ {\vphantom {\mu  \varepsilon }} \right. \kern-\nulldelimiterspace} \varepsilon }}$ is the intrinsic impedance in vacuum; (ii) $P_0$ is the average power density (i.e., the average power per unit area) of the incident EM field; (iii) $\theta_i$ and $\theta_r$ are the angles of incidence and reflection, respectively, with respect to the $z$-axis; and (iv) $A_r$ is the amplitude of the reflected radio wave, which is a scaling factor independent of $(x,y)$. For simplicity, $A_r$ is assumed to be a real positive number. The amplitude $A_r$ can be chosen according to different criteria. In the present paper, we focus our attention on setups for $A_r$ that account for the power flows (obtained from the Poynting vector) of the incident and reflected EM fields. This choice allows us to analyze the power efficiency of a metasurface, i.e., the amount of power that is reflected towards the desired direction given the amount of power that is incident from a given direction, and the design guidelines and constraints for engineering metasurfaces with high power efficiency. The details about the conditions for balancing the incident and reflected power flows are discussed in the next paragraphs. It is worth nothing that, given the assumption of transverse electric plane waves, the surface EM fields are independent of $y$, i.e., $E_y^i\left( {x,y} \right) = E_y^i\left( x \right)$, $E_y^r\left( {x,y} \right) = E_y^r\left( x \right)$,  $H_x^i\left( {x,y} \right) = H_x^i\left( x \right)$, and $H_x^r\left( {x,y} \right) = H_x^r\left( x \right)$. 

\textbf{Benchmark EM fields transformation}. The analytical formulation of the EM fields in \eqref{Eq_FullReflection_PlaneWave} is referred to as \textit{benchmark} or \textit{desired} fields transformation, since it provides one with the ideal function that a perfect anomalous reflector is intended to realize: The metasurface transforms a single incident plane wave into a single reflected plane wave, without any spurious reflections or transmissions towards other (undesired) directions. The EM fields in \eqref{Eq_FullReflection_PlaneWave} constitute a theoretical tool that is used for the synthesis of a perfect anomalous reflector. In further text, we discuss different possibilities for representing the resulting metasurface structure, which are convenient to use in wireless communications and electromagnetics.

\textbf{Surface susceptibility function}. Based on \eqref{Eq_SusceptibilityExample}, the metasurface structure that realizes the desired EM fields transformation can be obtained by an appropriate design of the surface susceptibility functions. Under the considered assumptions, in particular, a single surface susceptibility function needs to be studied, e.g., $\chi _{{\rm{ee}}}^{yy}\left( {x,y} \right)$, since the other surface susceptibility functions can be obtained from it. In particular, the following holds true:
\begin{equation} \label{Eq_FullReflection_Susceptibility}
\chi _{{\rm{ee}}}^{yy}\left( {x,y} \right) = - \frac{2}{{j\omega \varepsilon }}\frac{{H_x^i\left( {x,y} \right) + H_x^r\left( {x,y} \right)}}{{E_y^i\left( {x,y} \right) + E_y^r\left( {x,y} \right)}}
\end{equation}
\noindent where it is taken into account that zero transmission implies $E_x^t\left( {x,y} \right) = E_y^t\left( {x,y} \right) = 0$ and $H_x^t\left( {x,y} \right) = H_y^t\left( {x,y} \right) = 0$.

\textbf{Explicit analytical formulation and period of a metasurface structure}. By inserting \eqref{Eq_FullReflection_PlaneWave} in \eqref{Eq_FullReflection_Susceptibility}, the surface susceptibility function can be formulated as follows:
\begin{equation} \label{Eq_FullReflection_Susceptibility__ClosedFormReflection}
\chi _{{\rm{ee}}}^{yy}\left( x \right) = \frac{{2j}}{k}\frac{{\cos \left( {{\theta _i}} \right) - {A_r}\cos \left( {{\theta _r}} \right)\Phi \left( x \right)}}{{1 + {A_r}\Phi \left( x \right)}}
\end{equation}
\noindent where $\Phi \left( x \right) = \exp \left( { - jk\left( {\sin \left( {{\theta _r}} \right) - \sin \left( {{\theta _i}} \right)} \right)x} \right)$ is a shorthand notation. By direct inspection of \eqref{Eq_FullReflection_Susceptibility__ClosedFormReflection}, we evince that $\chi _{{\rm{ee}}}^{yy}\left( x \right)$ is a periodic function in $x$, i.e., $\chi _{{\rm{ee}}}^{yy}\left( x \right) = \chi _{{\rm{ee}}}^{yy}\left( x + \Pi \right)$, where $\Pi$ is the period. In particular, the period is determined by the function $\Phi \left( x \right)$ and it is equal to:
\begin{equation} \label{Eq_FullReflection_Period}
\Pi  = \frac{{2\pi }}{{k\left( {\sin \left( {{\theta _r}} \right) - \sin \left( {{\theta _i}} \right)} \right)}}
\end{equation}

It is worth mentioning that \eqref{Eq_FullReflection_Susceptibility__ClosedFormReflection} provides one with a general analytical representation of a metasurface structure that operates as a perfect anomalous reflector. The specific properties of the resulting metasurface, including its power efficiency and implementation complexity, highly depend on the particular choice of the parameter $A_r$. The impact of $A_r$ is elaborated in the next paragraphs.

\textbf{Net power flow of the metasurface structure}. One of the main features of nearly-passive metasurfaces consists of being passive during the normal operation phase. This implies that the power flow entering the metasurface structure (incoming power flow) needs to be less than or equal to the power flow leaving the metasurface structure (outgoing power flow). The two power flows coincide if the metasurface structure has unitary power efficiency. Given the (desired) analytical formulation of the EM fields in \eqref{Eq_FullReflection_PlaneWave}, the balance between the incoming and outgoing power flows can be ensured by appropriately optimizing the parameter ${A_r}$. Since the metasurface structure is periodic, the net power flow can be formulated for a single period. More precisely, the net power flow, i.e. the sum of the incoming and outgoing power flows, per unit period of the metasurface structure can be obtained from the Poynting vector, as follows:
\begin{equation} \label{Eq_PowerBudget__1}
\begin{split}
& {P_{{\rm{net}}}}\left( x \right) =  - \left( {{1 \mathord{\left/
 {\vphantom {1 2}} \right.
 \kern-\nulldelimiterspace} 2}} \right){\mathop{\rm Re}\nolimits} \left\{ {{E_\parallel }\left( x \right)H_\parallel ^*\left( x \right)} \right\}\\
& {P_{{\rm{net}}}} = \frac{1}{\Pi} \int\nolimits_0^\Pi  {{P_{{\rm{net}}}}\left( x \right)dx} 
\end{split}
\end{equation}
\noindent where ${E_\parallel }\left( {x} \right) = {E_\parallel }\left( {x,y} \right) = E_y^i\left( {x,y} \right) + E_y^r\left( {x,y} \right)$ and ${H_\parallel }\left( {x} \right) = {H_\parallel }\left( {x,y} \right) = H_x^i\left( {x,y} \right) + H_x^r\left( {x,y} \right)$ are the tangential components of the electric and magnetic field, respectively. It is worth mentioning that it is sufficient to consider a single period because the size of a metasurface is usually set equal to an integer number of periods and the net power flow is usually normalized to the total size of the metasurface structure.

By inserting \eqref{Eq_FullReflection_PlaneWave} in \eqref{Eq_PowerBudget__1}, and computing the resulting integral, we obtain the following expression:
\begin{equation} \label{Eq_PowerBudget__2}
\begin{split}
&{P_{{\rm{net}}}}\left( x \right) = {P_0}A_r^2\cos \left( {{\theta _r}} \right) - {P_0}\cos \left( {{\theta _i}} \right)\\
& \hspace{1.15cm} + {P_0}{A_r}\left( {\cos \left( {{\theta _r}} \right) - \cos \left( {{\theta _i}} \right)} \right)\cos \left( {\phi \left( x \right)} \right)\\
& {P_{{\rm{net}}}} = {P_0}\left( {A_r^2\cos \left( {{\theta _r}} \right) - \cos \left( {{\theta _i}} \right)} \right)
\end{split}
\end{equation}
\noindent where $\phi \left( x \right) = k\left( {\sin \left( {{\theta _r}} \right) - \sin \left( {{\theta _i}} \right)} \right)x$ is a shorthand notation.

\textbf{Locally and globally passive metasurfaces}. The net power flow of a metasurface structure can be defined in a local and global sense. From a local standpoint, one is interested in analyzing the surface power flow ${P_{{\rm{net}}}}\left( x \right)$. From a global standpoint, on the other hand, one is interested in analyzing the integral power flow ${P_{{\rm{net}}}}$. More specifically, the local and global power flows in \eqref{Eq_PowerBudget__1} can be negative and positive. If they are negative, it implies that the incident power flow is greater than the reflected power flow. If they are positive, on the other hand, the opposite holds true. The physical meaning of negative and positive power flows can be understood by introducing the definition of locally and globally passive metasurfaces. A metasurface is defined to be locally passive if ${P_{{\rm{net}}}}\left( x \right) \le 0$ for every $x$. Likewise, a metasurface is defined to be globally passive if ${P_{{\rm{net}}}} \le 0$. It is worth mentioning that a locally non-passive metasurface, i.e., ${P_{{\rm{net}}}}\left( x \right) \ge 0$ for some $x$, can be globally passive, i.e., ${P_{{\rm{net}}}} \le 0$. This occurs if the metasurface is characterized by regions where there are power absorptions and regions where there are power gains that compensate each other along the entire metasurface structure. This concept is elaborated in further text, since it is instrumental for realizing metasurfaces with a high power efficiency.

\textbf{Globally passive metasurfaces with unitary power efficiency}. By definition, globally passive metasurfaces with a unitary power efficiency, i.e., the total reflected power coincides with the total incident power, are obtained by imposing the constraint ${P_{{\rm{net}}}}=0$. From \eqref{Eq_PowerBudget__2}, this can be ensured if:
\begin{equation} \label{Eq_Ar_1}
{A_r} = \sqrt {\frac{{\cos \left( {{\theta _i}} \right)}}{{\cos \left( {{\theta _r}} \right)}}}
\end{equation}

In further text, the physical meaning that originates by imposing that a perfect anomalous reflector is globally passive by design is discussed. Furthermore, the design guidelines for engineering globally passive metasurface structures that realize perfect anomalous reflectors with unitary power efficiency are elaborated. It is worth nothing that, by setting $A_r$ as in \eqref{Eq_Ar_1}, ${P_{{\rm{net}}}}\left( x \right)$ in \eqref{Eq_PowerBudget__2} is not identically equal to zero for every $x$, but it is an oscillating function in $x$ that takes positive and negative values. This implies that a perfect anomalous reflector may, in general, exhibit absorptions and gains along the surface. It is interesting to note, however, that a perfect specular reflector, i.e., $\theta _r = \theta _i$, can be made locally and globally passive by setting $A_r=1$. In this case, in particular, the incident and reflected power flows coincide and the metasurface structure has a unitary power efficiency.

\textbf{Surface-averaged reflection coefficient}. The EM fields in \eqref{Eq_FullReflection_PlaneWave} are necessary and sufficient (benchmark) equations for the synthesis of a metasurface structure with full-reflection capability, which is able of bending a radio wave from a generic direction $\theta_i$ towards a generic direction $\theta_r$. The corresponding surface-averaged susceptibility function is given in \eqref{Eq_FullReflection_Susceptibility__ClosedFormReflection}. In particular, \eqref{Eq_FullReflection_Susceptibility__ClosedFormReflection} is a convenient tool for the synthesis of a metasurface. Often, however, researchers are more used to characterize the properties of structures that behave as reflectors by using the concept of reflection coefficient. It is important to note that the definition of (surface-averaged) reflection coefficient is slightly different and serves for a different purpose in the context of wireless communications and in the context of metasurface design. More precisely, the following two definitions hold true.
\begin{itemize}
\item \textbf{\textit{Wireless design: Reflection coefficient based on surface EM fields}}. Let us consider that, departing from \eqref{Eq_FullReflection_Susceptibility__ClosedFormReflection}, one has designed and realized the corresponding metasurface structure by using the synthesis procedure described in previous sections. Wireless researchers are interested in introducing equivalent representations of metasurface structures that yield an explicit and direct relation between the incident and reflected radio waves. In this context, the reflection coefficient is defined as the ratio between the reflected electric field and the incident electric field. From an engineering standpoint, it can be viewed as a ``global'' response of the metasurface structure to some incident  EM fields. As detailed in further text, this definition of reflection coefficient is useful for evaluating the performance of RIS-empowered wireless networks, since a direct link between incident and reflected EM fields is established. In the present paper, the reflection coefficient defined based on surface EM fields is denoted by ${R_{{\rm{EM}}}}\left( x,y \right) = {R_{{\rm{EM}}}}\left( x \right)$.
\item \textbf{\textit{Metasurface design: Reflection coefficient based on surface impedances}}. Let us consider that, departing from \eqref{Eq_FullReflection_Susceptibility__ClosedFormReflection} (or its corresponding surface-averaged impedance, which is  the reciprocal of \eqref{Eq_FullReflection_Susceptibility__ClosedFormReflection} in the considered example), one has to design and manufacture the corresponding metasurface structure that realizes the desired anomalous reflection. Metasurface scientists are interested in obtaining initial insights from \eqref{Eq_FullReflection_Susceptibility__ClosedFormReflection}, in order to identify appropriate physical structures that realize the desired functionality. Broadly speaking, the reflection coefficient based on surface EM fields that is usually employed in wireless communications provides one with information on ``what'' needs to be realized, but it does not yield design guidelines on ``how'' to realize it, which is, on the other hand, the main objective of the designers of metasurfaces. To this end, it is more useful to define the reflection coefficient as a parameter that quantifies how different the ``local'' response of a metasurface structure is as compared with a suitable reference. In the context of anomalous reflection, a suitable choice as a reference is a metasurface structure that behaves as a specular reflector. Therefore, the reflection coefficient is more conveniently defined as a parameter that quantifies the mismatch between the surface-averaged impedances of anomalous reflection and specular reflection \cite{Alu_Reflection2016}, \cite{Asadchy_2017}. This concept is detailed in further text. For clarity, we anticipate that the reflection coefficient based on surface impedances cannot be used to related the incident and reflected EM fields. It yields, however, fundamental information on the design of metasurfaces, e.g., how to realize a globally passive metasurface with unitary efficiency but without using active elements. In the present paper, the reflection coefficient defined based on surface impedances is denoted by ${R_{{\rm{Z}}}}\left( x,y \right) = {R_{{\rm{Z}}}}\left( x \right)$.
\end{itemize}

\textbf{Surface-averaged reflection coefficient: Unified notation}. An objective of the present paper is to analyze and compare against each other the two reflection coefficients ${R_{{\rm{EM}}}}\left( x \right)$ and ${R_{{\rm{Z}}}}\left( x \right)$. To this end, we adopt a unified notation that encompasses both definitions of reflection coefficient and that allows us to unveil the differences and similarities between ${R_{{\rm{EM}}}}\left( x \right)$ and ${R_{{\rm{Z}}}}\left( x \right)$. In the adopted unified notation, we depart from the EM fields in \eqref{Eq_FullReflection_PlaneWave}. We emphasize, once again, that ${R_{{\rm{Z}}}}\left( x \right)$ cannot be used, in general, to directly related the reflected and incident fields. The employed unified notation is, however, useful for unveiling the analytical relation between ${R_{{\rm{EM}}}}\left( x \right)$ and ${R_{{\rm{Z}}}}\left( x \right)$, and for explaining the physical meaning of ${R_{{\rm{Z}}}}\left( x \right)$. In particular, we introduce the following notation and mathematical definition for ${R_{{\rm{EM}}}}\left( x \right)$ and ${R_{{\rm{Z}}}}\left( x \right)$:
\begin{equation} \label{Eq_FullReflection_LocalReflectionCoefficient}
\begin{split}
& E_y^r\left( {x,y} \right) = E_y^r\left( x \right) = R\left( x \right)E_y^i\left( x \right)\\
& H_x^r\left( {x,y} \right) = H_x^r\left( x \right) =  - r_{i,r} R\left( x \right)H_x^i\left( x \right)
\end{split}
\end{equation}
\noindent where: 
\begin{equation} \label{Eq_FullReflection_LocalReflectionCoefficient}
{r_{i,r}} = \begin{cases}
\frac{{\cos \left( {{\theta _r}} \right)}}{{\cos \left( {{\theta _i}} \right)}}\quad \quad \quad & {\rm{if}} \quad R\left( x \right) = {R_{{\rm{EM}}}}\left( x \right)\\
1\quad \quad \quad & {\rm{if}} \quad R\left( x \right) = {R_{\rm{Z}}}\left( x \right)
\end{cases}
\end{equation}

In the sequel, thanks to the unified definition in \eqref{Eq_FullReflection_LocalReflectionCoefficient}, the surface-averaged reflection coefficient is denoted by $R\left( x \right)$.

\textbf{Reformulation of the EM fields}. As a function of the reflection coefficient in \eqref{Eq_FullReflection_LocalReflectionCoefficient}, the non-zero components of the electric and magnetic fields in \eqref{Eq_FullReflection_PlaneWave} can be written as follows:
\begin{equation} \label{Eq_FullReflection_R}
\begin{split}
& E_y^i\left( {x} \right) = \sqrt {{2 \eta P_0}} \exp \left( { - jk\sin \left( {{\theta _i}} \right)x} \right)\\
& E_y^r\left( {x} \right) = \sqrt {{2 \eta P_0}} R\left( x \right)\exp \left( { - jk\sin \left( {{\theta _i}} \right)x} \right)\\
& H_x^i\left( {x} \right) =  \sqrt {\frac{{2{P_0}}}{\eta }} \cos \left( {{\theta _i}} \right)\exp \left( { - jk\sin \left( {{\theta _i}} \right)x} \right)\\
& H_x^r\left( {x} \right) = - \sqrt {\frac{{2{P_0}}}{\eta }} \cos \left( {{\theta _i}}  \right)r_{i,r}R\left( x \right)\exp \left( { - jk\sin \left( {{\theta _i}} \right)x} \right)
\end{split}
\end{equation}

It is worth emphasizing that, in contrast with $A_r$ in \eqref{Eq_FullReflection_PlaneWave}, the surface-averaged reflection coefficient $R(x)$ depends on $x$. This is necessary because the reflected EM fields are expressed in terms of incident EM fields. This is consistent with the analytical formulation given in, e.g., \cite{Holloway_Nov2005}.

\textbf{$\bf{R_{{\rm{\bf EM}}}}\left( x \right)$ and $\bf{R_{{\rm{\bf Z}}}}\left( x \right)$: Some initial remarks}. By direct inspection of \eqref{Eq_FullReflection_R}, some preliminary remarks can be made.
\begin{itemize}
\item \textbf{\textit{Reflected electric and magnetic fields}}. From \eqref{Eq_FullReflection_R}, we obtain the following identities:
\begin{equation} \label{Comparing_ReflectionCoefficients__1}
{R_{{\rm{EM}}}}\left( x \right) = \frac{{E_y^r\left( x \right)}}{{E_y^i\left( x \right)}} \ne  - \frac{{H_y^r\left( x \right)}}{{H_y^i\left( x \right)}}
\end{equation}
\begin{equation} \label{Comparing_ReflectionCoefficients__2}
{R_{\rm{Z}}}\left( x \right) = \frac{{E_y^r\left( x \right)}}{{E_y^i\left( x \right)}} =  - \frac{{H_y^r\left( x \right)}}{{H_y^i\left( x \right)}}
\end{equation}

The identity in \eqref{Comparing_ReflectionCoefficients__1} is in agreement with the widely used definition of reflection coefficient as the ratio between the reflected electric field and the incident electric field. With the exception of the special case ${\cos \left( {{\theta _r}} \right) = \cos \left( {{\theta _i}} \right)}$, i.e., specular reflection, the same does not hold true for the corresponding magnetic fields. The identities in \eqref{Comparing_ReflectionCoefficients__2} are, on the other hand, in agreement with the typical definition of reflection coefficient that is employed in electric circuits, according to which the ratio of the reflected and incident voltages (i.e., the electric fields) is, except for a negative sign, equal to the ratio of the reflected and incident currents (i.e., the magnetic fields).
\item \textbf{\textit{Impedances of incident and reflected EM fields}}. From \eqref{Eq_FullReflection_R}, we obtain the following identities:
\begin{equation} \label{Comparing_ReflectionCoefficients__3}
{Z_i} = \frac{{E_y^i\left( x \right)}}{{H_x^i\left( x \right)}} = \frac{\eta }{{\cos \left( {{\theta _i}} \right)}}
\end{equation}
\begin{equation} \label{Comparing_ReflectionCoefficients__4}
{Z_r} = \frac{{E_y^r\left( x \right)}}{{H_x^r\left( x \right)}} = \begin{cases}
 - \frac{\eta }{{\cos \left( {{\theta _r}} \right)}}\quad & {\rm{if}}\; R\left( x \right) = {R_{{\rm{EM}}}}\left( x \right)\\
 - \frac{\eta }{{\cos \left( {{\theta _i}} \right)}}\quad & {\rm{if}}\; R\left( x \right) = {R_{\rm{Z}}}\left( x \right)
\end{cases}
\end{equation}

By leveraging the analogy of the electric and magnetic fields in vacuum with the voltages and currents in transmission lines, respectively, \eqref{Comparing_ReflectionCoefficients__3} and \eqref{Comparing_ReflectionCoefficients__4} give the impedances of the incident radio wave and the reflected radio wave, respectively. By viewing a metasurface as an equivalent electric circuit, \eqref{Comparing_ReflectionCoefficients__3} and \eqref{Comparing_ReflectionCoefficients__4} can be interpreted as the input and output impedance that is viewed by the radio wave entering and leaving the equivalent electric circuit, respectively. Based on the definition of  ${{R_{{\rm{EM}}}}\left( x \right)}$, the input and output radio waves view two different impedances. Based on the definition of  ${{R_{{\rm{Z}}}}\left( x \right)}$, the input and output radio waves view the same impedance. This is the main difference between the definitions of ${{R_{{\rm{EM}}}}\left( x \right)}$ and ${{R_{{\rm{Z}}}}\left( x \right)}$, which is elaborated in further text.
\item \textbf{\textit{Explicit expression for $\bf{R_{{\rm{\bf EM}}}}\left( x \right)$}}. By direct inspection and comparison between \eqref{Eq_FullReflection_PlaneWave} and \eqref{Eq_FullReflection_R}, the following simple expression for ${{R_{{\rm{EM}}}}\left( x \right)}$ can be obtained:
\begin{equation} \label{EM_ReflectionCoefficient}
{R_{{\rm{EM}}}}\left( x \right) = {A_r}\exp \left( { - jk\left( {\sin \left( {{\theta _r}} \right) - \sin \left( {{\theta _i}} \right)} \right)x} \right)
\end{equation}

In the considered example of a perfect anomalous reflector, it is, therefore, not difficult to identify an explicit analytical expression for ${{R_{{\rm{EM}}}}\left( x \right)}$. This is, in general, not always the case. By analyzing ${{R_{{\rm{EM}}}}\left( x \right)}$ carefully, we observe that the metasurface structure introduces a phase shift that compensates the difference between the angles of incidence and reflection, as well as it scales the amplitude of the incident and reflected fields by a factor $A_r$. In the literature, it is usually assumed that $A_r = 1$ in order to ensure that the metasurfaces are passive \cite{MDR_OverheadAware}. This is necessary, however, only if one wishes that the metasurface structure is locally passive. In general, $A_r$ can be chosen according to different criteria, and it is, in general, dependent on the angles of incidence and reflection. Based on \eqref{Eq_Ar_1}, for example, this is obtained for a globally passive perfect anomalous reflector with a unitary power efficiency. If one is interested in smart surfaces that are realized through tiny discrete antenna elements, e.g., the RFocus prototype in Fig. \ref{Fig_2}, on the other hand, the concepts of locally and globally passive are (almost) the same, since advanced methods to realize non-locally passive but globally passive metasurfaces without using active elements may not be employed \cite{Asadchy_2017}. This concept is better elaborated in further text. At this point, it is sufficient to keep in mind that globally passive metasurfaces do not necessarily imply that the amplitude of the reflection coefficient ${{R_{{\rm{EM}}}}\left( x \right)}$, i.e., $A_r$ in \eqref{EM_ReflectionCoefficient}, is equal to or less than one.
\end{itemize}

\textbf{Relating the surface-averaged susceptibility function to the surface-averaged reflection coefficient}. In order to gain deeper understanding of and insight on the specific features of the perfect anomalous reflector under design, the first step consists of finding an explicit relation between the surface susceptibility function and the surface reflection coefficient $R(x)$. The analytical derivation, in particular, applies to ${R_{{\rm{EM}}}}\left( x \right)$ and ${R_{{\rm{Z}}}}\left( x \right)$. The analytical relation is obtained by inserting \eqref{Eq_FullReflection_R} in \eqref{Eq_FullReflection_Susceptibility}, which yields the following:
\begin{equation} \label{Eq_FullReflection_Relation1}
\chi _{{\rm{ee}}}^{yy}\left( {x,y} \right) = \chi _{{\rm{ee}}}^{yy}\left( x \right) = \frac{{2j}}{k}\cos \left( {{\theta _i}} \right)\frac{{1 - r_{i,j} R\left( x \right)}}{{1 + R\left( x \right)}}
\end{equation}
\noindent or, equivalently, the following:
\begin{equation} \label{Eq_FullReflection_Relation2}
R\left( x \right) = \frac{{1 + j\frac{k}{{2\cos \left( {{\theta _i}} \right)}}\chi _{{\rm{ee}}}^{yy}\left( x \right)}}{{r_{i,j} - j\frac{k}{{2\cos \left( {{\theta _i}} \right)}}\chi _{{\rm{ee}}}^{yy}\left( x \right)}}
\end{equation}

\textbf{Explicit formulation of the surface-averaged reflection coefficient}. An explicit, as a function of the angles of incidence and reflection, relation of the surface-averaged reflection coefficient $R(x)$ can be obtained by inserting \eqref{Eq_FullReflection_Susceptibility__ClosedFormReflection} in \eqref{Eq_FullReflection_Relation2}. The resulting analytical expression is the following:
\begin{equation} \label{Eq_FullReflection_ExplicitFormulaReflection}
\begin{split}
& R\left( x \right) = \frac{{1 - \Gamma \left( x \right)}}{{r_{i,j} + \Gamma \left( x \right)}}\\
& \Gamma \left( x \right) = \frac{{\cos \left( {{\theta _i}} \right) - {A_r}\cos \left( {{\theta _r}} \right)\Phi \left( x \right)}}{{\cos \left( {{\theta _i}} \right)\left( {1 + {A_r}\Phi \left( x \right)} \right)}}
\end{split}
\end{equation}

It is worth mentioning that $A_r$ in \eqref{Eq_FullReflection_ExplicitFormulaReflection} is an arbitrary coefficient that is not necessarily equal to \eqref{Eq_Ar_1}. In the next paragraphs, in particular, we discuss the impact of $A_r$ on the power efficiency and associated implementation complexity for realizing the resulting metasurface structure, e.g., if the structure is locally passive, globally passive, or if it requires specific (advanced) implementation designs.

\textbf{Surface-averaged reflection coefficient: Net power flow and conditions for globally passive metasurfaces}. Similar to the analytical formulation for the transformations of the EM fields in \eqref{Eq_FullReflection_PlaneWave}, we are interested in computing the net power flow of a metasurface structure that is given by the analytical reformulation in \eqref{Eq_FullReflection_R}. Also, we are interested in formulating the conditions for ensuring that a metasurface structure is globally passive as a function of the surface-averaged reflection coefficient $R(x)$, and in comparing the two definitions of reflection coefficient ${R_{{\rm{EM}}}}\left( x \right)$ and ${R_{{\rm{Z}}}}\left( x \right)$. In detail, the net power flow can be computed by inserting the EM fields in \eqref{Eq_FullReflection_R} into \eqref{Eq_PowerBudget__1}, which yield the following analytical expressions:
\begin{equation} \label{Eq_PowerBudget__3}
\begin{split}
& {P_{{\rm{net}}}}\left( x \right) = {P_0}\cos \left( {{\theta _i}} \right) \\ & \hspace{1.20cm} \times \left( {{{r_{i,j}\left| {R\left( x \right)} \right|}^2} - 1 - (1-r_{i,j}){\mathop{\rm Re}\nolimits} \left\{ {R\left( x \right)} \right\}} \right)\\
& {P_{{\rm{net}}}} = \frac{{{P_0}r_{i,j}\cos \left( {{\theta _i}} \right)}}{\Pi }\int\nolimits_0^\Pi  {{{\left| {R\left( x \right)} \right|}^2}dx}  - {P_0}\cos \left( {{\theta _i}} \right)
\end{split}
\end{equation}

Once again, we observe that ${P_{{\rm{net}}}}\left( x \right)$ is, in general, an oscillating function in $x$, which, depending on the parameters, can be positive or negative, and it is not identically equal to zero. It is apparent from \eqref{Eq_PowerBudget__3} that the metasurface structure is locally passive if ${P_{{\rm{net}}}}\left( x \right) \le 0$ for every $x$. Likewise, the metasurface structure can be made globally passive with unitary power efficiency by imposing the design constraint ${P_{{\rm{net}}}} = 0$, which yields:
\begin{equation} \label{Eq_PowerBudget__4}
\frac{1}{\Pi }\int\nolimits_0^\Pi  {{{\left| {R\left( x \right)} \right|}^2}dx}  = \frac{1}{r_{i,j}}
\end{equation}

The difference between ${R_{{\rm{EM}}}}\left( x \right)$ and ${R_{{\rm{Z}}}}\left( x \right)$ can be understood by comparing them in terms of the net power flow given in \eqref{Eq_PowerBudget__3} and in terms of the condition in \eqref{Eq_PowerBudget__4} for the design of globally passive metasurfaces with unitary power efficiency.

\textbf{Reflected EM fields and net power flow from $\bf{R_{{\rm{\bf EM}}}}\left( x \right)$}. In this case, $r_{i,j} = {{{\cos \left( {{\theta _r}} \right)} \mathord{\left/ {\vphantom {{\cos \left( {{\theta _r}} \right)} {\cos \left( {{\theta _i}} \right)}}} \right. \kern-\nulldelimiterspace} {\cos \left( {{\theta _i}} \right)}}}$. By inserting $r_{i,r}$ in \eqref{Eq_FullReflection_ExplicitFormulaReflection}, we obtain, as expected, \eqref{EM_ReflectionCoefficient}, since ${1 \mathord{\left/ {\vphantom {1 {{r_{i,j}} = A_r^2}}} \right.
\kern-\nulldelimiterspace} {{r_{i,j}} = A_r^2}}$. Therefore, \eqref{Eq_PowerBudget__3} reduces to \eqref{Eq_PowerBudget__2} and \eqref{Eq_PowerBudget__4} is in agreement with \eqref{Eq_Ar_1}. This confirms that ${R_{{\rm{EM}}}}\left( x \right)$ is inherently suitable for applications to wireless communications, since its definition ensures that the reflected EM fields corresponds to physical fields that yield a power budget in agreement with the desired field transformations.

\textbf{Reflected EM fields from $\bf{R_{{\rm{\bf Z}}}}\left( x \right)$}. In this case, $r_{i,j} = 1$. Under this setup, we obtain results that are not consistent with the analytical expression of the desired EM fields in \eqref{Eq_FullReflection_PlaneWave}. By inserting $r_{i,j} = 1$ in \eqref{Eq_FullReflection_ExplicitFormulaReflection}, in particular, the resulting reflection coefficient $R(x)={R_{{\rm{Z}}}}\left( x \right)$ would be the following:
\begin{equation} \label{Eq_FieldsBack__R}
{R_{\rm{Z}}}\left( x \right) = \frac{{{A_r}\left( {\cos \left( {{\theta _r}} \right) + \cos \left( {{\theta _i}} \right)} \right)\Phi \left( x \right)}}{{2\cos \left( {{\theta _i}} \right) - {A_r}\left( {\cos \left( {{\theta _r}} \right) - \cos \left( {{\theta _i}} \right)} \right)\Phi \left( x \right)}}
\end{equation}

By inserting \eqref{Eq_FieldsBack__R} in \eqref{Eq_FullReflection_R}, we evince that the resulting electric and magnetic reflected fields are different from the desired ones that are given in \eqref{Eq_FullReflection_PlaneWave}. In particular, the EM fields obtained from \eqref{Eq_FieldsBack__R} and \eqref{Eq_FullReflection_R} may not be a solution of Maxwell's equations either. We observe, notably, that ${R_{\rm{Z}}}\left( x \right) = {R_{\rm{EM}}}\left( x \right)$ only if $\theta _r=\theta _i$, i.e., if specular reflection is considered, as remarked in previous text already.

\textbf{Net power flow from $\bf{R_{{\rm{\bf Z}}}}\left( x \right)$}. By inserting $r_{i,j} = 1$ in \eqref{Eq_FullReflection_ExplicitFormulaReflection} and by then plugging the resulting reflection coefficient in \eqref{Eq_PowerBudget__4}, in addition, we would obtain the following analytical condition for the design of a globally passive perfect anomalous reflector with unitary power efficiency:
\begin{equation} \label{Eq_PowerBudget__5}
{\left| {\frac{{{{\left( {1 - \frac{{\cos \left( {{\theta _r}} \right)}}{{\cos \left( {{\theta _i}} \right)}}} \right)}^2}}}{{{{\left( {1 + \frac{{\cos \left( {{\theta _r}} \right)}}{{\cos \left( {{\theta _i}} \right)}}} \right)}^2}}} - \frac{4}{{A_r^2{{\left( {1 + \frac{{\cos \left( {{\theta _r}} \right)}}{{\cos \left( {{\theta _i}} \right)}}} \right)}^2}}}} \right|^{ - 1}} = 1
\end{equation}

The design constraint in \eqref{Eq_PowerBudget__5} could be guaranteed by appropriately optimizing the parameter $A_r$ in \eqref{Eq_FullReflection_ExplicitFormulaReflection}. In particular, by solving the equation in \eqref{Eq_PowerBudget__5} as a function of $A_r$, we would obtain the following:
\begin{equation} \label{Eq_Ar_2}
{A_r} = \sqrt {\frac{2}{{1 + {{\left( {{{\cos \left( {{\theta _i}} \right)} \mathord{\left/
 {\vphantom {{\cos \left( {{\theta _i}} \right)} {\cos \left( {{\theta _r}} \right)}}} \right.
 \kern-\nulldelimiterspace} {\cos \left( {{\theta _r}} \right)}}} \right)}^2}}}}
\end{equation}

\textbf{$\bf{R_{{\rm{\bf EM}}}}\left( x \right)$ vs. $\bf{R_{{\rm{\bf Z}}}}\left( x \right)$: Comparison in terms of net power flow}. By comparing \eqref{Eq_Ar_2} with \eqref{Eq_Ar_1}, we evince that, in general, two different values for $A_r$ are obtained. The two values of $A_r$ coincide under the assumption of perfect specular reflection, i.e., $\theta _r = \theta _i$. In this case, in fact, both formulas simplify to $A_r=1$. If $r_{i,r}=1$, more in general, the net power flow in \eqref{Eq_PowerBudget__3} is different from the (benchmark) net power flow in \eqref{Eq_PowerBudget__2}. As an example, we can insert \eqref{Eq_Ar_1} in \eqref{Eq_PowerBudget__3} and \eqref{Eq_Ar_2} in \eqref{Eq_PowerBudget__2}, respectively, and can compute the two expressions of the corresponding net power flows. We denote the two resulting formulas by $P_{{\rm{net}}}^{\left( {{\rm{Z}}} \right)}$ and $P_{{\rm{net}}}^{\left( {{\rm{benchmark}}} \right)}$, respectively. It can be readily proved that $P_{{\rm{net}}}^{\left( {{\rm{Z}}} \right)} \ge 0$ and $P_{{\rm{net}}}^{\left( {{\rm{benchmark}}} \right)} \le 0$ for any angle of incidence and for any angle of reflection. In different terms, a metasurface structure that is expected to be globally passive and with unitary power efficiency under the desired field transformations in \eqref{Eq_FullReflection_PlaneWave}, i.e., by setting $A_r$ equal to \eqref{Eq_Ar_1}, turns out to be globally active if ${R_{{\rm{Z}}}}\left( x \right)$ is interpreted as a field reflection coefficient.

\textbf{What is the role and meaning of $\bf{R_{{\rm{\bf Z}}}}\left( x \right)$}? The comparison between ${R_{{\rm{EM}}}}\left( x \right)$ and ${R_{{\rm{Z}}}}\left( x \right)$ allows us to conclude that ${R_{{\rm{Z}}}}\left( x \right)$ cannot be employed as a field reflection coefficient and that ${R_{{\rm{EM}}}}\left( x \right)$ is the correct reflection coefficient to use if one is interested in the response of a metasurface structure in terms of input-output EM fields. Therefore, a fundamental point to clarify is what the meaning of ${R_{{\rm{Z}}}}\left( x \right)$ is why it is routinely employed for the design and optimization of metasurfaces (see, e.g., \cite{Alu_Reflection2016}, \cite{Asadchy_2017}). To this end, it is instructive to formulate a perfect anomalous reflector by using the homogenized model based on surface-averaged impedances.

\textbf{Anomalous reflection modeled via surface-averaged impedances}. A perfect anomalous reflector can be represented in terms of surface-averaged impedances by applying the equivalent transmission line representation for a metasurface structure that is depicted in Fig. \ref{Fig_SurfaceImpedances}. Based on the desired EM field transformations in \eqref{Eq_FullReflection_PlaneWave}, the following voltages and currents, at the two sides of the metasurface structure, can be obtained:
\begin{equation} \label{Eq_Impedances__1}
\begin{split}
& {v^ + }\left( x \right) = E_y^i\left( x \right) + E_y^r\left( x \right)\\
& {v^\_}\left( x \right) = 0\\
& {i^ + }\left( x \right) = H_x^i\left( x \right) + H_x^r\left( x \right)\\
& {i^\_}\left( x \right) = 0
\end{split}
\end{equation}

By applying Kirchhoff's circuit laws to the equivalent transmission line in Fig. \ref{Fig_SurfaceImpedances}, the following system of equations can be obtained:
\begin{equation} \label{Eq_Impedances__2}
\left\{\begin{split}
& {v^ + }\left( x \right) = \left( {{Z_1}\left( x \right) + {Z_3}\left( x \right)} \right){i^ + }\left( x \right) + {Z_3}\left( x \right){i^ - }\left( x \right)\\
& {v^ - }\left( x \right) = {Z_3}\left( x \right){i^ + }\left( x \right) + \left( {{Z_2}\left( x \right) + {Z_3}\left( x \right)} \right){i^ - }\left( x \right)
\end{split} \right.
\end{equation}

By inserting \eqref{Eq_Impedances__1} in \eqref{Eq_Impedances__2}, we evince that a solution of the system of equations, i.e., an explicit expression of the surface-averaged impedances as a function of the transformations of the EM fields to be realized, is the following:
\begin{equation} \label{Eq_Impedances__3}
\begin{split}
& {Z_1}\left( x \right) = {{{v^ + }\left( x \right)} \mathord{\left/
 {\vphantom {{{v^ + }\left( x \right)} {{i^ + }\left( x \right)}}} \right.
 \kern-\nulldelimiterspace} {{i^ + }\left( x \right)}}\\
& {Z_2}\left( x \right) = {\rm{any \; function }}\\
& {Z_3}\left( x \right) = 0
\end{split}
\end{equation}

Therefore, ${Z_1}\left( x \right)$ is the surface-averaged impedance that determines the properties and EM response of the metasurface structure under design. By inserting \eqref{Eq_Impedances__1} in \eqref{Eq_Impedances__3} and using  \eqref{Eq_FullReflection_PlaneWave}, in particular, the following benchmark analytical formulation for ${Z_1}\left( x \right)$ can be obtained (for any value of $A_r$):
\begin{equation} \label{Eq_Impedances__4}
\begin{split}
{Z_1}\left( x \right) &=  \frac{{E_y^i\left( x \right) + E_y^r\left( x \right)}}{{H_x^i\left( x \right) + H_x^r\left( x \right)}}\\
& \hspace{-0.5cm} = \eta \frac{{1 + {A_r}\Phi \left( x \right)}}{{\cos \left( {{\theta _i}} \right) - {A_r}\cos \left( {{\theta _i}} \right)\Phi \left( x \right)}}\\
& \hspace{-0.5cm} \mathop  = \limits^{\left( a \right)} \frac{\eta }{{\sqrt {\cos \left( {{\theta _i}} \right)\cos \left( {{\theta _r}} \right)} }}\frac{{\sqrt {\cos \left( {{\theta _r}} \right)}  + \sqrt {\cos \left( {{\theta _i}} \right)} {\Phi}\left( x \right)}}{{\sqrt {\cos \left( {{\theta _i}} \right)}  - \sqrt {\cos \left( {{\theta _r}} \right)} {\Phi}\left( x \right)}}
\end{split}
\end{equation}
\noindent where (a) holds true if $A_r$ is set equal to \eqref{Eq_Ar_1}.

\textbf{Comparison of homogenized models}. By direct inspection of \eqref{Eq_Impedances__4}, we observe that: (i) ${Z_1}\left( x \right)$ is, except for a scaling factor, the same as the surface-averaged susceptibility function in \eqref{Eq_FullReflection_Susceptibility}. This proves the equivalence of the homogenized models based on susceptibilities and impedances; (ii) ${Z_1}\left( x \right)$ coincides with the analytical formulations reported in \cite{Alu_Reflection2016}, \cite{Asadchy_2017}, \cite{Asadchy_2016}, \cite{Diaz-Rubio_2017}. We note, in particular, that ${Z_1}\left( x \right)$ is consistent with the advanced design in \cite[Eq. (3)]{Diaz-Rubio_2017}, for which a practical implementation of metasurface based on the physical principle of leaky-wave antennas exists; and (iii) ${Z_1}\left( x \right)$ is a complex number whose real part can be positive and negative. This brings to our attention that the resulting metasurface structure is non-locally passive but it can be made globally passive if $A_r$ is set equal to \eqref{Eq_Ar_1}. An example of realization of a non-locally passive but globally passive metasurface that behaves as a perfect anomalous reflector is elaborated in further text based on the recent results reported in \cite{Diaz-Rubio_2017}.

\textbf{Definition of $\bf{R_{{\rm{\bf Z}}}}\left( x \right)$ based on surface-averaged impedances}. In the context of designing and engineering metasurfaces, the usual \text{definition} of surface-averaged reflection coefficient is the following \cite{Alu_Reflection2016}, \cite{Asadchy_2017}, \cite{Asadchy_2016}, \cite{Diaz-Rubio_2017}:
\begin{equation} \label{Eq_Impedances__R}
r\left( x \right) = \frac{{{Z_1}\left( x \right) - \frac{\eta }{{\cos \left( {{\theta _i}} \right)}}}}{{{Z_1}\left( x \right) + \frac{\eta }{{\cos \left( {{\theta _i}} \right)}}}} \mathop  = \limits^{\left( a \right)} \frac{{{Z_1}\left( x \right) - {Z_i}\left( x \right)}}{{{Z_1}\left( x \right) + {Z_i}\left( x \right)}}
\end{equation}
\noindent where (a) is obtained by using \eqref{Comparing_ReflectionCoefficients__3}.

By inserting \eqref{Eq_Impedances__4} in \eqref{Eq_Impedances__R}, we obtain $r\left( x \right) = {R_{\rm{Z}}}\left( x \right)$, where ${R_{\rm{Z}}}\left( x \right)$ is the surface-averaged reflection coefficient in \eqref{Eq_FieldsBack__R}. We conclude that ${R_{\rm{Z}}}\left( x \right)$ is the surface-averaged reflection coefficient that is usually found in physics-related papers, and that is typically employed for getting insights for designing metasurface structures with the desired reflection capabilities. Notably, ${R_{\rm{Z}}}\left( x \right)$ fulfills the identities in \eqref{Comparing_ReflectionCoefficients__2}, which are typically employed as a definition of reflection coefficient in the context of circuits design and analysis.

\textbf{Physical interpretation and usefulness of $\bf{R_{{\rm{\bf Z}}}}\left( x \right)$}. Even though, as mentioned, ${R_{\rm{Z}}}\left( x \right)$ is not appropriate for relating the reflected EM fields to the incident EM fields, it provides one with important information for designing perfect anomalous reflectors. This can be inferred by a careful analysis of the equality (a) in \eqref{Eq_Impedances__R}. In a nutshell, $r\left( x \right) = {R_{\rm{Z}}}\left( x \right)$ provides one with information on how different the local (or point-wise) responses of a perfect anomalous reflector and a perfect specular reflector are. In general terms, according to \eqref{Comparing_ReflectionCoefficients__3} and \eqref{Comparing_ReflectionCoefficients__4}, the impedances of the incident and reflected EM waves are equal to ${Z_i} = {\eta  \mathord{\left/ {\vphantom {\eta  {\cos \left( {{\theta _i}} \right)}}} \right. \kern-\nulldelimiterspace} {\cos \left( {{\theta _i}} \right)}}$ and ${Z_r} = {\eta  \mathord{\left/ {\vphantom {\eta  {\cos \left( {{\theta _r}} \right)}}} \right. \kern-\nulldelimiterspace} {\cos \left( {{\theta _r}} \right)}}$, respectively. The larger the difference between the angle of reflection and the angle of reflection is, therefore, the larger the difference between ${Z_i}$ and ${Z_r}$ is. This implies that the transformation of the EM fields applied by the metasurface structure needs to be more distinct and marked as the difference between the angles of reflection and incidence increases. The reflection coefficient $r\left( x \right) = {R_{\rm{Z}}}\left( x \right)$ is a parameter that quantifies the local distinctive response of the metasurface structure. For a deeper understanding, let us assume that $A_r$ is equal to \eqref{Eq_Ar_1}. The reflection coefficient $r\left( x \right) = {R_{\rm{Z}}}\left( x \right)$ of a specular reflector can be obtained from \eqref{Eq_Impedances__R} by setting ${{\theta _r}}= {{\theta _i}}$, which yields $r\left( x \right) = {R_{\rm{Z}}}\left( x \right) = 1$ for every $x$. For a specular reflector, in other words, $r\left( x \right) = {R_{\rm{Z}}}\left( x \right)$ is constant in $x$. If ${{\theta _r}}\ne {{\theta _i}}$, on the other hand, the amplitude  of $r\left( x \right) = {R_{\rm{Z}}}\left( x \right)$, i.e., $\left| {{R_{\rm{Z}}}\left( x \right)} \right|$, exhibits spatial variations in $x$, which becomes more pronounced as the difference between the angle of reflection and the angle of incidence increases. By inspecting $r\left( x \right) = {R_{\rm{Z}}}\left( x \right)$, we obtain information on the mismatch of the local spatial responses of the metasurface structures that correspond to anomalous and specular reflection. The specular reflector is a meaningful reference because it corresponds to the usual response of conventional surfaces and because it is characterized by an $r\left( x \right) = {R_{\rm{Z}}}\left( x \right)$ that is a uniform function in $x$. It is worth noting that the reflection coefficient ${R_{\rm{EM}}}\left( x \right)$ is not capable of unveiling these properties. From \eqref{EM_ReflectionCoefficient}, in particular, we evince that the amplitude of ${R_{\rm{EM}}}\left( x \right)$ exhibits no spatial variations along the metasurface structure, i.e., $\left| {{R_{\rm{EM}}}\left( x \right)} \right| = A_r$ is independent of $x$ for any angles of incidence and reflection. This confirms, once more, that ${R_{\rm{EM}}}\left( x \right)$ and ${R_{\rm{Z}}}\left( x \right)$ have different physical meanings and serve for different purposes. In the next paragraphs, numerical illustrations are used in order to further elucidate the properties, differences, and similarities of ${R_{\rm{EM}}}\left( x \right)$ and ${R_{\rm{Z}}}\left( x \right)$.

\textbf{Canonical case: Normal incidence}. In the rest of this sub-section, for simplicity but without loss of generality, we focus our attention on the canonical case of EM fields with normal incidence, i.e., $\theta_i = 0$. In this case, ${R_{\rm{EM}}}\left( x \right)$ in \eqref{EM_ReflectionCoefficient} and ${R_{\rm{Z}}}\left( x \right)$ in \eqref{Eq_FieldsBack__R} reduce, respectively, to the following:
\begin{equation} \label{Eq_FullReflection_NormalIncidence__EM}
{R_{{\rm{EM}}}}\left( x \right) = {A_r}\exp \left( { - jk\sin \left( {{\theta _r}} \right)x} \right)
\end{equation}
\begin{equation} \label{Eq_FullReflection_NormalIncidence__Z}
{R_{\rm{Z}}}\left( x \right) = \frac{{{A_r}\left( {1 + \cos \left( {{\theta _r}} \right)} \right)\exp \left( { - jk\sin \left( {{\theta _r}} \right)x} \right)}}{{2 + {A_r}\left( {1 - \cos \left( {{\theta _r}} \right)} \right)\exp \left( { - jk\sin \left( {{\theta _r}} \right)x} \right)}}
\end{equation}

In the following paragraph, as a case study, we describe some numerical illustrations in order to gain important engineering insights on the main features and properties of the reflection coefficients in \eqref{Eq_FullReflection_NormalIncidence__EM} and \eqref{Eq_FullReflection_NormalIncidence__Z} for various system setups and optimization criteria. It is worth emphasizing that \eqref{Eq_FullReflection_NormalIncidence__EM} and \eqref{Eq_FullReflection_NormalIncidence__Z} hold true for general values of $A_r$. The specific choice of $A_r$ determines, however, the ultimate performance, the power efficiency, and the implementation constraints for realizing, in practice, the resulting metasurface structure.

\begin{figure}[!t]
\begin{centering}
\includegraphics[width=\columnwidth]{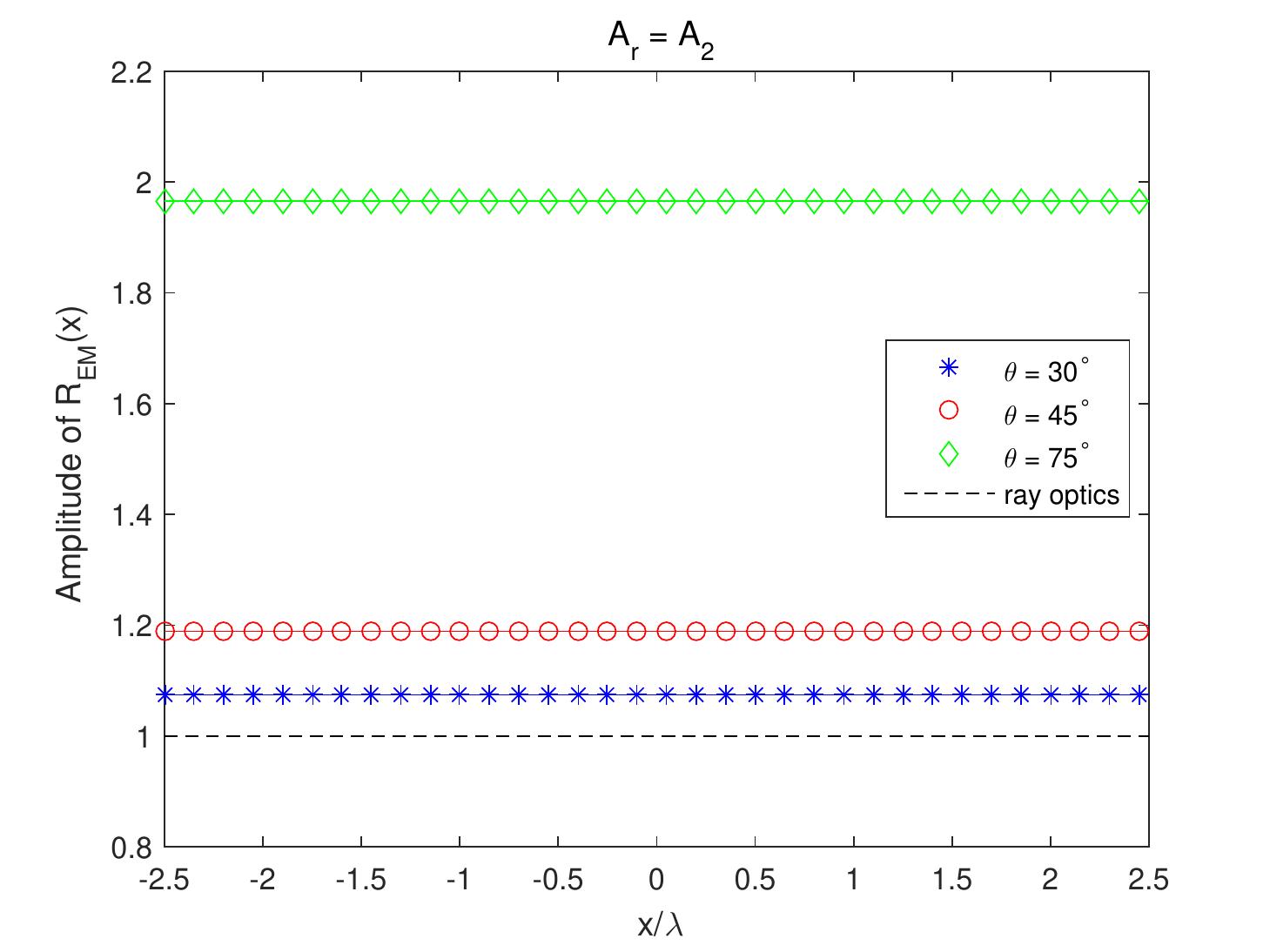}
\caption{Amplitude of the reflection coefficient $R_{\rm{EM}}(x)$ in \eqref{Eq_FullReflection_NormalIncidence__EM} for $\theta_r = \theta$. Dashed lines: Ray optics approximation (i.e., $A_r=1$).}
\label{Fig_15a}
\end{centering} 
\end{figure}
\begin{figure}[!t]
\begin{centering}
\includegraphics[width=\columnwidth]{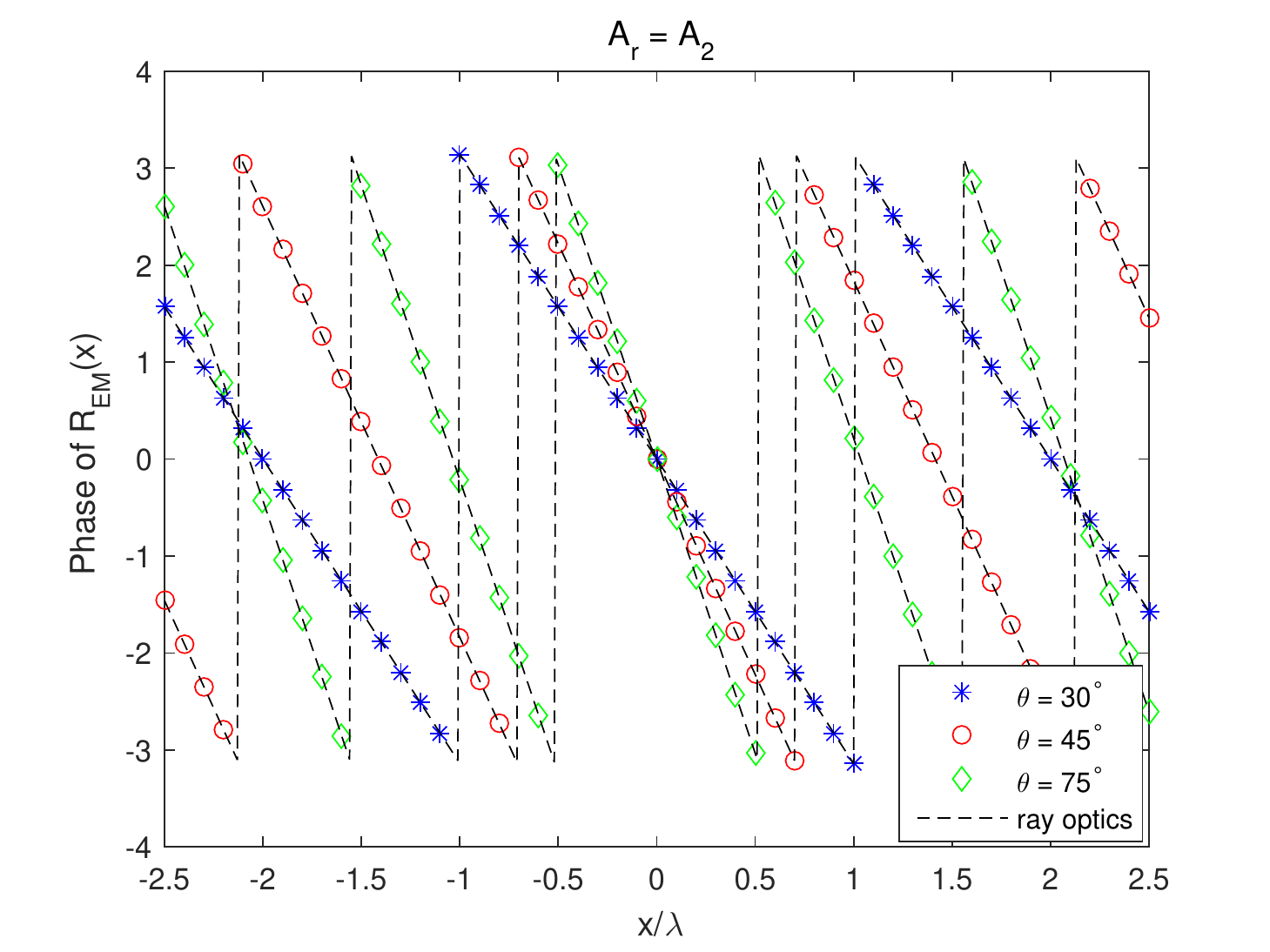}
\caption{Phase of the reflection coefficient $R_{\rm{EM}}(x)$ in \eqref{Eq_FullReflection_NormalIncidence__EM} for $\theta_r = \theta$. Dashed lines: Ray optics approximation (i.e., $A_r=1$).}
\label{Fig_15b}
\end{centering} 
\end{figure}
\textbf{Numerical illustrations: Understanding anomalous reflection}. Based on the closed-form expressions in \eqref{Eq_FullReflection_NormalIncidence__EM} and \eqref{Eq_FullReflection_NormalIncidence__Z}, the amplitude and phase of the surface-averaged reflection coefficient $R_{\rm{EM}}(x)$ are depicted in Figs. \ref{Fig_15a} and \ref{Fig_15b}, and the amplitude and phase of the surface-averaged reflection coefficient $R_{\rm{Z}}(x)$ are depicted in Figs. \ref{Fig_15a_1}-\ref{Fig_15b_2}, respectively, for different values of the angle of reflection $\theta_r$ and for two different values of $A_r$: (i) $A_r=A_1=1$ and (ii) $A_r=A_2$ equal to \eqref{Eq_Ar_1} by setting $\theta_i = 0$ in both cases. The considered case studies are relevant in order to understand the impact of $A_r$ on the design of metasurfaces, and to clarify the differences and similarities between $R_{\rm{EM}}(x)$ and $R_{\rm{Z}}(x)$. By direct inspection of \eqref{Eq_FullReflection_NormalIncidence__EM}, \eqref{Eq_FullReflection_NormalIncidence__Z}, and Figs. \ref{Fig_15a}-\ref{Fig_15b_2}, the following important observations can be made.
\begin{itemize}
\item \textit{\textbf{Designs based on ray optics are approximated}}. The surface-averaged reflection coefficients $R_{\rm{EM}}(x)$ and $R_{\rm{Z}}(x)$ of a perfect anomalous reflector are complex numbers whose amplitude and phase are, in general, different from those obtained by using ray optics approximation, i.e., $R_{\rm{ray \, optics}}\left( x \right) = \exp \left( { - jk\sin \left( {{\theta _r}} \right)x} \right)$ \cite{Capasso_Science2011}. In general, however, ray optics approximation is sufficiently accurate for low-medium angles of reflection, $\theta_r$, that need to be realized. It is worth nothing, in particular, that $R_{\rm{EM}}(x)$ and $R_{\rm{Z}}(x)$ tends, for every $x$, to one if  $\theta _r=0$, i.e., if specular reflection is considered. We observe, however, a distinct difference between $R_{\rm{EM}}(x)$ and $R_{\rm{Z}}(x)$. As far as $R_{\rm{EM}}(x)$ is concerned, Figs. \ref{Fig_15a} and \ref{Fig_15b} show that the phase response of $R_{\rm{EM}}(x)$ coincide with ray optics approximation but the amplitude of $R_{\rm{EM}}(x)$ is different under the assumption that $A_r$ is optimized for achieving unitary power efficiency (in the global sense). We observe, however, that the amplitude of $R_{\rm{EM}}(x)$ exhibits no spatial variations and it is constant in $x$. As far as $R_{\rm{Z}}(x)$ is concerned, Figs. \ref{Fig_15a_1}-\ref{Fig_15b_2} unveil a different behavior as a function of $x$. The main difference stems from the amplitude of $R_{\rm{Z}}(x)$, which exhibits a highly varying spatial response whose fluctuations are more pronounced as the angle of reflection increases (with respect to the angle of incidence). The highly non-uniform local behavior of $R_{\rm{Z}}(x)$ unveils the inherent difficulty of realizing perfect anomalous reflectors with large angles of deflection and high power efficiency.
\item \textit{\textbf{Amplitude and phase responses are correlated}}. For a given angle of reflection, $\theta_r$, the amplitude and phase of $R_{\rm{EM}}(x)$ and $R_{\rm{Z}}(x)$ are, in general, not independent of each other. The term ``correlation'' has, however, a different meaning for $R_{\rm{EM}}(x)$ and $R_{\rm{Z}}(x)$. As far as $R_{\rm{EM}}(x)$ is concerned, the term correlation refers to the fact that the amplitude and phase of $R_{\rm{EM}}(x)$ depend on the angle of reflection if $A_r$ is equal to \eqref{Eq_Ar_1}. In order to maximize the power efficiency of a metasurface structure, this implies that the amplitude and phase of $R_{\rm{EM}}(x)$ cannot be set independently of each other. This amplitude-phase interplay is not captured by ray optics approximation. However, no spatial coupling, i.e., as a function of $x$, is observed because the amplitude of $R_{\rm{EM}}(x)$ is independent of $x$. As far as $R_{\rm{Z}}(x)$ is concerned, on the other hand, the amplitude and phase of $R_{\rm{Z}}(x)$ are tightly (spatially) correlated along the surface, i.e., as a function of $x$. It is not possible, in general, to optimize the amplitude and the phase of $R_{\rm{Z}}(x)$ independently of each other. Similar to $R_{\rm{EM}}(x)$, the amplitude and phase of $R_{\rm{Z}}(x)$ depend on the angle of reflection. The highly non-uniform spatial coupling between the amplitude and the phase of $R_{\rm{Z}}(x)$ unveils the inherent difficulty of realizing perfect anomalous reflectors with large angles of deflection and high power efficiency. 
\item \textit{\textbf{The surface reflection coefficients are periodic functions}}. Figures \ref{Fig_15a}-\ref{Fig_15b_2} confirm the periodicity of the surface-averaged reflection coefficients $R_{\rm{EM}}(x)$ and $R_{\rm{Z}}(x)$ and, therefore, of the corresponding surface-averaged susceptibility function and surface-averaged impedance. This allows one to realize a perfect anomalous reflector based on the periodic repetition of a super-cell structure and to employ periodic boundary conditions for its design and optimization.
\begin{figure}[!t]
\begin{centering}
\includegraphics[width=\columnwidth]{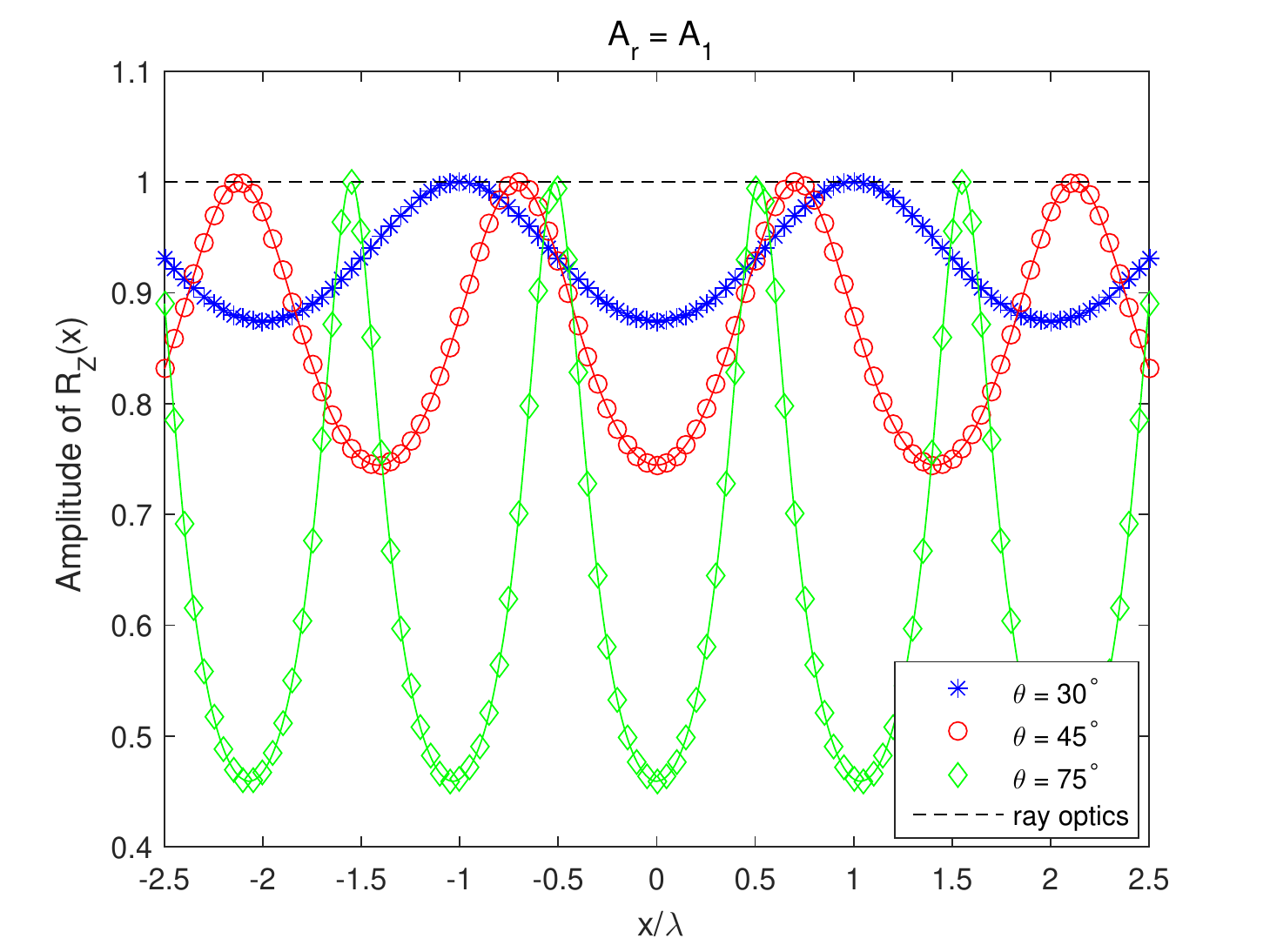}
\caption{Amplitude of the reflection coefficient $R_{\rm{Z}}(x)$ in \eqref{Eq_FullReflection_NormalIncidence__Z} for $\theta_r = \theta$ and $A_r = A_1$. Dashed lines: Ray optics approximation.}
\label{Fig_15a_1}
\end{centering} 
\end{figure}
\begin{figure}[!t]
\begin{centering}
\includegraphics[width=\columnwidth]{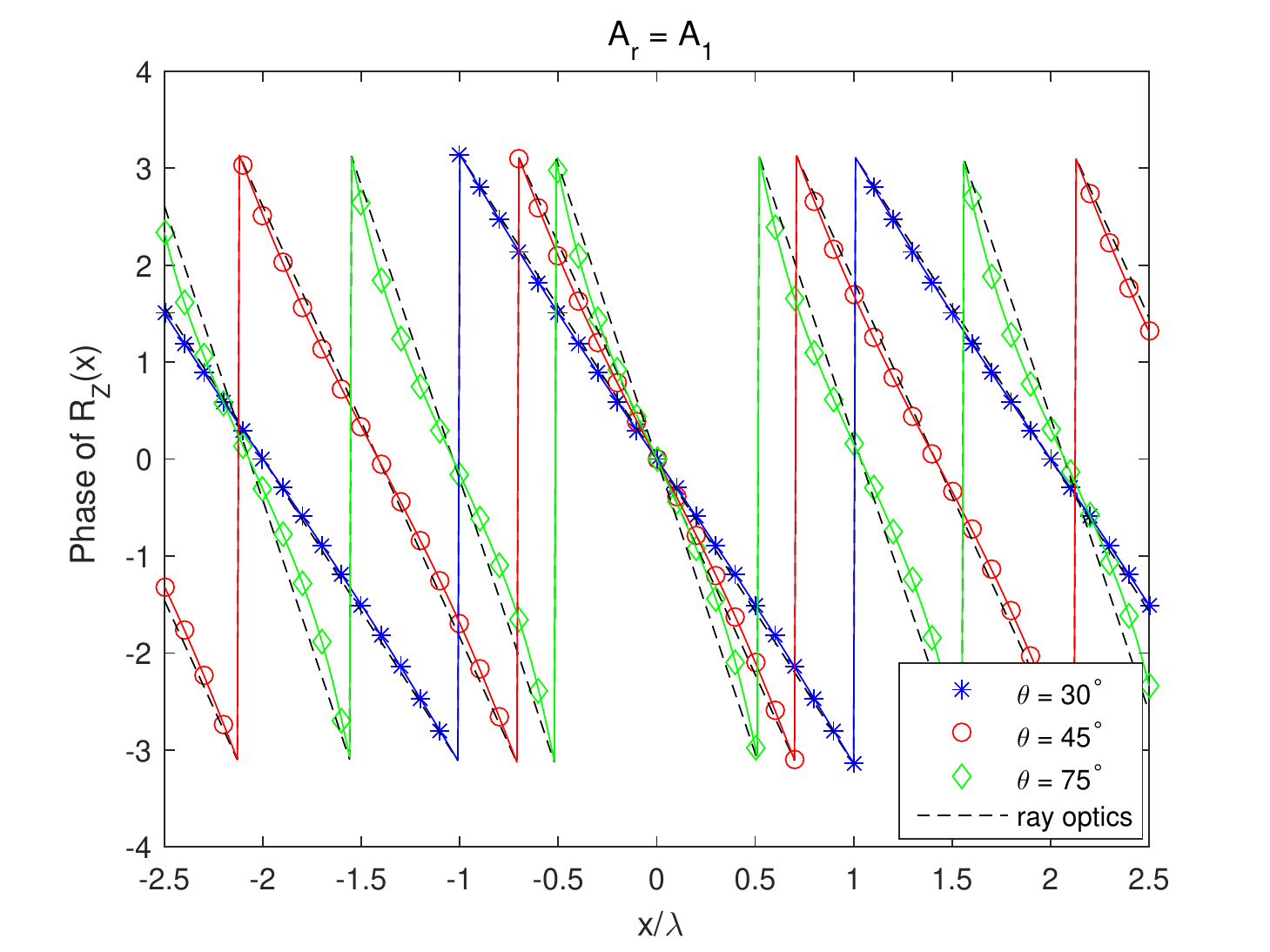}
\caption{Phase of the reflection coefficient $R_{\rm{Z}}(x)$ in \eqref{Eq_FullReflection_NormalIncidence__Z} for $\theta_r = \theta$ and $A_r = A_1$. Dashed lines: Ray optics approximation.}
\label{Fig_15b_1}
\end{centering} 
\end{figure}
\begin{figure}[!t]
\begin{centering}
\includegraphics[width=\columnwidth]{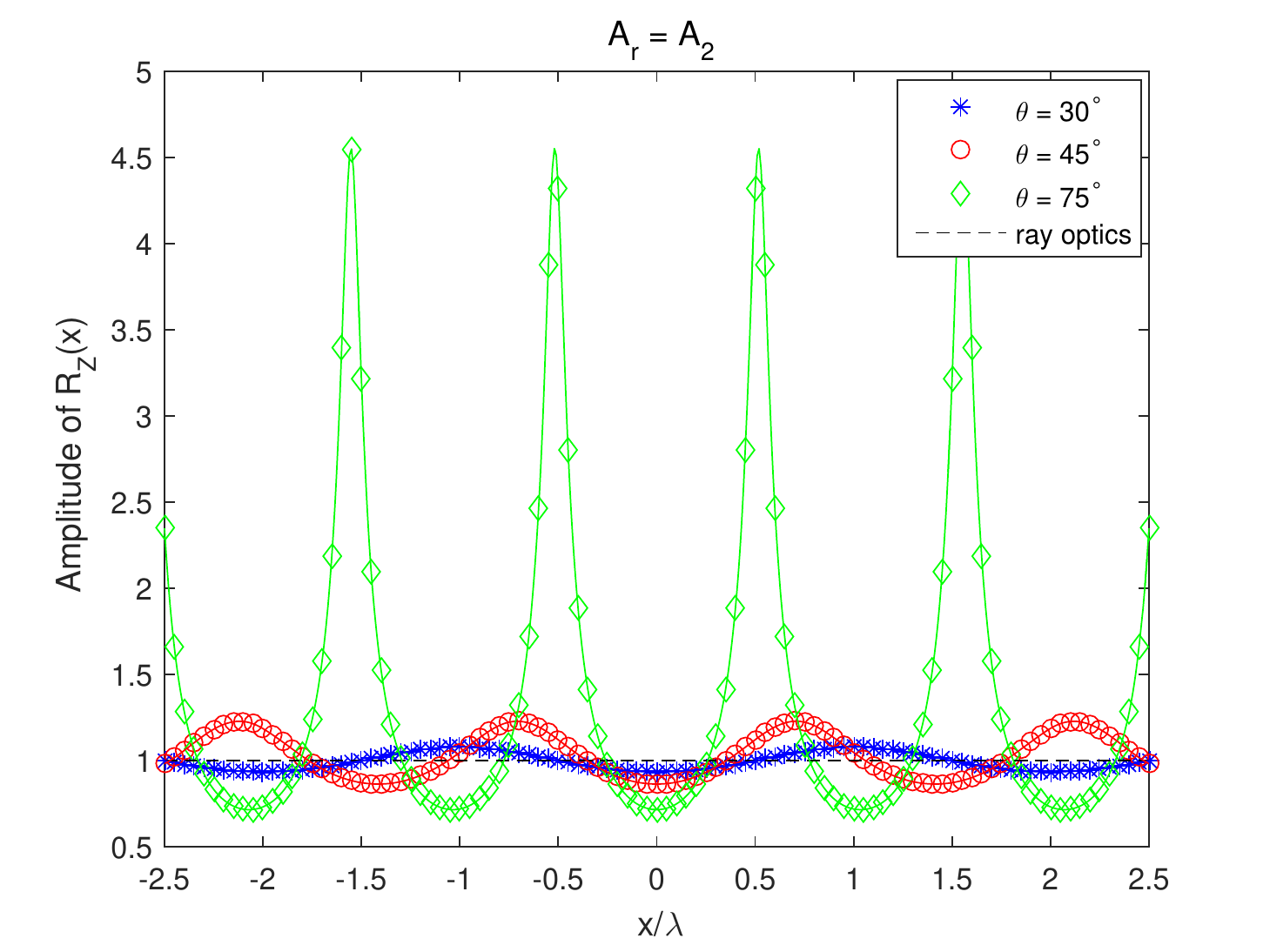}
\caption{Amplitude of the reflection coefficient $R_{\rm{Z}}(x)$ in \eqref{Eq_FullReflection_NormalIncidence__Z} for $\theta_r = \theta$ and $A_r = A_2$. Dashed lines: Ray optics approximation.}
\label{Fig_15a_2}
\end{centering} 
\end{figure}
\begin{figure}[!t]
\begin{centering}
\includegraphics[width=\columnwidth]{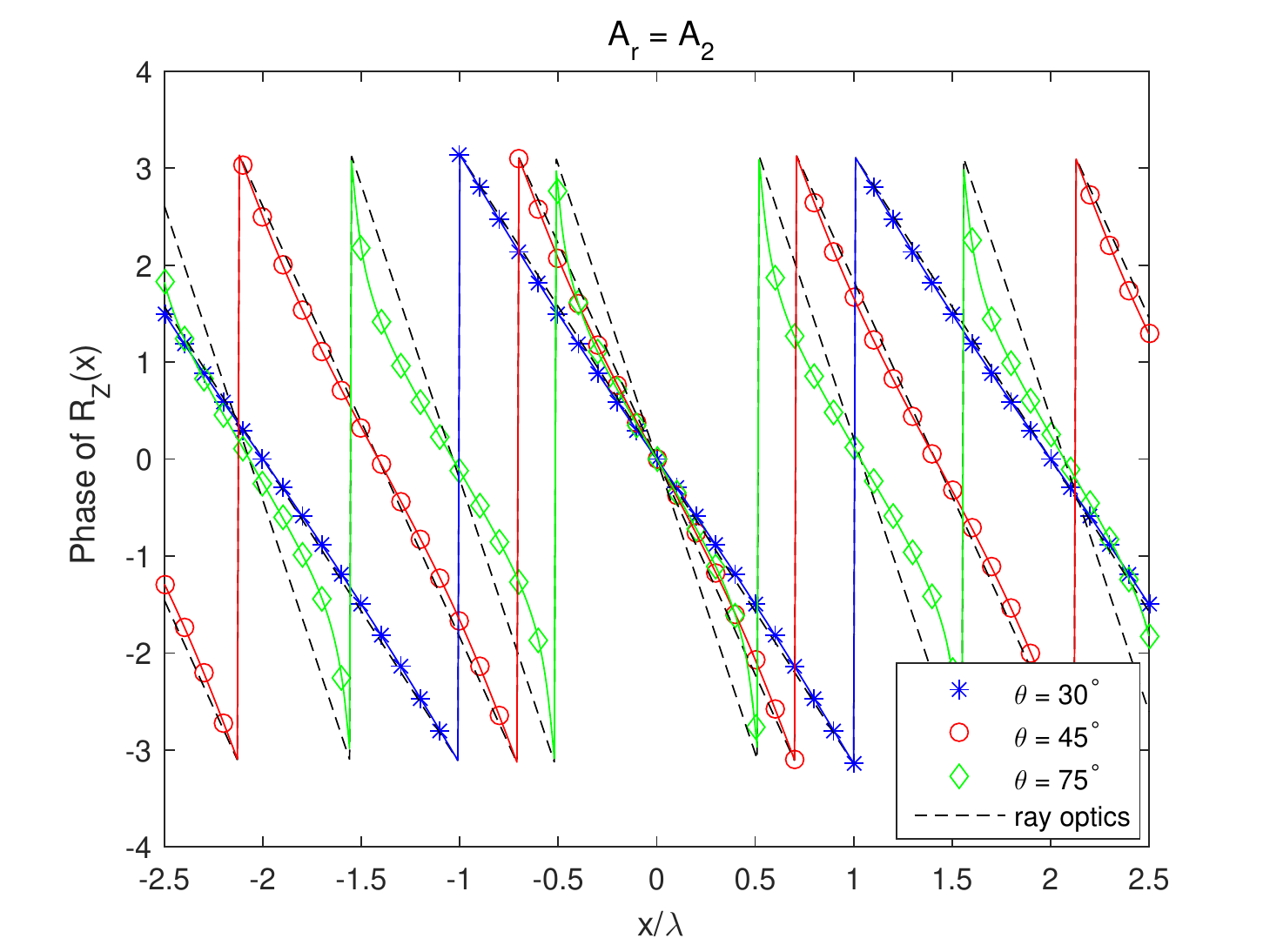}
\caption{Phase of the reflection coefficient $R_{\rm{Z}}(x)$ in \eqref{Eq_FullReflection_NormalIncidence__Z} for $\theta_r = \theta$ and $A_r = A_2$. Dashed lines: Ray optics approximation.}
\label{Fig_15b_2}
\end{centering} 
\end{figure}
\item \textit{\textbf{Perfect anomalous reflectors exhibit local power absorptions and gains}}. This important property that characterize perfect anomalous reflectors can be unveiled by analyzing $R_{\rm{Z}}(x)$. It is more difficult to unveil it by studying $R_{\rm{EM}}(x)$, since the amplitude of $R_{\rm{EM}}(x)$ is independent of $x$. More specifically, let us study the amplitude of $R_{\rm{Z}}(x)$ as a function of $A_r$. If $A_r = 1$, we observe that a metasurface structure that operates as a perfect anomalous reflector is locally passive, i.e., ${P_{{\rm{net}}}}\left( x \right) \le 0$, since the amplitude of $R_{\rm{Z}}(x)$ is less than or equal to one for every $x$. This implies that the metasurface structure exhibits power absorptions, since the reflected power flow is (locally) less than the incident power flow. If $A_r$ is, on the other hand, equal to the optimal setup for realizing the globally passive metasurface structure according to \eqref{Eq_Ar_1}, we observe that the amplitude of $R_{\rm{Z}}(x)$ cannot be equal to or smaller than one for every $x$. This implies that, if the metasurface structure is made globally passive, a perfect anomalous reflector needs to exhibit (virtual) local power absorptions and (virtual) local power gains along the surface of the metasurface structure. Even though a perfect anomalous reflector is designed to be globally passive, therefore, it is not, in general, locally passive. An simple design approach would consist of using active elements for ensuring $P_{{\rm{net}}} \ge 0$. This is, however, not the best solution if the ultimate target is to realize nearly-passive metasurfaces. Since nearly-passive metasurfaces are not intended to rely on power amplifiers (or active elements) during the normal operation phase, one may wonder how the non-local power gains can be obtained. This is discussed in, e.g., \cite{Asadchy_2017}, \cite{Asadchy_2016}, \cite{Diaz-Rubio_2017}, \cite{Epstein_2016}. The implementation recently reported in \cite{Diaz-Rubio_2017}, which assumes the optimal setup of $A_r$ in \eqref{Eq_Ar_1}, solves this issue by proposing a metasurface design that is realized by exciting evanescent (surface) waves (or leaky modes) that travel along the surface of the metasurface and that carry power from the effectively lossy regions to the effectively active regions of the metasurface structure. In other words, the metasurface design in \cite{Diaz-Rubio_2017} excites auxiliary EM fields, besides the propagating EM fields transformation in \eqref{Eq_FullReflection_PlaneWave}, so that (i) the energy of the impinging radio wave travels along the surface, from the (virtual) lossy regions to the (virtual) active regions; (ii) the metasurface structure is kept globally passive and has (nearly) unitary power efficiency; and (iii) no or minimal spurious reflections are generated towards unwanted directions. Evanescent fields are, more precisely, oscillating electric and magnetic fields that do not propagate but whose energy is spatially concentrated in the vicinity of the surface of the metasurface structure. The evanescent waves decay exponentially and are appropriately designed so as not to interfere with the main functionality of the metasurface structure at distances of a few (typically two or three) wavelengths from the metasurface. As far as the practical implementation in \cite{Diaz-Rubio_2017} is concerned, the fields transformation in \eqref{Eq_FullReflection_PlaneWave} constitutes, therefore, an approximated model, since in the very close vicinity of the metasurface structure the EM fields are not given by the summation of two simple incident and reflected plane waves, but other evanescent EM fields need to be considered. For wireless applications, these evanescent EM fields can be ignored as a first approximation, since the distances of interest are usually larger than two-three wavelengths, both in the radiative near-field and in the far-field. This point is elaborated in further text. Figures \ref{Fig_15a_1}-\ref{Fig_15b_2} show, in particular, that the larger the reflection angle (with respect to the incident angle) is, the larger the (virtual) power absorptions and gains along the metasurface structure need to be. The optimal setup of $A_r$ in \eqref{Eq_Ar_1}, results, in addition, in larger fluctuations of the amplitude of $R_{\rm{Z}}(x)$, which may be more difficult to realize in practice. No power fluctuation along the surface of the metasurface structure is obtained only if the angle of reflection coincides with the angle of incidence. 
\begin{figure}[!t]
\begin{centering}
\includegraphics[width=\columnwidth]{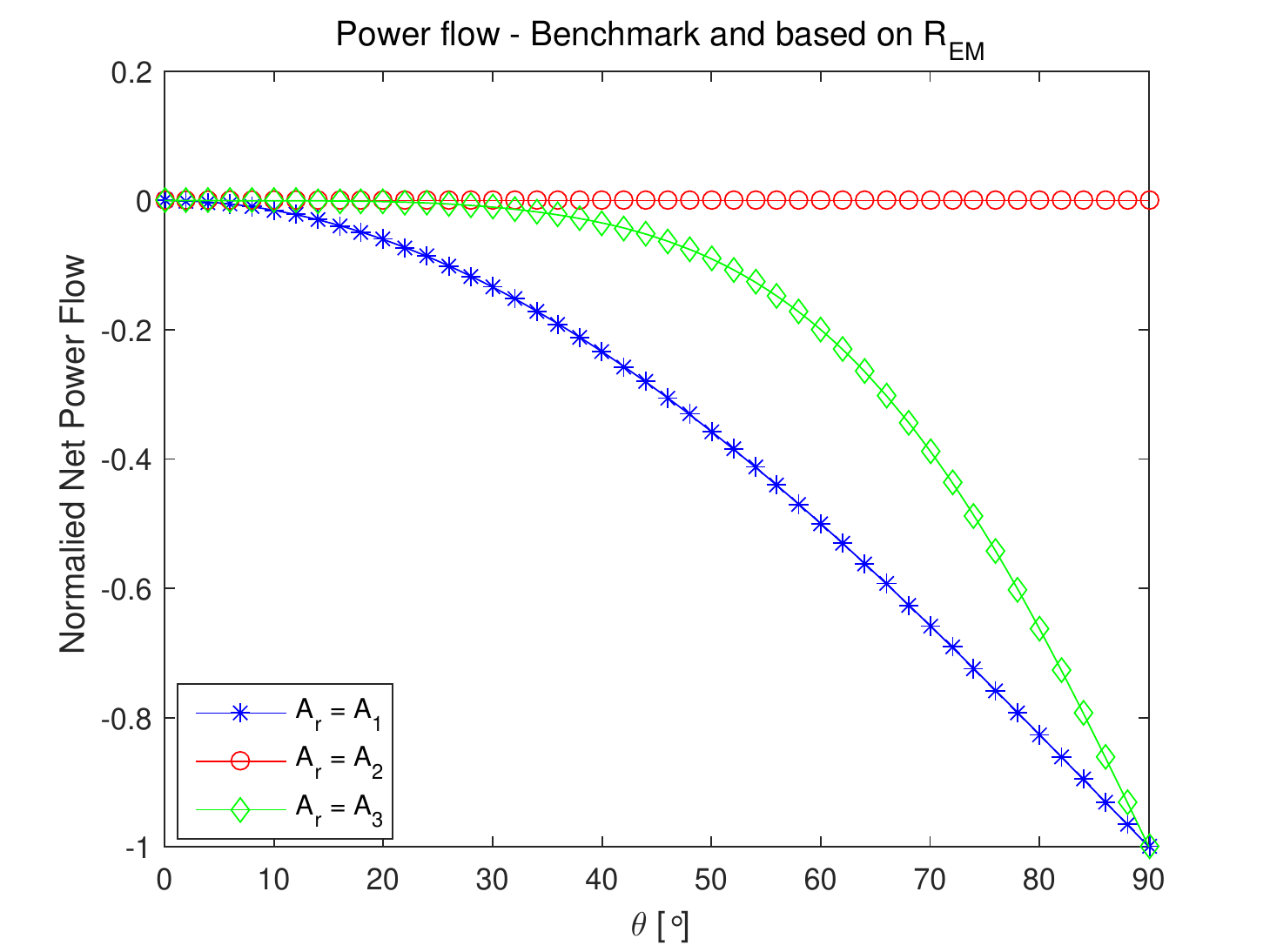}
\caption{Normalized net power flow (${P_{{\rm{net}}}}/{P_0}$) in \eqref{Eq_PowerBudget__2} and \eqref{Eq_PowerBudget__3}, by setting $r_{i,r}={\cos \left( {{\theta _r}} \right)}/{\cos \left( {{\theta _i}} \right)}$, for $\theta_r = \theta$ and $A_r = \{A_1, A_2, A_3\}$.}
\label{Fig_16a}
\end{centering} 
\end{figure}
\begin{figure}[!t]
\begin{centering}
\includegraphics[width=\columnwidth]{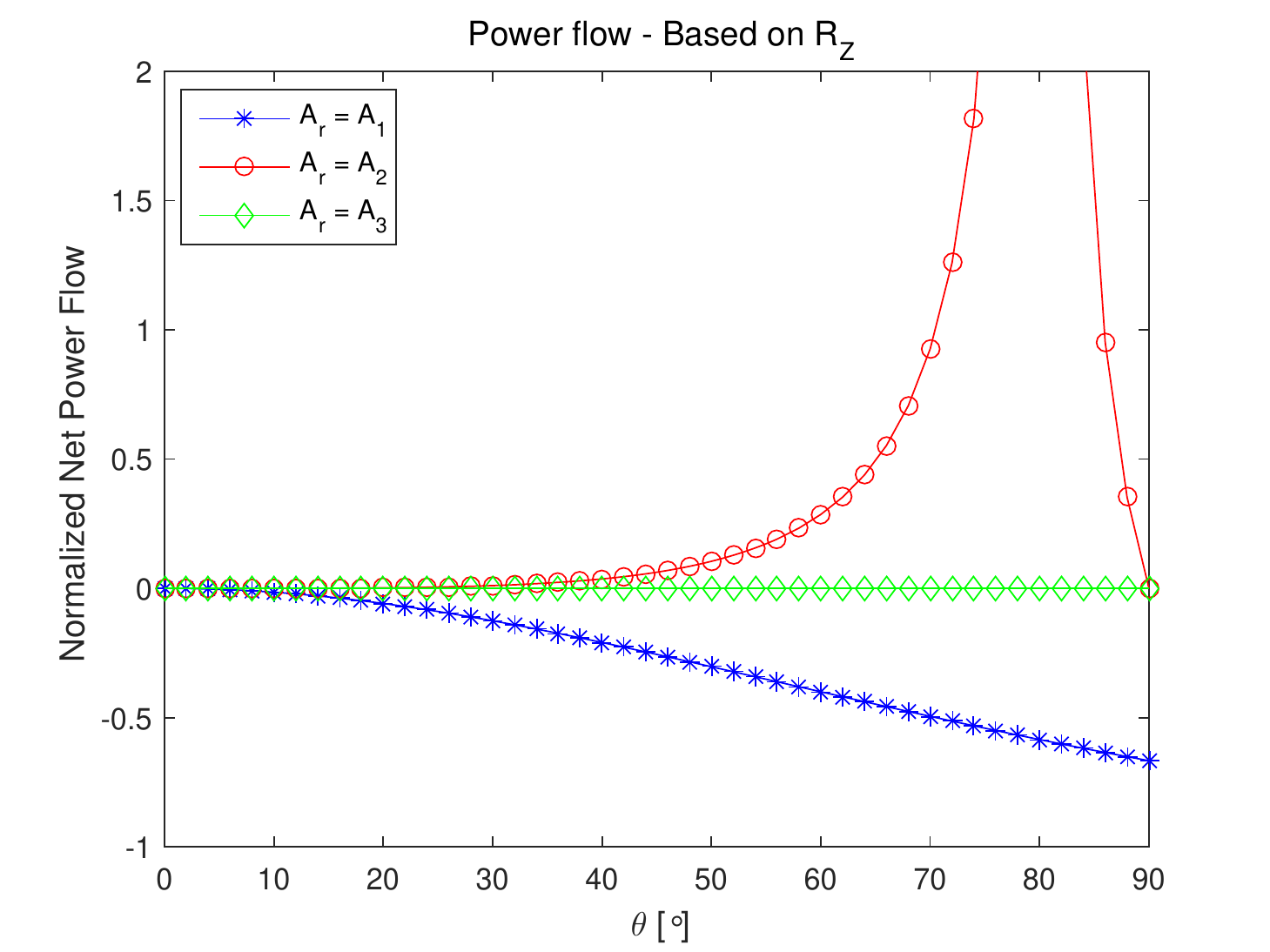}
\caption{Normalized net power flow (${P_{{\rm{net}}}}/{P_0}$) in \eqref{Eq_PowerBudget__3}, by setting $r_{i,r}=1$, for $\theta_r = \theta$ and $A_r = \{A_1, A_2, A_3\}$.}
\label{Fig_16b}
\end{centering} 
\end{figure}
\item \textit{\textbf{On the global efficiency of perfect anomalous reflectors}}. The importance of realizing metasurfaces with a surface-averaged reflection coefficient $R_{\rm{EM}}(x)$ whose amplitude is greater than one and with a surface-averaged reflection coefficient $R_{\rm{Z}}(x)$ whose amplitude exhibits local absorptions and gains along the surface can be unveiled by analyzing the normalized net power flows reported in Figs. \ref{Fig_16a} and \ref{Fig_16b} for different values of $A_r$. In particular, Figs. \ref{Fig_16a} and \ref{Fig_16b} are obtained by employing the two analytical formulations of the net power flows in \eqref{Eq_PowerBudget__2} and \eqref{Eq_PowerBudget__3}, respectively, for three different values of $A_r$: $A_r=A_1=1$, $A_r=A_2$ in \eqref{Eq_Ar_1}, and $A_r=A_3$ in \eqref{Eq_Ar_2}. Negative and positive values of the normalized net power flow imply that the metasurface structure exhibits global power absorptions and gains, respectively. Figure \ref{Fig_16a} shows, in particular, that globally passive metasurfaces can be realized, based on \eqref{Eq_PowerBudget__2}, for all considered values of $A_r$. If $A_r$ is set equal to \eqref{Eq_Ar_1}, the metasurface structure is, as expected, globally passive and has a unitary power efficiency. In general, however, the power efficiency decreases as the angle of reflection (with respect to the angle of incidence) increases. From these results, we evince that highly efficient metasurface structures need to be non-locally passive. By setting $A_r=1$, in fact, one can obtain a locally passive metasurface structure but its power efficiency is relatively low, especially for large values of the angle of reflection with respect to the angle of incidence. By analyzing Fig. \ref{Fig_16b}, similar conclusions can be made. However, Fig. \ref{Fig_16b} shows that the normalized net power flow computed with \eqref{Eq_PowerBudget__3} is greater than one if $A_r$ is set equal to \eqref{Eq_Ar_1}. This results confirms that $R_{\rm{Z}}(x)$ is not an appropriate definition for the surface-averaged reflection coefficient if one is interested in calculating the reflected EM fields from the incident EM fields. The corresponding power budget is, for example, not consistent with the benchmark transformation of the EM fields in \eqref{Eq_FullReflection_PlaneWave}, which leads to different formulas for the net power flow, as exemplified by comparing \eqref{Eq_PowerBudget__2} with \eqref{Eq_PowerBudget__3} by setting $r_{i,r} =1$.
\end{itemize}

\textbf{$\bf{R_{{\rm{\bf EM}}}}\left( x \right)$ vs. $\bf{R_{{\rm{\bf Z}}}}\left( x \right)$: Main takeaway message}. Based on the theoretical analysis and the numerical comparison between ${R_{{\rm{EM}}}}\left( x \right)$ and ${R_{{\rm{Z}}}}\left( x \right)$, it is apparent that they are two different parameters that serve for different purposes. In general, ${R_{{\rm{EM}}}}\left( x \right)={R_{{\rm{Z}}}}\left( x \right)$ only for perfect specular reflectors. ${R_{{\rm{EM}}}}\left( x \right)$ is intended to be used in the of context wireless communications in order to characterize the response of a metasurface structure in terms of the phase and amplitude difference between the incident and reflected EM fields. ${R_{{\rm{Z}}}}\left( x \right)$ is intended to be used in the context of metasurface design in order to characterize the response of a metasurface structure in terms of the mismatch between the impedances of the incident and reflected EM waves. Both ${R_{{\rm{EM}}}}\left( x \right)$ and ${R_{{\rm{Z}}}}\left( x \right)$ are, however, surface-averaged, and, therefore, macroscopic parameters. As a consequence, it is not appropriate to interpret either ${R_{{\rm{EM}}}}\left( x_0 \right)$ or ${R_{{\rm{Z}}}}\left( x_0 \right)$ as the response of an individual unit cell of the metasurface structure that is located in the point $x_0$ on the surface. This is because ${R_{{\rm{EM}}}}\left( x \right)$ and ${R_{{\rm{Z}}}}\left( x \right)$ are defined and calculated based on surface-averaged EM fields and are not based on acting (local) EM fields, as discussed in previous text. \\

\noindent \underline{\textbf{\textit{Takeaway messages for application to wireless networks}}}. \textbf{Contrasting current models for RISs and the synthesis of metasurfaces}. Several recently published papers (see, e.g., \cite{MDR_OverheadAware} and references therein) typically assume that an RIS can be modeled as a structure that is made of a discrete number of individually tunable scatterers whose properties, e.g., the phase and amplitude of the reflection coefficient (i.e., ${R_{{{\rm  EM}}}}\left( x \right)$), can be arbitrary and individually controlled. In general, the amplitude of the reflection coefficient of each scatterer is assumed to be unitary in order to ensure that the RIS is nearly-passive. Furthermore, the spatial correlation among the individual scatterers is usually ignored. These assumptions may be considered to be sufficiently accurate if an RIS is made of discrete tiny antenna elements, similar to the prototype in Fig. \ref{Fig_2}, that are sufficiently distant from each other. The procedure for the synthesis of metasurfaces described in this sub-section highlights, however, that these conventional assumptions may not necessarily be appropriate for modeling metasurfaces. 

\textbf{Key points to remember}. In particular, the following key points need to be kept in mind when designing metasurface-based RIS-empowered wireless networks.
\begin{itemize}
\item \textit{\textbf{The spatial coupling is not ignored}}. It is general belief that it is possible to individually control the reflection coefficient of each unit cell without taking into accounting the mutual coupling among them. This may not necessarily be correct. The synthesis procedure described in this sub-section unveils that the mutual coupling is not explicitly apparent thanks to the macroscopic description of the metasurfaces through the surface-averaged susceptibility functions (or the surface-averaged impedances). The macroscopic description of metasurfaces originates, however, from a microscopic description that explicitly accounts for the mutual coupling via the local fields. From an implementation standpoint, a metasurface is realized by jointly designing and optimizing the unit cells of the super-cell structure. This procedure inherently accounts for the mutual coupling among the unit cells that constitute the super-cell.
\item \textit{\textbf{Globally passive does not imply locally passive}}. Nearly-passive RISs may not necessarily be locally passive. In general, highly-efficient metasurfaces may exhibit some (virtual) absorptions and gains along the surface.
\item \textit{\textbf{Amplitude and phase responses are intertwined}}. The amplitude and phase of the radio waves impinging upon an RIS may not be modified independently of each other, but they may be coupled depending on the specific wave transformations to be realized and the constraints imposed to realize locally passive or globally passive metasurface structures.
\end{itemize}

\textbf{How to use the surface-averaged reflection coefficient}. As mentioned in previous text, the surface reflection coefficient for modeling the scattering from objects is a widespread tool in the context of wireless communications. Therefore, the analytical formulation in \eqref{Eq_FullReflection_R} may be particular convenient to use. In the context of designing wireless communication systems and networks, it is important to emphasize that one is interested in the surface-averaged reflection coefficient ${R_{{\rm{EM}}}}\left( x \right)$, which provides one with the response of a metasurface structure in terms of incident and reflected EM fields. The surface-averaged reflection coefficient ${R_{{\rm{EM}}}}\left( x \right)$ needs, however, not to be confused with the surface-averaged reflection coefficient ${R_{{\rm{Z}}}}\left( x \right)$, which is often used for designing and for gaining insight for implementing metasurface structures. Similar comments may be made if an RIS is employed for perfect anomalous transmission, or if it is designed to realize joint anomalous reflection and transmission. Based on these considerations, it is important to clarify how the surface-averaged reflection coefficient ${R_{{\rm{EM}}}}\left( x \right)$ should be used and should be interpreted in the context of wireless networks design. In this regard, the following comments can be made.
\begin{itemize}
\item \textit{\textbf{The surface-averaged reflection coefficient $\bf{R_{{\rm{\bf EM}}}}\left( x \right)$ is not a local parameter}}. ${R_{{\rm{EM}}}}\left( x \right)$ is a surface-averaged parameter and, therefore, it is not appropriate to discretize it and to interpret the resulting sampled values as the response of individual unit cells of the metasurface structure. This is because ${R_{{\rm{EM}}}}\left( x \right)$ is defined and calculated based on surface-averaged EM fields and not based on acting (local) EM fields. In this regard, ${R_{{\rm{EM}}}}\left( x \right)$ can be viewed as a ``global'' parameter that can be employed for analyzing RIS-empowered wireless networks. In other words, the representation of an RIS based on ${R_{{\rm{EM}}}}\left( x \right)$ inherently hides the implementation details of the corresponding metasurface structure, but it provides one with a tool to treat RISs as ``black boxes'' that can be used for designing and optimizing complex wireless communication systems and networks.
\item \textit{\textbf{$\bf{A_r}$ can be viewed as an optimization variable}}. The analytical formulation of the surface-averaged reflection coefficient ${R_{{\rm{EM}}}}\left( x \right)$ is valid for any values of $A_r$. Therefore, one may restrict the range of values of $A_r$ for the specific application of interest. More precisely, the parameter $A_r$ can be viewed as an optimization parameter that allows one to account for the interplay between power efficiency and implementation complexity. If $A_r=1$, in fact, the metasurface structure is easier to realize as compared with the setup of $A_r$ equal to \eqref{Eq_Ar_1}, since sophisticated metasurface structures that exhibit virtual power losses and virtual power gains are needed in the latter case. It is worth mentioning that $A_r$ is, in general, a function of the angles of incidence and reflection, as confirmed by \eqref{Eq_Ar_1}. It is important to bear in mind, however, that the EM fields in \eqref{Eq_FullReflection_R} needs to be solutions of Maxwell's equations. $A_r$ and the corresponding reflected fields need to obey Maxwell's equations, and, therefore, some (physical) constraints on their choice exist. 
\item \textit{\textbf{$\bf{A_r}$ can be greater than one}}. Conventional methods for modeling RISs in wireless communications are based on representing an RIS through a diagonal matrix with individually tunable reflecting elements whose amplitudes are all unitary and are independent of the applied phase shifts \cite{MDR_OverheadAware}. This modeling approach corresponds to using ${R_{{\rm{EM}}}}\left( x \right)$ in \eqref{EM_ReflectionCoefficient} by setting $A_r=1$. The resulting metasurface structure is inherently locally passive but sub-optimal from the power efficiency point of view. In general, $A_r$ is greater than one in globally passive metasurfaces and it depends on the angles of incidence and reflection.
\end{itemize}

\textbf{``Black box'' modeling of a perfect anomalous reflector}. Based on the described synthesis procedure and recipe for application to wireless networks, therefore, we can model an RIS that is designed to operate as a perfect anomalous reflector as a ``black box'' whose output field is obtained by multiplying the input field and the surface-averaged reflection coefficient ${R_{{\rm{EM}}}}\left( x \right)$ in \eqref{EM_ReflectionCoefficient}, as formulated in \eqref{Eq_FullReflection_R}. This simple procedure allows one to account for several practical aspects of a metasurface, which include the mutual coupling among individual unit cells and the fact that, with the exception of the configuration network, the RIS is globally passive even though it is not necessarily locally passive. Further details on the analysis of general metasurfaces are given in the next sub-section. It is worth anticipating, however, that modeling a perfect anomalous reflector by means of a surface reflection coefficient relies on the assumption that the incident EM fields can be approximated to be locally-plane waves at the scale of an individual unit cell (not necessarily at the scale of the entire metasurface). Usually, this can be considered to be true at distances from the metasurface structure that are greater than a few (two or three) wavelengths. From the physics standpoint, this is necessary since the surface reflection coefficient defined in the present paper does not account for the evanescent EM waves in close vicinity of the metasurface, which may be needed to realize non-local passive structures. This concept is illustrated in Fig. \ref{Fig_18} and it is further elaborated in the next sub-section.

\textbf{Relevance of the surface-averaged reflection coefficient for wireless applications}. The analytical formulation in terms of surface-averaged reflection coefficient ${R_{{\rm{EM}}}}\left( x \right)$ is especially useful for wireless applications. The reason is that wireless researchers are used to model systems as black boxes that perform some specific functions, but, at the same time, they wish to hide unnecessary details related to the practical implementation and realization of the systems themselves. This is necessary for analytical tractability, in order to model and optimize complex wireless networks, in which a single metasurface is one of the many network elements. With this in mind, the representation of EM fields as a function of surface-averaged susceptibility functions, surface-averaged impedances, or surface-averaged reflection coefficient ${R_{{\rm{Z}}}}\left( x \right)$ do not yield an explicit analytical formulation of the output fields as a function of the input fields. These latter homogenized representations, however, explicitly depend on the physical structure of the metasurface itself. This is useful for the synthesis of metasurfaces, since important insights on practical structures of unit cells can be explicitly obtained, and design guidelines on realizing practical structures can be obtained (e.g., from ${R_{{\rm{Z}}}}\left( x \right)$). The representation in terms of surface-averaged reflection coefficient ${R_{{\rm{EM}}}}\left( x \right)$ provides, on the other hand, information only on the function realized by a metasurface from the global point of view while hiding implementation-related details. Concretely, ${R_{{\rm{EM}}}}\left( x \right)$ provides one with an explicit relation that depends only on the angle of incidence and reflection of the radio waves, while hiding the physical structure of the metasurface. At the same time, the analytical formulation of ${R_{{\rm{EM}}}}\left( x \right)$ in \eqref{EM_ReflectionCoefficient} overcomes the limitations of current models used for wireless applications, according to which ${R_{{\rm{EM}}}}\left( x_1 \right)$ and ${R_{{\rm{EM}}}}\left( x_2 \right)$ are optimized independently of each other for any $x_1 \ne x_2$ \cite{MDR_OverheadAware}.

\subsubsection{Analysis of a Metasurface}
\textbf{EM fields scattered by a given metasurface}. In the previous sub-section, we have reported a general procedure for the synthesis of a metasurface that needs to realize some specified transformations on the impinging radio waves. In this sub-section, on the other hand, we consider that the metasurface is given and we are interested in computing a general and explicit expression of the reflected and transmitted electric and magnetic fields for some given incident electric and magnetic fields. In particular, we focus our attention on the EM fields that are at the two sides of a metasurface, i.e., at $z=0^+$ and  $z=0^-$, which are the result of the EM discontinuity introduced by the metasurface structure. They are referred to as \textit{surface EM fields}. The computation of the electric and magnetic fields at any arbitrary point of a given volume where a metasurface is deployed are, on the other hand, discussed in the next sub-section.

\textbf{Surface EM fields}. The analytical expressions of the surface EM fields that we are interested in calculating can be obtained from the GSTCs in \eqref{Eq_GSTCs_Explicit_H} and \eqref{Eq_GSTCs_Explicit_E}. In order to obtain an explicit expression of every tangential component of the electric and magnetic fields, we introduce the following matrices (for ease of writing, the dependency on $(x,y)$ is not explicitly reported):
\begin{equation} \label{Eq_Analysis__Def1}
\begin{split}
&{\textbf{\textsf A}}\left( {x,y} \right) = \left[ {\begin{array}{*{20}{c}}
{{{\rm{A}}_1}}&{{{\rm{A}}_2}}\\
{{{\rm{A}}_3}}&{{{\rm{A}}_4}}
\end{array}} \right] = \frac{{j\omega \varepsilon }}{2}\left[ {\begin{array}{*{20}{c}}
{\chi _{{\rm{ee}}}^{xx}}&{\chi _{{\rm{ee}}}^{xy}}\\
{\chi _{{\rm{ee}}}^{yx}}&{\chi _{{\rm{ee}}}^{yy}}
\end{array}} \right] \\
& {\textbf{\textsf B}}\left( {x,y} \right) = \left[ {\begin{array}{*{20}{c}}
{{{\rm{B}}_1}}&{{{\rm{B}}_2}}\\
{{{\rm{B}}_3}}&{{{\rm{B}}_4}}
\end{array}} \right] = \frac{{j\omega \sqrt {\mu \varepsilon } }}{2}\left[ {\begin{array}{*{20}{c}}
{\chi _{{\rm{em}}}^{xx}}&{\chi _{{\rm{em}}}^{xy}}\\
{\chi _{{\rm{em}}}^{yx}}&{\chi _{{\rm{em}}}^{yy}}
\end{array}} \right]\\
& {\textbf{\textsf C}}\left( {x,y} \right) = \left[ {\begin{array}{*{20}{c}}
{{{\rm{C}}_1}}&{{{\rm{C}}_2}}\\
{{{\rm{C}}_3}}&{{{\rm{C}}_4}}
\end{array}} \right] = \frac{{j\omega \mu }}{2}\left[ {\begin{array}{*{20}{c}}
{\chi _{{\rm{mm}}}^{xx}}&{\chi _{{\rm{mm}}}^{xy}}\\
{\chi _{{\rm{mm}}}^{yx}}&{\chi _{{\rm{mm}}}^{yy}}
\end{array}} \right]\\
& {\textbf{\textsf D}}\left( {x,y} \right) = \left[ {\begin{array}{*{20}{c}}
{{{\rm{D}}_1}}&{{{\rm{D}}_2}}\\
{{{\rm{D}}_3}}&{{{\rm{D}}_4}}
\end{array}} \right] = \frac{{j\omega \sqrt {\mu \varepsilon } }}{2}\left[ {\begin{array}{*{20}{c}}
{\chi _{{\rm{me}}}^{xx}}&{\chi _{{\rm{me}}}^{xy}}\\
{\chi _{{\rm{me}}}^{yx}}&{\chi _{{\rm{me}}}^{yy}}
\end{array}} \right]
\end{split}
\end{equation}
\begin{equation} \label{Eq_Analysis__Def2}
{\textbf{\textsf Z}} = {\left[ {\begin{array}{*{20}{l}}
{E_x^r}&{E_x^t}&{E_y^r}&{E_y^t}&{H_x^r}&{H_x^t}&{H_y^r}&{H_y^t}
\end{array}} \right]^T}
\end{equation}
\begin{equation} \label{Eq_Analysis__Def3}
{\textbf{\textsf M}} = \left[ {\begin{array}{*{20}{c}}{{{{\textbf{\textsf M}}}_1}}&{{{{\textbf{\textsf M}}}_2}}\end{array}} \right] 
\end{equation}
\begin{equation} \nonumber
\begin{split}
& {{\textbf{\textsf M}}_1} = \left[ {\begin{array}{*{20}{c}}
{{{\rm{A}}_1}}&{{{\rm{A}}_1}}&{{{\rm{A}}_2}}&{{{\rm{A}}_2}}\\
{{{\rm{A}}_3}}&{{{\rm{A}}_3}}&{{{\rm{A}}_4}}&{{{\rm{A}}_4}}\\
{{{\rm{D}}_1}}&{{{\rm{D}}_1}}&{{{\rm{D}}_2} + 1}&{{{\rm{D}}_2} - 1}\\
{{{\rm{D}}_3} - 1}&{{{\rm{D}}_3} + 1}&{{{\rm{D}}_4}}&{{{\rm{D}}_4}}
\end{array}} \right] \\
& {{\textbf{\textsf M}}_2} = \left[ {\begin{array}{*{20}{c}}
{{{\rm{B}}_1}}&{{{\rm{B}}_1}}&{{{\rm{B}}_2} - 1}&{{{\rm{B}}_2} + 1}\\
{{{\rm{B}}_3} + 1}&{{{\rm{B}}_3} - 1}&{{{\rm{B}}_4}}&{{{\rm{B}}_4}}\\
{{{\rm{C}}_1}}&{{{\rm{C}}_1}}&{{{\rm{C}}_2}}&{{{\rm{C}}_2}}\\
{{{\rm{C}}_3}}&{{{\rm{C}}_3}}&{{{\rm{C}}_4}}&{{{\rm{C}}_4}}
\end{array}} \right]
\end{split}
\end{equation}
\begin{equation} \label{Eq_Analysis__Def4}
{\textbf{\textsf W}} = \left[ {\begin{array}{*{20}{c}}
{ - {{\rm{A}}_1}}&{ - {{\rm{A}}_2}}&{ - {{\rm{B}}_1}}&1-{{\rm{B}}_2}\\
{ - {{\rm{A}}_3}}&{ - {{\rm{A}}_4}}&-1-{{\rm{B}}_3} &{ - {{\rm{B}}_4}}\\
{ - {{\rm{D}}_1}}&-1-{{\rm{D}}_2}&{ - {{\rm{C}}_1}}&{ - {{\rm{C}}_3}}\\
1-{{\rm{D}}_3}&{ - {{\rm{D}}_4}}&{ - {{\rm{C}}_3}}&{ - {{\rm{C}}_4}}
\end{array}} \right]
\end{equation}
\begin{equation} \label{Eq_Analysis__Def5}
{\textbf{\textsf F}} = {\left[ {\begin{array}{*{20}{c}}
{E_x^i}&{E_y^i}&{H_x^i}&{H_y^i}
\end{array}} \right]^T}
\end{equation}

\noindent With the aid of these definitions, \eqref{Eq_GSTCs_Explicit_H} and \eqref{Eq_GSTCs_Explicit_E} can be re-written in a single matrix form as follows:
\begin{equation} \label{Eq_Analysis__Matrix}
{\textbf{\textsf {MZ}}} = {\textbf{\textsf {WF}}}
\end{equation}

\textbf{Explicit analytical formulation}. An explicit expression for the reflected and transmitted surface EM fields can be obtained by solving the system of equations in \eqref{Eq_Analysis__Matrix}. To this end, the matrix $\textbf{\textsf {M}}$ is formulated in its singular value decomposition (SVD) form. In particular, let us introduce the following notation: (i) $\textbf{\textsf {V}}$ is the matrix whose columns are the eigenvectors of ${{\textbf {\textsf{M}}}^H}{\textbf {\textsf{M}}}$; (ii) $\textbf {\textsf{U}}$ is the matrix whose columns are the eigenvectors of ${\textbf {\textsf{M}}}{{\textbf {\textsf{M}}}^H}$; (iii) ${ \textbf {\textsf{ S}}}$ is a diagonal matrix whose entries are zero or are the square root of the non-zero eigenvalues of ${{\textbf {\textsf{M}}}^H}{\textbf {\textsf{M}}}$ or ${\textbf {\textsf{M}}}{{\textbf {\textsf{M}}}^H}$; and (iv) ${ \textbf {\textsf{ S}}} ^\dag$ is a diagonal matrix whose entries are either zero or are the reciprocal of the non-zero entries of ${ \textbf {\textsf{ S}}}$, i.e., ${ \textbf {\textsf{ S}}} ^\dag$ is the pseudo-inverse of ${ \textbf {\textsf{ S}}}$. Therefore, ${\textbf {\textsf{M}}} = {\textbf {\textsf{U S}}} {{\textbf {\textsf{V}}}^H}$ and ${\textbf {\textsf{Z}}}$ in \eqref{Eq_Analysis__Matrix} can be explicitly written as follows:
\begin{equation} \label{Eq_Analysis__Solution}
{\textbf {\textsf{Z}}} = \left( \textbf {\textsf{V}}{{\textbf {\textsf{S}}} ^\dag }{{\textbf {\textsf{U}}}^H}{\textbf {\textsf{W}}} \right) {\textbf {\textsf{F}}} = {\textbf {\textsf{PF}}}
\end{equation}
\noindent where ${\textbf {\textsf{P}}} = {\textbf {\textsf{V}}}{{\textbf {\textsf{S}}} ^\dag }{{\textbf {\textsf{U}}}^H}{\textbf {\textsf{W}}}$ is a matrix that depends only on the metasurface structure, i.e., the surface susceptibility functions, while it is independent of the incident, reflected, and transmitted fields. It is worth mentioning that, for any given matrix, the pseudo-inverse always exists and is unique. This implies that, given a metasurface structure, i.e., the surface susceptibility functions and the incident EM fields, one can always find the corresponding reflected and transmitted EM fields.

\textbf{Functional structure of the surface EM fields}. The matrix ${\textbf {\textsf{P}}}$ in \eqref{Eq_Analysis__Solution} offers an explicit analytical relation between the incident (surface) EM fields, i.e., the vector ${\textbf {\textsf{F}}}$, and the reflected and transmitted (surface) EM fields. By direct inspection of \eqref{Eq_Analysis__Solution}, in particular, it is apparent that the reflected and transmitted surface EM fields can be explicitly formulated as:
\begin{equation} \label{Eq_Analysis__Final}
\begin{split}
E_{x,y}^{r,t}\left( {x,y} \right) &\propto \varsigma _{E,x}^{r,t}\left( {x,y} \right)E_x^i\left( {x,y} \right) \\ & + \varsigma _{E,y}^{r,t}\left( {x,y} \right)E_y^i\left( {x,y} \right)\\
& + \varsigma _{H,x}^{r,t}\left( {x,y} \right)H_x^i\left( {x,y} \right)  \\ &+ \varsigma _{H,y}^{r,t}\left( {x,y} \right)H_y^i\left( {x,y} \right) \\
H_{x,y}^{r,t}\left( {x,y} \right) &\propto \xi _{E,x}^{r,t}\left( {x,y} \right)E_x^i\left( {x,y} \right) \\ & + \xi _{E,y}^{r,t}\left( {x,y} \right)E_y^i\left( {x,y} \right)\\
& + \xi _{H,x}^{r,t}\left( {x,y} \right)H_x^i\left( {x,y} \right) \\ &+ \xi _{H,y}^{r,t}\left( {x,y} \right)H_y^i\left( {x,y} \right)
\end{split}
\end{equation}
\noindent where $\varsigma _{\left\{ {E,H} \right\},\left\{ {x,y} \right\}}^{r,t}\left( {x,y} \right)$ and $\xi _{\left\{ {E,H} \right\},\left\{ {x,y} \right\}}^{r,t}\left( {x,y} \right)$ are coefficients that are obtained from ${\textbf {\textsf{P}}}$ and that allow one to express each tangential component of the surface EM fields on the two sides of a metasurface structure as a linear combination of the tangential components of the incident surface EM fields.

\textbf{Examples: Perfect anomalous reflection and perfect anomalous transmission}. In order to better understand the general formalism in \eqref{Eq_Analysis__Final}, it is instructive to explicitly analyze the case studies of metasurface structures that realize perfect anomalous reflection and perfect anomalous transmission. To this end, we can consider the constitutive relations in \eqref{Eq_SusceptibilityExample}.
\begin{itemize}
\item \textbf{\textit{Perfect anomalous reflection}}. Under the assumption of perfect anomalous reflection, we have $E_x^t\left( {x,y} \right) = E_y^t\left( {x,y} \right) = 0$ and $H_x^t\left( {x,y} \right) = H_y^t\left( {x,y} \right) = 0$. As an example, let us consider that calculation of $E_y^r\left( {x,y} \right)$ and $H_x^r\left( {x,y} \right)$ in \eqref{Eq_SusceptibilityExample}. Similar analytical expressions can be obtained for $E_x^r\left( {x,y} \right)$ and $H_y^r\left( {x,y} \right)$. Therefore, $\chi _{{\rm{ee}}}^{yy}\left( {x,y} \right)$ and $\chi _{{\rm{mm}}}^{xx}\left( {x,y} \right)$ are the surface-averaged susceptibility functions of interest. Since, $\chi _{{\rm{ee}}}^{yy}\left( {x,y} \right)\chi _{{\rm{mm}}}^{xx}\left( {x,y} \right) =  - {4 \mathord{\left/ {\vphantom {4 {{k^2}}}} \right. \kern-\nulldelimiterspace} {{k^2}}}$, the two corresponding constitutive relations in \eqref{Eq_SusceptibilityExample} are not linearly independent. This implies that multiple EM field transformations fulfill the constitutive relations in \eqref{Eq_SusceptibilityExample}. As a case study, let us consider solutions of the kind $H_x^r\left( {x,y} \right) = e\left( {x,y} \right)E_y^r\left( {x,y} \right)$, where $e\left( {x,y} \right)$ is a given function that fulfills Maxwell's equations. For example, this solution is in agreement with \eqref{Eq_FullReflection_R} for plane waves. Under these assumptions, we obtain the following explicit expression of the reflected EM fields as a function of the incident EM fields:
\begin{equation} \label{Eq_Example__1}
\begin{split}
& E_y^r\left( {x,y} \right) = \frac{{ - \frac{{j\omega \varepsilon }}{2}\chi _{{\rm{ee}}}^{yy}\left( {x,y} \right)}}{{e\left( {x,y} \right) + \frac{{j\omega \varepsilon }}{2}\chi _{{\rm{ee}}}^{yy}\left( {x,y} \right)}}E_y^i\left( {x,y} \right)\\
& \hspace{1.3cm} - \frac{1}{{e\left( {x,y} \right) + \frac{{j\omega \varepsilon }}{2}\chi _{{\rm{ee}}}^{yy}\left( {x,y} \right)}}H_x^i\left( {x,y} \right)\\
&H_x^r\left( {x,y} \right) = \frac{{ - \frac{{j\omega \varepsilon }}{2}e\left( {x,y} \right)\chi _{{\rm{ee}}}^{yy}\left( {x,y} \right)}}{{e\left( {x,y} \right) + \frac{{j\omega \varepsilon }}{2}\chi _{{\rm{ee}}}^{yy}\left( {x,y} \right)}}E_y^i\left( {x,y} \right)\\
& \hspace{1.3cm} - \frac{{e\left( {x,y} \right)}}{{e\left( {x,y} \right) + \frac{{j\omega \varepsilon }}{2}\chi _{{\rm{ee}}}^{yy}\left( {x,y} \right)}}H_x^i\left( {x,y} \right)
\end{split}
\end{equation}
\item \textbf{\textit{Perfect anomalous transmission}}. Under the assumption of perfect anomalous transmission, we have $E_x^r\left( {x,y} \right) = E_y^r\left( {x,y} \right) = 0$ and $H_x^r\left( {x,y} \right) = H_y^r\left( {x,y} \right) = 0$. As an example, let us consider that calculation of $E_y^t\left( {x,y} \right)$ and $H_x^t\left( {x,y} \right)$. Similar analytical expressions can be obtained for $E_x^t\left( {x,y} \right)$ and $H_y^t\left( {x,y} \right)$. Therefore, $\chi _{{\rm{ee}}}^{yy}\left( {x,y} \right)$ and $\chi _{{\rm{mm}}}^{xx}\left( {x,y} \right)$ are the surface-averaged susceptibility functions of interest. Under these assumptions, we obtain the following explicit expression of the transmitted EM fields as a function of the incident EM fields:
\begin{equation} \label{Eq_Example__2}
\begin{split}
& E_y^t\left( {x,y} \right) = \frac{{1 + \frac{{j\omega \varepsilon }}{2}\chi _{{\rm{ee}}}^{yy}\left( {x,y} \right)\frac{{j\omega \mu }}{2}\chi _{{\rm{mm}}}^{xx}\left( {x,y} \right)}}{{1 - \frac{{j\omega \varepsilon }}{2}\chi _{{\rm{ee}}}^{yy}\left( {x,y} \right)\frac{{j\omega \mu }}{2}\chi _{{\rm{mm}}}^{xx}\left( {x,y} \right)}}E_y^i\left( {x,y} \right)\\
& \hspace{1.3cm} + 2\frac{{\frac{{j\omega \mu }}{2}\chi _{{\rm{mm}}}^{xx}\left( {x,y} \right)}}{{1 - \frac{{j\omega \varepsilon }}{2}\chi _{{\rm{ee}}}^{yy}\left( {x,y} \right)\frac{{j\omega \mu }}{2}\chi _{{\rm{mm}}}^{xx}\left( {x,y} \right)}}H_x^i\left( {x,y} \right)\\
& H_x^t\left( {x,y} \right) = 2\frac{{\frac{{j\omega \varepsilon }}{2}\chi _{{\rm{ee}}}^{yy}\left( {x,y} \right)}}{{1 - \frac{{j\omega \varepsilon }}{2}\chi _{{\rm{ee}}}^{yy}\left( {x,y} \right)\frac{{j\omega \mu }}{2}\chi _{{\rm{mm}}}^{xx}\left( {x,y} \right)}}E_y^i\left( {x,y} \right)\\
& \hspace{1.3cm} + \frac{{1 + \frac{{j\omega \varepsilon }}{2}\chi _{{\rm{ee}}}^{yy}\left( {x,y} \right)\frac{{j\omega \mu }}{2}\chi _{{\rm{mm}}}^{xx}\left( {x,y} \right)}}{{1 - \frac{{j\omega \varepsilon }}{2}\chi _{{\rm{ee}}}^{yy}\left( {x,y} \right)\frac{{j\omega \mu }}{2}\chi _{{\rm{mm}}}^{xx}\left( {x,y} \right)}}H_x^i\left( {x,y} \right)
\end{split}
\end{equation}
\end{itemize}

In both case studies, we note that the transmitted electric and magnetic fields are expressed as a linear combination of both the incident electric and magnetic fields. In general, in other words, cross-coupling among the EM fields exist.

\textbf{``Black box'' modeling of metasurfaces}. Based on these general considerations and specific examples, the main conclusion that can be drawn from \eqref{Eq_Analysis__Solution} is that a metasurface can be modeled as a ``black box'' that takes the electric and magnetic (surface-averaged) EM fields in ${\textbf {\textsf{F}}}$ at its input and yields as an output the (surface-averaged) EM fields in ${\textbf {\textsf{Z}}}$. The example of perfect anomalous reflector analyzed in the previous sub-section is a special case of the general analytical formulation in \eqref{Eq_Analysis__Solution}, in which many entries of the surface susceptibility matrices are equal to zero by design. From a conceptual point of view, $\varsigma _{\left\{  \cdot  \right\},\left\{  \cdot  \right\}}^{r, \cdot }\left( {x,y} \right)$ and $\xi _{\left\{  \cdot  \right\},\left\{  \cdot  \right\}}^{r, \cdot }\left( {x,y} \right)$ can be interpreted as surface-averaged reflection coefficients (i.e., ${R_{{\rm{EM}}}}\left( x \right)$) and $\varsigma _{\left\{  \cdot  \right\},\left\{  \cdot  \right\}}^{t, \cdot }\left( {x,y} \right)$ and $\xi _{\left\{  \cdot  \right\},\left\{  \cdot  \right\}}^{t, \cdot }\left( {x,y} \right)$ as surface-averaged transmission coefficients for specific pairs of tangential components of the surface-averaged electric and magnetic fields on the two sides of the metasurface. As remarked in the previous sub-section, these coefficients implicitly account for the mutual coupling among the unit cells of the metasurface, as well as for the inherent coupling between the phase response and the amplitude response of the surface susceptibility functions. It is important to point out, however, that the surface-averaged EM fields under consideration can, in principle, be measured at distances from the metasurface that are larger than of two-three times the size of a unit cell (see also Fig. \ref{Fig_18} discussed in further text). The EM fields in close vicinity of the metasurface may contain, on the other hand, significant evanescent-field content that is carefully optimized in the design of the super-cell structure. As far as the calculation of the EM fields scattered by a metasurface is concerned, however, it is sufficient to consider the surface-averaged EM fields that are obtained from the homogenized modeling of metasurfaces and that are formulated in terms of surface susceptibility functions or surface impedances. This modeling assumption is employed in the next sub-section to formulate the EM field scattered by a phase gradient metasurface.

\textbf{Relation between synthesis and analysis}. It is worth mentioning and emphasizing that the synthesis and analysis of metasurfaces are strongly intertwined. As far as the synthesis is concerned, the metasurface structure is designed and optimized based on some given EM fields and desired radio wave transformations. This implies that, once the metasurface structure is manufactured and deployed, it will realize the desired response if the impinging radio waves have the same properties as those assumed during the design and synthesis. Otherwise, the efficiency of the metasurface structure will be reduced and the actual response of the metasurface structure will be different from the desired response. Similar comments apply to the analysis and performance evaluation of wireless networks for a given metasurface structure. The surface susceptibility functions, in fact, correspond to incident, reflected, and transmitted EM fields with specific properties, e.g., the radio waves are plane, spherical, have a specified polarization, or simply are supposed to impinge upon the metasurface at only some angles of incidence. These constraints need to be taken into account when analyzing and optimizing RIS-empowered SREs.

\textbf{Beyond surface EM fields}. The general functional relations in \eqref{Eq_Analysis__Solution} constitute the departing point in order to compute the electric and magnetic fields at any point of a given volume, e.g., far away from a metasurface, where an RIS may be deployed. This issue is investigated in the next sub-section.

\subsection{On Modeling Radio Wave Propagation in the Presence of Metasurfaces} \label{WavePropagation}
\textbf{From surface EM fields to EM fields in volumes}. In the previous sub-section, we have shown that a metasurface can be described in terms of surface susceptibility functions and that the electric and magnetic (surface) fields scattered by the metasurface can be formulated in an algebraic form under sufficiently general modeling assumptions. In this sub-section, we show that the knowledge of the EM field at the two sides of a metasurface, i.e., only at $z=0^+$ and  $z=0^-$, is sufficient for computing the EM field at any point of a given volume. 

\begin{figure}[!t]
\begin{centering}
\includegraphics[width=\columnwidth]{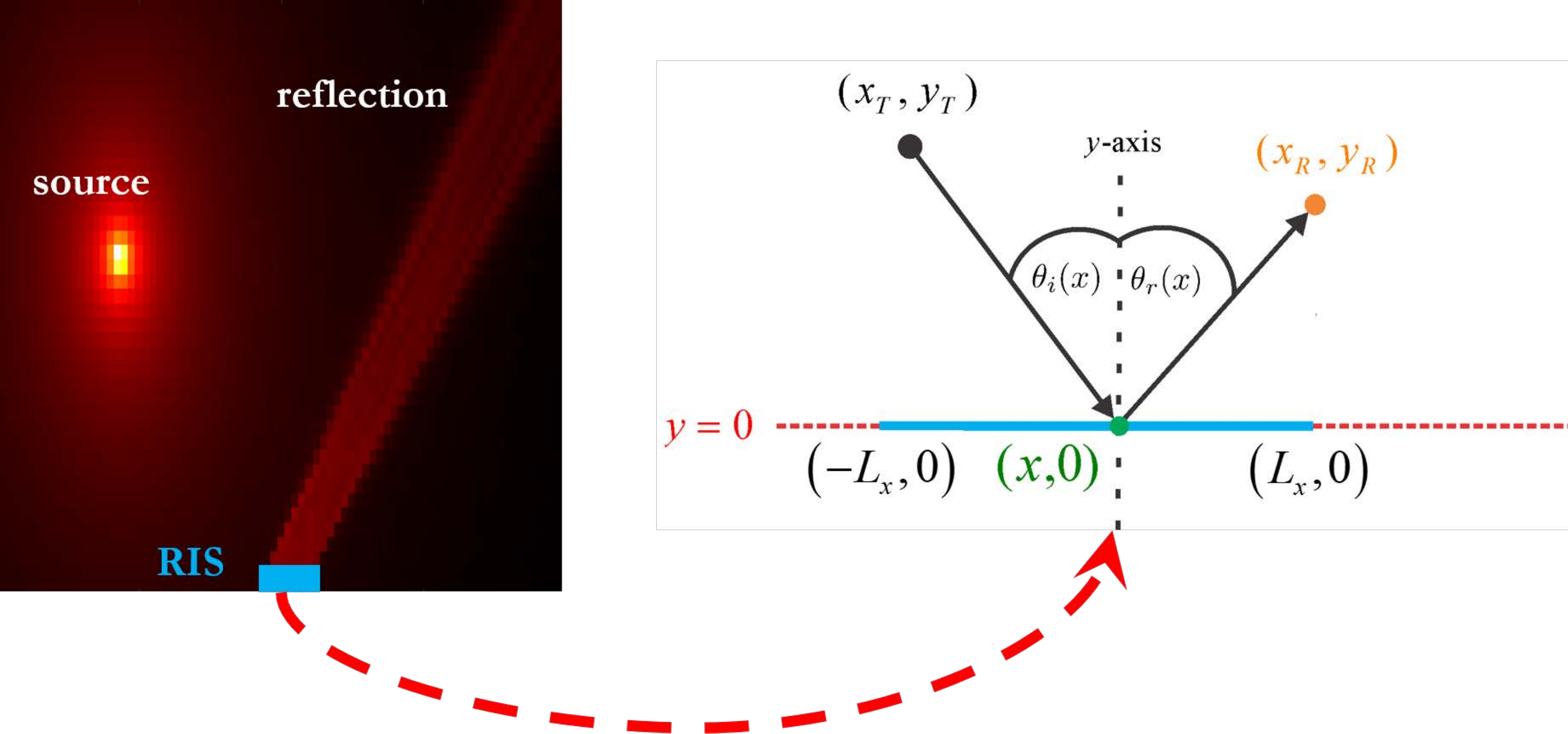}
\caption{System model for modeling the propagation of radio waves in the presence of metasurfaces.}
\label{Fig_17}
\end{centering} 
\end{figure}
\textbf{Reference system model}. We summarize, in particular, the analytical approach recently reported in \cite{MDR_SPAWC2020}. We highlight only the most important aspects of \cite{MDR_SPAWC2020}, some of which are not extensively elaborated in the original paper. For simplicity, in addition, we restrict the analysis to a one-dimensional metasurface and to a two-dimensional space (i.e., the plane). The general case study that encompasses two-dimensional metasurfaces in a three-dimensional space can be found in \cite{MDR_PathLossFadil}. For ease of illustration, the system model under analysis is sketched in Fig. \ref{Fig_17}, which depicts an infinitesimal small source that radiates isotropically in the plane and a one-dimensional metasurface. Further details about the system model are given in further text.

\begin{figure}[!t]
\begin{centering}
\includegraphics[width=\columnwidth]{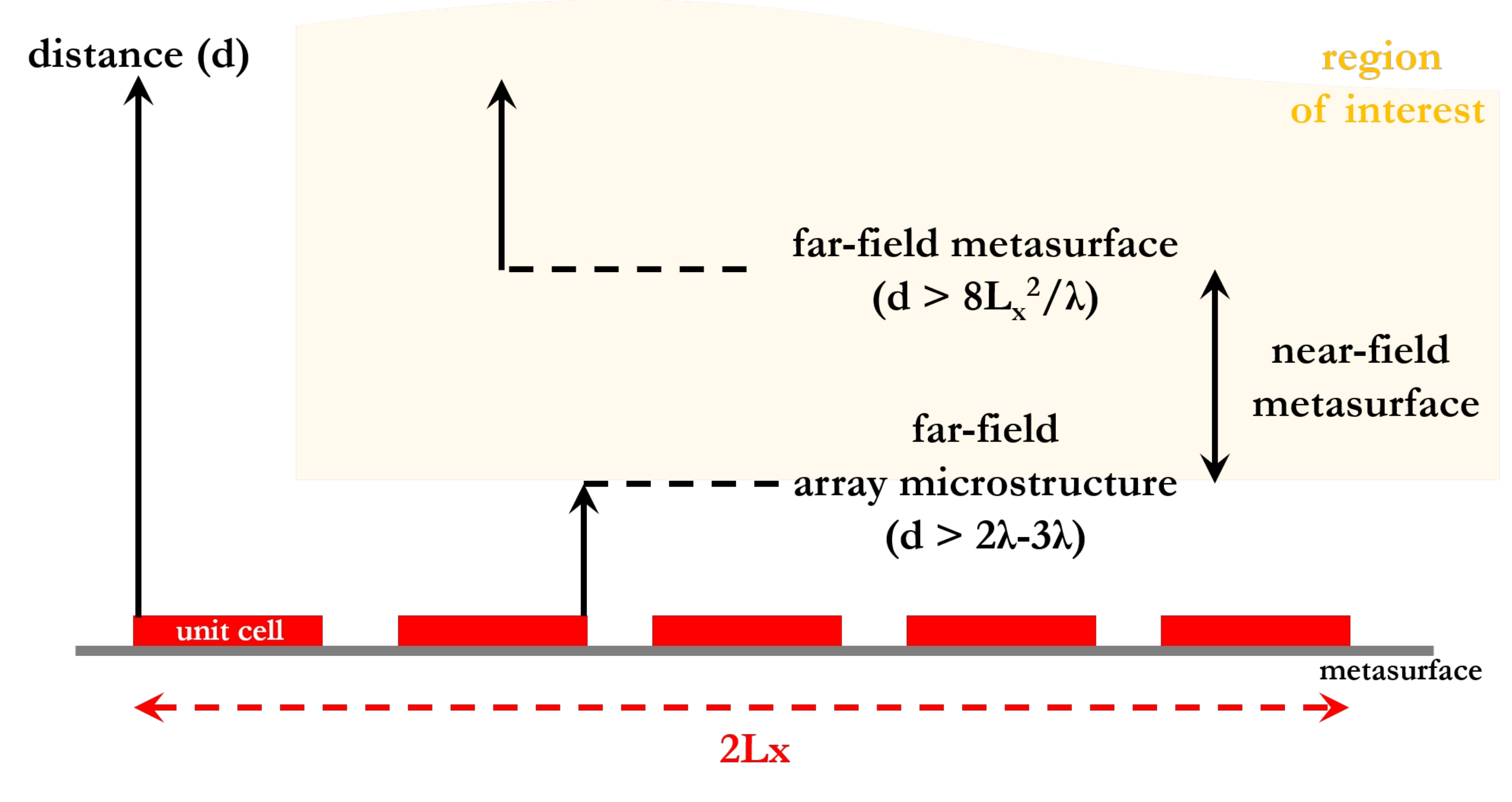}
\caption{Near-field vs. far-field of a metasurface and far-field of the array microstructure. The far-field boundary of the metasurface is obtained from the Fraunhofer distance as illustrated in Figs. \ref{Fig_7} and \ref{Fig_8}.}
\label{Fig_18}
\end{centering} 
\end{figure}

\textbf{Reference operating regimes}. Before introducing the main recipe for obtaining the EM field scattered by a metasurface, we feel important to introduce some relevant operating regimes and concepts. To help us clarify them, the illustrations in Figs. \ref{Fig_18} and \ref{Fig_19} are utilized. 
\begin{itemize}
\item \textit{\textbf{Near-field vs. far-field}}. As depicted in Fig. \ref{Fig_4}, a metasurface is made of unit cells whose size is usually smaller than the wavelength of the impinging radio waves. At distances from the metasurface that are larger than two-three times the size of a unit cell, which is still small compared with the wavelength, the EM fields of the evanescent surface modes of the metasurface structure, which may be present, can be considered to be negligible already. From the macroscopic point of view, in particular, the radio waves impinging upon a metasurface can be assumed to be locally plane waves regardless of the actual characteristics of the source (e.g., the source may not necessarily emit plane waves). In simple terms, this implies that, e.g., a spherical wavefront impinging upon a unit cell can be approximated by its tangent provided that the size of the unit cell is sufficiently small that the phase of the impinging radio wave does not change along the unit cell. Since we are interested in EM fields that are evaluated at transmission distances from the metasurface that are larger than two-three times the size of a unit cell, we can ignore the evanescent fields that may be present along the metasurface structure, and we can utilize the surface-averaged reflection coefficient for locally plane waves. In the present paper, this region is referred to as ``the far-field of the array microstructure''. Physically, as mentioned, this region corresponds to distances from the metasurface structure at which the evanescent EM fields are negligible. If we consider the example of an anomalous perfect reflector, this implies that the EM field scattered by the metasurface is approximately equal to the sum of two plane waves (as modeled in the previous example). At these distances, one can define and use the surface-averaged reflection coefficient ${R_{{\rm{EM}}}}\left( x \right)$ as defined in the previous example. As far as the whole metasurface structure is concerned, on the other hand, the operating regime is not unique. The transverse size of a metasurface can, in fact, be tens or hundreds (or more) times larger than the wavelength of the impinging radio waves. This implies that it is not possible to ignore, in general, the phase change of the impinging radio waves along the entire metasurface. Depending on the actual network geometry, the size of the metasurface, the operating wavelength, therefore, a metasurface can operate either in the near-field regime or in the far-field regime. For clarity, these operating regimes are conceptually illustrated in Fig. \ref{Fig_18}.
\begin{figure}[!t]
\begin{centering}
\includegraphics[width=\columnwidth]{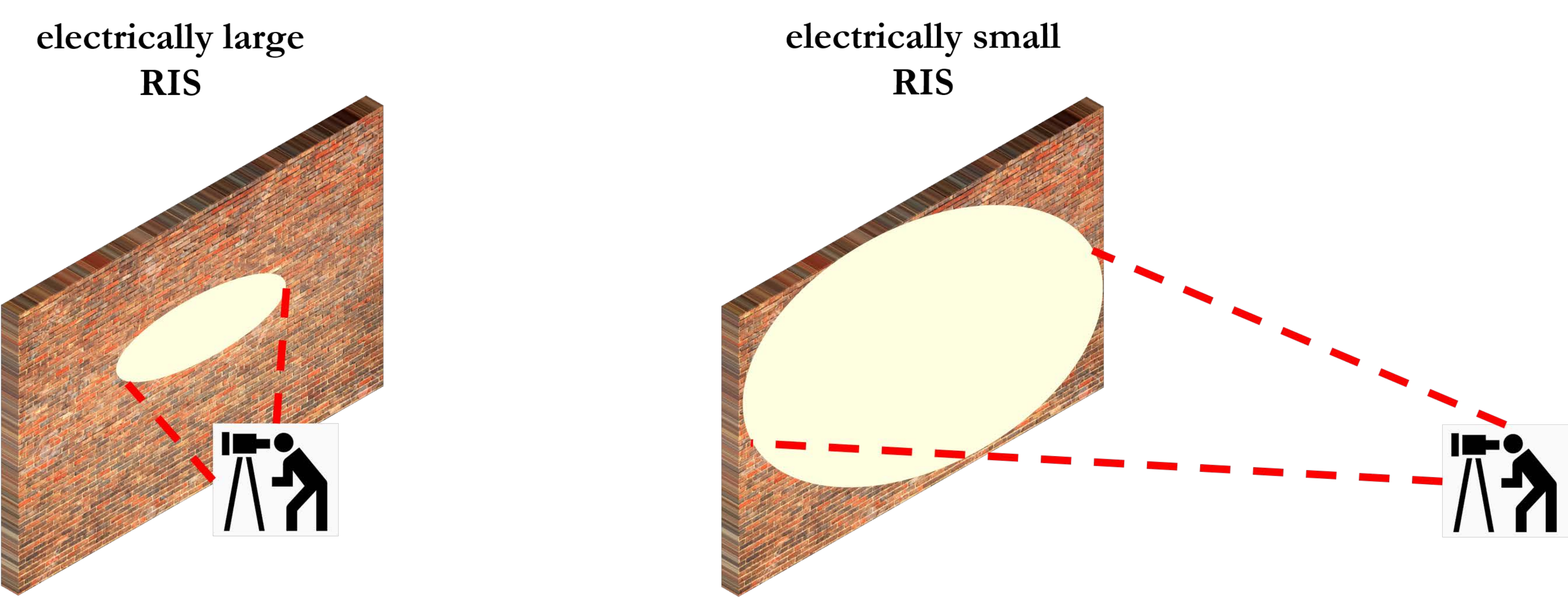}
\caption{Electrically large vs. electrically small metasurfaces.}
\label{Fig_19}
\end{centering} 
\end{figure}
\item \textit{\textbf{Electrically large and electrically small metasurfaces}}. As mentioned, we cannot assume, a priori, that a metasurface operates in the far-field regime. This implies that the signal models and the corresponding analytical frameworks need to be sufficiently general to take this into account. This can be considered to be a not so usual situation in mainstream wireless communications, where the far-field operating regime is usually \textit{de facto} implied, even though often not explicitly stated. In the present paper, we do not attempt to give a formal definition of near-field and far-field for a metasurface as a whole. As an illustrative example, and based on the conventional definition for active antennas, Figs. \ref{Fig_7} and \ref{Fig_8} report the boundary of the radiative near-field and far-field as computed from the Fraunhofer distance. We provide, on the other hand, arguments to distinguish these two operating regimes from a practical standpoint. Instead of using the terms near-field and far-field, we introduce the concepts of electrically large and electrically small metasurfaces. For clarity, these two concepts are illustrated in Fig. \ref{Fig_19}. A metasurface is referred to as electrically large if its geometric size is large enough as compared with the wavelength of the radio waves, and as compared with the transmission distance from the source to the metasurface and from the metasurface to the observation point. If this is not the case, then the metasurface is referred to as electrically small. The term ``large enough'' is deliberately left a bit vague, since a formal definition would depend on a large number of system parameters. In simple terms, based on Fig. \ref{Fig_19}, a metasurface can be considered to be electrically large if the size of a metasurface and the transmission distances are such that one sees the metasurface as being infinitely large, and, therefore, one cannot see the edges of the metasurface structure. If the opposite holds true, a metasurface can be considered to be electrically small. By considering a metasurface of a given and fixed size, with a similar line of thought, a metasurface can likely be considered to be electrically large if the transmission distances are sufficiently short. If the transmission distances are, on the other hand, sufficiently long, then a metasurface can likely be considered to be electrically small. In further text, we show that the scaling laws as a function of key system parameters, e.g., the size of the metasurface and the transmission distances, are usually different for electrically large and electrically small metasurfaces.
\end{itemize}
\begin{figure}[!t]
\begin{centering}
\includegraphics[width=\columnwidth]{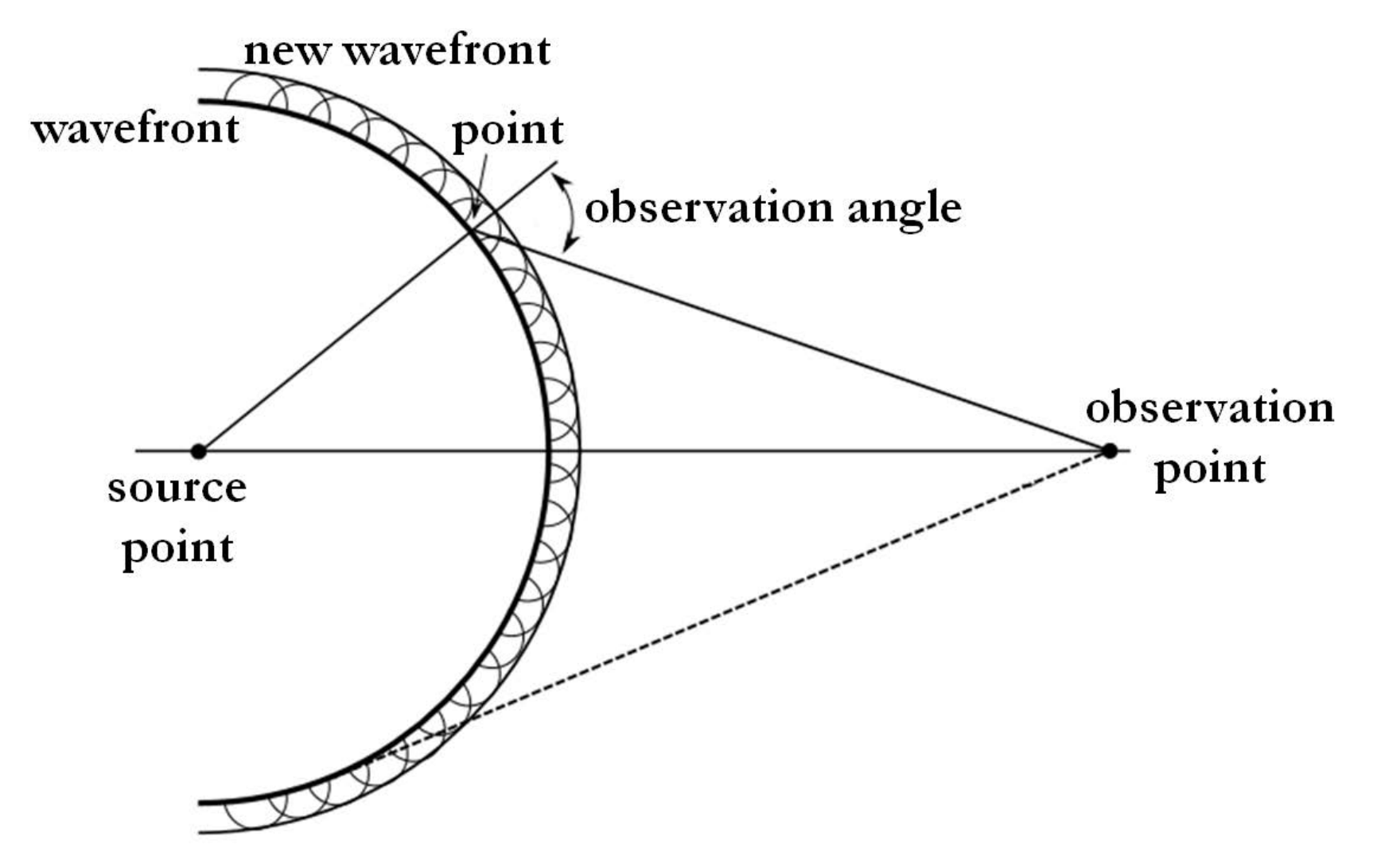}
\caption{The Huygens-Fresnel principle.}
\label{Fig_20}
\end{centering} 
\end{figure}
\textbf{Theory of diffraction and Huygens-Fresnel principle}. With these preliminary definitions at hand, we can now discuss how to appropriately model the propagation of radio waves in the presence of a metasurface. As recently proved in \cite{MDR_SPAWC2020} and \cite{MDR_PathLossFadil}, the EM field at any point of a volume and in the presence of a metasurface can be obtained by invoking the theory of diffraction and the Huygens-Fresnel principle. In general terms, the theory of diffraction provides one with the mathematical tools for modeling the bending of the radio waves when they encounter an object or a discontinuity, which in our case is a metasurface characterized by its specific effective parameters. In classical physics, more precisely, diffraction-based phenomena, including the reflection and transmission from objects, are described by the Huygens-Fresnel principle. According to this principle, every point of a propagating wavefront is viewed as a collection of individual spherical wavelets. In particular, every point on a wavefront is itself the source of spherical wavelets, and the secondary wavelets emanated from different points interfere with each other. The sum of the emanated spherical wavelets forms a new wavefront. This concept is sketched in Fig. \ref{Fig_20}.

\textbf{Green's theorem}. When the radio waves emitted by a source impinge upon a metasurface, based on the Huygens-Fresnel principle, the wavefront of the impinging radio wave in correspondence of a metasurface, i.e., at each point of a metasurface, becomes the source of secondary wavelets that determine the EM field at any point of the volume of interest. Based on the divergence theorem (or, its more general formulation, the Green theorem \cite{Green_1828}), which states that the surface integral of a vector field over a closed surface is equal to the volume integral of the divergence of the vector field over the region inside the surface, the EM field of interest is uniquely determined by the EM fields on the two sides of a metasurface, i.e., the surface EM fields. The readers are referred to \cite{MDR_SPAWC2020}, \cite{MDR_PathLossFadil} for the analytical details. As far as the present paper is concerned, it is sufficient to understand that this implies that the EM field at any point of a volume can be obtained from the GSTCs computed in the previous sub-section, which, in fact, formulate the tangential components of the EM (surface) fields at the two sides of a metasurface. Formally, this is reported in \eqref{Eq_Analysis__Final} for a general metasurface. 

\textbf{The atomic case study}. In the rest of the present paper, we assume, for simplicity, that the metasurface under analysis is characterized by a single non-zero reflection coefficient $\varsigma _{\left\{  \cdot  \right\},\left\{  \cdot  \right\}}^{r, \cdot }\left( {x,y} \right) = \Delta \left( {x,y} \right)\exp \left( {jk\Upsilon \left( {x,y} \right)} \right)$, where $\Delta \left( {x,y} \right) = \left| {\varsigma _{\left\{  \cdot  \right\},\left\{  \cdot  \right\}}^{r, \cdot }\left( {x,y} \right)} \right|$ and $k\Upsilon \left( {x,y} \right) = \arg \left( {\varsigma _{\left\{  \cdot  \right\},\left\{  \cdot  \right\}}^{r, \cdot }\left( {x,y} \right)} \right)$ are the amplitude and the phase of $\varsigma _{\left\{  \cdot  \right\},\left\{  \cdot  \right\}}^{r, \cdot }\left( {x,y} \right)$. Also, we assume $\Delta \left( {x,y} \right) = \Delta \left( {x} \right)$ and $\Upsilon \left( {x,y} \right) = \Upsilon \left( {x} \right)$. If other coefficients in \eqref{Eq_Analysis__Final} are not equal to zero, the electric or magnetic tangential components of the field of interest can be formulated as a linear combination of integrals similar to those reported in further text. It is worth emphasizing that the term surface-averaged reflection coefficient is always referred to ${R_{{\rm{EM}}}}\left( x \right)$ (not to ${R_{{\rm{Z}}}}\left( x \right)$) if one is interested in analyzing the actual EM fields reflected from an RIS.

\subsubsection{Analytical Formulation of the EM Field} \label{FieldFormulation}
\textbf{An approach based on Green's functions}. By capitalizing on the Huygens-Fresnel principle and by assuming an infinitesimally small source whose complex amplitude is the Green function in a two-dimensional plane, the authors of \cite{MDR_SPAWC2020} have given the following expression for a generic component of the EM field reflected by a metasurface located at $y=0$ and $L_x \le x \le L_x$:
\begin{equation} \label{Eq_Field_Integral}
\begin{split}
& Z\left( {{x_R},{y_R}} \right) \approx \frac{1}{{8\pi }}\int\nolimits_{ - {L_x}}^{ + {L_x}} {{\mathcal{I}}\left( x \right)\exp \left( { - jk{\mathcal{P}}\left( x \right)} \right)dx} \\
& {\mathcal{P}}\left( x \right) = {d_T}\left( x \right) + {d_R}\left( x \right) - \Upsilon \left( x \right)\\
& {\mathcal{I}}\left( x \right) = \frac{{\Delta \left( x \right)}}{{\sqrt {{d_T}\left( x \right){d_R}\left( x \right)} }}\left( {\frac{{{y_T}}}{{{d_T}\left( x \right)}} + \frac{{{y_R}}}{{{d_R}\left( x \right)}}} \right)
\end{split}
\end{equation}
\noindent where $2 L_x$ is the size (length) of the metasurface, $\left( {{x_T},{y_T}} \right)$ is the location of the source, $\left( {{x_R},{y_R}} \right)$ is the location of the observation point (e.g., where an intended receiver is located), and ${{d_T}\left( x \right)}$ and ${{d_R}\left( x \right)}$ are the transmission distances from the source to the point $x$ of the metasurface and from the point $x$ of the metasurface to the observation point, respectively:
\begin{equation} \label{Eq_Distances}
\begin{split}
& {d_T}\left( x \right) = \sqrt {{{\left( {{x_T} - x} \right)}^2} + y_T^2} \\
& {d_R}\left( x \right) = \sqrt {{{\left( {{x_R} - x} \right)}^2} + y_R^2} 
\end{split}
\end{equation}

It is worth mentioning that the integral (instead of a sum of EM fields scattered by individual scatterers) formulation in \eqref{Eq_Field_Integral} is a consequence of employing a homogenized approach for modeling metasurfaces, which is consistent with the sub-wavelength inter-distance between adjacent unit cells and the GSTCs introduced in the previous section.

\textbf{Interpretation of the reflected field}. The analytical formulation in \eqref{Eq_Field_Integral}, based on the Huygens-Fresnel principle, can be interpreted by rewriting the integrand function as follows: 
\begin{equation} \label{Eq_Interpretation}
\begin{split}
{\mathcal{I}}\left( x \right)\exp \left( { - jk{\mathcal{P}}\left( x \right)} \right) & = \frac{{\exp \left( { - jk{d_T}\left( x \right)} \right)}}{{\sqrt {{d_T}\left( x \right)} }}\\
& \times \Delta \left( x \right)\exp \left( {jk\Upsilon \left( x \right)} \right)\\
& \times \frac{{\exp \left( { - jk{d_R}\left( x \right)} \right)}}{{\sqrt {{d_R}\left( x \right)} }}\\
& \times \left( {\frac{{{y_T}}}{{{d_T}\left( x \right)}} + \frac{{{y_R}}}{{{d_R}\left( x \right)}}} \right)
\end{split}
\end{equation}
\noindent where: (i) the first term corresponds to the radio wave impinging upon the metasurface (i.e., the Green function); (ii) the second term corresponds to the surface response of the metasurface that is obtained from the homogenized model; (iii) the third term corresponds to the radio wave that is re-emitted by the metasurface upon reception of the impinging radio wave (i.e., the Green function); and (iv) the fourth term represents the sum of the cosines of the inclination angles under which the metasurface is viewed by the source and by the observation point. Finally, the integral originates from the fact that a metasurface can be homogeneized and by applying the divergence theorem, which, in simple terms, allows one to formulate a vector field in a given volume as a function of the corresponding surface integral. It is worth mentioning that ${\mathcal{I}}\left( x \right)$ depends on the square root of the transmission distances because a bi-dimensional system model is considered, and the amplitude of the corresponding Green function, in the absence of a metasurface, decays with the square root of the transmission distance.

\subsubsection{Configuration of the Metasurface} \label{Configuration}
\textbf{Beamsteering vs. focusing}. From \eqref{Eq_Field_Integral}, it is apparent that a metasurface can be appropriately configured and can perform different functions depending on the specific choice of $\Delta \left( x \right)$ and ${\Upsilon \left( x \right)}$. As illustrative examples, we consider three case studies that have important applications in wireless communications: (i) specular reflector (or specular beamsteering); (ii) anomalous reflector (or anomalous beamsteering); and (iii) focusing lens (or beamforming). We assume, for simplicity, $\Delta \left( x \right) =1$, which corresponds to a benchmark for the three configurations of the metasurface under analysis.
\begin{itemize}
\item \textit{\textbf{Specular reflector}}. The specified wave transformation that the metasurface needs to realize is reflection by modifying the phase of the impinging radio wave. This operation can be obtained by using a uniform metasurface whose phase response per unit length is as follows: 
\begin{equation} \label{Eq_Specular}
\Upsilon \left( x \right) = {\phi _0}/k
\end{equation}
\noindent where $\phi _0 \in [0,2\pi)$ is a fixed phase shift. If $\phi _0 =\pi$, for example, a perfect electric conductor (PEC) is obtained, whose reflection coefficient is $-1$.
\item \textit{\textbf{Anomalous reflector}}. The specified wave transformation that the metasurface needs to realize is reflection by modifying the angle of reflection and the phase of the impinging radio wave. This operation can be obtained by using a phase gradient metasurface whose phase response per unit length is as follows:
\begin{equation} \label{Eq_Anomalous}
\Upsilon \left( x \right) = \left( {{{\bar \phi }_T} - {{\bar \phi }_R}} \right)x + {\phi _0}/k
\end{equation}
\noindent where $\phi _0 \in [0,2\pi)$ is a fixed phase shift, and ${{\bar \phi }_T}$ and ${{\bar \phi }_R}$ are two parameters to be optimized in order to steer the impinging radio wave towards a specified angle of reflection. In \cite{MDR_SPAWC2020}, for example, ${{\bar \phi }_T}$ and ${{\bar \phi }_R}$ are optimized as a function of the angles under which the source and the receiver view the center of the metasurface. It is worth mentioning that the design of the phase gradient in \eqref{Eq_Anomalous} can be considered to be an approximation of the optimal reflector in \eqref{Eq_FullReflection_ExplicitFormulaReflection} that originates from ray optics arguments. The numerical results in \cite{MDR_SPAWC2020} show, however, that the desired function is appropriately realized, even though the efficiency of the metasurface, i.e., the ratio between the reflected and incident powers, may not necessarily be optimized.
\item \textit{\textbf{Focusing lens}}. The specified wave transformation that the metasurface needs to realize is focusing the impinging radio wave towards a specified location. This operation can be obtained by using a phase gradient metasurface whose phase response per unit length is as follows:
\begin{equation} \label{Eq_Focusing}
\Upsilon \left( x \right) = \sqrt {{{\left( {x - {x_T}} \right)}^2} + y_T^2}  + \sqrt {{{\left( {x - {{\bar x}_R}} \right)}^2} + \bar y_R^2}
\end{equation}
\noindent where $\left( {{{\bar x}_R},{{\bar y}_R}} \right)$ is the location where the radio wave needs to be focused.
\end{itemize}
\begin{figure}[!t]
\begin{centering}
\includegraphics[width=\columnwidth]{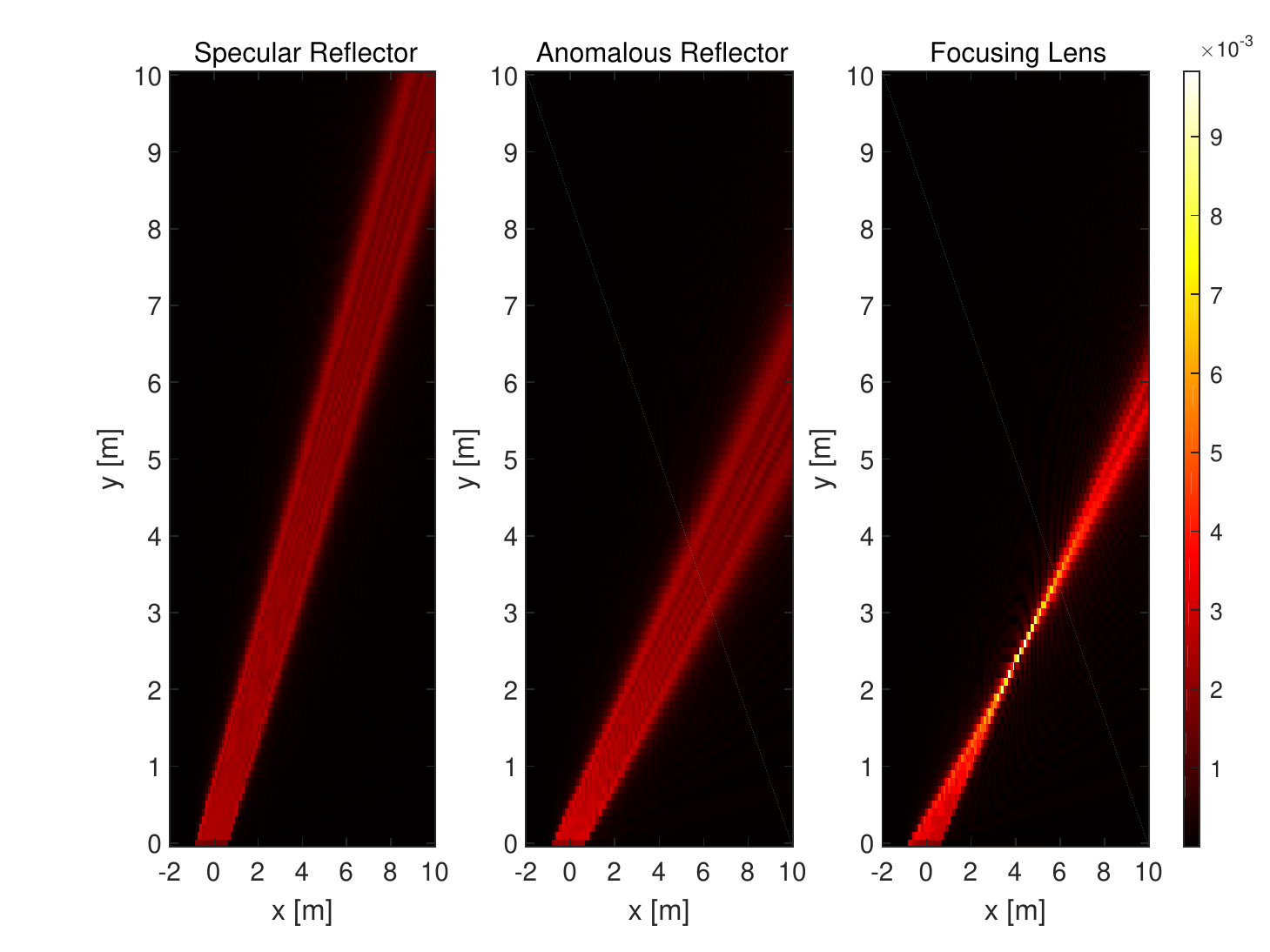}
\caption{Comparison among specular reflector, anomalous reflector, and focusing lens (reproduced from \cite{MDR_SPAWC2020}). The source is not reported for clarity.}
\label{Fig_21}
\end{centering} 
\end{figure}
\textbf{Optimization criteria for beamsteering and focusing}. We observe, therefore, that the phase response of the metasurface structure under analysis is quite different depending on the specified function to be realized. As an example, we can analyze the differences between the two phase responses in \eqref{Eq_Anomalous} and \eqref{Eq_Focusing}. In simple terms, the criterion employed for obtaining \eqref{Eq_Anomalous} consists of ensuring that, as a first-order approximation, the first-order derivative of $\mathcal{P}(x)$ in \eqref{Eq_Field_Integral} is equal to zero in correspondence of those points, $x$, of the metasurface for which the angle of reflection is the desired one (assuming that the angle of incidence is given and fixed). The criterion employed for obtaining \eqref{Eq_Focusing} consists, on the other hand, of ensuring that $\mathcal{P}(x)$ in \eqref{Eq_Field_Integral} is equal to zero in correspondence of the location where the radio wave needs to be focused to. By using this criterion, in fact, the complex terms as a function of $x$ in \eqref{Eq_Field_Integral} are all co-phased at the location of interest, thus yielding the desired focusing effect. In conclusion, we can state that \eqref{Eq_Anomalous} is chosen by considering the stationary points of $\mathcal{P}(x)$, while \eqref{Eq_Focusing} is chosen by considering the zeros of $\mathcal{P}(x)$. An example of the different behavior of the considered metasurface structure based on the configurations specified by the phase responses in \eqref{Eq_Specular}, \eqref{Eq_Anomalous}, and \eqref{Eq_Focusing} is illustrated in Fig. \ref{Fig_21}. 

\textbf{Comparing metasurfaces obtained from different optimization criteria}. Based on these considerations, we conclude that, in general, the performance of a metasurface is different depending on the criterion employed for optimizing $\mathcal{P}(x)$. This implies that it may not be relevant or interesting to first optimize a metasurface for operating as, e.g., a focusing lens, and then to study when and if the same metasurface can operate as an anomalous reflector. This is because the two metasurfaces are designed based on different criteria and in order to perform different functions. Furthermore, the comparison may not necessarily be fair since different designs usually rely on a different amount of channel state information and on different implementation complexities. As an example, let us consider the realization and practical operation of a metasurface with the phase response in \eqref{Eq_Anomalous} and a metasurface with the phase response in \eqref{Eq_Focusing}. In the first case,  one needs to know only the angle of incidence and the angle of reflection for optimizing the phase response. This implies that the design needs not to be changed if, e.g., a receiver moves along the same direction. This could be true if, e.g., a user moves along a street or a hallway. In the second case, on the other hand, the optimization of \eqref{Eq_Focusing} necessitates the exact location of the receiver. This implies that the metasurface needs to be reconfigured every time that the user moves, even though the direction of motion is kept the same. This imposes some practical constraints on the rate at which a metasurface needs to be reconfigured, and the rate at which the environmental information needs to be estimated in order to achieve good performance.

\subsubsection{Electrically Large vs. Electrically Small Regimes} \label{largeVSsmall}
\textbf{Scaling laws and performance trends}. In this sub-section, we analyze in further detail the conditions under which a metasurface can be assumed to be electrically large and electrically small. In particular, we discuss the different scaling laws, as a function of many system parameters, that emerge in these two operating regimes. For simplicity, we consider only the configuration of a metasurface that operates as a specular reflector. This is the simplest case study, which allows us to draw several interesting and general conclusions on the performance trends that can be expected, and, at the same time, allows us to validate the consistency of the analytical formulation in \eqref{Eq_Field_Integral} in a canonical case study. The interested readers can find a similar analysis for metasurfaces that operate as an anomalous reflector and as a focusing lens in \cite{MDR_SPAWC2020} and \cite{MDR_PathLossFadil}.

\textbf{Electrically large regime}. If the metasurface structure can be assumed to be electrically large, e.g., given its size and operating frequency, the transmission distances are sufficiently short that the metasurface is viewed as infinitely large, the intensity of the field in \eqref{Eq_Field_Integral} can be approximated as follows:
\begin{equation} \label{Eq_Large}
\left| {{Z}\left( {{x_R},{y_R}} \right)} \right| \approx \frac{1}{{\sqrt {8\pi k} }}\frac{1}{{\sqrt {{d_T}\left( {{x_s}} \right) + {d_R}\left( {{x_s}} \right)} }}
\end{equation}
\noindent where ${x_s} \in \left[ {-L_x,L_x} \right]$ is the unique solution, if it exists, of the equation:
\begin{equation} \label{Eq_Xs}
\frac{{{x_s} - {x_T}}}{{{d_T}\left( {{x_s}} \right)}} - \frac{{{x_R} - {x_s}}}{{{d_R}\left( {{x_s}} \right)}} = 0
\end{equation}
\noindent which corresponds to the stationary point of $\mathcal{P}(x)$.

\textbf{Electrically small regime}. If, on the other hand, the metasurface structure can be assumed to be electrically small, e.g., given its size and operating frequency, the transmission distances are sufficiently long that the metasurface is viewed as a small scatterer, the intensity of the field in \eqref{Eq_Field_Integral} can be approximated as follows:
\begin{equation} \label{Eq_Small}
\begin{split}
\left| {{Z}\left( {{x_R},{y_R}} \right)} \right| &\approx \frac{L_x}{{4\pi }}\left| \frac{{\cos \left( {{\theta _{T0}}} \right) + \cos \left( {{\theta _{R0}}} \right)}}{{\sqrt {{d_{T0}}{d_{R0}}} }}\right|\\
& \times \left|\frac{{\sin \left( {k L_x\left( {\sin \left( {{\theta _{T0}}} \right) - \sin \left( {{\theta _{R0}}} \right)} \right)} \right)}}{{k L_x\left( {\sin \left( {{\theta _{T0}}} \right) - \sin \left( {{\theta _{R0}}} \right)} \right)}}\right|
\end{split}
\end{equation}
\noindent where, for $Q=\{T, R \}$, ${d_{Q0}} = \sqrt {x_Q^2 + y_Q^2}$ is the transmission distance between the source and the center of the metasurface (i.e., the origin) if $Q=T$ and is the transmission distance between the center of the metasurface and the observation point if $Q=R$, and $\cos \left( {{\theta _{Q0}}} \right) = {{{y_Q}} \mathord{\left/ {\vphantom {{{y_Q}} {{d_{Q0}}}}} \right. \kern-\nulldelimiterspace} {{d_{Q0}}}}$ is the cosine of the angle that the direction of propagation of the incident (if $Q=T$) and reflected (if $Q=R$) radio wave forms with the normal to the metasurface, i.e., the $z$-axis. 

\begin{figure}[!t]
\begin{centering}
\includegraphics[width=\columnwidth]{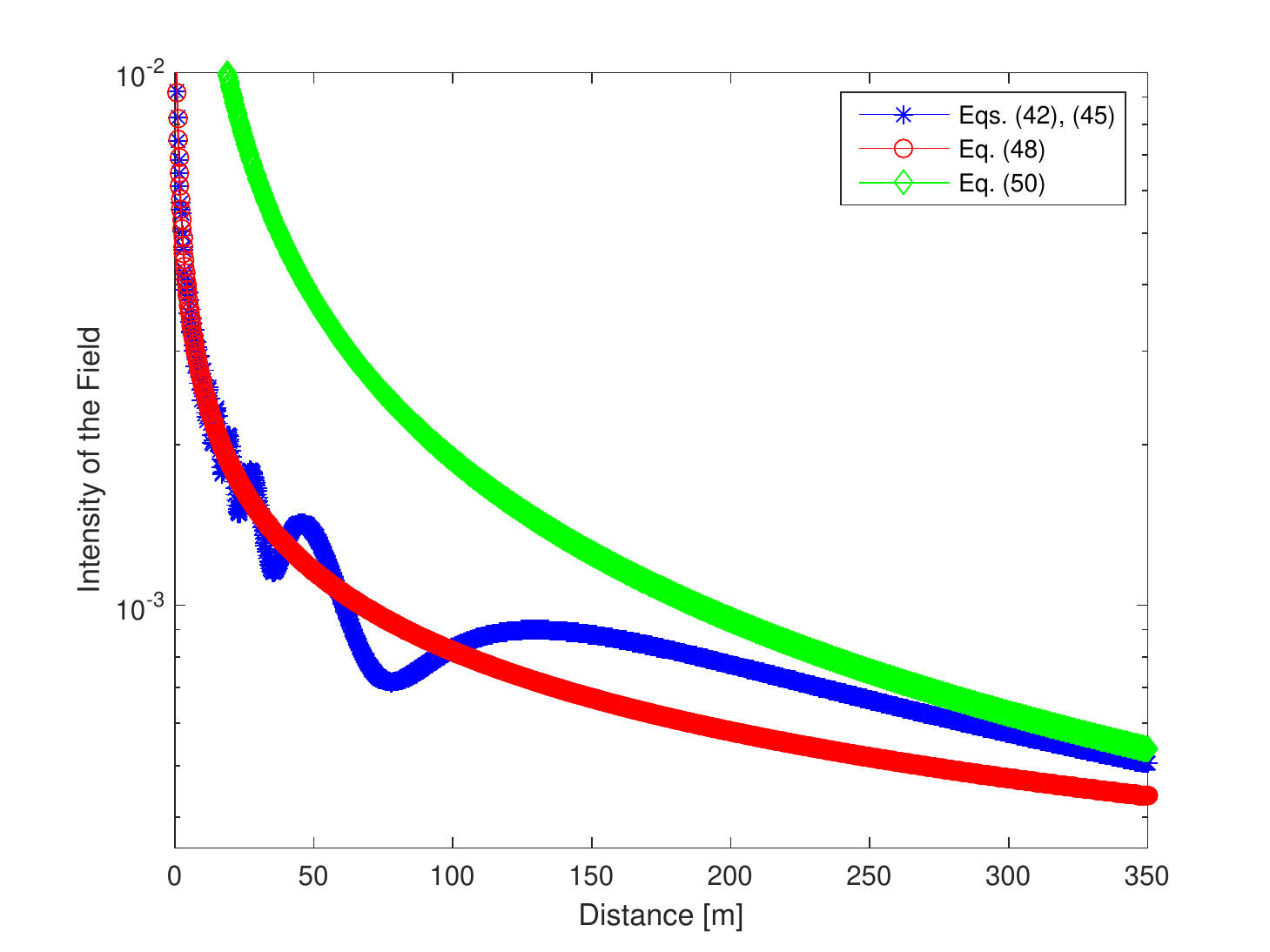}
\caption{Intensity of the field as a function of the transmission distance (reproduced from \cite{MDR_SPAWC2020}). The distance is the sum of the distances between the source-to-RIS and RIS-to-destination, which are assumed to be equal.}
\label{Fig_22}
\end{centering} 
\end{figure}
\textbf{Comparing the two regimes}. In simple terms, the main difference between \eqref{Eq_Large} and \eqref{Eq_Small} is that the latter equation is obtained by exploiting the parallel rays approximation of the radio wave emitted by the source thanks to the long transmission/observation distances. In the electrically small regime, in other words, it is possible to capitalize on the plane wave approximation of the impinging radio waves for the whole metasurface. The accuracy of the approximations in \eqref{Eq_Large} and \eqref{Eq_Small} is analyzed in Fig. \ref{Fig_22} for a metasurface that operates in the millimeter-wave frequency band (28 GHz) and whose size is 1.5 meters. In the considered case study, the metasurface structure is viewed as electrically large for distances of the order of tens of meters. The range of distances for which the approximation in \eqref{Eq_Large} is accurate depends, however, among other parameters, on the size of the metasurface structure and the operating frequency. In general, the larger the size of the metasurface structure  is and the higher the operating frequency is, the more accurate the approximation in \eqref{Eq_Large} becomes, i.e., it can be used for longer transmission distances.

\textbf{Design insights and performance trends}. From the obtained closed-form analytical expressions of the intensity of the EM field in \eqref{Eq_Large} and \eqref{Eq_Small}, several important conclusions can be drawn. The most important are summarized as follows.
\begin{itemize}
\item \textit{\textbf{Specular mirror}}. In the electrically large operating regime, (e.g., for short transmission distances), the metasurface structure behaves as a specular mirror. From \eqref{Eq_Large}, in particular, we observe that the (end-to-end, i.e., from the source to the observation point) intensity of the field reflected from the metasurface decays as a function of the reciprocal of the square root of the sum of the transmission distances from the source to the stationary point in \eqref{Eq_Xs} and from the stationary point in \eqref{Eq_Xs} to the observation point. The presence of the square root originates from the assumption of two-dimensional space as elaborated in previous text. 
\item \textit{\textbf{Method of images}}. The scaling law as a function of the transmission distances in \eqref{Eq_Large} is in agreement with the method of (mirror) images, which justifies the denomination ``specular mirror''. In particular,  \eqref{Eq_Large} can be regarded as an approximation of \eqref{Eq_Field_Integral} under the assumption of geometric optics propagation. More precisely, \eqref{Eq_Large} unveils that the intensity of the field is approximately the same as the one obtained from a single ray (i.e., the direction of propagation of the wavefront of the EM wave) that corresponds to the piece-wise linear function made of the line segment the connects the source with with stationary point in \eqref{Eq_Xs} and the line segment the connects the stationary point in \eqref{Eq_Xs} with the observation point. Therefore, the stationary point in \eqref{Eq_Xs} is often referred to as the reflection point for a given source and observation point.
\item \textit{\textbf{Law of reflection}}. Given a point $x$ of the metasurface structure, the angle of incidence, ${\theta _T}\left( x \right)$, and the angle of reflection, ${\theta _R}\left( x \right)$, fulfill the identities ${{\left( {{x} - {x_T}} \right)} \mathord{\left/ {\vphantom {{\left( {{x} - {x_T}} \right)} {{d_T}\left( {{x}} \right)}}} \right. \kern-\nulldelimiterspace} {{d_T}\left( {{x}} \right)}} = \sin \left( {{\theta _T}\left( {{x}} \right)} \right)$ and ${{\left( {{x_R} - {x}} \right)} \mathord{\left/ {\vphantom {{\left( {{x_R} - {x}} \right)} {{d_R}\left( {{}} \right)}}} \right. \kern-\nulldelimiterspace} {{d_R}\left( {{x}} \right)}} = \sin \left( {{\theta _R}\left( {{x}} \right)} \right)$, respectively. From \eqref{Eq_Xs}, this implies that the angle of incidence and the angle of reflection coincide at the reflection (stationary) point ${x_s} \in \left[ {-L_x,L_x} \right]$, i.e., ${\theta _T}\left( x_s \right) = {\theta _R}\left( x_s \right)$. In other words, \eqref{Eq_Large} and \eqref{Eq_Xs} allow us to retrieve the law of reflection.
\item \textit{\textbf{Scatterer or diffuser}}. In the electrically small operating regime, (e.g., for long transmission distances), the metasurface structure does not behave as a specular mirror. Thanks to the parallel ray approximation that is employed to obtain \eqref{Eq_Large}, we observe that the intensity of the field depends, predominantly, on the distance, ${d_{T0}}$, from the source to the center (i.e., the origin) of the metasurface and on the distance, ${d_{R0}}$, from the center of the metasurface to the observation point. In particular, the scaling law that governs the intensity of the field as a function of the transmission distances is not straightforward to unveil. To shed light on the impact of the transmission distances, i.e., ${d_{T0}}$ and ${d_{R0}}$ in \eqref{Eq_Large}, we can consider the case study in which the source and the observation point move along two straight lines such that the angles $\theta_{T0}$ and $\theta_{R0}$ are kept constant (the two angles do not have to be necessarily the same), but ${d_{T0}}$ and ${d_{R0}}$ are different. In this case, we evince from \eqref{Eq_Large} that the intensity of the field decays as a function of the square root of the product of the transmission distances ${d_{T0}}$ and ${d_{R0}}$. In the electrically small operating regime, therefore, a metasurface is better modeled as a scatterer rather than a mirror, since its size is relatively small in comparison with the transmission distances ${d_{T0}}$ and ${d_{R0}}$.
\item \textit{\textbf{Directivity}}. The intensity of the field in the electrically large and electrically small operating regimes is quite different not only as a function of the transmission distances, but as a function of the angular response as well. By direct inspection of \eqref{Eq_Small}, we retrieve the ``sinc-like'' behavior that is typical of the far-field regime, since ${\rm{sinc}}\left( {ax} \right) = {{\sin \left( {ax} \right)} \mathord{\left/ {\vphantom {{\sin \left( {ax} \right)} {\left( {ax} \right)}}} \right. \kern-\nulldelimiterspace} {\left( {ax} \right)}}$. Assuming that the angle of incidence, $\theta_{T0}$, of the impinging radio wave is fixed, this implies that the intensity of the field (i) achieves its maximum in the direction of observation that fulfills the condition $\theta_{R0} =\theta_{T0}$ and (ii) has an oscillating behavior whose envelope decays as the difference between $\theta_{R0}$ and $\theta_{T0}$ increases. This implies that, even in the electrically small operating regime, the metasurface structure has a behavior that resembles a specular reflector, but it is different from a specular mirror. In general, in addition, the main lobe of the ${\rm{sinc}}\left( {ax} \right)$ function becomes narrower as $a$ increases. With a slight abuse of terminology, we could say that the metasurface structure becomes more directive. In \eqref{Eq_Small}, $a = {{{L_x}} \mathord{\left/ {\vphantom {{{L_x}} k}} \right. \kern-\nulldelimiterspace} k} = {{\lambda {L_x}} \mathord{\left/ {\vphantom {{\lambda {L_x}} {\left( {2\pi } \right)}}} \right. \kern-\nulldelimiterspace} {\left( {2\pi } \right)}}$. This implies that the end-to-end response of the metasurface structure becomes more directive as the product ${\lambda {L_x}}$ increases, i.e., if either the operating frequency decreases or the size of the metasurface increases. The comparison between \eqref{Eq_Large} and \eqref{Eq_Small} informs us that the scaling laws that originate from \eqref{Eq_Small} hold true only if the metasurface structure operates in the electrically small regime, which implies that, given the transmission distances, the size of the metasurface cannot be increased without any limits. If the size of the metasurface exceeds a given threshold, then the parallel rays approximation cannot be used anymore, and \eqref{Eq_Large} and \eqref{Eq_Small} become a more accurate and a less accurate approximation for \eqref{Eq_Field_Integral}, respectively. By direct inspection of \eqref{Eq_Large}, we observe that the electrically large operating regime is not characterized by the ``sinc-like'' angular response that, on the other hand, characterizes \eqref{Eq_Small}. The intensity of the field in \eqref{Eq_Large} still depends, however, on the observation angle, since \eqref{Eq_Large} holds true only if there exists a point on the metasurface structure that fulfills \eqref{Eq_Xs}. In contrast with \eqref{Eq_Small}, the probability that there exists a point on the metasurface structure  that fulfills \eqref{Eq_Xs} increases as the size of the metasurface increases. We conclude, therefore, that the scaling laws that emerge in the electrically large and electrically small operating regimes are quite different. Depending on the system setup, the correct approximation needs to be used.
\item \textit{\textbf{Size vs. power conservation principle}}. The scaling law as a function of the size, $2 L_x$, of the metasurface structure is different in the electrically large and electrically small operating regimes. For ease of comparison, let us consider \eqref{Eq_Small} in correspondence of its maximum value that is obtained for $\theta_{R0} =\theta_{T0}$. We observe that \eqref{Eq_Large} is independent of $L_x$ while \eqref{Eq_Small} increases linearly with $L_x$. This scaling law as a function of $L_x$ can be intuitively justified by direct inspection of Fig. \ref{Fig_19}. In the electrically large operating regime, we cannot ``see'' the edges of the metasurface structure. Therefore, the intensity of the field cannot depend on the size of the metasurface. In the electrically small operating regime, the opposite holds true, and we expect that the intensity of the field depends on the size of the metasurface. This different scaling law is important, since it allows us to prove that the proposed approach based on the Huygens-Fresnel principle is consistent with the power conservation principle. For ease of exposition and for ease of comparison with other research works available in recent papers \cite{Wankai_PathLoss}, let us consider a discretized version for the metasurface structure. A metasurface can be discretized in $N$ unit cells whose inter-distance is ${\lambda  \mathord{\left/ {\vphantom {\lambda  \mathcal{D}}} \right. \kern-\nulldelimiterspace} \mathcal{D}}$ with $\mathcal{D} > 1$, such that $L_x = N \lambda/\mathcal{D}$. From \eqref{Eq_Small}, this implies, in agreement with some recent research papers \cite{Wankai_PathLoss}, that the power of the received field increases quadratically with the number $N$ of unit cells (by assuming that $\mathcal{D}$ is fixed). From \eqref{Eq_Large}, on the other hand, the power of the received field is independent of $N$. Based on \eqref{Eq_Small}, the received power would grow infinitely large as $N$ tends to infinity. This observation usually leads many researchers to conclude that the considered system model is physically incorrect, since it is not in agreement with the power conservation principle. The truth is, on the other hand, that the scaling law and the considered model are physically correct, since \eqref{Eq_Small} can be applied only in the electrically small operating regime. If $N$ increases tending to infinity, then \eqref{Eq_Small} cannot be utilized anymore. As $N$ increases towards infinity, \eqref{Eq_Small} becomes the less and less accurate, and it needs to be replaced by \eqref{Eq_Large} that well captures the regime of $N$ tending to infinity. Based on \eqref{Eq_Large}, we observe that the power does not grow infinitely large as the size of the  metasurface structure tends to infinity, but it reaches an asymptotic limit that is independent of the size of the metasurface. Once again, we conclude that the correct approximation needs to be utilized in order to draw meaningful conclusions, and caution needs to be taken on arguing the inconsistency of models that are, on the other hand, correct in the operating regimes in which they are computed.
\end{itemize}

\textbf{Beyond uniform metasurfaces}. The remarks made in this section hold true for a metasurface that is uniform and operates as a specular reflector. A similar study can be performed for metasurfaces that are not uniform and are configured to operate as anomalous reflectors or as focusing lenses. As recently proved in \cite{MDR_SPAWC2020} and \cite{MDR_PathLossFadil}, the performance trends as a function of the system parameters can be different as compared with a specular reflector. In addition, the performance of each configuration can be different and needs to be judged based on the complexity and the amount of environmental information that is needed for an appropriate operation. A metasurface that operates as a focusing lens, for example, is expected to maximize the received power at a given location. Therefore, it is expected to outperform a metasurface that operates as a specular or as an anomalous reflector. As mentioned in previous text, however, the side information that is needed for operating a metasurface as a reflector and as a lens may be different, which has an impact on the overhead necessary for their configuration and, eventually, on their ultimate performance, \cite{MDR_OverheadAware}, \cite{MDR_SPAWC2020}, \cite{MDR_PathLossFadil}.

\subsection{Modeling and Analyzing Reconfigurable Intelligent Surfaces: Recipe for Wireless Researchers} \label{RecipeWireless}
\textbf{Two case studies}. Based on the approaches described for the synthesis and analysis of RISs in wireless networks, we summarize in this sub-section the recipe that wireless researchers could employ in order to incorporate analytically tractable but sufficiently accurate models for the metasurfaces in their signal and system models. In particular, we distinguish two case studies: (i) to analyze and optimize the performance of wireless networks by assuming to employ a metasurface structure that operates in a predetermined manner; and (ii) to identify the structure of the optimal metasurface structure that maximizes some desired performance metrics. These two case studies are analyzed in the following two sub-sections.

\subsubsection{The Metasurface Structure is Given} In this case study, the structure of the metasurface is decided a priori, i.e., we assume to have a metasurface that transforms the impinging radio waves in specified ways. The following procedure for application in wireless networks can be used.
\begin{itemize}
\item \textit{\textbf{Input}}: The incident, reflected, and transmitted electric and magnetic fields in \eqref{Eq_Analysis__Def2} and \eqref{Eq_Analysis__Def5} are given and known.
\item \textit{\textbf{Step 1}}: From \eqref{Eq_Analysis__Def2} and \eqref{Eq_Analysis__Def5}, the surface susceptibility functions of the metasurface structure that realizes the specified wave transformations are obtained from the constitutive relations in \eqref{Eq_GSTCs_Explicit_H} and \eqref{Eq_GSTCs_Explicit_E}.
\item \textit{\textbf{Step 2}}: From the constitutive relations in \eqref{Eq_GSTCs_Explicit_H} and \eqref{Eq_GSTCs_Explicit_E}, the representation of the EM fields in terms of surface reflection and transmission coefficients is obtained from \eqref{Eq_Analysis__Solution}.
\item \textit{\textbf{Step 3}}: From \eqref{Eq_Analysis__Solution}, the EM fields in a generic observation point of a given volume are obtained from the analytical formulation of the Huygens-Fresnel principle in \eqref{Eq_Field_Integral}.
\item \textit{\textbf{Output}}: The performance metric of interest is formulated as a function of the EM fields, and it is used for system optimization.
\end{itemize}

It is worth mentioning that the surface susceptibility functions (in Step 1) may be available without any explicit formulation of the incident, reflected, and transmitted electric and magnetic fields. In this case, the procedure starts at Step 2. Also, it is worth mentioning that the analytical formulation in Step 3 needs to be consistent with the input fields.

\subsubsection{The Metasurface Structure is Optimized} In this case study, the structure of the metasurface is not decided a priori but needs to be chosen in order to maximize some performance metrics, e.g., the spectral efficiency and the energy efficiency. The following procedure for application in wireless networks can be used.
\begin{itemize}
\item \textit{\textbf{Input}}: The surface susceptibility functions are the unknown of the problem.
\item \textit{\textbf{Step 1}}: The EM fields are formulated in terms of generic surface susceptibility functions by using \eqref{Eq_Analysis__Solution}.
\item \textit{\textbf{Step 2}}: From the generic analytical formulation in \eqref{Eq_Analysis__Solution}, the EM fields at a generic observation point of a given volume are obtained from the analytical formulation of the Huygens-Fresnel principle in \eqref{Eq_Field_Integral}.
\item \textit{\textbf{Step 3}}: The performance metric of interest is formulated in terms of EM fields, and, therefore, as a function of generic surface susceptibility functions.
\item \textit{\textbf{Output}}: The optimal surface susceptibility functions that maximize the performance metric of interest are calculated, and the corresponding metasurface structure is identified.
\end{itemize}

\textbf{Appropriate models need to be used}. We conclude this section by reminding the readers that, regardless of the specific problem as hand, it is important that realistic models for the metasurfaces and for the propagation of radio waves are employed. This is necessary in order to obtain results that are in agreement with the fundamental laws of physics. The methods described in this section, even though are based on some assumptions, may allow one to solve a broad class of research problems in wireless networks, and, more importantly, may help one understand the potential benefits and the limitations of employing RISs for wireless applications.

\section{The State of Research} \label{SOTA}
In this section, we report a comprehensive summary of the research activities on RIS-empowered SREs that are available to date. The research papers are split into main categories and a brief description for each paper is given.

\subsection{Tutorials and Surveys}
During the last two years, several survey and tutorial papers on the application of RISs to the design of wireless networks have been published \cite{MDR_AI_VTM}, \cite{MDR_AM_TCOM}, \cite{MDR_AI_RIS}, \cite{Liaskos_COMMAG}, \cite{Liaskos_ACM}, \cite{MDR_Survey_EURASIP}, \cite{MDR_Survey_Access}, \cite{MDR_RISs_Relays}, \cite{Liang2019}, \cite{MDR_Survey_Holos}, \cite{RuiZhang_Survey}, \cite{JunZhao_Survey_1}, \cite{Angela_Survey}, \cite{Niyato_Survey}. In \cite{MDR_Survey_EURASIP}, the concept of SREs is comprehensively discussed along with a historical perspective of initial contributions in this field of research. In \cite{MDR_Survey_Access}, the applications of RISs as anomalous reflectors and singe-RF transmitters are elaborated, and the bit error probability offered by both implementations is analyzed and compared. In \cite{MDR_RISs_Relays}, RISs and relays are compared against each other. It is shown that RISs are capable of providing a similar average rate as relays, but without necessitating a power amplifier and provided that they are sufficiently large. In \cite{MDR_AI_VTM}, \cite{MDR_AM_TCOM}, \cite{MDR_AI_RIS}, the potential need and applications of machine learning and AI to SREs are investigated.

\subsection{Collaborative Research Projects}
Besides the upsurge of research activities in academia and industry, several collaborative research projects have been funded during the last two years. In particular, the European Commission has recently funded three major research projects on the application of RISs to future wireless networks. The \textbf{\textit{VisorSurf}} project \cite{VisorSurf} began in January 2017. VisorSurf is aimed to develop a full stack of hardware and software components for smart and interconnected planar objects with a programmable electromagnetic behavior. The \textbf{\textit{ARIADNE}} project \cite{Ariadne} began in November 2019. ARIADNE is aimed to develop innovative technologies and protocols for providing enhanced connectivity to wireless networks that operate in the D-band (110-170 GHz), by coating environmental objects with RISs and turning them into controllable reflectors. The \textbf{\textit{PathFinder}} project \cite{PathFinder} will begin in May 2021. PathFinder is aimed to build the theoretical and algorithmic foundation of RIS-empowered wireless networks by introducing cross-disciplinary methodologies that capitalize on tools from communication theory, physics, and electromagnetism.

\subsection{Physics and Electromagnetism}
Research on metamaterials and metasurfaces has a very long history and countless research contributions can be found in the open technical literature. The readers interested in having background information on this wide research topic are invited to consult the many survey papers, tutorial papers, and books that are available in the literature. For example, background information on \textbf{\textit{metamaterials}} can be found in the books \cite{Book_Caloz}, \cite{Book_Engheta}, \cite{Book_Capolino}, \cite{Book_TieJunCui}, \cite{Book_Maier}. Notably, a historical perspective on metamaterials can be found in \cite{Book_Capolino}. The theory and applications of \textbf{\textit{metasurfaces and surface electromagnetics}} can be found in the book \cite{Book_SurfaceElectromagnetics}, and in the recent survey papers \cite{Survey_Tretyakov} and \cite{Survey_Others}. Readers interested in the \textbf{\textit{analytical modeling of applied electromagnetics and scattering theory}} are referred to the books \cite{Book_Tretyakov} and \cite{Book_Osipov}. The concept of \textbf{\textit{coding metamaterials}}, which can be viewed as the foundation of RISs, can be found in \cite{Coding_Cui}. Examples of realizations of metasurfaces with perfect control of the reflection and transmission of radio waves can be found in \cite{Asadchy_2017}, \cite{Asadchy_2016}, \cite{Epstein_2016}, \cite{Diaz-Rubio_2017}. A recent approach to design metasurfaces with multiple reconfigurable functions can be found in \cite{Tretyakov_Multifunction}. A recent short survey paper on the state-of-the-art of reconfigurable metasurfaces can be found in \cite{VisorSurf_Survey}. A comprehensive survey paper on reconfigurable reflectarrays and array lenses for dynamic antenna beam control is available in \cite{Perruisseau-Carrier}.

\textbf{Main references of this paper}. In the present paper, the focus of Sections II and IV is on the development of computational methods for modeling and analyzing metasurfaces for application to the optimization of wireless communications and networks. The approach reported in Sections II and IV is, in particular, focused on a macroscopic homogenized abstraction modeling based on surface susceptibility functions and the corresponding surface reflection and transmission coefficients. Readers interested in further information on the fundamental theory behind this approach are invited to consult \cite{Caloz_GSTCs2015}, \cite{Caloz_Synthesis2018}, \cite{Caloz_Computational2018} and the first four chapters of the book \cite{Book_SurfaceElectromagnetics}. Readers interested in the implementation of metasurfaces with perfect anomalous reflection and transmission capabilities are invited to consult \cite{Asadchy_2017}, \cite{Asadchy_2016}, \cite{Epstein_2016}, \cite{Diaz-Rubio_2017}. In these latter papers, in particular, the implementation challenges of realizing locally and globally passive metasurfaces with high power efficiency are extensively discussed and early references on the topic are given.

\subsection{Smart Radio Environments Empowered by Reconfigurable Intelligent Surfaces: Initial Works}
Tracing back the history of an emerging research field is not a simple task. Going back to the years 2003-2014, we were able to find a few early papers in which the concepts of ``smart surfaces'' and ``smart environments" were advocated. 

In \cite{Holloway_Oct2003} and \cite{Holloway_Nov2005} published in 2003 and 2005, respectively, one can read ``\textit{The GSTCs are expected to have wide application to the design and analysis of antennas, reflectors, and other devices where controllable scatterers are used
to form a ``smart'' surface}''; ``\textit{This could in principle enable us to realize controllable surfaces, able for instance to switch electronically between reflecting and transmitting states}''; and ``\textit{We see that if a scatterer can be chosen such that the electric and/or magnetic polarizability is changeable on demand, then a smart or controllable surface would be realized}''.

In the tree papers \cite{Subrt2010}, \cite{Subrt2012a}, \cite{Subrt2012} published in 2010 and 2012, respectively, one can read ``\textit{This paper deals with a new approach to the control of the propagation environment in indoor scenarios using intelligent walls. The intelligent wall is a conventional wall situated inside a building, but equipped with an active frequency selective surface and sensors. The intelligent wall can be designed as a self-configuring and self-optimizing autonomous part of a collaborative infrastructure working within a high-capacity mobile radio system. The paper shows how such surfaces can be used to adjust the electromagnetic characteristics of the wall in response to changes in traffic demand, monitored using a network of sensors, thereby controlling the propagation environment inside the building}'' and ``\textit{We have proposed a new method of controlling signal coverage which can be used inside buildings both for coverage controlling and avoiding of interference between transmitters. The method is based on changing of electromagnetic properties of intelligent walls which are conventional walls equipped with an active frequency selective surface and sensors. The walls are part of a collaborative autonomous infrastructure and they are able to adjust their properties, e.g., on the basis of data obtained by sensing of users or cognitive network sensing. The results show that intelligent walls can be strong instrument to control signal coverage and subsequently to control available level of service. The controlling of coverage is very useful when mobile users change their position in an indoor scenario, because then intelligent walls provide the efficient way to change propagation environment rapidly}''.

In \cite{Kaina2014} published in 2014, the authors demonstrate that spatial microwave modulators are capable of shaping, in a passive way, complex microwave fields in complex wireless environments, by using only binary phase state tunable metasurfaces. These findings are obtained through the design of a 0.4 $m^2$ spatial microwave modulator that consists of 102 controllable EM reflectors and  that operates at a working frequency of 2.47 GHz. The 102 reflectors are controlled by using two Arduino 54-channel digital controllers. In the abstract of the paper, more specifically, one can read ``\textit{In this article, we propose to use electronically tunable metasurfaces as spatial microwave modulators. We demonstrate that like spatial light modulators, which have been recently proved to be ideal tools for controlling light propagation through multiple scattering media, spatial microwave modulators can efficiently shape in a passive way complex existing microwave fields in reverberating environments with a non-coherent energy feedback. We expect that spatial microwave modulators will be interesting tools for fundamental physics and will have applications in the field of wireless communications}''. 

During the last five years (2015-2020), and, in particular, during the last two years, the research contributions on RIS-empowered SREs have increased exponentially. The research papers available to date are discussed in the following thematic sub-sections.

\subsection{Path-Loss and Channel Modeling}
A major research issue for analyzing the ultimate performance limits, optimizing the operation, and assessing the advantages and limitations of RIS-empowered SREs is the development of simple but sufficiently accurate models for the power received at a given location in space when a transmitter emits radio waves that illuminate an RIS. This is an open and challenging research issue to tackle. Recently, however, a few initial research works have attempted to shed light on modeling the path-loss of RISs. 

In \cite{Wankai_PathLoss}, the authors perform a measurement campaign in an anechoic chamber and show that the power reflected from an RIS follows a scaling law that depends on many parameters, including the size of the RIS, the mutual distances between the transmitter/receiver and the RIS (i.e., near-field vs. far-field conditions), and whether the RIS is used for beamforming or broadcasting applications. 

In \cite{garcia2019reconfigurable}, the authors employ antenna theory to compute the electric field in the near-field and far-field of a finite-size RIS, and prove that an RIS is capable of acting as an anomalous mirror in the near-field of the array. The results are obtained numerically and no explicit analytical formulation of the received power as a function of the distance is given. Similar findings are reported in \cite{ellingson2019path}. 

In \cite{Khawaja-Coverage}, the authors study, through experimental measurements, the power scattered from passive reflectors that operate in the millimeter-wave frequency band. Also, the authors compare the obtained results against ray tracing simulations. By optimizing the area of the surface that is illuminated, it is shown that a finite-size passive reflector can act as an anomalous mirror. 

In \cite{ozdogan2019intelligent}, the authors study the path-loss of RISs in the far-field regime by leveraging antenna theory. The obtained results are in agreement with those reported in \cite{garcia2019reconfigurable} and \cite{ellingson2019path} under the assumption of far-field propagation.

In \cite{MDR_SPAWC2020}, the authors use the scalar theory of diffraction and the Huygens-Fresnel principle in order to model the path-loss in both the near-field and far-field of RISs, which are modeled as homogenized sheets of electromagnetic
material with negligible thickness. By using the stationary phase method, the authors unveil the regimes under which the path-loss depends on the sum and the product of the distances between the RIS and the transmitter, and the RIS and the receiver. The proposed analytical approach is shown to be sufficiently general for application to uniform reflecting surfaces, anomalous reflectors, and focusing reflecting lenses.

\subsection{Surface-Based Modulation and Encoding}
One of the emerging and promising applications of RISs is to employ them for modulating and encoding data into their individual reconfigurable elements. This application of RISs can be viewed as an instance of spatial modulation and index modulation \cite{MDR_SM_COMMAG}, \cite{MDR_SM_PIEEE}, \cite{MDR_SM_CST2015}, \cite{MDR_SM_CST2016}, \cite{MDR_SM_MILCOM}, \cite{MDR_IM_Access}, \cite{MDR_SM_JSAC}. In particular, recent research activities on this topic constitute a generalization of the concept of spatial modulation based on reconfigurable antennas, which was introduced in \cite{MDR_SM_MILCOM} and was recently engineered and realize in \cite{PhanHuy2019}. The bit error probability of spatial modulation based on the reconfigurable antennas is studied in \cite{MDR2019_Dung}, by using radiation patterns obtained from a manufactured reconfigurable antenna. The bit error probability of RIS-based spatial modulation is studied in \cite{MDR_Survey_Access}.

In \cite{Basar2019}, the author studies the error probability offered by space shift keying and spatial modulation schemes that employ RISs. The author proposes maximum energy-based suboptimal and exhaustive search-based optimal detectors, and analyzes their performance with the aid of analytical frameworks and numerical simulations. The obtained results show that RIS-based spatial modulation is capable of providing high data rates at low error rates. A similar study can be found in \cite{canbilen2020reconfigurable}.

In \cite{yan2019WCL}, the authors apply the principle of spatial modulation by modulating information onto the ON/OFF states of the reflecting elements of RISs. In addition, passive beamforming is obtained by adjusting the phase shifts of the activated reflecting elements. To optimize passive beamforming, the authors formulate an optimization problem and solve it by using semidefinite relaxation. To retrieve the information data from the transmitter and the RIS, the authors introduce a two-step detection algorithm based on compressed sensing and matrix factorization. The tradeoff between the passive beamforming gain
and the information rate provided by RISs is investigated.

In \cite{Guo19RefMod}, the authors propose a reflecting modulation scheme for RIS-based communications, where both the reflecting patterns and transmit signals carry information. To enhance the reliability of transmission, the authors devise a discrete optimization-based joint signal mapping, shaping, and reflecting design that minimizes the bit error probability for a given transmit signal candidate set and a given reflecting pattern candidate set. Numerical results show that the proposed optimized schemes outperform existing solutions in terms of bit error probability.

In \cite{Hanzo__2020}, the authors propose three different architectures based on RISs for beam-index modulation in millimeter wave communication, which circumvent the line-of-sight blockage of millimeter wave frequencies. The authors derive the optimal maximum likelihood detector and a low-complexity compressed sensing detector for the proposed schemes. The proposed solutions are evaluated through extensive simulations and analytical bounds.

In \cite{MDR_ISIT2020}, the authors conduct an information-theoretic analysis of RIS-empowered wireless networks. An RIS-aided communication link is considered, in which the transmitter can control the state of an RIS via a finite-rate control link. The authors derive information-theoretic limits and demonstrate that the capacity is achieved by a scheme that jointly encodes information onto the transmitted signal and onto the RIS configuration. In addition, a novel signaling strategy based on layered encoding is proposed, which enables practical successive cancellation-type decoding at the receiver. Numerical experiments demonstrate that conventional schemes that fix the configuration of RISs in order to maximize the signal-to-noise ratio are strictly suboptimal, since they are outperformed by the proposed strategies.

\subsection{Channel Estimation}
Channel estimation is a fundamental issue to tackle in SREs wherein nearly-passive RISs are employed. In contrast to other communication systems, e.g., relays, which are equipped with sufficiently powerful signal processing capabilities, nearly-passive RISs are equipped with minimal on-board signal processing units that are not intended to be used during their normal operation phase. Therefore, new algorithms and protocols are necessary in order to perform channel estimation, while keeping the complexity of RISs as low as possible and avoiding on-board signal processing operations as much as possible.

In \cite{Nadeem2019b}, the authors investigate channel estimation in RIS-based multiple-input-single-output (MISO) links. The implementation and performance advantages of RIS-based communications over traditional relay-based and MIMO systems are discussed. A minimum mean square error channel estimation protocol is developed, and, based on the estimated channels, the phase shifts of the RIS are optimized by employing a gradient ascent algorithm.

In \cite{Jensen2019}, the authors introduce optimal methods for estimating the cascaded channel from the transmitter to the receiver, i.e., the combined channel that includes the link from the transmitter to the RIS and the link from the RIS to the receiver. The proposed method is based on activating the RIS elements based on a sequence of patterns, which are selected based on the minimum variance unbiased estimation principle. It is found that the channel estimation phase mimics a series of discrete Fourier transforms. The authors show that the estimation variance of the proposed algorithms is one order of magnitude smaller compared with traditional approaches.

In \cite{Lin2019}, the authors formulate the channel estimation problem in RIS-based systems as a constrained estimation error minimization problem, which is tackled by using the methods of Lagrange multipliers and a dual ascent-based algorithm. The Cramer-Rao lower bounds are determined for benchmarking purposes. It is shown that the derived method yields better accuracy, in the low signal-to-noise ratio regime, compared with least square methods.

In \cite{Chen2019b}, the authors develop an algorithm for estimating the cascaded channel. The proposed approach leverages compressive sensing methods, by leveraging the sparsity available in the cascaded channel. In addition, the authors extend the approach to the joint estimation of multiple channels in a multi-user network setup. In order to improve the estimation performance, the optimization of the training reflection sequences is considered.

In \cite{Ning2019}, the authors tackle the channel estimation problem by means of beam training. The associated quantization error is characterized and evaluated.  As a low-complexity alternative for channel estimation, a hierarchical search codebook design is proposed. Based on the proposed channel estimation methods, the performance of RIS-based wireless networks is studied.

In \cite{Wang2019d}, the authors propose a three-phase protocol for estimating large numbers of channel coefficients, for application to RIS-assisted uplink systems, by using a small number of pilot symbols. Building upon the observation that each RIS element reflects the signals from all the users to the transmitter via the same channel, the authors quantify the time that is necessary to estimate all required channels. The authors show that massive MIMO may play an important role in reducing the channel estimation overhead in RIS-based communication systems.

In \cite{Xia2019}, the authors consider the uplink of a network with massive connectivity and investigate the joint problem of channel estimation and activity detection. The problem is formulated as a sparse matrix factorization, matrix completion, and multiple measurement vector problem, and a three-stage algorithm is proposed for solving it, by leveraging message passing methods.

In \cite{cui2019efficient}, the authors propose a low-complexity channel information acquisition method for RIS-assisted communication systems that exploit the sparsity of millimeter-wave channels and the topology of RISs. Compared with state-of-the-art methods, the proposed approach requires less time-frequency resources for acquiring the channel information. 

In \cite{You2019}, the authors tackle the problem of joint channel estimation and sum-rate maximization in the uplink of a single-user RIS-based system. In particular, the phase shifts of RISs are assumed to be discretized. A practical but sub-optimal channel estimation method based on the discrete Fourier transform and on a truncation of Hadamard matrices is developed. Based on the estimated channel, rate optimization is tackled by designing the discrete phase shifts at the RIS. To this end, a low-complexity successive refinement algorithm is developed. By using numerical simulations, the authors show that the performance of the proposed method depends on the quality of the initialization point.

In \cite{liu2019matrix}, the authors formulate the channel estimation problem of an RIS-assisted multiuser MIMO system as a matrix-calibration based matrix factorization task. This is exploited in order to reduce the length of the training sequences for estimating the cascaded channel, which includes the transmitter-to-RIS and the RIS-to-receiver links. By assuming slow-varying channel components and channel sparsity, the authors propose a message-passing based algorithm that factorizes the cascaded channel. The authors present an analytical framework to characterize the theoretical performance of the proposed estimator in the large-system limit. 

In \cite{wei2020parallel}, the authors propose a method for channel estimation that is based on the parallel factor decomposition  algorithm in order to unfold the cascaded channel from the transmitter to the receiver. The proposed method is based on an alternating least squares algorithm that iteratively estimates the channel between the transmitter and the RIS as well as the channel between the RIS and the users. Simulation results show that the proposed iterative channel estimation method outperforms benchmark schemes based on genie-aided information. 

In \cite{wan2020broadband}, the authors propose a compressed-sensing-based channel estimation algorithm in which the angular channel sparsity of large-scale arrays that operate at millimeter wave frequencies is exploited to perform channel estimation at a reduced pilot overhead. The authors design pilot signals based on prior knowledge of the line-of-sight-dominated transmitter-to-RIS channel as well as prior information on the high-dimensional transmitter-to-receiver and RIS-to-receiver channels, by using compressed sensing methods. A distributed orthogonal matching pursuit algorithm is exploited in order to capitalize on the channel sparsity.

In \cite{wang2019compressed}, the authors develop a joint channel estimation and beamforming design for RIS-assisted MISO systems in the millimeter-wave frequency band. In order to reduce the training overhead, the inherent sparsity of millimeter-wave channels is exploited. The authors first find a sparse representation of the cascaded channel, and then develop a compressed-sensing-based channel estimation method. Also, joint beamforming design is performed based on the estimated channel.

In \cite{he2020channel}, the authors propose a two-stage channel estimation scheme by using iterative re-weighted methods for estimating the channel parameters in a sequential optimization loop. Simulation results show that the proposed method provides better performance than the two-stage orthogonal matching pursuit approach, in terms of channel estimation error and spectral efficiency gain.

In \cite{mishra2019channel}, the authors introduce a channel estimation protocol for RIS-assisted energy transfer over MISO Rician fading channels. In order to account for the practical limitations of RISs and energy harvesting users, all computations are carried out at the transmitter. The authors derive near-optimal and closed-form expressions for active beamforming at the transmitter and for passive beamforming at the RIS.

In \cite{de2020parafac}, the authors propose two novel channel estimation methods by assuming a structured time-domain pattern of pilots and RIS phase shifts. The authors show that the received signal follows a parallel factor tensor model that can be exploited to estimate the involved communication channels in closed-form or iteratively. Numerical results corroborate the effectiveness of the proposed channel estimation methods and highlight the existing tradeoffs.

In \cite{ning2019channel}, the authors introduce a channel estimation and transmission design for low scattering channels that is based on hybrid beamforming architectures. The authors design two different codebooks for efficiently realizing the proposed estimation procedure via a three-tree hierarchical search algorithm. The two codebooks are referred to as tree dictionary codebook and phase shifter deactivation codebook. By using the estimation information, the authors compute two closed-form transmission designs that maximize the overall spectral efficiency.

In \cite{you2019progressive}, the authors propose a hierarchical training reflection design for progressively estimating the channels based on an RIS-elements grouping and partition method. Specifically, a new discrete Fourier transform Hadamard-based basis training reflection matrix design is proposed in order to minimize the mean square channel estimation error. Moreover, the authors propose two subgroup partition schemes and design efficient subgroup training reflection matrices to progressively resolve the subgroup aggregated channels of each group over the blocks. Based on the resolved subgroup aggregated channels,  passive beamforming is used for improving the achievable rate, by taking into account the derived channel estimation error covariance matrix. Also, a low-complexity successive refinement algorithm with properly designed initializations is proposed in order to obtain high-quality suboptimal solutions.

In \cite{mirza2019channel}, the authors introduce a two stage channel estimation method. Based on the proposed algorithm, the direct channel between the access point and the users is estimated during the first stage. The RIS channels are estimated during the second stage by using a bilinear adaptive vector approximate message passing algorithm. A phase shift design for the RIS elements, which approximately maximizes the channel gain at the user, is proposed. 

In \cite{hu2019two}, the authors propose a two-timescale channel estimation framework. The key idea of the framework is to exploit the property that the transmitter-RIS channel is high-dimensional but quasi-static, while the RIS-receiver channel is mobile but low dimensional. To estimate the quasi-static transmitter-RIS channel, the authors propose a dual-link pilot transmission scheme, in which the transmitter emits downlink pilots and receives uplink pilots reflected by the RIS. The authors propose a
coordinate descent-based algorithm to recover the transmitter-RIS channel. With the aid of numerical results, the authors show that the proposed two-timescale channel estimation framework can achieve accurate channel estimation with low pilot overhead.

In \cite{zheng2019channel}, the authors study the problem of channel estimation for ambient backscattering. The authors design an estimator consisting of an initial estimation step and a subsequent iteration phase in order to acquire the channel parameters. To further evaluate the performance of the designed estimator, the authors derive Cramer-Rao lower bounds for the channel estimates. 

In \cite{XiaojunYuan}, the authors introduce a general framework for the estimation of the transmitter-RIS and RIS-receiver cascaded channel, by leveraging a combined bilinear spare matrix factorization and matrix completion. In particular, the authors present a two-stage algorithm that includes a generalized bilinear message passing algorithm for matrix factorization and a
Riemannian manifold gradient-based algorithm for matrix completion. Simulation results illustrate that the proposed method achieves accurate channel estimation performance for application to RIS-assisted systems.

In \cite{alexandropoulos2020hardware}, the authors present an architecture for RISs that comprises an arbitrary number of passive reflecting elements, a simple controller for their adjustable configuration, and a single RF chain for baseband measurements. By capitalizing on the proposed architecture and by assuming sparse wireless channels in the beamspace domain, the authors introduce an alternating optimization approach for the explicit estimation of the channel gains by relaying on a single RF chain. Simulation results demonstrate the channel estimation accuracy and achievable performance of the proposed solution as a function of the  number of training symbols and the number of RIS reflecting elements.

\subsection{Performance Evaluation}
Since their inception, RISs have been under study for unveiling their fundamental performance limits and the impact of the imperfect knowledge of various systems parameters on their achievable performance. In this context, several exact, approximate, and asymptotic analytical frameworks have been developed in order to quantify the advantages and limitations of RISs in different network scenarios.

In \cite{Zhang2019e}, the authors study the impact of the finite resolution of the RIS phase shifts on the achievable performance. In particular, the authors introduce an approximated expression of the achievable data rate of an RIS-aided communication system, and derive the minimum number of phase shifts that is necessary in order to guarantee a data rate degradation constraint. 

In \cite{Hu2018b}, the authors study the capacity degradation that stems from potential hardware impairments in RIS-aided systems. In the absence of hardware impairments, the first-order derivative of the capacity with respect to the surface-area is shown to be proportional to the inverse of it. In the presence of hardware impairments, the latter performance metric is degraded. The authors introduce a model for the hardware impairments, and derive the effective noise density and the capacity loss in closed-form. The authors investigate distributed implementations for RIS-aided wireless systems as well, and show that the impact of hardware impairments can be reduced due to the smaller size of each RIS.

In \cite{Jung2018}, the authors study the asymptotic rate of an RIS-based uplink channel in which an RIS is equipped with a large number of reflecting elements. Also, the authors take channel estimation errors and spatially correlated multi-user interference in account. It is shown that channel hardening effects occur, similar to massive MIMO systems. Moreover, it is shown that the impairments due to thermal noise, interference, and channel estimation errors become negligible as the number of RIS reflectors increases. This leads to results comparable with those of massive MIMO systems, but at a reduced deployment area and costs. 

In \cite{Jung2019}, the authors study the asymptotic optimality of the achievable rate of a downlink RIS-aided system. In order to increase the achievable system sum-rate, the authors propose a modulation scheme that can be used in an RIS-aided system without interfering with existing users. The authors study the average symbol error probability and introduce a resource allocation algorithm that jointly optimizes the scheduling of users and the power control at the transmitter. The proposed asymptotic analysis is validated with the aid of numerical simulations.

In \cite{Nadeem2019}, the authors study the downlink channel of a single-cell multi-user system in which a base station  equipped with multiple antennas communicates with several single-antenna users through an RIS. The minimum signal-to-interference+noise-ratio (SINR) among the mobile users is asymptotically analyzed, by considering the case studies in which the line-of-sight channel matrix between the base station and the RIS has unit rank and full rank. In the unit rank case, the minimum SINR is shown to be bounded by a quantity that goes to zero as the number of users increases. In the full rank case, on the other hand, deterministic approximations are derived and are used to optimize the RIS phase shifts by using the gradient ascent method.

In \cite{Zhang2019c}, the authors consider a single-user multi-RISs link communication system The RIS phase shifts are adapted to the line-of-sight components of the propagation channels. The outage probability of this system model is derived and it is shown that it decreases as the number of RISs and as the number of reflecting elements at each RIS increase. Moreover, the outage probability is characterized in the high signal-to-noise-ratio regime, and it is proved that it decreases with the power of the line-of-sight links.

In \cite{basar2019transmission}, the author introduces a general mathematical framework for the calculation of the symbol error probability of network scenarios in which an RIS is used as a reflector and as a transmitter. With the aid of numerical simulations, the author compares RIS-based systems with and without the knowledge of channel phases. 

In \cite{jung2019reliability}, the authors study the distribution of the uplink sum-rate of an RIS-based system. Leveraging the derived asymptotic expression of the rate, the authors introduce an analytical framework for computing the outage probability. The authors show that the proposed asymptotic analysis is in close agreement with numerical simulations, under the assumption of a large number of antennas and devices.

In \cite{kudathanthirige2020performance}, the authors compute the outage probability, the average symbol error probability, and the achievable rate of an RIS-aided system. It is proved that the accuracy of the obtained analytical expressions becomes tighter as the number of RIS elements grows large. The achievable diversity order is quantified by deriving single-polynomial approximations in the high signal-to-noise-ratio regime. The diversity order is shown to be equal to the number of RIS elements.

In \cite{badiu2019communication}, the authors study the transmission through an RIS in the presence of phase errors. The authors show that the RIS-based composite channel is equivalent to a point-to-point Nakagami fading channel. This equivalent representation allows the authors to perform theoretical analysis of the performance and to study the interplay between the performance, the distribution of phase errors, and the number of reflectors. Numerical evaluation of the error probability for a limited number of reflectors confirms the theoretical prediction and shows that the performance of RISs is remarkably robust against the phase errors.

In \cite{psomas2019low}, the authors present an analytical framework for the performance evaluation of random rotation-based RIS-aided communications. Under this framework, the authors propose four low-complexity and energy efficient techniques based on two approaches: A coding-based and a selection-based approach. Both approaches depend on random phase rotations and require no channel state information. In particular, the coding-based schemes use time-varying random phase rotations in order to produce a time-varying channel. The selection-based schemes select a partition of the RIS elements at each time slot based on the received signal power at the destination. Analytical expressions for the outage probability and energy efficiency of each scheme are derived. The authors demonstrate that all schemes can provide significant performance gains and full diversity order.

In \cite{Guo20CL}, the authors study an RIS-assisted MISO system in which a base station equipped with multiple antennas arranged in a uniform rectangular array serves a single-antenna user with the help of an RIS equipped with multiple elements arranged in a uniform rectangular array. The authors consider a Rician fading model, where the non-line-of-sight components vary slowly and the line-of-sight components do not change. To reduce the overhead for channel estimation and phase adjustment, the authors adopt a fixed maximum-ratio transmission scheme at the base station, fixed phase shifts at the RIS, and a constant rate transmission. The authors obtain analytical expressions of the outage probability and optimize the RIS phase shifts based on them. Based on the obtained analytical framework, the authors analyze the impact of several key system parameters on the optimal outage probability.

In \cite{basar2019reconfigurable}, the authors revisit the multipath fading phenomenon in the context of RISs. In particular, the authors investigate the feasibility of eliminating or mitigating the multipath fading effect, which originates from the movement of mobile receivers, with the aid of RISs. The authors show that the rapid fluctuations in the received signal strength due to the Doppler effect can be effectively reduced by using real-time tunable RISs. Also, the authors prove that the multipath fading effect can be eliminated if all available reflectors are coated with RISs, as well as that the presence of a few RIS-coated reflectors can significantly reduce the Doppler spread and the deep fades in the received signal.

In \cite{ozdogan2020using}, the authors study RISs in order to improve the rank of wireless channels. One of the classical bottlenecks of point-to-point MIMO communications is that the capacity gains provided by spatial multiplexing are large in the high signal-to-noise-ratio regime. However, the latter regime usually occurs in line-of-sight scenarios in which the channel matrix has a low rank, and, therefore, it does not support spatial multiplexing. The authors demonstrate that an RIS can be used and optimized in line-of-sight scenarios in order to increase the rank of the channel matrix, which leads to substantial capacity gains.

In \cite{xu2019discrete}, the authors study an RIS-aided downlink communication system. The RIS is equipped with $B$-bit discrete phase shifters and the base station exploits low-resolution digital-to-analog converters. Without knowledge of the channel state information, the authors propose a practical phase shift design method whose computational complexity depends on $B$, but it is independent of the number of RIS reflecting elements. The authors compute a tight lower bound for the asymptotic rate of the users. As the number of reflecting elements increases, the authors prove that the asymptotic rate saturates because the received signal power and the quantization noise of the digital-to-analog converters increase. Compared with the optimal continuous phase shift design with perfect channel state information, the authors prove that the proposed design asymptotically approaches the ideal benchmark performance for moderate to high values of $B$. 

In \cite{cao2019capacity}, the authors analyze the performance of a distributed RIS-aided large-scale antenna system by formulating an upper bound expression of the ergodic capacity. The authors provide an asymptotic optimal phase shift design in order to maximize the ergodic capacity, and they propose a novel greedy scheduling scheme that takes the scheduling fairness into account. With the aid of simulations, the authors show that the analytical upper bounds are sufficiently accurate, and they validate that the proposed scheduling scheme achieves almost the same ergodic capacity as classical greedy scheduling algorithms.

In \cite{hu2019spherical}, the authors consider three-dimensional RISs that are deployed as spherical surfaces. Compared with two-dimensional RISs, the authors prove that spherical RISs are capable of offering wider coverage, simpler positioning techniques, and a more flexible deployment.

In \cite{zhang2020intelligent}, the authors investigate the capacity region of a multiple access channel in which two users send independent messages to an access point that is aided by $M$ reflecting elements. The authors study two practical deployment strategies: A distributed deployment in which the $M$ reflecting elements form two RISs that are deployed in the vicinity of two users; and a centralized deployment in which the $M$ reflecting elements are deployed in the vicinity of the access point. As for the distributed deployment, the authors derive the capacity region in closed-form. As for the centralized deployment, the authors derive a capacity region outer bound and propose an efficient rate-profile based method to characterize an achievable rate region. Based on the obtained analytical frameworks, the authors show that the centralized deployment outperforms the distributed deployment under practical channel setups, and that the capacity gain is higher when the user rates are asymmetric.

In \cite{lyu2020spatial}, the authors characterize the spatial throughput of a single-cell multiuser system aided by multiple RISs that are randomly deployed in the cell. By using analysis and simulations, the authors prove that an RIS-aided system outperforms full-duplex relaying in terms of spatial throughput, provided that the number of RIS elements exceeds a certain value. Moreover, the authors show that different deploying strategies for RISs and relays should be adopted for their respective throughput maximization. Finally, it is shown that, for a given total number of reflecting elements, the system spatial throughput increases if fewer RISs, each having a larger number of reflecting elements, are deployed. However, this comes at the cost of higher spatially varying user rates.

In \cite{williams2019communication}, the authors consider an RIS that is made of active elements and that is used as a transmitter. The RIS is modeled as a collection of tiny closely spaced antenna elements. Due to the proximity of the elements, mutual coupling arises. The authors introduce analytical expressions for the mutual coupling of two types of planar arrays. The singular values of the mutual coupling matrix is shown to approach zero as the antenna element density increases. When the spacing between elements becomes small (smaller than half a wavelength), the authors show that the directivity surpasses the conventional directivity equal to the number of antennas, as well as the gain that is obtained when modeling the surface as a continuous one. The authors show that the gain is theoretically unbounded as the element density increases for a constant aperture.

In \cite{wang2020intelligent}, the authors study the feasibility of adopting RISs for improving the beamforming gain and throughput of uplink massive MIMO communications. The authors show that the favorable propagation property of conventional massive MIMO systems no longer holds in the presence of RISs. This is because each RIS element reflects the signals from all the users to the base station via the same channel, thus making the channels correlated. As a result, a maximal ratio combining receive beamforming leads to strong inter-user interference and low user rates. To tackle this challenge, the authors propose a novel zero-forcing beamforming strategy that efficiently nulls the interference. By using this approach, the authors show that RIS-assisted massive MIMO systems can achieve a higher throughput compared with their counterpart that does not use RISs.

\subsection{Stochastic Geometry Based Analysis}
Random spatial processes are considered to be the most suitable analytical tool to shed light on the ultimate performance limits of innovative technologies when applied in wireless networks and to guide the design of optimal algorithms and protocols for attaining such ultimate limits. In particular, random spatial processes have been successfully applied to the analysis of cellular networks, multi-tier cellular networks, millimeter-wave cellular networks, multi-antenna cellular networks, wireless information and power transfer, energy efficiency optimization, etc. \cite{SG_Jeff}, \cite{SG_MDR_1}, \cite{SG_MDR_2}, \cite{SG_MDR_3}, \cite{SG_MDR_4a}, \cite{SG_MDR_4}, \cite{SG_MDR_7}, \cite{SG_MDR_5}, \cite{SG_MDR_6}. Despite the many results available, fundamental issues remain open. In the current literature, in particular, the environmental objects are modeled as entities that can only attenuate the signals, by making the links either line-of-sight or non-line-of-sight. Modeling anything else is acknowledged to be difficult. Modeling reflections, for example, either according to the conventional law of reflection or to the generalized law of reflection, can be considered to be an open and challenging research issue to tackle. Due to the importance of quantifying the performance gains offered by RISs in large-scale wireless networks, some researchers have recently attempted to fill this fundamental research gap.

In \cite{MDR2019_Eurasip}, the authors model the environmental objects with a modified random line process of fixed length and with random orientations and locations. Based on the proposed modeling approach, the authors introduce the first analytical framework that provides one with the probability that a randomly distributed object that is coated with an RIS acts as a reflector for a given pair of transmitter and receiver. In contrast to the conventional network setup where the environmental objects are not coated with RISs, the authors prove that the probability that the typical random object acts as a reflector is independent of the length of the object itself. 

In \cite{Hou2019}, the authors consider a MIMO system that is intended to provide wireless services to randomly roaming users, and they analyze the achievable performance by utilizing stochastic geometry. Each user is assumed to receive the superposed signals reflected by multiple RISs. The objective is to serve multiple users by jointly designing the beamforming weights at the RISs and the receiver weights vectors at the users, respectively. The authors derive closed-form expressions of the outage probability and the ergodic rate of the users, and their corresponding diversity orders. The analytical results demonstrate that the specific fading environments between the RISs and users have almost no impact on the diversity order attained. 

In \cite{kishk2020exploiting}, the authors use tools from stochastic geometry in order to study the effect of large-scale deployment of RISs on the performance of cellular networks. In particular, the authors model the blockages using a line Boolean model. The authors focus their attention on how equipping a subset of the blockages with RISs can enhance the performance of a cellular network. The authors derive the ratio of the blind-spots to the total area, and the probability that a typical mobile user associates with a base station in the presence of an RIS. Also, they derive the probability distribution of the path-loss between the typical user and its associated base station. Based on the proposed analytical framework, the main conclusion drawn by the authors unveils that the deployment of RISs is capable of highly improving the coverage regions of cellular base stations.

\subsection{Optimization and Resource Allocation}
Optimization of the system radio resources is one of the most researched topics in the context of RIS-based systems. A significant share of research articles has considered the problem of passive and active beamforming in MISO systems, i.e., the optimization of the transmit (active) beamforming and the (passive) RIS phase shifts, by resorting to the popular tool of alternating maximization in order to handle the non-convexity of the optimization problem. These papers are reviewed in the first sub-section. In the second sub-section, we review the use of other optimization techniques for  solving the problem of active and passive beamforming in MISO systems. Research contributions on other resource allocation problems are discussed in the third sub-section. In the fourth sub-section, we summarize the notable scenario in which RISs can only apply discrete/quantized phase shifts. In the fifth sub-section, we report some recent research directions in the context of resource optimization for RIS-based networks.

\textit{1) Passive / Active Beamforming in MISO Systems by Using Alternating Optimization}\label{Sec:RA1}\hfill\\

\textbf{Rate and Energy Efficiency}.
In \cite{Huang_ICASSP,ZapTWC2019}, the authors consider the maximization of the spectral and energy efficiency in a multi-user MISO downlink network, by alternatively optimizing the base station beamforming and the RIS phase shifts. The power allocation problem is optimally solved by a water-filling-like solution, while two methods are developed for optimizing the RIS phase shifts. The first method employs a gradient search approach, while the second method exploits the framework of sequential optimization. Both methods provide phase shifts allocations that are stationary points of the spectral and energy efficiency. Moreover, numerical results show that the use of RISs significantly improves the energy efficiency compared to the use of amplify-and-forward relays, while a gap is observed with respect to the spectral efficiency. This is due to the fact that RISs are passive devices, while amplify-and-forward relays are able to apply signal amplification. 

In \cite{Abeywickrama2019}, the authors develop a phase shift model that relates the RIS phase shifts to the amplitude of the reflection coefficient. A single-user MISO link is considered, and, based on the new phase shift model, the problem of maximizing the system achievable rate with respect to the transmit beamforming and to the RIS phase shifts is formulated. In order to cope with the non-convexity of the problem, alternating optimization is employed to obtain a low-complexity algorithm, in which the transmit beamforming and the RIS phase shifts are iteratively optimized. 

In \cite{Yxu2020resource}, the authors consider the downlink of an RIS-enabled heterogeneous network, in which a multiple-antenna base station serves single-antenna users. In this context, the authors develop an algorithm that jointly allocates the base station transmit power and the RIS phase shifts, in order to maximize the SINR and the sum-rate at the small cell users, subject to a minimum SINR constraint on the rate of the macro-cell users. An algorithm based on alternating maximization coupled with semi-definite relaxation is developed to tackle the sum-rate maximization problem, by trading off complexity and optimality. 

In \cite{xie2019max}, the authors investigate an RIS-aided multi-cell MISO system, with the objective of maximizing the minimum weighted SINR at all users by jointly optimizing the base station beamforming and the RIS phase shifts, subject to individual power constraints at the base station. The authors propose an algorithm to tackle this problem based on the alternating optimization framework, wherein the base station beamforming and the RIS phase shifts are alternatively optimized. The beamforming sub-problem is formulated as a second-order-cone program, while the optimization of the RIS phase shifts is tackled by leveraging semi-definite relaxation and successive convex approximation. 

In \cite{gao2020reconfigurable}, the authors consider the problem of maximizing the spectral efficiency in RIS-aided MISO systems under nonlinear proportional rate constraints. The problem is tackled by alternatively optimizing the transmit power at the base station and the reflecting phase shifts at the RIS, which enables the authors to develop a practical algorithm even though no optimality properties can be guaranteed. Numerical simulations based on the proposed alternating optimization algorithm show that RISs are beneficial for improving the network data rate.

In \cite{Pan2019b}, the authors study the downlink of a MIMO multi-cell system, wherein an RIS is deployed at the boundary between multiple cells. In this context, the authors tackle the problem of weighted sum-rate maximization, by alternating optimization of the base station beamformer and the RIS phase shifts. The beamformer optimization subproblem is obtained in closed form, for fixed RIS phase shifts. Instead, two efficient, but sub-optimal, algorithms are proposed for optimizing the RIS phase shifts, for fixed beamformers, by leveraging the sequential optimization and the complex circle manifold methods. Both algorithms yield first-order optimality.

In \cite{Guo2019}, the authors address the problem of weighted sum-rate maximization in a multi-user RIS-based network. The problem is tackled by resorting to alternating optimization, in which the beamforming a the base station is optimized for fixed RIS phase shifts, and then the RIS phase shifts are optimized by keeping fixed the base station beamforming. Lagrangian duality is coupled with fractional programming methods for beamforming optimization at the base station, while three closed-form allocations of the RIS phase shifts are proposed.

In \cite{Zhou2019b}, the authors consider a MISO multi-group multicast system in which an RIS is used to assist the communication. The authors study the problem of optimizing the base station precoding matrix and the RIS reflection coefficients for sum-rate maximization. The resulting optimization problem is first approximated by means of a concave lower bound of the objective function, which is then optimized by alternating optimization between the BS precoding and the RIS phase shifts. Both subproblems are second-order cone programming problems, which can be solved by convex optimization theory. Moreover, the use of sequential optimization is utilized in order to formulate approximate problems that can be solved in closed-form.

In \cite{Yang2019}, the authors study an RIS-aided system model and propose a practical transmission protocol and channel estimation algorithm, by assuming an orthogonal frequency division multiplexing (OFDM) transmission scheme and a frequency-selective channel. The problem of maximizing the achievable rate by jointly optimizing the transmit power allocation and the RIS reflection coefficients is analyzed. The resulting optimization problem is non-convex, and the authors propose an efficient algorithm by solving the power allocation problem and the optimization of the phase shifts in an iterative manner. Simulation results show that RISs significantly improve the performance of OFDM systems.

In \cite{Li2019c}, the authors study an RIS-enhanced wide-band multiuser OFDM-based MISO system. The transmit beamformer and the RIS phase shifts are optimized, in order to maximize the average sum-rate over all subcarriers. The problem is tackled by exploiting the connection between the sum-rate maximization and the mean square error minimization, which is used to equivalently reformulate the problem and to tackle it by using alternating optimization.

In \cite{Yang2019c}, the authors consider an RIS-aided OFDM-based system and tackle the problem of maximizing the system rate by optimizing the reflection coefficients over different time slots within each channel coherence block. The resulting optimization problem is shown to be non-convex, and it is tackled by using alternating optimization with respect to the resource block assignment, the transmit power over each resource block, and the RIS phase shifts. Numerical results show that RISs are capable of increasing the system rate.

\textbf{Other performance metrics}.
In \cite{Han2019}, the authors address the problem of power control for physical layer broadcasting in the downlink of a multi-user network. In particular, the authors design the base station transmit beamforming and the RIS phase shifts by using alternating optimization, in order to minimize the base station power consumption subject to a minimum signal-to-noise ratio constraint for the mobile users. Given the sub-optimality of alternating optimization, a lower-bound of the optimal solution is derived for benchmarking purposes. Simulation results show that the proposed method approaches the lower bound as the number of RIS reflecting elements increases.

In \cite{Wu2019b}, the authors address  the problem of minimizing the base station power consumption with respect to the base station beamformer and the RIS phase shifts, subject to minimum SINR constraints at the receivers. The resulting problem is tackled by using alternating optimization in order to obtain a practical optimization method, at the price of sacrificing optimality. A power scaling law for RIS-based links is derived, and it is shown that the received power scales with the square of the number of RIS reflecting elements.

In \cite{zhou2020framework}, the authors consider an RIS-aided multi-user MISO system and investigate the problem of robust beamforming under the assumption of imperfect knowledge of the cascaded channel from the base station to the users. Alternating optimization is adopted to minimize the transmit power subject to worst-case rate constraints under the assumption of a bounded channel error model, and subject to rate outage probability constraints under a statistical channel state information error model. The results reveal that the number of RIS elements may have a negative impact on the system performance if the cascaded channel is not properly estimated.

In \cite{shao2020minimum}, the authors study the symbol error probability of an RIS-aided multiuser MISO downlink network. An expression of the symbol error probability is derived and alternating optimization is used to optimize the symbol-level base station precoder and the RIS phase shifts. The resulting algorithm is guaranteed to yield a stationary solution for the non-convex symbol error probability minimization problem. Simulation results demonstrate that the use of RISs reduces the bit error probability, especially if a large number of RIS elements is deployed.

In \cite{liu2019joint}, the authors consider the problem of joint symbol-level precoding and reflection coefficients design in RIS-enhanced multi-user MISO systems. The authors propose iterative algorithms to solve the power consumption minimization and quality of service balancing problems. Symbol-level precoding and RIS phase shifts are designed by using alternating optimization and the gradient projection and Riemannian conjugate gradient algorithms, in order solve the resulting sub-problems. The results confirm that the employment of RISs in systems in which symbol-level precoding is used can provide more efficient multi-user interference exploitation by intelligently manipulating the multi-user channels.

In \cite{jiang2019over}, the authors propose to employ RISs with the aim of improving the performance of over-the-air computation in wireless multi-user networks, thereby achieving ultra-fast data aggregation. The authors propose an alternating minimization approach to optimize the transceiver and the RIS phase shifts. The resulting non-convex quadratically constrained quadratic sub-problem is formulated as a rank-one constrained matrix optimization problem, which is tackled via the matrix lifting method.

In \cite{Zhou2019}, the authors perform the robust design of an RIS-aided multiuser MISO system wherein only imperfect channel state information is available for optimization purposes. The considered problem consists of minimizing the transmit power with respect to the base station beamformer and the RIS phase shifts, under the constraint of achievable rate guarantees for each user and for all possible channel error realizations. By leveraging some approximations, the considered resource allocation problem is transformed into a more tractable form, which is then tackled by solving a sequence of semidefinite programming subproblems that can be efficiently solved.

\textit{2) Design of Passive / Active Beamforming in MISO Systems by Using Other Optimization Techniques}\label{Sec:RA2}\\ \hfill
In \cite{guo2019model}, the authors consider a single-user MISO downlink wireless communication system, wherein multiple RISs are deployed to aid the communication. The authors tackle the problem of optimizing the phase shift matrices of all the RISs by assuming that only statistical channel state information is available. The resulting optimization problem is solved by means of the stochastic successive convex approximation method. Numerical results show that the proposed algorithms significantly outperform the random phase shift scheme, especially when the channel randomness is limited.

In \cite{li2020weighted}, the authors consider a wireless network wherein multiple RISs are deployed to cooperatively assist the communication between a multi-antenna base station and multiple single-antenna cell-edge users. The goal is to maximize the weighted sum rate of all cell-edge users by jointly optimizing the base station transmit beamforming and the RISs phase shifts. Optimal beamforming at the base station is found by using the Lagrangian method, while the RIS phase shifts are obtained based on the Rienmann manifold conjugate gradient method.

In \cite{Yu2019}, the authors maximize the spectral efficiency of an RIS-aided point-to-point MISO system. Transmit beamforming and RIS phase shifts are optimized by resorting to iterative methods and manifold optimization techniques. Two suboptimal but low-complexity algorithms are developed. Numerical results reveal that deploying large-scale RISs in wireless systems provides higher spectral and energy efficiency than increasing the size of the transmit antenna-array at the base station. This shows how RIS can be effectively used to reduce  the number of antennas deployed at the transmitter and receiver, which is an expensive part of a communication system.

In \cite{Li2019b}, the authors consider a massive MIMO system, in which multiple RISs equipped with a large number of reflectors are deployed to assist a base station that serves a small number of single-antenna users. The problem of maximizing the minimum SINR at the users is tackled by jointly optimizing the transmit precoding vector and the RISs phase shifts. The problem is converted into an RIS-user association problem, for which an exhaustive search scheme and a greedy-based search scheme are proposed.

In \cite{jia2020analysis}, the authors consider the design of an RIS-assisted system in which a multi-antenna base station serves a single-antenna user with the help of an RIS in the presence of interference generated by another multi-antenna base station serving its own single-antenna user. A tractable expression of the ergodic rate is obtained and it is used as the objective function for optimizing the RIS phase shifts. The optimization is performed by exploiting the parallel coordinate descent algorithm, which, in general, converges to a stationary point of the considered resource allocation problem.

In \cite{razavizadeh20203d}, the authors discuss the problem of using three-dimensional beamforming in RIS-empowered wireless networks. The authors, in particular, propose a new scheme that provides more degrees of freedom in designing and deploying RIS-based networks. The authors consider a base station equipped with a full dimensional array of antennas, which optimizes its radiation pattern in the three dimensional space in order to maximize the received signal-to-noise-ratio at a target user. The tilt and azimuth angles of the base station and the phase shifts at the RIS are jointly optimized. The authors study the impact of the number of reflecting elements at the RIS, by taking into account the effect of the angle of incidence of signal received by the RIS.

In \cite{yan2019passive}, the authors investigate the problem of passive beamforming and information transfer in the context of an RIS-aided MIMO system. An RIS is used for enhancing the primary communication while at the same time transmitting a second data stream by adjusting the state of the reflecting elements. The RIS phase shifts are optimized for sum-capacity maximization, by formulating the problem as a two-step stochastic program that is solved by leveraging the stochastic average approximation algorithm. 

In \cite{yu2020optimal}, the authors investigate the design of an RIS-assisted single-user MISO wireless communication system, by jointly optimizing the beamformer at the base station and the phase shifts at the RIS. The resulting non-convex optimization problem is tackled by resorting to the branch-and-bound algorithm, which enables the authors to handle the non-convex unit modulus constraints that need to be enforced on the RIS phase shifts, at the expense of an exponential complexity of the overall optimization algorithm. 

\textit{3) Design of Other RIS-Based Wireless Scenarios}\label{Sec:RA3}\\ \hfill
In \cite{gao2020reflection}, the authors consider the problem of joint transmit power allocation and RIS phase shifts optimization in RIS-aided peer-to-peer communication networks. The objective is to jointly select a sparse passive beamforming matrix and allocate the transmit power such that the minimum SINR is maximized, subject to a maximum transmit power constraint and to RIS reflection coefficients constraints. The parallel alternating direction method of multipliers is employed to tackle the resulting non-convex resource optimization problem.

In \cite{zhang2020joint}, the authors introduce the concept of RIS-aided cell-free network, which is aimed at improving the network capacity at a low cost and power consumption. In a wideband scenario, the authors formulate the problem of optimizing the base station beamforming and the RIS phase shifts, with the aim of maximizing the weighted sum-rate under a transmit power constraint at the base station and under unit-modulus phase shifts constraints at the RIS. A general joint precoding framework is proposed to solve the resulting problem.

In \cite{zhou2019ris}, the authors propose to use RISs for off-shore wireless communications. In particular, the authors assume that an RIS is placed off-shore to facilitate the communication between single-antenna ships and a multi-antenna coastal base station. In this context, the article considers the problem of optimizing the system sum-rate with respect to the base station beamforming, the RIS phase shifts, and the service time allocated to each ship. The resulting non-convex problem is tackled by means of alternating optimization, whose performance is numerically analyzed to reveal that it yields comparable performance to the case in which fully digital antenna-arrays are used. 

In \cite{he2020coordinated}, the authors consider a distributed RIS-empowered communication network architecture, in which multiple source-destination pairs communicate through multiple distributed RISs. In this scenario, the authors study the problem of maximizing the network achievable sum-rate as a function of the transmit power vector at the sources and the phase shift matrix at the distributed RISs. The resulting non-convex problem is tackled by employing alternating optimization, in which each sub-problem is cast as a fractional programming problem.

In \cite{xu2020resource}, the authors consider an RIS-based cognitive radio communication system, and focus their attention on the problem of optimizing the sum-rate of the secondary system as a function of the downlink transmit beamformers at the base station and the phase shifts at the RIS, and under quality of serice constraints for the primary system. The resulting optimization problem is non-convex due to the unit-modulus constraints imposed to the RIS reflection coefficients. A practical iterative optimization method is developed by using the alternating optimization algorithm with respect to the base station beamformer and the RIS phase shifts. 

In \cite{atapattu2020reconfigurable}, the authors consider a wireless system in which two-way communications take place between a pair of users, with the help of an RIS. Reciprocal and non-reciprocal channels are considered, and the RIS phase shifts are optimized in order to maximize the SINR at the two receivers. In the reciprocal case, the optimal RIS phase shifts are determined. In the non-reciprocal case, the problem turns out to be more difficult and is tackled by means of semidefinite relaxation. In both cases, the special case of an RIS with a single element is investigated, and optimal solutions and expressions of the system outage probability are calculated. 

In \cite{cao2020sum}, the authors consider a device-to-device communication system in which an RIS is employed to enhance the communication between the device-to-device users who underlay the cellular users. The authors formulate the problem of sum-rate maximization of the device-to-device users as a function of the transmit powers, the receive linear filters, and the RIS phase shifts. These three blocks of variables are optimized iteratively by using the alternating maximization method. Closed-form solutions for the transmit powers and receive filters are provided, while the semi-definite relaxation method is exploited for optimizing the RIS phase shifts. Numerical results demonstrate that the use of RISs significantly improves the system sum-rate.

In \cite{du2019multiple}, the authors consider a wireless system in which a multiple-antenna base station sends a multicast transmission to single-antenna mobile users, with the help of an RIS. The authors address the problem of maximizing the system capacity as a function of the transmit covariance matrix and the RIS phase shifts. Sub-gradient and gradient descent methods are adopted to tackle the resulting optimization problem. In addition, asymptotic analysis is performed in order to derive the order of growth of the maximum capacity in the regime of an infinite number of RIS reflecting elements, base station antennas, and mobile users.

In \cite{yuan2019intelligent}, the authors consider an RIS-aided cognitive radio network, wherein multiple RISs are used to improve the achievable rate of secondary users without disturbing the existing primary users in the network. The transmit beamforming vector at the secondary user transmitter and the reflecting coefficients at each RIS are jointly optimized under a total transmit power constraint at the secondary user and under an interference temperature constraints at the primary user. The impact of perfect and imperfect channel state information is analyzed.

In \cite{ye2019joint}, the authors consider the joint design of the RIS phase shifts and the full rank precoder at the transmitter in a point-to-point RIS-assisted MIMO system. The problem of minimizing the error probability is tackled by employing the alternating maximization method, wherein the two sub-problems are globally solved. In order to develop resource allocation algorithms with lower complexity, approximate solutions for the two sub-problems are derived. Numerical results indicate that the lower-complexity methods can still provide satisfactory performance. 

In \cite{Ning2019b}, the authors study the maximization of the spectral efficiency in an RIS-assisted MIMO link. The transmit beamforming and the RIS phase shifts are optimized by converting the spectral efficiency maximization problem into the simpler problem of maximizing the sum of the gains of the different communication paths. This simplified problem is then tackled by resorting to the alternating direction method of multipliers.

In \cite{sanchez2019iterative}, the authors address the uplink detection problem in the communication from multiple single-antenna users to an RIS that is connected to a central processing unit that performs data detection. The central processing unit acts as a receiver and applies a linear filter to the received signal. The authors determine the system sum-rate and derive an algorithm that yields the equalizer that maximizes the system sum-rate. Numerical results indicate that the considered RIS-based architecture outperforms  traditional approaches based on matched filtering.

In \cite{Hu2018}, the authors consider an RIS-aided wireless communication system, in which each RIS element has a separate signal processing unit and is connected to a central processing unit that coordinates the behavior of all RIS elements. The user assignments are optimized to maximize the sum-rate and the minimal user-rate. By constructing a cost matrix based on the received signal strength at each RIS element for each user, a weighted bipartite graph between the users and the RIS elements is obtained. Thanks to the constructed bipartite graph, the problem of optimal user assignments for sum-rate maximization is transformed into a linear sum assignment problem, and the problem of minimum user-rate maximization is transformed into a linear bottleneck assignment problem.

\textit{4) Discrete Phase Shifts and Phase Quantization}\label{Sec:RA4} \\ \hfill
In \cite{wu2019beamforming}, the authors address the problem of beamforming optimization under practical discrete phase shift constraints at the RIS. The transmit precoder at the access point and the discrete phase shifts at the RIS are jointly optimized to minimize the transmit power at the access point while meeting the individual SINR targets at the users. Optimal solutions and suboptimal solutions based on successive convex approximations are proposed. The authors  show that RISs with  one bit phase shifters are able to achieve the same asymptotic squared power gain as continuous phase shifts, but they are subject to a constant power loss.

In \cite{omid2020irs}, the authors consider a large-scale MIMO system in which an that employ discrete phase shifts aids the downlink transmission. Passive precoding methods are used at the base station and at the RIS. The objective is to minimize the sum power of multi-user interference by jointly optimizing the RIS beamforming and the base stations precoding vectors. A trellis-based joint RIS and base station precoding design is introduced, according to which the base station precoding in each cell is performed individually. By applying stochastic optimization, a low-overhead trellis-based optimization technique is introduced, and performance improvements are obtained by minimizing the inter-cell and intra-cell interference. In addition, semi-definite relaxation is applied to develop benchmark methods for comparison. 

In \cite{han2019large}, the authors evaluate the ergodic spectral efficiency of an RIS-assisted large-scale antenna system. An upper-bound for the ergodic spectral efficiency is computed and it is utilized for optimizing the RIS phase shifts. In particular, the authors analyze the case study in which the RIS is capable of applying only discrete phase shifts and they evaluate the minimum number of quantization bits that are required in order to ensure an acceptable spectral efficiency. Numerical results show that a two bit quantizer is sufficient to ensure an ergodic capacity degradation of at most one bit/s/Hz. 

In \cite{li2019sum}, the authors consider an RIS-aided multi-user MISO downlink system. The objective is to maximize the system sum rate by jointly optimizing the transmit beamformer at the base station and the  phase shifts at the RIS. The case studies with continuous and discrete RIS phase shifts are addressed. In order to tackle the resulting optimization problem, a smooth approximation of the objective function is derived and a block coordinate accelerated projected gradient algorithm is developed. Numerical simulations demonstrate that the proposed design achieves satisfactory performance, especially if the RIS is equipped with a large number of reflecting elements. 

In \cite{zhao2019intelligent}, the authors consider an RIS-aided multi-user system and design a two-timescale joint active and passive beamforming optimization approach. The RIS phase shifts  are optimized leveraging the statistical channel state information of all links, while beamforming at the access point is performed based on the instantaneous channel state information of the effective channels of the users, which account for the optimized RIS phase shifts. This approach is used to maximize the system weighted sum-rate, under the assumption of discrete phase shift at the RIS. A penalty dual decomposition based algorithm is developed for the single-user and multi-user case studies.

In \cite{BoyaDi20}, the authors study an RIS-based downlink multi-user multi-antenna system, in which the base station transmits signals to users via an RIS with limited-resolution phase shifters. The authors consider the practical case study in which only the large-scale fading gain is known at the transceivers. A hybrid beamforming scheme is proposed for sum-rate maximization, in which digital beamforming is performed at the base station and analog beamforming is employed at the RIS. The resulting sum-rate maximization problem is tackled by using alternating optimization.

In \cite{Liu2019}, the authors address the problem of precoder design in an RIS-based multi-user MISO wireless system, in which the RIS phase shifts have a low resolution. The information to be transmitted is encoded into the configuration of the RIS phase shifts and the problem of designing a symbol-level RIS precoder that minimizes the maximum symbol error rate of single-antenna receivers is considered. Under the assumption that the symbol-level precoding vector has a one bit resolution, the resulting optimization problem is tackled by using relaxation methods coupled with the Riemannian conjugate gradient algorithm and the branch-and-bound method.

In \cite{Wu2019c}, the authors consider a single-user system, in which an RIS with finite-resolution phase shifters aids the communication from a multi-antenna transmitter to a single-antenna receiver. The problem of transmit power minimization at the transmitter is analyzed, under a minimum signal-to-noise ratio constraint. The problem is tackled by using alternating maximization, by taking turn in optimizing the transmit beamforming and the RIS phase shifts. The authors show that the power gain at the receiver scales quadratically with the number of phase shifts of the RIS, but only in the limit of a large number of RIS reflecting elements. 

In \cite{Di2019}, the authors consider a downlink multi-user MISO system, in which an RIS is used to help a base station communicate with the downlink users. Under the assumption that the RIS can apply only discrete phase shifts, the sum-rate maximization problem is addressed by optimizing the digital beamforming at the base station and the discrete phase shifts at the RIS. The resulting resource allocation problem is tackled by an iterative algorithm based on alternating optimization.

In \cite{mu2020capacity}, the authors investigate an RIS-assisted multi-user communication system, in which a single-antenna access point sends independent information to multiple single-antenna users with the aid of an RIS that is capable of applying discrete phase shifts. The authors consider the bi-objective maximization of the system ergodic capacity and the delay-limited capacity, and characterize the corresponding Pareto boundary. The authors show that the ergodic capacity can be achieved by using an alternating transmission strategy in which the RIS phase shifts are dynamically changed. In order to achieve the delay-limited capacity, the authors show that the RIS phase shift matrix needs to be fixed at a specific value.

\textit{5) Overhead for Estimation and Control}\label{Sec:Perspectives} \\ \hfill
All available works on resource optimization for RIS-based networks are focused only on the communication phase and assume that channel state information (perfect or imperfect) has been acquired already. This is the conventional approach in the context of resource allocation in wireless networks. In RIS-based systems, however, channel estimation and feedback affect the performance of resource allocation schemes in a more fundamental way than in other instances of wireless networks. For this reason, the design of RIS-based wireless networks requires, more than in any other wireless systems, the development of algorithms for joint channel estimation/feedback and system optimization. A first contribution in this direction has recently appeared in \cite{MDR_OverheadAware}. Therein, the authors derive an expression of the rate and energy efficiency of a single-user RIS-based MIMO system that explicitly accounts for the overhead due to channel estimation and feedback. Also, the authors introduce an optimization framework for the design of the transmit beamforming, the receive filter, the RIS phase shifts, and the power and bandwidth that are allocated for both channel feedback and data transmission phases.

\subsection{Physical Layer Security}
The premise of physical-layer security is to exploit the physical properties of wireless channels in order to enhance the communication security through appropriate coding and signal processing schemes. The analysis of the potential gains and advantages of RISs for enhancing the physical-layer security of communication links has recently attracted research interests and attention.

In \cite{Chu2019}, the authors consider an RIS-aided wiretap channel model, and study the problem of minimizing the power consumption under a minimum secrecy rate constraint and by optimizing the transmit power and the RIS phase shifts. The resulting optimization problem is shown to be non-convex, and is tackled by means of semi-definite relaxation and alternating optimization between the transmit powers and the RIS phase shifts. A closed-form allocation strategy for the transmit beamformer is derived.

In \cite{Cui2019}, the authors analyze the problem of secrecy rate optimization in an RIS-aided wiretap channel, in which a multi-antenna access point communicates with a single-antenna receiver in the presence of a single-antenna eavesdropper. The authors assume that exists a strong spatial correlation among the channels of the legitimate link and the eavesdroppers. The transmit beamforming and the RIS phase shifts are optimized by resorting to alternating optimization and semidefinite relaxation methods.

In \cite{Feng2019b}, the authors consider a network scenario in which an RIS is used to help a base station equipped with a uniform linear antenna-array that transmits confidential signals to a single-antenna legitimate user in the presence of a single-antenna eavesdropper. The problem of secrecy rate maximization is addressed by optimizing the transmit beamforming and the RIS phase shifts. The authors introduce a sub-optimal but low-complexity optimization algorithm by leveraging quadratic transformations and manifold optimization techniques, whose convergence is theoretically proved.

In \cite{Shen2019}, the authors study the problem of secrecy rate maximization in an RIS-assisted wiretap channel. The authors, assume a maximum transmit power constraint at the transmitter and unit modulus constraints for the RIS phase shifts. The problem is tackled by using alternating optimization, and, in particular, the transmit power control sub-problem is solved in closed-form, and the RIS phase shifts allocation sub-problem is solved in semi-closed form. The convergence of the proposed algorithm is guaranteed theoretically.

In \cite{Feng2019}, the authors study the problem of secure transmissions in an RIS-aided wiretap channel while minimizing the system energy consumption. The authors analyze the case studies in which the channel between the transmitter and the RIS has unitary and full rank. The optimal transmit beamforming is obtained by using eigenvalue-based allocation, while the semi-definite relaxation algorithm and the projected gradient algorithm are used to compute the optimal RIS phase shifts.

In \cite{Yu2019c}, the authors study an RIS-assisted wiretap channel, in which a multi-antenna transmitter communicates with a single-antenna receiver in the presence of an eavesdropper. The problem of secrecy rate maximization is considered with respect to the transmit beamformer and the RIS phase shifts. The resulting optimization problem is tackled by using alternating optimization and sequential optimization, which leads to an efficient but, in general, sub-optimal algorithm. 

In \cite{Guan2019}, the authors analyze a wiretap channel in which an RIS is used to aid a multi-antenna transmitter to communicate with a single-antenna receiver in the presence of multiple single-antenna eavesdroppers. The authors employ artificial noise for improving the system achievable secrecy rate. Alternating optimization is used for optimizing the transmit beamforming, the artificial noise jammer, and the RIS phase shifts.

In \cite{Yu2019b}, the authors use artificial noise for ensuring secure communication in an RIS-assisted network. In particular, a multi-antenna access point exploits an RIS in order to serve multiple single-antenna legitimate users in the presence of multiple multi-antenna potential eavesdroppers, whose channel state information is not perfectly known at the access point. The access point makes use of artificial noise to jam the eavesdroppers and maximize the system secrecy rate. To solve this optimization problem, the authors develop an algorithm based on alternating optimization, successive convex approximation, and semidefinite relaxation.

In \cite{Lu2019}, the authors study an RIS-based system in which the secrecy of the communication link is enhanced by making the signal transmission covert to potential eavesdroppers. This is achieved by using an RIS that focuses the transmit signal only towards the direction of the intended receiver. The authors substantiate the proposed approach in a system scenario in which a single-antenna transmitter communicates with a single-antenna receiver. In particular, the signal leakage towards a single-antenna eavesdropper is minimized by optimizing the transmit power and the RIS phase shifts.

In \cite{Chen2019}, the authors study a downlink MISO system, in which the base station transmits independent data streams to multiple legitimate receivers in the presence of multiple eavesdroppers. By jointly optimizing the beamformers at the base station and the RIS phase shifts, the authors maximize the minimum secrecy rate under the assumption of continuous and discrete RIS phase shifts. The optimization problem is solved by using alternating optimization and the path-following algorithm.

In \cite{hong2020artificial}, the authors consider a MIMO wireless communication system that employs physical layer security. The transmitter sends artificial noise in order to maximize the system secrecy rate. This is achieved by optimizing the precoding matrix at the transmitter, the covariance matrix of the artificial noise, and the phase shifts at the RIS under the assumption of a maximum transmit power limit and unit modulus constrains for the phase shifts. The optimization problem is tackled by using the block coordinate descent algorithm.

In \cite{dong2020secure}, the authors consider an RIS-assisted Guassian MIMO wiretap channel, in which a multi-antenna transmitter communicates with a multi-antenna receiver in the presence of a multi-antenna eavesdropper. The secrecy rate is maximized by using an alternating optimization algorithm, which jointly optimizes the transmit covariance and the phase shifts at the RIS. The successive convex approximation algorithm is adopted to solve the optimization sub-problems.

In \cite{jiang2020intelligent}, the authors study an RIS-based MIMO network, in which a multi-antenna access point communicates with a multi-antenna legitimate user in the presence of a multi-antenna eavesdropper. The authors investigate the joint optimization of the transmit covariance matrix at the access point and the phase shifts at the RIS, in order to maximize the system secrecy rate under the assumption of continuous and discrete RIS phase shifts. The optimization problem is tackled by applying alternating optimization to the transmit covariance matrix at the access point and to the phase shifts at the RIS.

In \cite{wang2020angle}, the authors study the physical-layer security in a massive MIMO multicasting system. The authors propose an angle-aware user cooperation scheme that avoids direct transmission to the attacked user and relies on other users for cooperative relaying. The proposed scheme requires only the angle information about the eavesdropper. By considering an angular secrecy model to formulate the average secrecy rate of the attacked system, the authors use successive convex optimization to design the beamformer.

In \cite{xu2019resource}, the authors consider the secure communication in RIS-aided MISO communication systems. To enhance physical layer security, artificial noise is transmitted from the base station in order to impair the eavesdroppers. The problem of jointly optimizing the phase shifts at the RIS and the beamforming vectors and artificial noise covariance matrix at the base station is investigated. The secrecy rate maximization problem is tackled by using non-convex optimization. In addition, a suboptimal algorithm based on alternating optimization, successive convex approximation, semidefinite relaxation, and manifold optimization is proposed.

In \cite{lyu2020irs}, the authors consider a jammer that attacks a legitimate communication without using any internal energy to generate jamming signals. The jammer uses an RIS that controls the reflected signals in order to decrease the SINR at the legitimate receiver. Simulation results show that, although no energy is used, the proposed RIS-based jammer outperforms conventional active jamming attacks.

In \cite{makarfi2019physical}, the authors study the physical layer security in a vehicular network that employs an RIS. Two network models are investigated: A vehicle-to-vehicle communication in which the transmitter employs an RIS-equipped access point, and a vehicular ad-hoc network in which an RIS-equipped relay in deployed in a building. Both system setups assume the presence of an eavesdropper and the average secrecy capacity is investigated. Numerical results show that the secrecy capacity is affected by the location of the RIS and the size of the RIS.

\subsection{Non-Orthogonal Multiple Access}
Non-orthogonal multiple access (NOMA) has received major attention for the design of radio access techniques in future wireless networks. The basic concept behind NOMA is to serve more than one user in the same resource block, e.g., time slot, subcarrier, spreading code, space. Based on this approach, NOMA promotes massive connectivity, lowers latency, improves user fairness and spectral efficiency, and increases reliability compared to orthogonal multiple access (OMA) techniques. The potential advantages and limitations of RISs in the context of NOMA-based networks have been investigated in a few recent papers.

In \cite{Fu2019}, the authors analyze the use of RISs in the context of NOMA systems. Specifically, an RIS is used to enhance the downlink communication from a base station to mobile users. The base station beamforming and the RIS phase shifts are optimized in order to minimize the system total power consumption under the constraint of given rate requirements at the receivers. The problem is tackled by using alternating optimization and the difference of convex functions programming. The resulting algorithm is shown to be provably convergent but, in general, suboptimal.

In \cite{Li2019}, the authors consider an RIS-aided NOMA-based downlink MISO channel. The authors minimize the total transmit power by optimizing the transmit beamforming vectors at the base station and the reflection coefficient vector at the RIS. This optimization problem is tackled by using semidefinite programming, second-order cone programming, and the alternating direction method of multipliers. A low-complexity algorithm based on zero-forcing transmission is proposed.

In \cite{Mu2019}, the authors analyze the sum-rate optimization of an RIS-aided NOMA-based downlink MISO channel. In particular, the system sum-rate is maximized with respect to the base station beamformer and the RIS phase shifts. The problem is solved by using alternating optimization and by applying successive convex approximation methods. The case study of discrete phase shifts at the RIS is studied and a quantization-based scheme is proposed.

In \cite{Yang2019b}, the authors maximize the minimum rate among the users of an RIS-assisted downlink NOMA system. The optimization problem is formulated in terms of the transmit beamforming vector and the RIS phase shifts. A user ordering scheme based on the signal strength of the combined channel is introduced, in order to decouple the user-ordering design and the  beamforming design. The resulting non-convex problem is tackled by using alternating optimization and semi-definite relaxation methods. Single-antenna and multi-antenna system setups at the base station are analyzed.

In \cite{zheng2020intelligent}, the authors compare the performance between NOMA and OMA. The authors optimize the RIS reflection coefficients, by assuming a discrete resolution for the phase shifts, by minimizing the transmit power of the access point under the assumption of some specified constraints for the user rates. It is shown that NOMA may underperform OMA in some network configurations.

In \cite{ding2019simple}, the authors propose an RIS-based NOMA transmission scheme in which the number of users that can be served in each orthogonal spatial direction is larger compared with spatial division multiple access schemes. It is shown that, by employing an RIS, the directions of the channel vectors of the users can be effectively aligned and that this facilitates the implementation of NOMA.

In \cite{ding2020impact}, the authors study the performance of two phase shifting designs, i.e., random phase shifting and coherent phase shifting, for application to RIS-assisted NOMA systems. Simulation results prove the accuracy of the obtained analytical results and are used to compare RIS-NOMA with respect to conventional relaying and RIS-OMA.

In \cite{fu2019intelligent}, the authors study the downlink transmit power minimization problem in RIS-empowered NOMA networks, by jointly optimizing the transmit beamformer and the RIS phase shifts. The optimization problem is tackled by using alternating minimization. In particular, the underlaying subproblems are shown to be non-convex quadratic programs, and they are tackled by using difference of convex programming methods.

In \cite{yue2020performance}, the authors consider a NOMA-based system in which a base station transmits a multiplex of signals to multiple users by means of an RIS. The performance of the system model is studied by assuming perfect and imperfect successive interference cancellation and a one bit coding scheme. Exact and asymptotic expressions of the system outage probability and ergodic rate are derived. From the obtained analytical frameworks, the authors show that RIS-NOMA outperforms RIS-OMA and conventional relaying schemes, and that the performance of RIS-NOMA improves if the number of reflecting elements increases.

In \cite{zuo2020resource}, the authors study the downlink of an RIS-assisted NOMA system. In order to maximize the system throughput, the authors formulate an optimization problem as a function of the channel assignment, the decoding order of NOMA users, the transmit power, and the RIS reflection coefficients. The resulting optimization problem is non-convex, and a three-step alternating optimization algorithm is proposed to tackle it. In particular, the channel assignment problem is solved by using a many-to-one matching algorithm, the decoding order problem is solved by introducing a low complexity algorithm, and, finally, given the channel assignment and decoding order, the joint power allocation and reflection coefficient design is solved. The three sub-problems are solved iteratively until convergence.

In \cite{zhu2019power}, the authors optimize the beamforming vectors and the RIS phase shifts in an RIS-assisted NOMA system. The authors propose an improved quasi-degradation condition under which RIS-NOMA achieves the same performance as dirty paper coding. The authors characterize the optimal beamforming of RIS-assisted NOMA and RIS-assisted zero-forcing schemes, by assuming a given RIS phase shift matrix. It is shown that NOMA always outperforms zero-forcing if the improved quasi-degradation condition is considered and the same RIS phase shift matrix is used.

In \cite{hou2019reconfigurable}, the authors propose a priority-oriented design for enhancing the spectral efficiency of an RIS-aided NOMA network. Accurate and approximated closed-form expressions for the outage probability and the ergodic rate are derived. Numerical results show that the diversity order of the system can be enhanced by increasing the number of phase shifts of the RIS.

\subsection{Internet of Things and Backscattering}
In the broadest sense, the Internet of Things (IoT) encompasses everything connected to the Internet, but it is often used to define objects that talk to each other. The IoT is made up of massive deployments of devices, encompassing sensors, smart phones, wearables, etc., which are all connected together. One of the major open challenges for the IoT is the limited network lifetime, which is due to the massive deployment of devices that are powered by batteries of finite capacities. In this context, backscatter communication (or simply backscattering), which enables the transmission of information through the passive reflection and modulation of an incident radio wave, has emerged as a promising technology for enabling the connectivity of massive deployments of devices at a low power consumption and at a low complexity. In \cite{MDR_Survey_EURASIP}, the authors have elaborated on the potential applications and benefits of employing RISs for IoT applications. Since then, several authors have started researching on the applications of RISs in this context. 

In \cite{park2020intelligent}, the authors propose an RIS-aided phase shift backscatter communication system. The authors study the scenario in which the RIS is employed for transmitting secondary information data while relaying primary information data. This is obtained by employing the RIS to relay the primary information data through beamforming and by encoding the secondary information data onto the phase shifts. The spectral efficiency of the secondary system subject to a minimum spectral efficiency requirement for the primary system is analyzed, and it is shown that RIS-aided backscattering is capable of outperforming the system performance.

In \cite{zhang2020large}, the authors propose an RIS-assisted symbiotic radio system. The joint design and optimization of the transmit beamformer and the reflecting phases at the RIS is studied, by minimizing the total transmit power under a rate constraint for the primary transmission and a signal-to-noise-ratio constraint for the IoT communication. Alternating optimization is adopted to tackle the resulting optimization problem. 

In \cite{ZhaoPerformanceAnalysis20}, the authors study an RIS-aided backscatter system in order to support ultra-reliable communications for IoT applications. By considering random and optimized phase shifts at the RIS, the authors compute the error probability, by using the moment generating function approach. Tight upper bounds are obtained for both independent and correlated channels. Numerical results unveil the impact on the achievable error performance of the number of reflecting elements of the RIS.

In \cite{makarfi2019reconfigurable}, the authors analyze and RIS-aided IoT system over composite fading and shadowing channels. Exact and approximate expressions for the average capacity, average error probability, and outage probability are given. The impact of the transmit power, the fading and shadowing severity, and the size of the RIS is analyzed.

\subsection{Aerial Communications}
Over the last years, the use of Unmanned Aerial Vehicles (UAVs), commonly referred to as drones, has received significant attention from various industry sectors and research communities. This is mostly attributed to the broad range of UAV-enabled use cases in, e.g., agriculture, public protection, disaster relief, logistics, media production, mapping, etc. The transformation of UAVs into distributed aerial communication and computing platforms is gathering pace by leveraging research advancements in 5G and 6G technologies, edge computing, and machine learning. This transformation promises exciting and high impact use cases and innovations far beyond legacy UAV solutions. A few recent papers have started investigating the possible applications of RISs in the context of UAV-based communications.

In \cite{Li2019d}, the authors study the joint design of the trajectory of an UAV and the beamforming design of an RIS in order to maximize the average achievable rate. To tackle the resulting non-convex optimization problem, the authors divide the optimization into two sub-problems that consist of optimizing the beamforming at the RIS and the trajectory of the UAV. The authors derive a closed-form solution for the beamforming vector for a given trajectory. A sub-optimal solution for the trajectory is obtained by using successive convex approximation methods. Numerical results demonstrate that the use of RISs enhances the average achievable rate.

In \cite{Ma2019}, the authors consider the deployment of RISs on the walls of buildings and assume that they can be remotely configured by cellular base stations in order to coherently direct the reflected radio waves towards specific UAVs for increasing the received signal strength. By using standardized ground-to-air channel models, the authors analyze the signal gain at the UAVs, as a function of the height of the UAVs, and the size, altitude, and distance from base station of the RIS. The authors identify the optimal location (height and distance from the base stations) of the RIS that maximizes the performance.

In \cite{lu2020enabling}, the authors introduce a three dimensional networking architecture, enabled by UAVs and RISs, in order to achieve panoramic signal reflections from the sky. The authors study the problem of maximizing the worst-case signal-to-noise ratio in a given coverage area, by jointly optimizing the transmit beamforming, and the placement and phase shifts of RIS-mounted UAVs. The resulting problem is non-convex and the optimization variables are coupled. To tackle this problem, the authors develop an efficient sub-optimal solution that exploits the similarity between optimization of the RIS phase shifts and the optimization of the analog beamforming of phased arrays. 

In \cite{Zhang2019d}, the authors consider a UAV-carried RIS in order to enhance the performance of millimeter-wave networks. In particular, the UAV-carried RIS is used to intelligently reflect signals from a base station towards a mobile outdoor user, while harvesting energy from the signals so as to power the RIS. To maintain a line-of-sight channel, a reinforcement learning approach is used to model the propagation environment, so that the location and reflection coefficient of the UAV-carried RIS are optimized to maximize the downlink transmission capacity. Simulation results show the advantages of using a UAV-carried RIS as compared with a conventional RIS, in terms of average data rate and achievable downlink line-of-sight connectivity.

\subsection{Wireless Power Transfer}
Wireless power transfer is the transmission of electrical energy without using wires as a physical link. In a wireless power transmission system, a transmitter (driven by electric power from a power source) generates a time-varying EM field towards a receiver that extracts power from the field and supplies it to an electrical load. The technology of wireless power transmission can eliminate the use of wires and batteries, thus increasing the mobility, convenience, and safety of an electronic device. Wireless power transfer has received major attention from the research community, and several authors are currently investigating the potential uses and applications of RISs in this context.

In \cite{ZhengIRS20}, the authors consider an RIS that assists the transmission of a two-user cooperative wireless-powered communication network. The authors study the problem of maximizing the throughput, by jointly optimizing the phase shifts of the RIS, the transmission time, and the power allocation strategy. Numerical results show that the use of RISs can effectively improve the throughput of cooperative transmissions.

In \cite{Pan2019}, the authors study a downlink MISO wireless network with simultaneous wireless and information power transfer, in which an RIS is employed for enhancing the wireless transfer capabilities of the system. The authors analyze the problem of maximizing the system weighted sum-rate with respect to  the transmit precoding matrices of the base station and the phase shifts of the RIS. Alternating optimization is employed to tackle the resulting optimization problem, and, in particular, the authors propose two low-complexity iterative algorithms that converge to a first-order optimal point of the respective optimization sub-problem.

In \cite{Tang2019}, the authors consider the downlink of a MISO system in which a base station sends information and energy signals to a set of receivers and an RIS is used to aid the information and energy transfer. In this setup, the base station beamformer and the RIS phase shifts are alternatively optimized with the aim of maximizing the minimum power received at all the energy-harvesting receivers, by fulfilling individual SINR constraints for the information receivers and a maximum transmit power constraint for the base station. The resulting optimization problem is tackled by using alternating optimization and semi-definite relaxation, thus obtaining a sub-optimal but efficient algorithm.

In \cite{Wu2019d}, the authors analyze the downlink of a MISO system in which the simultaneous transmission of information and power is considered. In the considered system model, multiple RISs are deployed in order to aid the information and power transfer. The authors consider the problem of minimizing the transmit power of the base station through the optimization of the transmit precoders and the phase shifts of the RISs. The use of alternating optimization methods to tackle the problem is shown to be unsuitable and a penalty-based algorithm is proposed instead. 

In \cite{shi2019enhanced}, the authors analyze the downlink of an RIS-aided MISO system where simultaneous wireless information and power transfer is employed. The authors study the design of secure transmit beamforming and RIS phase shifts in order to maximize the harvested power. In order to solve the resulting non-convex optimization problem, two alternating iterative algorithms based on semi-definite relaxation and successive convex approximation are proposed.

\subsection{Multiple-Access Edge Computing}
Multi-access edge computing (MEC) is a network architecture concept that enables cloud computing capabilities and an information technology service environment at the edge of any network, e.g., a cellular network. The basic idea behind MEC is that, by running applications and performing related processing tasks closer to the customers, network congestion is reduced and applications perform better. MEC technology is designed to be implemented at the cellular base stations or at other edge nodes. It enables flexible and rapid deployment of new applications and services for customers. Recently, the application of RISs in this context has been investigated by some researchers.

In \cite{Bai2019}, the authors consider a scenario in which single-antenna devices may decide to offload part of their computational tasks to the edge computing node via an RIS-aided multi-antenna access point. In this context, the authors formulate a latency minimization problem, by considering single-device and multi-device scenarios, in which constraints on the edge computing capabilities are imposed. Alternating optimization is used to tackle the optimization problem, by alternatively optimizing computing and communication operations. The computing sub-problem is solved optimally while only a first-order optimal solution is determined for the communication sub-problem.

In \cite{Hua2019}, the authors consider an RIS-aided system model in which inference tasks generated from mobile devices are uploaded to and cooperatively performed by multiple base stations. The objective is to minimize the network power consumption as a function of the set of tasks performed by each base station, the base stations transmit and receive beamforming vectors, the transmit power of the mobile devices, and the RIS phase shifts. The resulting combinatorial problem is addressed by exploiting the group sparsity structure of the beamforming vectors and by employing alternating maximization to decouple the optimization variables.

In \cite{cao2019intelligent}, the authors study an RIS-assisted MEC system model for application to the millimeter-wave frequency band. In particular, the RIS is employed to alleviate link blockage problems in an economical manner and to guarantee the real-time offloading of computing tasks. The objective is to minimize the uplink mobile power for all users while satisfying the offloading latency constraints, where the power of the individual devices, the multi-user detection matrix, and the beamforming coefficients of the RIS are jointly optimized. An alternating optimization framework is developed, in order to decompose the joint optimization of all coupled variables into tractable sub-problems. Simulation results highlight the benefits of using RISs in the context of MEC.

In \cite{liu2020intelligent}, the authors consider the problem of computation offloading for MEC in a wireless network equipped with RISs. In particular, mobile devices are assumed to offload computation tasks to an edge server located at the access point in order to reduce a cost function that is a weighted sum of time and energy. The edge server adjusts the phase shifts of the RIS in order to maximize its earning while maintaining the incentives of the mobile devices for offloading and guaranteeing each of them a customized information rate. This resulting optimization problem is tackled by using an iterative evaluation procedure that identifies the feasibility of the problem when confronting an arbitrary set of information rate requirements. Numerical results show that the use of RISs enables the access point to guarantee higher information rate to all mobile devices, while at the same time improving the earning of the edge server.

\subsection{Millimeter-Wave, Terahertz, and Optical Wireless Communications}
Wireless data traffic has been increasing at a high rate and this trend is expected to accelerate over the next decade. To address this demand, the wireless industry is designing future wireless transmission technologies and standards in order to unlock the potential opportunities offered by large and unused frequency bands, which include the millimeter-wave, the terahertz, and the visible light spectrum. Since the wavelengths shrink by order of magnitudes, compared to currently used microwave frequencies, at these high frequency bands, diffraction and material penetration will incur greater attenuation, thus elevating the importance of line-of-sight propagation, reflection, and scattering. RISs are considered to be an enabling technology for controlling the radio waves and for enhancing the connectivity of wireless networks that operate at these emerging high frequency bands.

In \cite{Cao2019}, the authors study the application of RISs in order to overcome signal path-loss and blind spots in millimeter-wave systems. The problem of maximizing the sum-rate in a multi-user network is addressed by jointly optimizing  the transmit beamforming and the RIS phase shifts. With the aid of alternating maximization, the resulting two optimization sub-problems are solved in closed-form, and the initial optimization problem is tackled by iterating between them until convergence. The impact of discrete phase shifts is investigated as well, and a projection method is developed to tackle the associated optimization problem.

In \cite{chaccour2020risk}, the authors consider a virtual reality network and study the problem of associating RISs to virtual reality users for application in the terahertz frequency band. In particular, the authors formulate a risk-based framework based on the entropic value-at-risk and optimize rate and reliability. The formulated problem accounts for the higher order statistics of the queue length, thus guaranteeing continuous reliability. Tools such as Lyapunov optimization, deep reinforcement learning, and recurrent neural network are adopted to tackle the optimization problem.

In \cite{ying2020gmd}, the authors propose geometric mean decomposition-based beamforming for application to RIS-assisted millimeter-wave hybrid MIMO systems. By exploiting the common angular-domain sparsity of millimeter-wave channels over different sub-carriers, an orthogonal matching pursuit algorithm is proposed to obtain the optimal beamforming vector. By leveraging the angle of arrival and angle of departure associated with line-of-sight channels, the authors design the RIS phase shifts by maximizing the array gain for line-of-sight channels. Simulation results show that the proposed scheme achieves better error performance.

In \cite{cao2019delay}, the authors consider an RIS-assisted millimeter-wave communication system in which an RIS is used to overcome the impact of spatial blockages. The objective is to jointly optimize the power of individual devices, the multi-user detection matrix, and the beamforming at the RIS. To this end, new methods for minimizing the power under delay requirements are proposed, and the optimization problem is solved by using alternating optimization. In particular, the configuration of the RIS is formulated as a sum-of-inverse minimization fractional programming problem, and it is solved by using the alternating direction method of multipliers. The proposed approach is shown to outperform semidefinite relaxation techniques.

In \cite{xiu2020irs}, the authors consider an RIS-aided downlink wireless system in the millimeter-wave frequency band and formulate a joint power allocation and beamforming design problem in order to maximize the weighted sum-rate. To solve the problem, the authors propose a novel alternating manifold optimization based beamforming algorithm. Simulation results show that the proposed optimization algorithm outperforms existing algorithms.

In \cite{chen2019sum}, the authors consider an indoor communication system for application to terahertz communications and study the application of RISs for improving the reliability of signal transmission. The authors study the problem of optimizing the sum-rate performance by selecting the optimal RIS phase shifts. Two algorithms based on a local search method and cross-entropy method are proposed and their performance analyzed with the aid of numerical simulations.

In \cite{jamali2019intelligent}, the authors analyze an RIS-aided millimeter-wave massive MIMO system and show that a proper positioning of the active antennas with respect to the RIS leads to a considerable improvement of the system spectral efficiency. It is revealed that conventional MIMO architectures are either energy inefficient or expensive and bulky if the number of transmit antennas is large. On the other hand, RIS-aided MIMO architectures are shown to be highly energy efficient and fully scalable in terms of number of transmit antennas.

In \cite{najafi2019intelligent}, the authors consider the application of RISs in order to alleviate the line-of-sight requirement in free space optical systems. The authors develop conditional and statistical channel models that characterize the impact of the physical parameters of an RIS, which include the size, position, and orientation, on the quality of the end-to-end channel. Furthermore, the authors propose a statistical channel model that accounts for the random movements of the RIS, the transmitter, and the receiver. With the aid of the proposed channel models, the authors analyze the advantages and limitations of RISs in different scenarios.

In \cite{Perovi2019}, the authors consider the application of RISs for improving the channel capacity of millimeter-wave indoor networks. More precisely, the authors propose two optimization schemes that exploit the customizing capabilities of the RIS reflection elements in order to maximize the channel capacity. The first optimization scheme exploits only the adjustability of the RIS reflection elements. For this scheme, approximate expressions of the channel capacity are derived and used to infer the connection between the channel capacity gains and the system parameters. The second optimization scheme jointly optimizes the RIS reflection elements and the transmit phase precoder. For this scheme, the authors propose a low-complexity technique called global co-phasing to determine the phase shift values for use at the RIS. Simulation results show that the optimization of the RIS reflection elements produces significant channel capacity gains, and that this gain increases with the number of RIS elements.

In \cite{Pradhan2019}, the authors study a millimeter-wave system in which a base station that employs hybrid precoding communicates with multiple single-antenna users through an RIS. The mean square error between the received symbols and the transmitted symbols is minimized by optimizing the hybrid precoder at the base-station and the RIS phase shifts. The optimization problem is tackled by resorting to the gradient projection method. The resulting algorithm is shown to be provably convergent to a point that fulfills the first-order optimality conditions of the considered optimization problem.

In \cite{Wang2019b}, the authors analyze a single-user downlink millimeter-wave system model in the presence of multiple RISs. The transmit beamforming and the RIS phase shifts are optimized by analyzing single-RIS and multi-RIS system setups. In the single-RIS case, the optimal solution that maximizes the signal-to-noise ratio is derived, under the assumption of rank-one channel. In the multi-RIS case, a sub-optimal solution based on the semidefinite relaxation method is proposed. The authors show that RISs provide an effective way for signal focusing that is potentially useful for applications in the millimeter-wave frequency band. In \cite{Wang2019}, the authors generalize the study by analyzing the impact of the finite resolution of the RIS phase shifts.

In \cite{wang2020performance}, the authors investigate the use of RISs for optical communications. In particular, the authors consider a system model in which multiple optical RISs are deployed in the environment and are used to build multiple artificial channels in oder to improve the system performance and to reduce the outage probability. The authors analyze the impact of three factors that affect the channel coefficients: beam jitter, jitter of the RISs, and the probability of obstruction. Based on the proposed model, the authors derive the probability density function of the channel coefficients, the asymptotic average bit error rate, and the outage probability for systems with single and multiple RISs. It is revealed that the probability density function contains an impulse function, which causes irreducible error rate and outage probability floors.

\subsection{Relays and Massive Multiple-Input-Multiple-Output Systems}
When new technologies come into the limelight, it is dutiful to critically investigate the potential benefits and limitations that they may provide as compared with similar and well established technologies. Therefore, it is sensible to compare RISs with transmission technologies that may be considered to be closely related to them. In the research literature, two technologies are often considered to be the most similar to RISs: Relays and massive MIMO. In general terms, RISs are different with respect to relays and massive MIMO because the former emerging technology is intended to be realized in a passive way with minimal power amplification and signal processing capabilities, while the latter two technologies rely upon the utilization of individual RF chains and powerful signal processing algorithms. The long-term vision of RIS-based SREs consists of realizing a multi-function platform in which the transmission, processing, and computation are, as much as possible, moved to the EM level. Relays and massive MIMO, have been envisioned, on the other hand, as platforms in which the EM signal are first converted into the digital domain and, after processing, are converted back to the EM domain. The large-scale experimental platforms reported in \cite{RFocus}, \cite{Jamieson_2017}, \cite{Jamieson_2019}, \cite{Bharadia_2020}, which are described in Section \ref{Testbeds}, clearly point this out. Despite these general considerations, it is important to identify a fair basis for the comparison of RISs against relays and massive MIMO. A few research papers have recently attempted to shed light on this issue. In \cite{Hu2017}, notably, the authors coin the term LIS and describe it as ``\textit{a new concept that conceptually goes beyond contemporary massive MIMO technology}''.

In \cite{MDR_RISs_Relays} and \cite{Bjornson2019}, the authors report a comparative study of RISs against decode and forward relaying. In \cite{MDR_RISs_Relays}, in particular, the authors discuss all the advantages and disadvantages of RISs and relays in a comprehensive manner. From a performance comparison point of view, both papers lead to similar conclusions: RISs may outperform relaying provided that the size of RISs is sufficiently large. If one considers the implementation of RISs by using a  large number of inexpensive antennas, this implies that several antenna elements are needed in order to achieve good performance without the need of using power amplification and signal regeneration. This finding was somehow expected. The interesting finding in \cite{MDR_RISs_Relays} is, however, that a sufficiently large RIS that is configured to operate as a simple anomalous reflector may outperform an ideal full-duplex single-antenna decode and forward relay in terms of achievable data rate. If one considers other performance metrics, such as the energy efficiency and the power consumption, the gains may be larger if RISs are realized without using active elements and power amplifiers. It is important to mention, however, that the signal manipulations that can be realized at the EM level are simpler than those that can be realized in the digital domain. In addition, the implementation of metasurface-based RISs with a high power efficiency requires the design of sophisticated metasurface structures, which are the state-of-the-art of research in the field.

In \cite{Dardari_DegreesFreedom} and \cite{bjornson2020power}, the authors report a comparative study of RISs against MIMO and massive MIMO. The author of \cite{Dardari_DegreesFreedom} analyzes the fundamental aspects of the communication between RISs from the EM standpoint, with a focus on the degrees of freedom that are offered by the system.  The departing point of the author is that the spatial multiplexing gains of MIMO systems, i.e., the channel degrees of freedom, are achieved in rich scattering environments and they are usually lost in favor of beamforming gains in rank-deficient channels. The main finding of the author is that the degrees of freedom offered by RISs are greater than one in propagation environments that are not characterized by rich scattering, e.g., in line-of-sight channels or for transmission at high frequencies (the millimeter-wave frequency range and beyond). Based on the findings of the author, this is, on the other hand, not possible in conventional MIMO systems. For this reason, the author concludes that RISs may outperform MIMO systems in those scenarios and may ultimately offer a large increase of the spatial capacity density. The authors of \cite{bjornson2020power} move from the findings in \cite{dardari2019communicating}, and analyze the scaling laws of RISs and massive MIMO as a function of the number of reflecting elements. The authors show that the scaling law is quadratic and linear in the number of antenna elements for RISs and massive MIMO, respectively. RISs are, however, expected to need a larger number of reflecting elements for achieving similar signal-to-noise ratios as massive MIMO, owing to the inherent assumption that RISs can apply only phase shifts and cannot amplify the incident EM fields if they are assumed to be made of locally or globally passive smart surfaces.

\subsection{Software Defined Networking Based Design and Nano-Communication Networks}
An important component for realizing the vision of SREs is the capability of controlling and programming the RISs in a software-based fashion \cite{Liaskos_COMMAG}, \cite{Liaskos_ACM}. More specifically, the long-term vision consists of realizing fully adaptive metasurfaces that are flexible enough to synthesize multiple concurrent functionalities at the EM level. Researchers of the European-funded project VisorSurf \cite{VisorSurf} have been researching towards the design and implementation of software-defined metamaterials for the last four years. Software-defined metamaterials are built upon descriptions of EM functionalities in reusable software modules that are specifically designed for changing the behavior of RISs by sending preset commands. In simple terms, the EM properties of RISs can be reconfigured at runtime by using a set of software primitives. In order to disseminate, interpret, and apply the commands, RISs need, however, the integration of a network of miniaturized controllers within their structure, which are capable of interacting locally and communicating globally to realize the desired EM behavior. The realization of truly software-defined metamaterials is, therefore, coupled with the realization of the network of miniaturized controllers, which are equipped with communication capabilities that can, e.g., be realized by exploiting nano-communication technologies. Several research papers on the design of software-defined metamaterials and nano-communication networks for application to metasurfaces have recently appeared in the literature. The research issues tacked in these papers encompass the definition of software-defined primitives, the analysis of communication and security issues, the design of medium access control protocols, routing protocols, and the network layer configuration and operation \cite{VisorSurf__6}, \cite{VisorSurf__7}, \cite{VisorSurf__4}, \cite{VisorSurf__5}, \cite{VisorSurf__3}, \cite{VisorSurf__1}, \cite{VisorSurf__2}. Interested readers are referred to \cite{VisorSurf} for further information.

\subsection{Machine Learning Based Design}
The field of machine learning has a long and extremely successful history. Machine learning has shown its overwhelming advantages in many areas, including computer vision, robotics, and natural language processing, where it is normally difficult to find a concrete mathematical model for feature representation. In those areas, machine learning has proved to be a powerful tool as it does not require a comprehensive specification of the model. Different from the aforementioned machine learning applications, the development of communications has vastly relied on theories and models, from information theory to channel modeling. These traditional approaches are currently showing some limitations, especially in view of the increased complexity of communication networks. Therefore, research on machine learning applied to communications, especially to wireless communications, is currently experiencing an incredible surge of interest. Recently, in addition, wireless researchers have started researching on the possibility of combining together model-based and data-driven approaches, in an attempt of capitalizing on their advantages while overcoming their inherent limitations. Machine learning is considered to be a potential enabler for realizing the vision of RIS-empowered SREs.

In \cite{MDR_AM_TCOM}, the authors put forth the connection between RISs and machine learning, arguing that both are needed to realize the vision of SREs. The authors discuss how the use of machine learning is essential to reduce the complexity related to the design of RIS-based wireless networks, which is anticipated to be higher than in conventional wireless networks. To this end, the authors propose a model-aided deep learning framework, which is further elaborated in \cite{MDR_AI_VTM}, that exploits transfer learning to simplify wireless network design by combining together model-based and data-driven methods. The same authors discuss the connection between RIS-empowered SREs and reinforcement learning in \cite{MDR_AI_RIS}.

In \cite{liaskos2019interpretable}, the authors introduce a neural-network-based approach for configuring the behavior of tiles in RIS-based environments. The wireless propagation is modeled as a custom, interpretable, back-propagating neural network, in which the RIS elements act as nodes and their cross-interactions as links. After a training phase, the neural network learns how to configure the RISs for improving the communication performance. By using a ray tracing simulation environment, the authors  show performance gain in agreement with state-of-the-art solutions, but with a distinct gain in reducing the total number of active tiles.

In \cite{Huang2019}, the authors employ an artificial neural network to maximize the received power in a RIS-based network. The downlink of a multi-user MISO system is considered, and the deep learning framework developed in \cite{MDR_AM_TCOM} is applied. An indoor scenario is considered, and a deep neural network is used as a function approximator to learn the map between the locations of the users and the corresponding configuration of the RIS reflecting elements that maximizes the signal-to-noise ratio at the intended receiver.

In \cite{gao2020unsupervised}, the authors employ deep learning techniques to reduce the design complexity of RIS-based wireless networks. In particular, the authors propose an unsupervised approach to optimize the RIS phase shifts. In the proposed approach, a customized deep neural network is trained offline by using the objective function to be optimized for training the parameters of the deep neural network. Simulation results show a gain over conventional approaches based on the use of semi-definite relaxation and alternating optimization, even though the considered approach remains heuristic in the sense that no optimality property can be claimed. 

In \cite{huang2020reconfigurable}, the authors propose a joint design of transmit beamforming and phase shifts in a RIS-assisted MIMO system, by using a deep reinforcement learning framework. The proposed deep reinforcement learning algorithm is shown to be scalable for accommodating various system settings. Instead of utilizing conventional alternating optimization techniques to obtain the transmit beamforming and the RIS phase shifts, the proposed algorithm obtains the joint design simultaneously at the output of a deep neural network whose training is refined in real-time.

In \cite{taha2020deep}, the authors propose a deep reinforcement learning framework with reduced training overhead that is able to tune the phase shifts of the RIS elements with reduced training overhead. The proposed approach paves the way to the deployment of distributed RISs, which are able of self-configuration and operation without the assistance of base stations or infrastructure nodes. The authors illustrate numerical results that show that the proposed online learning framework is able to approach the same rate as a benchmark system with perfect channel state information. 

In \cite{feng2020deep}, the authors optimize the design of the phase shifts of an RIS-aided downlink MISO wireless communication system, with the goal of maximizing the received signal-to-noise ratio. Because of the non-convexity of the considered resource allocation problem, the authors employ deep reinforcement learning in order to develop a practical phase shift design algorithm. Numerical results reveal that the developed algorithm can achieve near-optimal signal-to-noise ratio performance with relatively low complexity.

In \cite{liu2020ris}, the authors consider the problem of joint system deployment, RIS phase shift control, power allocation, and dynamic decoding order determination in a RIS-enhanced wireless system, while enforcing individual data rate requirements to the multiple users. To tackle this optimization problem, the authors use machine learning. In particular, the authors propose a novel long short-term memory based echo state network algorithm for predicting the future traffic demands of multiple users based on an empirical dataset. In addition, the authors propose a decaying double deep Q-network based position-acquisition and phase-control algorithm to determine the position and control policy of the RIS.

In \cite{Khan2019}, the authors use an artificial neural network for estimating and detecting symbols in an RIS-aided wireless system. A fully connected artificial neural network is employed to estimate the channels and phase angles from a reflected signal received through an RIS. This enables the authors to perform symbol detection without any dedicated pilot signaling, which significantly reduces the overhead required for the operation of RIS-aided networks. The proposed method is shown to achieve a lower bit error rate than traditional detectors.

In \cite{elbir2020deep}, the authors investigate the use of deep learning for channel estimation in a RIS-assisted massive MIMO system. In the proposed scheme, each user has an identical convolutional neural network which takes as input the received pilot signals and yields as output an estimate of the direct channel between the transmitter and receiver, and the cascaded channel from the transmitter to the receiver through the RIS. The approach is extended to a multi-user scenario, wherein each user has its own convolutional neural network and estimates its own channel. With the aid of numerical simulations, the approach is compared against state-of-the-art deep learning based techniques and performance gains are shown.

In \cite{taha2019enabling}, the authors propose an RIS-based architecture based on sparse channel sensors. All RIS elements are passive except for a few elements that are active and connected to the baseband of the RIS controller. The authors propose two solutions for the design of the RIS reflection matrices based on compressive sensing and deep learning. The deep learning based solution has the advantage of requiring a negligible training overhead, which makes it practical especially in the RIS context, since it reduces the overhead related to RIS control. 

In \cite{yang2020deep}, the authors study an RIS-aided wireless communication system for physical layer security, in which an RIS is deployed to adjust its surface reflecting elements in order to guarantee the secure communication of multiple legitimate users in the presence of multiple eavesdroppers. Aiming to improve the system secrecy rate, a design problem for jointly optimizing the base station beamforming and the RIS reflecting beamforming is formulated, under the assumption of different quality of service requirements and time-varying channel condition. As the system is highly dynamic and complex, and it is challenging to address the resulting non-convex optimization problem, the authors propose a deep reinforcement learning secure beamforming approach  that achieves the optimal beamforming policy. Furthermore, post-decision state and prioritized experience replay schemes are utilized to enhance the learning efficiency and secrecy performance. Simulation results demonstrate that the proposed secure beamforming approach can significantly improve the system secrecy rate and satisfaction probability.

\subsection{Localization, Positioning, and Sensing}
Future wireless networks are expected to offer more than allowing people, mobile devices, and objects to communicate with each other. Future wireless networks need to be turned into a distributed intelligent communication, sensing, and computing platform. Besides connectivity, more specifically, this envisioned 6G platform is expected to be capable of sensing the environment, as well as locally storing and processing information, in order to provide network applications and services with context-awareness capabilities. Such processing could accommodate the time critical, ultra-reliable, and energy-efficient delivery of data, and the accurate localization of people and devices. Besides enabling enhanced connectivity, therefore, RISs are envisioned to play an important role to accomplish several other tasks that complement and support communications. Thanks to the possibility of equipping smart surfaces with energy-harvesting sensors, RISs may be able to offer a dense and capillary network for sensing the environment, and for creating environmental maps supporting a variety of emerging applications. Thanks to the possibility of realizing large-size smart surfaces, RISs may offer a platform that provides localization and positioning services with high accuracy in outdoor and outdoor scenarios, as well as they may enable near-field highly focusing capabilities for supporting the communication of massive deployments of devices. Thanks to the possibility of performing algebraic operations and functions directly on the impinging radio waves, RISs may offer the opportunity to realize a fully EM-based computing platform unlocking the potential of reconfigurable backscatter communications. For all these reasons, several researchers have started investigating the potential opportunities offered by RISs for applications beyond enhanced connectivity. 

In \cite{Hu2018c}, the authors envision the use of RISs for positioning purposes. In particular, the authors analyze the Cram\'er-Rao lower bound for the localization accuracy offered by RISs. Closed-form expressions and accurate approximations for the Cram\'er-Rao lower bound are derived and are used to prove that the localization error offered by RISs decreases quadratically with their surface area. This holds true in general except for the case in which a terminal is located exactly on the central perpendicular line of the RIS. In this latter case, the localization error decreases linearly. The results show that a considerable performance improvement is possible over conventional approaches, even when an unknown phase shift is present in the RIS analog circuit. Moreover, the authors discuss different deployment architectures for RISs that can be used for positioning, and compare the case studies in which the deployment area is covered by a single large RIS or by multiple smaller RIS. Results show that none of the two approaches always outperforms the other. 

In \cite{AlegriaASILOMAR}, the authors derive the Cram\'er-Rao lower bound for the localization accuracy offered by RISs under the assumption of having discrete phase shifts and measuring amplitudes with a finite resolution. In addition, the authors compute analytical bounds for the Cram\'er-Rao lower bound related to the positioning error when the phase information is disregarded and the amplitude is measured with full resolution. Numerical results are used to analyze and quantify the Cram\'er-Rao lower bound loss with respect to the resolution of quantization. 

In \cite{he2019large}, the authors investigate the use of RISs for positioning and object tracking in millimeter-wave MIMO wireless networks. The authors show that RISs provide promising performance thanks to the very sharp beams that they can realize. In particular, the Cram\'er-Rao lower bound of the positioning error is derived, and the impact of the number of RIS reflecting elements on the localization performance is analyzed. Numerical results corroborate the theoretical analysis, and show that RIS-based positioning systems can achieve better performance than traditional schemes.

In \cite{He2019}, the authors investigate the design of millimeter-wave MIMO wireless networks with joint communication and localization capabilities. This authors show that the use of RISs is beneficial for both tasks, granting both accurate positioning and high data-rate transmission. The authors introduce an adaptive phase shifter design based on hierarchical codebooks and feedback from the mobile station. The proposed scheme does not require the deployment of active sensors and baseband processing units.

In \cite{wymeersch2019radio}, the authors analyze and discuss the fundamental technical and scientific challenges for the successful use of RISs for the localization of users and for creating maps of the coverage area in millimeter-wave wireless networks. In this context, the authors describe their vision on the use of RIS-based localization and illustrate initial results and solutions along this line of research. It is argued that the use of RISs for localization purposes has a great potential for improving the localization accuracy, thanks to the sharpness of the beams that RISs equipped with a large number of reflecting elements can realize.

\subsection{Experimental Assessment and Testbeds} \label{Testbeds}
From a technological point of view, RISs are a new and emerging paradigm in the context of wireless communications. Even though reflectarrays have been used for long time in several communication-related applications, the realization of metasurface-based RISs bring new implementation challenges, and necessitate the realization of proof-of-concept platforms and hardware testbeds in order to substantiate theoretical findings and to develop realistic channel and signal models for system design. Even though several research works on the design of metasurfaces with advanced functionalities can be found, research on the realization of hardware platforms for application to wireless communications is still at its infancy. However, a few relevant research works can be found already. As mentioned in Section II, the MIT's RFocus \cite{RFocus} and the NTT-DOCOMO \cite{Docomo_Glass} platforms are two notable examples of promising research activities towards the realization of RISs that are specifically designed for wireless applications. The world's first field trials performed by NTT DOCOMO and MetaWave in December 2018 on the application of RISs to the millimeter-wave frequency can be found in \cite{DocomoMetaWave}. Other examples available in the literature are briefly reported as follows.

In \cite{Tan_ICC2016} and \cite{Tan2018} the authors report the design of a smart reflectarray that is designed for enhancing indoor connectivity in the millimeter-wave frequency band. The reflector panel is fabricated to operate in the 60 GHz frequency band, and its dimensions are 337 mm, 345 mm, and 0.254 mm for its length, height, and thickness, respectively. The reflector panel is made of 224 reflector units, whose inter-distance is larger than one wavelength. Therefore, this design does not fall into the definition of metasurfaces. The authors show that the proposed solution enables the communication of multiple users in the same indoor area, over the same spectrum band, at the same time, and without suffering any significant interference, as well as it enables the reduction of the link outage probability. Recent research results of the same authors are available in \cite{Buffalo2019}.

Along the same lines as \cite{RFocus}, a few other research groups around the world have built prototypes for realizing smart surfaces that are made of large arrays of inexpensive antennas. Notable fully-functional experimental testbeds and experimental activities along this line of research include the PRESS (programmable radio environment for smart spaces) prototype \cite{Jamieson_2017}, the LAIA (large array of inexpensive antenna) prototype \cite{Jamieson_2019}, and the ScatterMIMO prototype \cite{Bharadia_2020}. As far as the realization of smart surfaces based on metasurfaces is concerned, a relevant example is the HyperSurface hardware platform: A software-defined, controllable, and programmable metasurface structure that is capable of shaping the EM waves in reconfigurable ways. The HyperSurface is under development by researchers of the  European-funded project VisorSurf \cite{VisorSurf}.

In \cite{dai2019reconfigurable}, the authors report a new type of high-gain yet low-cost RIS made of 256 elements. The proposed RIS combines the functions of phase shift and EM radiation on an electromagnetic surface where PIN diodes are used to realize two-bit phase shifts for beamforming. This proposed design forms the basis for the world's first wireless communication prototype that realize an RIS having 256 two-bit elements. The prototype consists of modular hardware and flexible software that encompass the hosts for parameter setting and data exchange, universal software radio peripherals for baseband and RF signal processing, as well as modules for signal transmission and reception. Experimental results confirm the feasibility and efficiency of RISs for wireless communications. In particular, at 2.3 GHz the proposed RIS can achieve a 21.7 dBi antenna gain. At the millimeter wave frequency of 28.5 GHz, it attains a 19.1 dBi antenna gain. Furthermore, it is shown that the developed prototype is capable of significantly reducing the power consumption.

In \cite{hu2019reconfigurable}, the authors design an RF sensing system for posture recognition based on RISs. The proposed system can actively customize the environment in order to provide the desirable propagation properties and diverse transmission channels. The prototype has a size of 69 cm $\times$ 69 cm $\times$ 0.52 cm, and it is composed of a two-dimensional array of
electrically controllable unit elements. Each row/column of the array contains 48 unit elements, which corresponds to a surface made of 2,304 unit elements. Each unit element has a size of 1.5 cm $\times$ 1.5 cm $\times$ 0.52 cm, and it is composed of four rectangle copper patches printed on a dielectric substrate. Any two adjacent copper patches are connected by a PIN diode that is controlled by applying appropriate voltages. By combing simulation and experimental results, the authors show that the posture recognition accuracy increases with the size of the RIS and with the number of independently controllable elements. 

In \cite{Wankai_ElectronLetterEditor}, \cite{Wankai_ElectronLetter}, \cite{Wankai_ChinaCommun}, \cite{Wankai_WCM}, \cite{Wankai_JSAC}, the authors report the design and some experimental results on multiple prototypes of reconfigurable metasurfaces that implement single-RF transmitters. In particular, the prototype reported in \cite{Wankai_ElectronLetter} is
designed to have a full 360 degrees phase response and a reflectivity of about 85\% at a working frequency of 4.25 GHz. The fabricated metasurface sample is made of $8 \times 32$ unit cells. The size of a unit cell is 12 mm $\times$ 12 mm $\times$ 5 mm, which approximately corresponds to 0.17$\lambda$ $\times$ 0.17$\lambda$ $\times$ 0.07$\lambda$ at 4.25 GHz. The capacitance of the unit cell is dominated by a varactor diode, which indicates that the phase response of the metasurface can be tuned through the bias voltage of the varactor diode. This bias voltage has an approximate linear relationship with the phase of the reflected wave and can achieve 360 degrees phase modulation. For example, the authors prove that the designed metasurface prototype is capable of mimicking the phase response of an eight phase shift keying modulation scheme.

In \cite{Wankai_PathLoss}, the authors report the world's first experimental measurements to characterize the path-loss experienced by EM waves in the presence of metasurfaces. The authors report experiments, which are conducted in an anechoic chamber, for three samples of manufactured metasurfaces with 10,200, 1,700, and 256 unit cells, and whose size is 1.02 m $\times$ 1 m at 10.5 GHz, 0.34 m $\times$ 0.5 m at 10.5 GHz, and 0.384 m $\times$ 0.096 m at 4.25 GHz, respectively. The authors prove that the path-loss as a function of the transmission distance depends on several parameters, which include the size of the metasurface, the function applied by the metasurface, and whether the system operates in the radiative near-field or in the far-field.

\section{The Road Ahead} \label{Beyond}
\textbf{Major research problems to tackle}. In this section, we elaborate on some major open research problems that we consider to be of great importance for unveiling the potential benefits and for assessing the advantages and limitations of RISs against other mature wireless technologies. The list of open research problems is, however, not intended to be exhaustive in light of the steady growth of fundamental and applied research on RIS-empowered SREs.

\textbf{EM-based circuital models}.
The homogenized and macroscopic model-based approach (either based on surface susceptibility functions or surface impedances) briefly summarized in the present paper is a powerful method for analyzing and optimizing RISs. It constitutes, in particular, a suitable interface between the physics phenomena underlying the operation of RISs and the level of abstraction that is needed for obtaining adequately accurate but sufficiently tractable models for application to wireless communications. This approach hides, however, some aspects related to the practical realization of RISs, which may be convenient to preserve in the analytical formulation. For analyzing and optimizing wireless networks, in particular, it may be appropriate to start from simple models for the unit cell of metasurfaces, e.g., simple dipoles, and to obtain equivalent EM-based circuital models for metasurface structures, which explicitly account for the mutual coupling among the unit cells as a function of their size, inter-distance, and spatial arrangements. This will allow us to better understand the potential benefits and limitations of sub-wavelength metasurface structures, and to better quantify the potential gains for wireless applications.

\textbf{Path-loss and channel modeling}.
In order to evaluate the performance limits of RISs in wireless networks, we need realistic models for the propagation of the signals scattered by metasurfaces. In addition, it is necessary to abandon the comfortable assumption of far-field propagation, since several promising applications and potential advantages of RISs may emerge in the radiative near-field regime. To this end, it is necessary to relay upon physics-based models for the propagation of EM fields in proximity of metasurfaces and to account for their circuital models in the problem formulation. Besides the development of accurate path-loss models for link budget analysis, it is necessary to develop fading models for sub-wavelength structures, both at the microscopic level, which account for individual unit cells, and at the macroscopic level, which may be more easily integrated into communication-theoretic frameworks.

\textbf{Fundamental performance limits}.
RISs can be employed for different purposes, which include beamsteering, beamforming, focusing, modulation, and joint modulation and encoding with the transmitter. Depending on the specific application and wave transformation applied, the ultimate performance limits are not known yet. Recent results have pointed out, in a simple point-to-point communication setup, that the channel capacity is achieved by performing joint encoding at the transmitter and RIS, whereas typical criteria that rely on maximizing the power of the received signal are sub-optimal. These findings are obtained, however, by considering simple models for the metasurface structure, which are in agreement with conventional assumptions in communication and information theory. The ultimate performance limits of RISs, how to best configure RISs to achieve them, and the actual gains with respect to other well established technologies are not known yet.

\textbf{Robust optimization and resource allocation}.
The simple examples reported in the present paper confirm that physics-based models for simple metasurfaces, e.g., perfect anomalous reflectors, are more complex that the typical models that communication theorists employ in formulating resource allocation problems. As an example, the amplitude and phase responses of a metasurface operating as a perfect anomalous reflector are not independent of each other, and the resulting metasurface structure is locally non-passive. It is necessary to integrate physics-based models for metasurface structures into resource allocation problems, and, in particular, to explicitly account for the impact of the sub-wavelength structure of RISs and the associated design constraints and shaping capabilities of the radio waves.

\textbf{Constrained system design and optimization}.
One of the potential novelties of RISs is the possibility of manipulating the radio waves impinging upon them without the need of relying on on-board power amplifiers and signal processing units. Some advantages of these assumptions lie in the opportunity of reducing the EM field exposure of human beings and in making the metasurfaces easier to recycle. These assumptions introduce, however, constraints on the operation of RISs when they are deployed in wireless networks. Usually, the restrictions on performing signal processing operations on the metasurfaces have an impact on the overhead that is associated with gathering the necessary environmental information for optimizing and configuring RISs. The fundamental trade-off between the complexity and power consumption of RISs, as well as the associated overhead needs to be carefully investigated and assessed.

\begin{figure}[!t]
\begin{centering}
\includegraphics[width=\columnwidth]{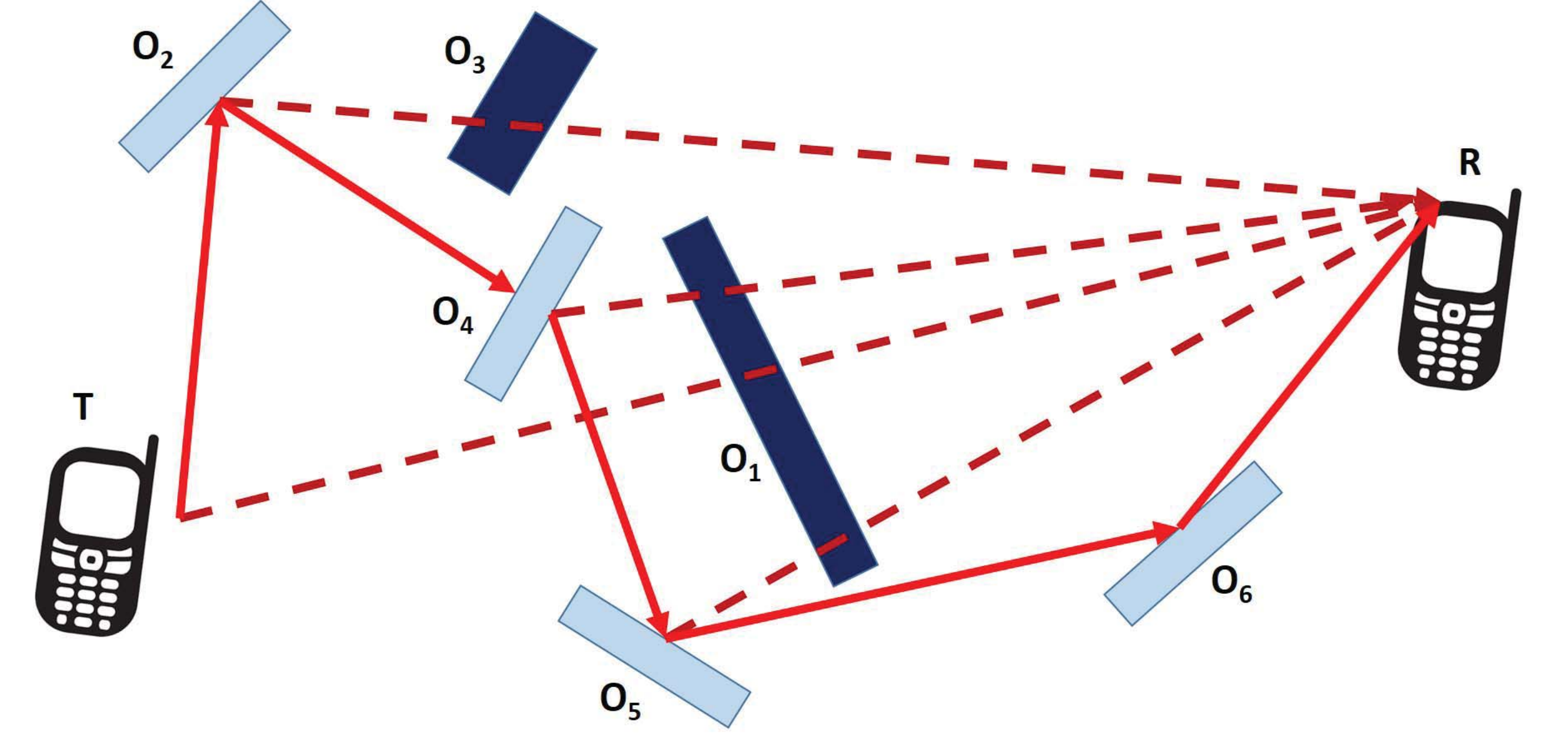}
\caption{Electromagnetic-based routing in smart radio environments (reproduced from \cite{MDR_Survey_EURASIP}).}
\label{Fig_23}
\end{centering} 
\end{figure}
\textbf{EM-based communications: ``Layer-0'' networking}. 
RISs may open the possibility of designing wireless networks in which the entire protocol stack for communication is realized through an appropriate shaping and manipulation of the radio waves at the EM level. This does not encompass only physical layer functionalities, which have been the main focus of research to date, but the medium access control layer and the routing layer as well. In scenarios with a high density of mobile terminals and devices, for example, RISs may be employed to create a large number of orthogonal wireless channels that may be employed for achieving high spatial capacities through EM-based medium access control protocols with reduced overhead and contention time, as well as for enabling spatial multiplexing gains in rank-deficient wireless environments. RIS-empowered SREs, in addition, may be viewed as an opportunity to realize EM-based routing protocols (see, e.g., Fig. \ref{Fig_23}), according to which the radio waves are routed from the transmitter to the receiver via controlled reflections and refractions at low latencies, low transmission delays/jitters, and high reliability, since all operations are realized directly at the EM level. Research on the potential uses, advantages, and limitations of capitalizing on RISs for the design of communication protocols beyond the physical layer is almost unexplored to date but deserves major attention and research efforts.

\textbf{Large-scale networks: Deployment, analysis, and optimization}.
RIS-empowered SREs have several potential applications, which encompass indoor and outdoor environments. The vast majority of research activities conducted so far have considered the analysis of optimization of ``small-size'' networks, usually networks that are made of a single RIS. This is the natural starting point for assessing the potential benefits of new technologies. In addition, the analysis of these simple scenarios is based on modeling assumptions that may not necessarily be sufficiency realistic for the assessment of sub-wavelength metasurfaces. It instrumental to quantify, however, the performance limits of SREs in large-scale deployments. For example, a relevant question may be to identify the density of RISs to be deployed for increasing the coverage probability or the energy efficiency in a large industrial factory or even in the downtown of a city. This is an almost unexplored research problem to be tackled.

\textbf{Ray tracing and system-level simulators}.
The deployment phase of new technologies is usually preceded by extensive system-level simulation studies and field test trials. In this context, the use of ray tracing and system-level simulators plays a crucial role for overcoming the limitations of simplified analytical studies, and for taking into account realistic spatial topologies, e.g., three-dimensional maps for territories, cities, streets, buildings, and floor plants of apartments, offices, etc. Current tools that are employed to this end are not suitable for application to RIS-empowered SREs because the wave transformations and the scattering models of metasurfaces are not integrated into them. In simple terms, the so-called generalized laws of reflection and refraction that can be viewed as approximations of the response of metasurfaces to impinging radio waves according to the assumptions of geometric optics are not available in commercial ray tracing tools and system-level simulators.

\textbf{Beyond far-field communications}. 
Most research works conducted in the recent present on RIS-empowered SREs rely upon far-field assumptions. RISs, however, may be made of geometrically large surfaces of the order of a few square meters. This implies that RISs may operate in the near-field regime in relevant application scenarios, e.g., in indoor environments. The use of geometrically large RISs opens the possibility of building new wireless networks that operate in the near-field regime, which is not a conventional design assumption in wireless communications. Research on the fundamental performance limits, design constraints, and potential applications and benefits of near-field communications in RIS-empowered SREs has not received significant attention so far. In light of the potential applications that may be unlocked, e.g., highly focusing capabilities, the near-field regime is worthed of further attention from the research community.

\textbf{Beyond communications}.
Most research works conducted in the recent present on RIS-empowered SREs are focused on applications that are related to communications. RISs, however, offer opportunity for research that go beyond communications. For example, the high focusing capability of electrically large RISs offers the opportunity of using them for high-precision radio localization and mapping (i.e., the construction of a model or map of the environment). The capabilities of RISs for these applications and the expected performance as a function of their size, sub-wavelength structure, and near-field vs. far-field operation constitute an open research problem. In addition, radio localization and mapping can be considered to be important enablers for realizing important communication-related tasks.

\textbf{Advantages and limitations}.
This can be considered to be the ultimate goal of conducting research on RIS-empowered SREs: ``\textit{Do RISs bring any (substantial) gains as compared with other well-established technologies in wireless networks}?'' At the time of writing, there is no precise answer to this question, and it will probably take a few years of multi-disciplinary research to uncover the tip of the iceberg.

\begin{figure}[!t]
\begin{centering}
\includegraphics[width=\columnwidth]{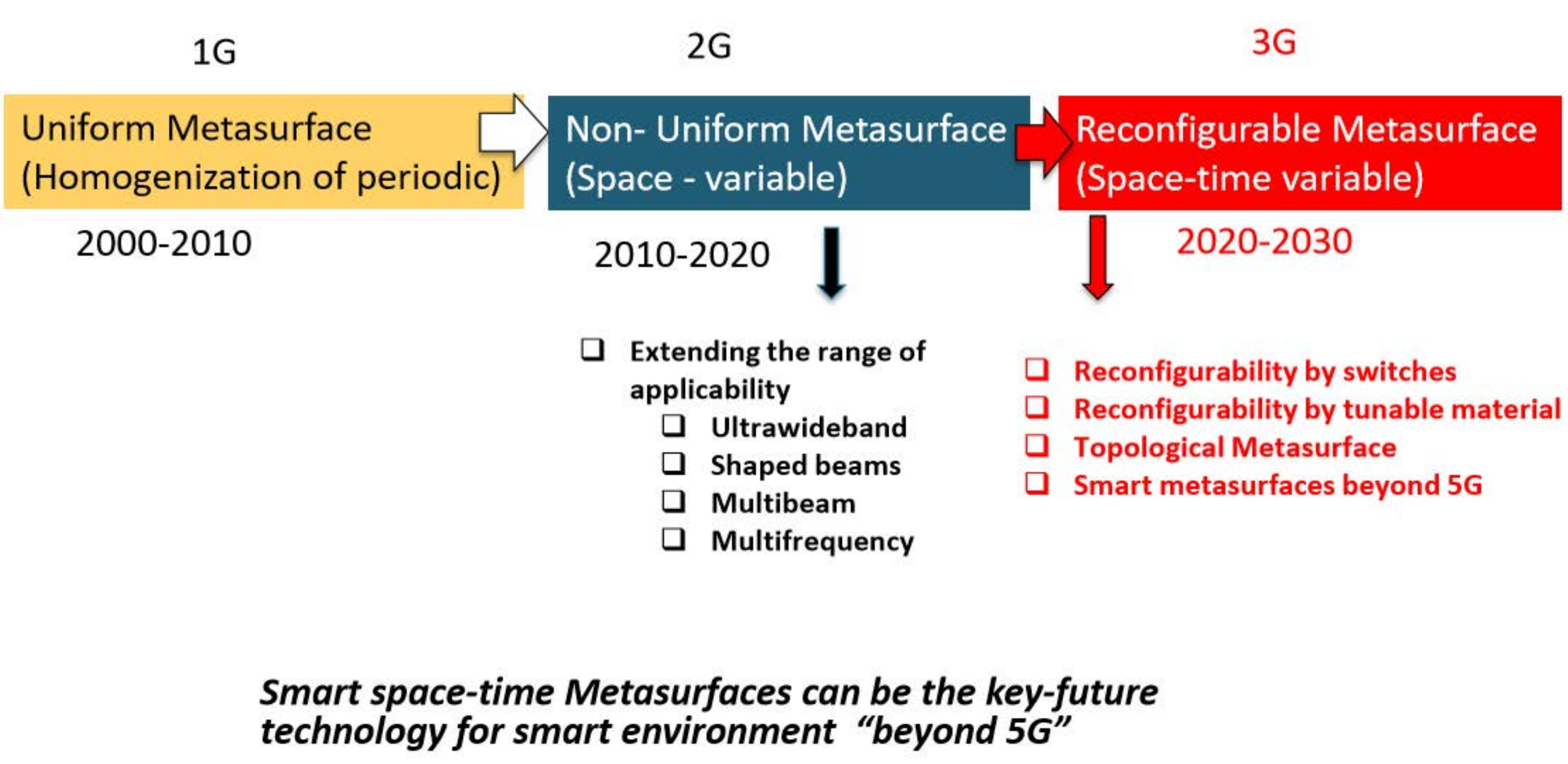}
\caption{Towards 3G metasurfaces (slide: Prof. Stefano Maci \cite{Maci_Huawei}).}
\label{Fig_Maci}
\end{centering} 
\end{figure}
\textbf{6G wireless vs. 3G metasurfaces}.
Finally, we conclude this section by emphasizing that the potential benefits of RIS-empowered SREs are conditioned upon the availability of reconfigurable multi-function metasurfaces with exceptional capabilities of manipulating EM waves. Research and development on this enabling technological field of SEM has just began, as illustrated in Fig. \ref{Fig_Maci}.

\section{Conclusion} \label{End}
For more than seven decades, communication and information theorists have built their theories and algorithms upon \textbf{C. E. Shannon}'s seminal work reported in the monograph ``\textit{A (The) Mathematical Theory of Communication}'' (1948). The present article constitutes a first preliminary attempt that elaborates on the need of reconciling and reuniting Shannon's mathematical theory of communication with \textbf{G. Green}'s landmark essay in the field of mathematical physics ``\textit{An Essay on the Application of Mathematical Analysis to the Theories of Electricity and Magnetism}'' (1828) and \textbf{J. C. Maxwell}'s fundamental treatise on the theory of electromagnetism ``\textit{A Dynamical Theory of the Electromagnetic Field}'' (1865). \\

We conclude the present article by quoting the last paragraph of the preface of \textbf{G. Green}'s essay: ``\textit{Should the present Essay tend in any way to facilitate the application of analysis to one of the most interesting of the physical sciences, the author will deem himself amply repaid for any labour he may have bestowed upon it; and it is hoped the difficulty of the subject will incline mathematicians to read this work with indulgence, more particularly when they are informed that it was written by a young man, who has been obliged to obtain the little knowledge he possesses, at such intervals and by such means, as other indispensable avocations which offer but few opportunities of mental improvement, afforded}".

\bibliographystyle{IEEEtran}
\bibliography{JSAC_RIS__BIB}

\begin{thebibliography}{100}
\providecommand{\url}[1]{#1}
\csname url@samestyle\endcsname
\providecommand{\newblock}{\relax}
\providecommand{\bibinfo}[2]{#2}
\providecommand{\BIBentrySTDinterwordspacing}{\spaceskip=0pt\relax}
\providecommand{\BIBentryALTinterwordstretchfactor}{4}
\providecommand{\BIBentryALTinterwordspacing}{\spaceskip=\fontdimen2\font plus
\BIBentryALTinterwordstretchfactor\fontdimen3\font minus
  \fontdimen4\font\relax}
\providecommand{\BIBforeignlanguage}[2]{{%
\expandafter\ifx\csname l@#1\endcsname\relax
\typeout{** WARNING: IEEEtran.bst: No hyphenation pattern has been}%
\typeout{** loaded for the language `#1'. Using the pattern for}%
\typeout{** the default language instead.}%
\else
\language=\csname l@#1\endcsname
\fi
#2}}
\providecommand{\BIBdecl}{\relax}
\BIBdecl

\bibitem{ITU_2015}
{Radiocommunication~Sector~of~ITU}, ``{IMT} traffic estimates for the years
  2020 to 2030,'' \emph{Report ITU-R M. 2370-0}, pp. 1--15, Jul. 2015.

\bibitem{Letaief_AI}
K.~B. Letaief, W.~Chen, Y.~Shi, J.~Zhang, and Y.~Zhang, ``The roadmap to {6G}:
  {AI} empowered wireless networks,'' \emph{IEEE Commun. Mag.}, vol.~57, no.~8,
  pp. 84--90, Aug. 2019.

\bibitem{RFocus}
V.~Arun and H.~Balakrishnan, ``{RFocus}: Beamforming using thousands of passive
  antennas,'' in \emph{USENIX Symposium on Networked Systems Design and
  Implementation}, Feb. 2020, pp. 1047--1061.

\bibitem{Docomo_Glass}
\BIBentryALTinterwordspacing
{NTT~DOCOMO}, ``{DOCOMO} conducts world's first successful trial of transparent
  dynamic metasurface,'' Jan. 2020. [Online]. Available:
  \url{https://www.nttdocomo.co.jp/english/info/media\_center/pr/2020/0117\_0
  0.html.}
\BIBentrySTDinterwordspacing

\bibitem{Book_SurfaceElectromagnetics}
F.~Yang and Y.~Rahmat-Samii, \emph{Surface Electromagnetics: With Applications
  in Antenna, Microwave, and Optical Engineering}.\hskip 1em plus 0.5em minus
  0.4em\relax Cambridge University Press, Jun. 2019.

\bibitem{MDR_SM_COMMAG}
M.~Di~Renzo, H.~Haas, and P.~M. Grant, ``Spatial modulation for
  multiple-antenna wireless systems: {A} survey,'' \emph{IEEE Commun. Mag.},
  vol.~49, no.~12, pp. 182--191, Dec. 2011.

\bibitem{MDR_SM_PIEEE}
M.~Di~Renzo, H.~Haas, A.~Ghrayeb, S.~Sugiura, and L.~Hanzo, ``Spatial
  modulation for generalized {MIMO}: Challenges, opportunities, and
  implementation,'' \emph{Proc. of the IEEE}, vol. 102, no.~1, pp. 56--103,
  Jan. 2014.

\bibitem{MDR_SM_CST2015}
P.~Yang, M.~Di~Renzo, Y.~Xiao, S.~Li, and L.~Hanzo, ``Design guidelines for
  spatial modulation,'' \emph{IEEE Commun. Surveys Tuts.}, vol.~17, no.~1, pp.
  6--26, First quarter 2015.

\bibitem{MDR_SM_CST2016}
P.~Yang, Y.~Xiao, Y.~L. Guan, K.~Hari, A.~Chockalingam, S.~Sugiura, H.~Haas,
  M.~Di~Renzo, C.~Masouros, Z.~Liu, L.~Xiao, S.~Li, and L.~Hanzo,
  ``Single-carrier {SM-MIMO}: A promising design for broadband large-scale
  antenna systems,'' \emph{IEEE Commun. Surveys Tuts.}, vol.~18, no.~3, pp.
  1687--1716, Third quarter 2016.

\bibitem{MDR_SM_MILCOM}
M.~Di~Renzo, ``Spatial modulation based on reconfigurable antennas - {A} new
  air interface for the {IoT},'' in \emph{IEEE Military Commun. Conf.}, Oct.
  2017, pp. 495--500.

\bibitem{MDR_IM_Access}
E.~Basar, M.~Wen, R.~Mesleh, M.~Di~Renzo, Y.~Xiao, and H.~Haas, ``Index
  modulation techniques for next-generation wireless networks,'' \emph{IEEE
  Access}, vol.~5, pp. 16\,693--16\,746, Aug. 2017.

\bibitem{MDR_SM_JSAC}
M.~Wen, B.~Zheng, K.~J. Kim, M.~Di~Renzo, T.~A. Tsiftsis, K.-C. Chen, and
  N.~Al-Dhahir, ``A survey on spatial modulation in emerging wireless systems:
  Research progresses and applications,'' \emph{IEEE J. Sel. Areas Commun.},
  vol.~37, no.~9, pp. 1949--1972, Sep. 2019.

\bibitem{Wankai_PathLoss}
\BIBentryALTinterwordspacing
W.~Tang, M.~Z. Chen, X.~Chen, J.~Y. Dai, Y.~Han, M.~Di~Renzo, Y.~Zeng, S.~Jin,
  Q.~Cheng, and T.~J. Cui, ``Wireless communications with reconfigurable
  intelligent surface: Path loss modeling and experimental measurement,''
  \emph{arXiv}, Nov. 2019. [Online]. Available:
  \url{https://arxiv.org/abs/1911.05326.}
\BIBentrySTDinterwordspacing

\bibitem{Wankai_ElectronLetterEditor}
W.~{Tang}, J.~{Dai}, M.~{Chen}, X.~{Li}, Q.~{Cheng}, S.~{Jin}, K.~{Wong}, and
  T.~J. {Cui}, ``Subject {E}ditor spotlight on programmable metasurfaces: The
  future of wireless?'' \emph{IET Electron. Lett.}, vol.~55, no.~7, pp.
  360--361, Apr. 2019.

\bibitem{Wankai_ElectronLetter}
W.~Tang, J.~Y. Dai, M.~Chen, X.~Li, Q.~Cheng, S.~Jin, K.-K. Wong, and T.~J.
  Cui, ``Programmable metasurface-based {RF} chain-free {8PSK} wireless
  transmitter,'' \emph{IET Electron. Lett.}, vol.~55, no.~7, pp. 417--420, Apr.
  2019.

\bibitem{Wankai_ChinaCommun}
W.~Tang, X.~Li, J.~Y. Dai, S.~Jin, Y.~Zeng, Q.~Cheng, and T.~J. Cui, ``Wireless
  communications with programmable metasurface: Transceiver design and
  experimental results,'' \emph{IEEE China Commun.}, vol.~16, no.~5, pp.
  46--61, May 2019.

\bibitem{Wankai_WCM}
\BIBentryALTinterwordspacing
W.~Tang, M.~Z. Chen, J.~Y. Dai, Y.~Zeng, X.~Zhao, S.~Jin, Q.~Cheng, and T.~J.
  Cui, ``Wireless communications with programmable metasurface: New paradigms,
  opportunities, and challenges on transceiver design,'' \emph{IEEE Wireless
  Commun Mag.}, 2020. [Online]. Available:
  \url{https://arxiv.org/abs/1907.01956.}
\BIBentrySTDinterwordspacing

\bibitem{Wankai_JSAC}
\BIBentryALTinterwordspacing
W.~Tang, J.~Y. Dai, M.~Z. Chen, K.-K. Wong, X.~Li, X.~Zhao, S.~Jin, Q.~Cheng,
  and T.~J. Cui, ``{MIMO} transmission through reconfigurable intelligent
  surface: System design, analysis, and implementation,'' \emph{IEEE J. Sel.
  Commun.}, 2020. [Online]. Available: \url{https://arxiv.org/abs/1912.09955.}
\BIBentrySTDinterwordspacing

\bibitem{MDR_ISIT2020}
\BIBentryALTinterwordspacing
R.~Karasik, O.~Simeone, M.~{Di Renzo}, and S.~Shamai, ``Beyond max-snr: Joint
  encoding for reconfigurable intelligent surfaces,'' \emph{IEEE Int. Symp.
  Inform. Theory}, Nov. 2019. [Online]. Available:
  \url{https://arxiv.org/abs/1911.09443.}
\BIBentrySTDinterwordspacing

\bibitem{NearFieldFocused_Antennas}
P.~Nepa and A.~Buffi, ``Near-field-focused microwave antennas: Near-field
  shaping and implementation,'' \emph{IEEE Antennas Propag. Mag.}, vol.~59,
  no.~3, pp. 42--53, Jun. 2017.

\bibitem{MDR_OverheadAware}
\BIBentryALTinterwordspacing
A.~Zappone, M.~Di~Renzo, F.~Shams, X.~Qian, and M.~Debbah, ``Overhead-aware
  design of reconfigurable intelligent surfaces in smart radio environments,''
  \emph{arXiv}, Mar. 2020. [Online]. Available:
  \url{https://arxiv.org/abs/2003.02538.}
\BIBentrySTDinterwordspacing

\bibitem{TIT_StateDependentChannels}
T.~Weissman, ``Capacity of channels with action-dependent states,'' \emph{IEEE
  Trans. Inform. Theory}, vol.~56, no.~11, pp. 5396--5411, Nov. 2010.

\bibitem{Dardari_DegreesFreedom}
\BIBentryALTinterwordspacing
D.~Dardari, ``Communicating with large intelligent surfaces: Fundamental limits
  and models,'' \emph{arXiv}, Dec. 2019. [Online]. Available:
  \url{https://arxiv.org/abs/1912.01719.}
\BIBentrySTDinterwordspacing

\bibitem{ORANGE_BackScattering_1}
Y.~Kokar, D.-T. Phan-Huy, R.~Fara, K.~Rachedi, A.~Ourir, J.~{de Rosny}, M.~{Di
  Renzo}, J.-C. Prevotet, and M.~Helard, ``First experimental ambient
  backscatter communication using a compact reconfigurable tag antenna,'' in
  \emph{IEEE Global Commun. Conf.}, Dec. 2019.

\bibitem{ORANGE_BackScattering_2}
R.~Fara, D.-T. Phan-Huy, and M.~{Di Renzo}, ``Ambient backscatters-friendly 5g
  networks: creating hot spots for tags and good spots for readers,'' in
  \emph{IEEE Wireless Commun. Netw. Conf.}, Apr. 2020.

\bibitem{MDR_AI_VTM}
A.~Zappone, M.~{Di~Renzo}, M.~Debbah, T.~T. Lam, and X.~Qian, ``Model-aided
  wireless artificial intelligence: Embedding expert knowledge in deep neural
  networks for wireless system optimization,'' \emph{IEEE Veh. Technol. Mag.},
  vol.~14, no.~3, pp. 60--69, Sep. 2019.

\bibitem{MDR_AM_TCOM}
A.~Zappone, M.~{Di~Renzo}, and M.~Debbah, ``Wireless networks design in the era
  of deep learning: Model-based, {AI}-based, or both?'' \emph{IEEE Trans.
  Commun.}, vol.~67, no.~10, pp. 7331--7376, Oct. 2019.

\bibitem{MDR_AI_RIS}
\BIBentryALTinterwordspacing
H.~Gacanin and M.~Di~Renzo, ``Wireless 2.0: Towards an intelligent radio
  environment empowered by reconfigurable meta-surfaces and artificial
  intelligence,'' \emph{arXiv}, Feb. 2020. [Online]. Available:
  \url{https://arxiv.org/abs/2002.11040.}
\BIBentrySTDinterwordspacing

\bibitem{Holloway_Oct2003}
E.~F. Kuester, M.~A. Mohamed, M.~Piket-May, and C.~L. Holloway, ``Averaged
  transition conditions for electromagnetic fields at a metafilm,'' \emph{IEEE
  Trans. Antennas Propag.}, vol.~51, no.~10, pp. 2641--2651, Oct. 2003.

\bibitem{Holloway_Nov2005}
C.~L. Holloway, M.~A. Mohamed, E.~F. Kuester, and A.~Dienstfrey, ``Reflection
  and transmission properties of a metafilm: With an application to a
  controllable surface composed of resonant particles,'' \emph{IEEE Trans.
  Electromagn. Compat.}, vol.~47, no.~4, pp. 853--865, Nov. 2005.

\bibitem{Caloz_GSTCs2015}
K.~Achouri, M.~A. Salem, and C.~Caloz, ``General metasurface synthesis based on
  susceptibility tensors,'' \emph{IEEE Trans. Antennas Propag.}, vol.~63,
  no.~7, pp. 2977--2991, Jul. 2015.

\bibitem{Caloz_Synthesis2018}
K.~Achouri and C.~Caloz, ``Design, concepts, and applications of
  electromagnetic metasurfaces,'' \emph{De Gruyter Nanophotonics}, vol.~7,
  no.~6, pp. 1095--1116, Jun. 2018.

\bibitem{Caloz_Computational2018}
Y.~Vahabzadeh, N.~Chamanara, K.~Achouri, and C.~Caloz, ``Computational analysis
  of metasurfaces,'' \emph{IEEE J. Multiscale Multiphys. Comput. Tech.},
  vol.~3, pp. 37--49, 2018.

\bibitem{ComputationImpedances}
M.~Albooyeh, R.~Alaee, C.~Rockstuhl, and C.~Simovski, ``Revisiting
  substrate-induced bianisotropy in metasurfaces,'' \emph{Phys. Rev. B},
  vol.~91, no.~19, pp. 1--11, 195304, May 2015.

\bibitem{CalozSergei_Mar2018}
G.~Lavigne, K.~Achouri, V.~S. Asadchy, S.~A. Tretyakov, and C.~Caloz,
  ``Susceptibility derivation and experimental demonstration of refracting
  metasurfaces without spurious diffraction,'' \emph{IEEE Trans. Antennas
  Propag.}, vol.~66, no.~3, pp. 1321--1330, 2018.

\bibitem{Alu_Reflection2016}
N.~M. Estakhri and A.~Al{\`u}, ``Wave-front transformation with gradient
  metasurfaces,'' \emph{Phys. Rev. X}, vol.~6, no.~4, pp. 1--17, 041008, Dec.
  2016.

\bibitem{Asadchy_2017}
V.~S. Asadchy, W.~Wickberg, A.~Diaz-Rubio, and M.~Wegener, ``Eliminating
  scattering loss in anomalously reflecting optical metasurfaces,'' \emph{ACS
  Photonics}, vol.~4, no.~5, pp. 1264--1270, Apr. 2017.

\bibitem{Asadchy_2016}
V.~S. Asadchy, M.~Albooyeh, S.~N. Tcvetkova, A.~D{\'\i}az-Rubio, Y.~Ra'di, and
  S.~Tretyakov, ``Perfect control of reflection and refraction using spatially
  dispersive metasurfaces,'' \emph{Phys. Rev. B}, vol.~94, no.~7, 075142, Aug.
  2016.

\bibitem{Diaz-Rubio_2017}
A.~D{\'\i}az-Rubio, V.~S. Asadchy, A.~Elsakka, and S.~A. Tretyakov, ``From the
  generalized reflection law to the realization of perfect anomalous
  reflectors,'' \emph{Science Advances}, vol.~3, no.~8, e1602714, Aug. 2017.

\bibitem{Capasso_Science2011}
N.~Yu, P.~Genevet, M.~A. Kats, F.~Aieta, J.-P. Tetienne, F.~Capasso, and
  Z.~Gaburro, ``Light propagation with phase discontinuities: generalized laws
  of reflection and refraction,'' \emph{Science}, vol. 334, no. 6054, pp.
  333--337, Oct. 2011.

\bibitem{Epstein_2016}
A.~Epstein and G.~V. Eleftheriades, ``Synthesis of passive lossless
  metasurfaces using auxiliary fields for reflectionless beam splitting and
  perfect reflection,'' \emph{Phys. Rev. Lett.}, 256103, 2016.

\bibitem{MDR_SPAWC2020}
\BIBentryALTinterwordspacing
M.~{Di Renzo}, F.~H. Danufane, X.~Xi, J.~de~Rosny, and S.~A. Tretyakov,
  ``Analytical modeling of the path-loss for reconfigurable intelligent
  surfaces - {A}nomalous mirror or scatterer ?'' \emph{ArXiv}, Jan. 2020.
  [Online]. Available: \url{https://arxiv.org/abs/2001.10862.}
\BIBentrySTDinterwordspacing

\bibitem{MDR_PathLossFadil}
F.~H. Danufane, M.~Di~Renzo, J.~de~Rosny, and S.~Tretyakov, ``On the path-loss
  of reconfigurable intelligent surfaces: An approach based on the theory of
  diffraction,'' 2020, (in writing).

\bibitem{Green_1828}
G.~Green, \emph{An essay on the application of mathematical analysis to the
  theories of electricity and magnetism}.\hskip 1em plus 0.5em minus
  0.4em\relax Wez{\"a}ta-Melins Aktiebolag, Nottingham, 1828.

\bibitem{Liaskos_COMMAG}
C.~{Liaskos}, S.~{Nie}, A.~{Tsioliaridou}, A.~{Pitsillides}, S.~{Ioannidis},
  and I.~{Akyildiz}, ``A new wireless communication paradigm through
  software-controlled metasurfaces,'' \emph{IEEE Commun. Mag.}, vol.~56, no.~9,
  pp. 162--169, Sep. 2018.

\bibitem{Liaskos_ACM}
C.~Liaskos, A.~Tsioliaridou, A.~Pitsillides, S.~Ioannidis, and I.~F. Akyildiz,
  ``Using any surface to realize a new paradigm for wireless communications,''
  \emph{ACM Commun.}, vol.~61, no.~11, pp. 30--33, Nov. 2018.

\bibitem{MDR_Survey_EURASIP}
M.~{Di Renzo}, M.~Debbah, D.~Phan-Huy, A.~Zappone, M.~Alouini, C.~Yuen,
  V.~Sciancalepore, G.~Alexandropoulos, J.~Hoydis, H.~Gacanin, and J.~De~Rosny,
  ``Smart radio environments empowered by {AI} reconfigurable meta-surfaces: An
  idea whose time has come,'' \emph{EURASIP J. Wireless Commun. Netw.}, vol.
  129, May 2019.

\bibitem{MDR_Survey_Access}
E.~Basar, M.~{Di~Renzo}, J.~{de~Rosny}, M.~Debbah, M.-S. Alouini, and R.~Zhang,
  ``Wireless communications through reconfigurable intelligent surfaces,''
  \emph{IEEE Access}, vol.~7, pp. 116\,753--116\,773, Aug. 2019.

\bibitem{MDR_RISs_Relays}
\BIBentryALTinterwordspacing
M.~{Di~Renzo}, K.~Ntontin, J.~Song, F.~Danufane, X.~Qian, F.~Lazarakis,
  J.~de~Rosny, D.-T. Phan-Huy, O.~Simeone, R.~Zhang, M.~Debbah, G.~Lerosey,
  M.~Fink, S.~Tretyakov, and S.~Shamai, ``Reconfigurable intelligent surfaces
  vs. relaying: Differences, similarities, and performance comparison,''
  \emph{arXiv}, Aug. 2019. [Online]. Available:
  \url{https://arxiv.org/abs/1908.08747.}
\BIBentrySTDinterwordspacing

\bibitem{Liang2019}
\BIBentryALTinterwordspacing
Y.-C. Liang, R.~Long, Q.~Zhang, J.~Chen, H.~V. Cheng, and H.~Guo, ``Large
  intelligent surface/antennas {(LISA):} making reflective radios smart,''
  \emph{arXiv}, Jun. 2019. [Online]. Available:
  \url{https://arxiv.org/abs/1906.06578.}
\BIBentrySTDinterwordspacing

\bibitem{MDR_Survey_Holos}
\BIBentryALTinterwordspacing
C.~Huang, S.~Hu, G.~C. Alexandropoulos, A.~Zappone, C.~Yuen, R.~Zhang, M.~{Di
  Renzo}, and M.~Debbah, ``Holographic {MIMO} surfaces for {6G} wireless
  networks: Opportunities, challenges, and trends,'' \emph{arXiv}, Nov. 2019.
  [Online]. Available: \url{https://arxiv.org/abs/1911.12296.}
\BIBentrySTDinterwordspacing

\bibitem{RuiZhang_Survey}
Q.~Wu and R.~Zhang, ``Towards smart and reconfigurable environment: Intelligent
  reflecting surface aided wireless network,'' \emph{IEEE Commun. Mag.},
  vol.~58, no.~1, pp. 106--112, Jan. 2020.

\bibitem{JunZhao_Survey_1}
\BIBentryALTinterwordspacing
J.~Zhao, ``A survey of intelligent reflecting surfaces ({IRSs}): Towards {6G}
  wireless communication networks,'' \emph{arXiv}, Jul. 2019. [Online].
  Available: \url{https://arxiv.org/abs/1907.04789.}
\BIBentrySTDinterwordspacing

\bibitem{Angela_Survey}
\BIBentryALTinterwordspacing
X.~Yuan, Y.-J. Zhang, Y.~Shi, W.~Yan, and H.~Liu,
  ``Reconfigurable-intelligent-surface empowered {6G} wireless communications:
  Challenges and opportunities,'' \emph{arXiv}, Jan. 2020. [Online]. Available:
  \url{https://arxiv.org/abs/2001.00364.}
\BIBentrySTDinterwordspacing

\bibitem{Niyato_Survey}
\BIBentryALTinterwordspacing
S.~Gong, X.~Lu, D.~T. Hoang, D.~Niyato, L.~Shu, D.~I. Kim, and Y.-C. Liang,
  ``Towards smart radio environment for wireless communications via intelligent
  reflecting surfaces: A comprehensive survey,'' \emph{arXiv}, Dec. 2019.
  [Online]. Available: \url{https://arxiv.org/abs/1912.07794.}
\BIBentrySTDinterwordspacing

\bibitem{VisorSurf}
``https://www.visorsurf.eu.''

\bibitem{Ariadne}
``https://www.ict-ariadne.eu.''

\bibitem{PathFinder}
``https://cordis.europa.eu.''

\bibitem{Book_Caloz}
C.~Caloz and T.~Itoh, \emph{Electromagnetic Metamaterials: Transmission Line
  Theory and Microwave Applications}.\hskip 1em plus 0.5em minus 0.4em\relax
  John Wiley \& Sons, Nov. 2005.

\bibitem{Book_Engheta}
N.~Engheta and R.~W. Ziolkowski, \emph{Metamaterials: Physics and Engineering
  Explorations}.\hskip 1em plus 0.5em minus 0.4em\relax John Wiley \& Sons,
  Jul. 2006.

\bibitem{Book_Capolino}
F.~Capolino, \emph{Theory and Phenomena of Metamaterials}.\hskip 1em plus 0.5em
  minus 0.4em\relax CRC press, Oct. 2009.

\bibitem{Book_TieJunCui}
T.~J. Cui, D.~Smith, and R.~Liu, \emph{Metamaterials: Theory, Design, and
  Applications}.\hskip 1em plus 0.5em minus 0.4em\relax Springer, Nov. 2009.

\bibitem{Book_Maier}
A.~Maier~S., \emph{Handbook of Metamaterials and Plasmonics}.\hskip 1em plus
  0.5em minus 0.4em\relax World Scientific, Dec. 2017.

\bibitem{Survey_Tretyakov}
S.~B. Glybovski, S.~A. Tretyakov, P.~A. Belov, Y.~S. Kivshar, and C.~R.
  Simovski, ``Metasurfaces: From microwaves to visible,'' \emph{Phys. Rep.},
  vol. 634, pp. 1--72, May 2016.

\bibitem{Survey_Others}
H.-T. Chen, A.~J. Taylor, and N.~Yu, ``A review of metasurfaces: physics and
  applications,'' \emph{Rep. Prog. Phys.}, vol.~79, no.~7, 076401, Jun. 2016.

\bibitem{Book_Tretyakov}
S.~Tretyakov, \emph{Analytical Modeling in Applied Electromagnetics}.\hskip 1em
  plus 0.5em minus 0.4em\relax Artech House, 2003.

\bibitem{Book_Osipov}
A.~V. Osipov and S.~A. Tretyakov, \emph{Modern Electromagnetic Scattering
  Theory with Applications}.\hskip 1em plus 0.5em minus 0.4em\relax John Wiley
  \& Sons, Feb. 2017.

\bibitem{Coding_Cui}
T.~J. Cui, M.~Q. Qi, X.~Wan, J.~Zhao, and Q.~Cheng, ``Coding metamaterials,
  digital metamaterials and programmable metamaterials,'' \emph{Light Sci.
  Appl.}, vol.~3, no.~10, e218, Oct. 2014.

\bibitem{Tretyakov_Multifunction}
F.~Liu, O.~Tsilipakos, A.~Pitilakis, A.~C. Tasolamprou, M.~S. Mirmoosa, N.~V.
  Kantartzis, D.-H. Kwon, M.~Kafesaki, C.~M. Soukoulis, and S.~A. Tretyakov,
  ``Intelligent metasurfaces with continuously tunable local surface impedance
  for multiple reconfigurable functions,'' \emph{Phys. Rev. Applied}, vol.~11,
  no.~4, 044024, Apr. 2019.

\bibitem{VisorSurf_Survey}
F.~Liu, A.~Pitilakis, M.~S. Mirmoosa, O.~Tsilipakos, X.~Wang, A.~C.
  Tasolamprou, S.~Abadal, A.~Cabellos-Aparicio, E.~Alarcon, C.~Liaskos, N.~V.
  Kantartzis, M.~Kafesaki, E.~N. Economou, C.~M. Soukoulis, and S.~Tretyakov,
  ``Programmable metasurfaces: State of the art and prospects,'' \emph{IEEE
  Int. Symp. Circuits and Systems}, May 2018.

\bibitem{Perruisseau-Carrier}
S.~V. Hum and J.~Perruisseau-Carrier, ``Reconfigurable reflectarrays and array
  lenses for dynamic antenna beam control: A review,'' \emph{IEEE Trans.
  Antennas Propag.}, vol.~62, no.~1, pp. 183--198, Ja. 2014.

\bibitem{Subrt2010}
L.~Subrt, D.~Grace, and P.~Pechac, ``Controlling the short-range propagation
  environment using active frequency selective surfaces,''
  \emph{Radioengineering}, vol.~19, no.~4, pp. 610--615, Dec. 2010.

\bibitem{Subrt2012a}
L.~Subrt and P.~Pechac, ``Controlling propagation environments using
  intelligent walls,'' \emph{IEEE European Conf. Antennas and Propagation}, pp.
  1--5, Mar. 2012.

\bibitem{Subrt2012}
------, ``Intelligent walls as autonomous parts of smart indoor environments,''
  \emph{IET Commun.}, vol.~6, no.~8, pp. 1004--1010, May. 2012.

\bibitem{Kaina2014}
N.~Kaina, M.~Dupre, G.~Lerosey, and M.~Fink, ``Shaping complex microwave fields
  in reverberating media with binary tunable metasurfaces,'' \emph{Sci. Rep.},
  vol.~4, pp. 1--8, Oct. 2014.

\bibitem{garcia2019reconfigurable}
\BIBentryALTinterwordspacing
J.~B. Garcia, A.~Sibille, and M.~Kamoun, ``Reconfigurable intelligent surfaces:
  Bridging the gap between scattering and reflection,'' \emph{arXiv}, Dec.
  2019. [Online]. Available: \url{https://arxiv.org/abs/1912.05344.}
\BIBentrySTDinterwordspacing

\bibitem{ellingson2019path}
\BIBentryALTinterwordspacing
S.~W. Ellingson, ``Path loss in reconfigurable intelligent surface-enabled
  channels,'' \emph{arXiv}, Dec. 2019. [Online]. Available:
  \url{https://arxiv.org/abs/1912.06759.}
\BIBentrySTDinterwordspacing

\bibitem{Khawaja-Coverage}
W.~Khawaja, O.~Ozdemir, Y.~Yapici, F.~Erden, M.~Ezuma, and I.~Guvenc,
  ``Coverage enhancement for {NLOS} mmwave links using passive reflectors,''
  \emph{IEEE Open J. Commun. Society}, vol.~1, pp. 263--281, Jan. 2020.

\bibitem{ozdogan2019intelligent}
{\"O}.~{\"O}zdogan, E.~Bj{\"o}rnson, and E.~G. Larsson, ``Intelligent
  reflecting surfaces: Physics, propagation, and pathloss modeling,''
  \emph{IEEE Wireless Commun. Lett.}, IEEE Early Access, 2020.

\bibitem{PhanHuy2019}
D.-T. {Phan-Huy}, Y.~{Kokar}, K.~{Rachedi}, P.~{Pajusco}, A.~{Mokh},
  T.~{Magounaki}, R.~{Masood}, C.~{Buey}, P.~{Ratajczak},
  N.~{Malhouroux-Gaffet}, J.~{Conrat}, J.~. {Prevotet}, A.~{Ourir}, J.~{de
  Rosny}, M.~{Crussière}, M.~{Helard}, A.~{Gati}, T.~{Sarrebourse}, and M.~{Di
  Renzo}, ``Single-carrier spatial modulation for the internet of things:
  Design and performance evaluation by using real compact and reconfigurable
  antennas,'' \emph{IEEE Access}, vol.~7, pp. 18\,978--18\,993, Jan. 2019.

\bibitem{MDR2019_Dung}
D.~{Nguyen Viet}, M.~{Di~Renzo}, V.~{Basavarajappa}, B.~{Bedia Exposito},
  J.~{Basterrechea}, and D.-T. {Phan-Huy}, ``Spatial modulation based on
  reconfigurable antennas: Performance evaluation by using the prototype of a
  reconfigurable antenna,'' \emph{EURASIP J. Wireless. Commun.}, pp. 1--17,
  Jun. 2019.

\bibitem{Basar2019}
E.~Basar, ``Reconfigurable intelligent surface-based index modulation: A new
  beyond {MIMO} paradigm for {6G},'' \emph{IEEE Trans. Commun.}, IEEE Early
  Access, 2020.

\bibitem{canbilen2020reconfigurable}
\BIBentryALTinterwordspacing
A.~E. Canbilen, E.~Basar, and S.~S. Ikki, ``Reconfigurable intelligent
  surface-assisted space shift keying,'' \emph{arXiv}, Jan. 2020. [Online].
  Available: \url{https://arxiv.org/abs/2001.11287.}
\BIBentrySTDinterwordspacing

\bibitem{yan2019WCL}
W.~Yan, X.~Yuan, and X.~Kuai, ``Passive beamforming and information transfer
  via large intelligent surface,'' \emph{IEEE Wireless Commun. Lett.}, vol.~9,
  no.~4, pp. 533--537, Apr. 2020.

\bibitem{Guo19RefMod}
\BIBentryALTinterwordspacing
S.~Guo, S.~Lv, H.~Zhang, J.~Ye, and Z.~Peng, ``Reflecting modulation,''
  \emph{arXiv}, Dec. 2019. [Online]. Available:
  \url{https://arxiv.org/abs/1912.08428.}
\BIBentrySTDinterwordspacing

\bibitem{Hanzo__2020}
\BIBentryALTinterwordspacing
S.~Gopi, S.~Kalyani, and L.~Hanzo, ``Intelligent reflecting surface assisted
  beam index-modulation for millimeter wave communication,'' \emph{arXiv}, Mar.
  2020. [Online]. Available: \url{https://arxiv.org/pdf/2003.12049.}
\BIBentrySTDinterwordspacing

\bibitem{Nadeem2019b}
\BIBentryALTinterwordspacing
Q.-U.-A. Nadeem, A.~Kammoun, A.~Chaaban, M.~Debbah, and M.-S. Alouini,
  ``Intelligent reflecting surface assisted multi-user {MISO} communication,''
  \emph{arXiv}, Jun. 2019. [Online]. Available:
  \url{https://arxiv.org/abs/1906.02360.}
\BIBentrySTDinterwordspacing

\bibitem{Jensen2019}
\BIBentryALTinterwordspacing
T.~L. Jensen and E.~D. Carvalho, ``On optimal channel estimation scheme for
  intelligent reflecting surfaces based on a minimum variance unbiased
  estimator,'' \emph{arXiv}, Sep. 2019. [Online]. Available:
  \url{https://arxiv.org/abs/1909.09440.}
\BIBentrySTDinterwordspacing

\bibitem{Lin2019}
\BIBentryALTinterwordspacing
J.~Lin, G.~Wang, R.~Fan, T.~A. Tsiftsis, and C.~Tellambura, ``Channel
  estimation for wireless communication systems assisted by large intelligent
  surfaces,'' \emph{arXiv}, Nov. 2019. [Online]. Available:
  \url{https://arxiv.org/abs/1911.02158.}
\BIBentrySTDinterwordspacing

\bibitem{Chen2019b}
\BIBentryALTinterwordspacing
J.~Chen, Y.-C. Liang, H.~V. Cheng, and W.~Yu, ``Channel estimation for
  reconfigurable intelligent surface aided multi-user {MIMO} systems,''
  \emph{arXiv}, Dec. 2019. [Online]. Available:
  \url{https://arxiv.org/abs/1912.03619.}
\BIBentrySTDinterwordspacing

\bibitem{Ning2019}
\BIBentryALTinterwordspacing
B.~Ning, Z.~Chen, W.~Chen, and Y.~Du, ``Channel estimation and transmission for
  intelligent reflecting surface assisted thz communications,'' \emph{arXiv},
  Nov. 2019. [Online]. Available: \url{https://arxiv.org/abs/1911.04719.}
\BIBentrySTDinterwordspacing

\bibitem{Wang2019d}
\BIBentryALTinterwordspacing
Z.~Wang, L.~Liu, and S.~Cui, ``Channel estimation for intelligent reflecting
  surface assisted multiuser communications,'' \emph{arXiv}, Nov. 2019.
  [Online]. Available: \url{https://arxiv.org/abs/1911.03084.}
\BIBentrySTDinterwordspacing

\bibitem{Xia2019}
\BIBentryALTinterwordspacing
S.~Xia and Y.~Shi, ``Intelligent reflecting surface for massive device
  connectivity: Joint activity detection and channel estimation,''
  \emph{arXiv}, Nov. 2019. [Online]. Available:
  \url{https://arxiv.org/abs/1911.12157.}
\BIBentrySTDinterwordspacing

\bibitem{cui2019efficient}
\BIBentryALTinterwordspacing
Y.~Cui and H.~Yin, ``An efficient {CSI} acquisition method for intelligent
  reflecting surface-assisted mmwave networks,'' \emph{arXiv}, Dec. 2019.
  [Online]. Available: \url{https://arxiv.org/abs/1912.12076.}
\BIBentrySTDinterwordspacing

\bibitem{You2019}
\BIBentryALTinterwordspacing
C.~You, B.~Zheng, and R.~Zhang, ``Intelligent reflecting surface with discrete
  phase shifts: Channel estimation and passive beamforming,'' \emph{arXiv},
  Nov. 2019. [Online]. Available: \url{https://arxiv.org/abs/1911.03916.}
\BIBentrySTDinterwordspacing

\bibitem{liu2019matrix}
\BIBentryALTinterwordspacing
H.~Liu, X.~Yuan, and Y.~Jun, ``Matrix-calibration-based cascaded channel
  estimation for reconfigurable intelligent surface assisted multiuser
  {MIMO},'' \emph{arXiv}, Dec. 2019. [Online]. Available:
  \url{https://arxiv.org/abs/1912.09025.}
\BIBentrySTDinterwordspacing

\bibitem{wei2020parallel}
\BIBentryALTinterwordspacing
L.~Wei, C.~Huang, G.~C. Alexandropoulos, and C.~Yuen, ``Parallel factor
  decomposition channel estimation in {RIS}-assisted multi-user {MISO}
  communication,'' \emph{arXiv}, Jan. 2020. [Online]. Available:
  \url{https://arxiv.org/abs/2001.09413.}
\BIBentrySTDinterwordspacing

\bibitem{wan2020broadband}
\BIBentryALTinterwordspacing
Z.~Wan, Z.~Gao, and M.-S. Alouini, ``Broadband channel estimation for
  intelligent reflecting surface aided mmwave massive {MIMO} systems,''
  \emph{arXiv}, Feb. 2020. [Online]. Available:
  \url{https://arxiv.org/abs/2002.01629.}
\BIBentrySTDinterwordspacing

\bibitem{wang2019compressed}
\BIBentryALTinterwordspacing
P.~Wang, J.~Fang, H.~Duan, and H.~Li, ``Compressed channel estimation and joint
  beamforming for intelligent reflecting surface-assisted millimeter wave
  systems,'' \emph{arXiv}, Nov. 2019. [Online]. Available:
  \url{https://arxiv.org/abs/1911.07202.}
\BIBentrySTDinterwordspacing

\bibitem{he2020channel}
\BIBentryALTinterwordspacing
J.~He, M.~Leinonen, H.~Wymeersch, and M.~Juntti, ``Channel estimation for
  {RIS}-aided mmwave {MIMO} channels,'' \emph{arXiv}, Feb. 2020. [Online].
  Available: \url{https://arxiv.org/abs/2002.06453.}
\BIBentrySTDinterwordspacing

\bibitem{mishra2019channel}
D.~Mishra and H.~Johansson, ``Channel estimation and low-complexity beamforming
  design for passive intelligent surface assisted {MISO} wireless energy
  transfer,'' \emph{IEEE Int. Conf. Acoustics, Speech and Signal Processing},
  pp. 4659--4663, May. 2019.

\bibitem{de2020parafac}
\BIBentryALTinterwordspacing
G.~T. de~Ara{\'u}jo and A.~L. de~Almeida, ``{PARAFAC}-based channel estimation
  for intelligent reflective surface assisted {MIMO} system,'' \emph{arXiv},
  Jan. 2020. [Online]. Available: \url{https://arxiv.org/abs/2001.06554.}
\BIBentrySTDinterwordspacing

\bibitem{ning2019channel}
\BIBentryALTinterwordspacing
B.~Ning, Z.~Chen, W.~Chen, Y.~Du, and J.~Fang, ``Channel estimation and hybrid
  beamforming for reconfigurable intelligent surfaces assisted {THz}
  communications,'' \emph{arXiv}, Dec. 2019. [Online]. Available:
  \url{https://arxiv.org/abs/1912.11662.}
\BIBentrySTDinterwordspacing

\bibitem{you2019progressive}
\BIBentryALTinterwordspacing
C.~You, B.~Zheng, and R.~Zhang, ``Progressive channel estimation and passive
  beamforming for intelligent reflecting surface with discrete phase shifts,''
  \emph{arXiv}, Dec. 2019. [Online]. Available:
  \url{https://arxiv.org/abs/1912.10646.}
\BIBentrySTDinterwordspacing

\bibitem{mirza2019channel}
\BIBentryALTinterwordspacing
J.~Mirza and B.~Ali, ``Channel estimation method and phase shift design for
  reconfigurable intelligent surface assisted {MIMO} networks,'' \emph{arXiv},
  Dec. 2019. [Online]. Available: \url{https://arxiv.org/abs/1912.10671.}
\BIBentrySTDinterwordspacing

\bibitem{hu2019two}
\BIBentryALTinterwordspacing
C.~Hu and L.~Dai, ``Two-timescale channel estimation for reconfigurable
  intelligent surface aided wireless communications,'' \emph{arXiv}, Dec. 2019.
  [Online]. Available: \url{https://arxiv.org/abs/1912.07990.}
\BIBentrySTDinterwordspacing

\bibitem{zheng2019channel}
H.~Zheng, Z.~Yang, G.~Wang, R.~He, and B.~Ai, ``Channel estimation for ambient
  backscatter communications with large intelligent surface,'' \emph{IEEE Int.
  Conf. Wireless Communications and Signal Processing}, pp. 1--5, Oct. 2019.

\bibitem{XiaojunYuan}
Z.-Q. He and X.~Yuan, ``Cascaded channel estimation for large intelligent
  metasurface assisted massive {MIMO},'' \emph{IEEE Wireless Commun. Lett.},
  vol.~9, no.~2, pp. 210--214, Feb. 2020.

\bibitem{alexandropoulos2020hardware}
\BIBentryALTinterwordspacing
G.~C. Alexandropoulos and E.~Vlachos, ``A hardware architecture for
  reconfigurable intelligent surfaces with minimal active elements for explicit
  channel estimation,'' \emph{arXiv}, Feb. 2020. [Online]. Available:
  \url{https://arxiv.org/abs/2002.10371.}
\BIBentrySTDinterwordspacing

\bibitem{Zhang2019e}
H.~Zhang, B.~Di, L.~Song, and Z.~Han, ``Reconfigurable intelligent surfaces
  assisted communications with limited phase shifts: How many phase shifts are
  enough?'' \emph{IEEE Trans. Veh. Technol.}, IEEE Early Access, 2020.

\bibitem{Hu2018b}
S.~Hu, F.~Rusek, and O.~Edfors, ``Capacity degradation with modeling hardware
  impairment in large intelligent surface,'' \emph{IEEE Conf. Global
  Communications}, pp. 1--6, Dec. 2018.

\bibitem{Jung2018}
M.~Jung, W.~Saad, Y.~R. Jang, K.~Gyuyeol, and C.~Sooyong, ``Performance
  analysis of large intelligence surfaces ({LISs}): Asymptotic data rate and
  channel hardening effects,'' \emph{IEEE Trans. Wireless Commun.}, vol.~19,
  no.~3, pp. 2052--2065, Jan. 2019.

\bibitem{Jung2019}
\BIBentryALTinterwordspacing
M.~Jung, W.~Saad, M.~Debbah, and C.~S. Hong, ``On the optimality of
  reconfigurable intelligent surfaces ({RISs}): Passive beamforming,
  modulation, and resource allocation,'' \emph{arXiv}, Oct. 2019. [Online].
  Available: \url{https://arxiv.org/abs/1910.00968.}
\BIBentrySTDinterwordspacing

\bibitem{Nadeem2019}
\BIBentryALTinterwordspacing
Q.-U.-A. Nadeem, A.~Kammoun, A.~Chaaban, M.~Debbah, and M.-S. Alouini,
  ``Asymptotic max-min {SINR} analysis of large intelligent surface assisted
  {MIMO} communications,'' \emph{arXiv}, Mar. 2020. [Online]. Available:
  \url{https://arxiv.org/abs/1903.08127.}
\BIBentrySTDinterwordspacing

\bibitem{Zhang2019c}
\BIBentryALTinterwordspacing
Z.~Zhang, Y.~Cui, F.~Yang, and L.~Ding, ``Analysis and optimization of outage
  probability in multi-intelligent reflecting surface-assisted systems,''
  \emph{arXiv}, Sep. 2019. [Online]. Available:
  \url{https://arxiv.org/abs/1909.02193.}
\BIBentrySTDinterwordspacing

\bibitem{basar2019transmission}
E.~Basar, ``Transmission through large intelligent surfaces: A new frontier in
  wireless communications,'' \emph{IEEE European Conf. Networks and
  Communications}, pp. 112--117, Jun. 2019.

\bibitem{jung2019reliability}
M.~Jung, W.~Saad, Y.~Jang, G.~Kong, and S.~Choi, ``Reliability analysis of
  large intelligent surfaces (liss): Rate distribution and outage
  probability,'' \emph{IEEE Wireless Commun. Lett.}, vol.~8, no.~6, pp.
  1662--1666, Aug. 2019.

\bibitem{kudathanthirige2020performance}
\BIBentryALTinterwordspacing
D.~Kudathanthirige, D.~Gunasinghe, and G.~Amarasuriya, ``Performance analysis
  of intelligent reflective surfaces for wireless communication,''
  \emph{arXiv}, Feb. 2020. [Online]. Available:
  \url{https://arxiv.org/abs/2002.05603.}
\BIBentrySTDinterwordspacing

\bibitem{badiu2019communication}
M.-A. Badiu and J.~P. Coon, ``Communication through a large reflecting surface
  with phase errors,'' \emph{IEEE Wireless Commun. Lett.}, vol.~9, no.~2, pp.
  184--188, Feb. 2020.

\bibitem{psomas2019low}
\BIBentryALTinterwordspacing
C.~Psomas, I.~Chrysovergis, and I.~Krikidis, ``Low-complexity random
  rotation-based schemes for intelligent reflecting surfaces,'' \emph{arXiv},
  Dec. 2019. [Online]. Available: \url{https://arxiv.org/abs/1912.10347.}
\BIBentrySTDinterwordspacing

\bibitem{Guo20CL}
C.~Guo, Y.~Cui, F.~Yang, and L.~Ding, ``Outage probability analysis and
  minimization in intelligent reflecting surface-assisted {MISO} systems,''
  \emph{IEEE Commun. Lett.}, IEEE Early Access, 2020.

\bibitem{basar2019reconfigurable}
\BIBentryALTinterwordspacing
E.~Basar and I.~F. Akyildiz, ``Reconfigurable intelligent surfaces for doppler
  effect and multipath fading mitigation,'' \emph{arXiv}, Dec. 2019. [Online].
  Available: \url{https://arxiv.org/abs/1912.04080.}
\BIBentrySTDinterwordspacing

\bibitem{ozdogan2020using}
\BIBentryALTinterwordspacing
{\"O}.~{\"O}zdogan, E.~Bj{\"o}rnson, and E.~G. Larsson, ``Using intelligent
  reflecting surfaces for rank improvement in {MIMO} communications,''
  \emph{arXiv}, Feb. 2020. [Online]. Available:
  \url{https://arxiv.org/abs/2002.02182.}
\BIBentrySTDinterwordspacing

\bibitem{xu2019discrete}
J.~Xu, W.~Xu, and A.~L. Swindlehurst, ``Discrete phase shift design for
  practical large intelligent surface communication,'' \emph{IEEE Pacific Rim
  Conf. Communications, Computers and Signal Processing}, pp. 1--5, Aug. 2015.

\bibitem{cao2019capacity}
F.~Cao, Y.~Han, Q.~Liu, C.-K. Wen, and S.~Jin, ``Capacity analysis and
  scheduling for distributed {LIS}-aided large-scale antenna systems,''
  \emph{IEEE Int. Conf. Communications in China}, pp. 659--664, Aug. 2019.

\bibitem{hu2019spherical}
\BIBentryALTinterwordspacing
S.~Hu, ``Spherical large intelligent surfaces,'' \emph{arXiv}, Jul. 2019.
  [Online]. Available: \url{https://arxiv.org/abs/1907.02699.}
\BIBentrySTDinterwordspacing

\bibitem{zhang2020intelligent}
\BIBentryALTinterwordspacing
S.~Zhang and R.~Zhang, ``Intelligent reflecting surface aided multiple access:
  Capacity region and deployment strategy,'' \emph{arXiv}, Feb. 2020. [Online].
  Available: \url{https://arxiv.org/abs/2002.07091.}
\BIBentrySTDinterwordspacing

\bibitem{lyu2020spatial}
\BIBentryALTinterwordspacing
J.~Lyu and R.~Zhang, ``Spatial throughput characterization for intelligent
  reflecting surface aided multiuser system,'' \emph{arXiv}, Jan. 2020.
  [Online]. Available: \url{https://arxiv.org/abs/2001.02447.}
\BIBentrySTDinterwordspacing

\bibitem{williams2019communication}
\BIBentryALTinterwordspacing
R.~J. Williams, E.~De~Carvalho, and T.~L. Marzetta, ``A communication model for
  large intelligent surfaces,'' \emph{arXiv}, Dec. 2019. [Online]. Available:
  \url{https://arxiv.org/abs/1912.06644.}
\BIBentrySTDinterwordspacing

\bibitem{wang2020intelligent}
\BIBentryALTinterwordspacing
Z.~Wang, L.~Liu, and S.~Cui, ``Intelligent reflecting surface assisted massive
  {MIMO} communications,'' \emph{arXiv}, Jan. 2020. [Online]. Available:
  \url{https://arxiv.org/abs/2001.05899.}
\BIBentrySTDinterwordspacing

\bibitem{SG_Jeff}
J.~G. Andrews, F.~Baccelli, and R.~K. Ganti, ``A tractable approach to coverage
  and rate in cellular networks,'' \emph{IEEE Trans. Commun.}, vol.~59, no.~11,
  pp. 3122--3134, Nov. 2011.

\bibitem{SG_MDR_1}
M.~{Di Renzo}, A.~Guidotti, and G.~E. Corazza, ``Average rate of downlink
  heterogeneous cellular networks over generalized fading channels - a
  stochastic geometry approach,'' \emph{IEEE Trans. Commun.}, vol.~61, no.~7,
  pp. 3050--3071, Jul. 2013.

\bibitem{SG_MDR_2}
M.~{Di Renzo}, ``Stochastic geometry modeling analysis of multi-tier millimeter
  wave cellular networks,'' \emph{IEEE Trans. Wireless Commun.}, vol.~14,
  no.~9, pp. 5038--5057, Sep. 2015.

\bibitem{SG_MDR_3}
W.~Lu and M.~{Di Renzo}, ``Stochastic geometry modeling of cellular networks:
  Analysis, simulation and experimental validation,'' \emph{ACM Int. Conf.
  Modeling, Analysis and Simulation of Wireless and Mobile Systems}, pp.
  179--188, Nov. 2015.

\bibitem{SG_MDR_4a}
M.~{Di Renzo} and P.~Guan, ``Stochastic geometry modeling and system-level
  analysis of uplink heterogeneous cellular networks with multi-antenna base
  stations,'' \emph{IEEE Trans. Commun.}, vol.~64, no.~6, pp. 2453--2476, Jun.
  2016.

\bibitem{SG_MDR_4}
M.~{Di Renzo}, W.~Lu, and P.~Guan, ``The intensity matching approach: A
  tractable stochastic geometry approximation to system-level analysis of
  cellular networks,'' \emph{IEEE Trans. Wireless Commun.}, vol.~15, no.~9, pp.
  5963--5983, Sep. 2016.

\bibitem{SG_MDR_7}
M.~{Di Renzo} and W.~Lu, ``System-level analysis and optimization of cellular
  networks with simultaneous wireless information and power transfer:
  Stochastic geometry modeling,'' \emph{IEEE Trans. Veh. Technol.}, vol.~66,
  no.~3, pp. 2251--2275, Aug. 2017.

\bibitem{SG_MDR_5}
M.~{Di Renzo}, A.~Zappone, T.~T. Lam, and M.~Debbah, ``System-level modeling
  and optimization of the energy efficiency in cellular networks - a stochastic
  geometry framework,'' \emph{IEEE Trans. Wireless Commun.}, vol.~17, no.~4,
  pp. 2539--2556, Apr. 2018.

\bibitem{SG_MDR_6}
M.~{Di Renzo}, S.~Wang, and X.~Xi, ``Modeling and analysis of cellular networks
  by using inhomogeneous poisson point processes,'' \emph{IEEE Trans. Wireless
  Commun.}, vol.~17, no.~8, pp. 5162--5182, Aug. 2018.

\bibitem{MDR2019_Eurasip}
D.~Marco and J.~Song, ``Reflection probability in wireless networks with
  metasurface-coated environmental objects: An approach based on random spatial
  processes,'' \emph{EURASIP J. Wireless Commun.}, no.~99, Apr. 2019.

\bibitem{Hou2019}
\BIBentryALTinterwordspacing
T.~Hou, Y.~Liu, Z.~Song, X.~Sun, Y.~Chen, and L.~Hanzo, ``{MIMO} assisted
  networks relying on large intelligent surfaces: A stochastic geometry
  model,'' \emph{arXiv}, Oct. 2019. [Online]. Available:
  \url{https://arxiv.org/abs/1910.00959.}
\BIBentrySTDinterwordspacing

\bibitem{kishk2020exploiting}
\BIBentryALTinterwordspacing
M.~A. Kishk and M.-S. Alouini, ``Exploiting randomly-located blockages for
  large-scale deployment of intelligent surfaces,'' \emph{arXiv}, Jan. 2020.
  [Online]. Available: \url{https://arxiv.org/abs/2001.10766.}
\BIBentrySTDinterwordspacing

\bibitem{Huang_ICASSP}
C.~{Huang}, A.~{Zappone}, M.~{Debbah}, and C.~{Yuen}, ``Achievable rate
  maximization by passive intelligent mirrors,'' \emph{IEEE Int. Conf.
  Acoustics, Speech and Signal Processing}, pp. 3714--3718, Apr. 2018.

\bibitem{ZapTWC2019}
C.~Huang, A.~Zappone, G.~C. Alexandropoulos, M.~Debbah, and C.~Yuen,
  ``Reconfigurable intelligent surfaces for energy efficiency in wireless
  communication,'' \emph{IEEE Trans. Wireless Commun.}, vol.~18, no.~8, pp.
  4157--4170, Aug. 2019.

\bibitem{Abeywickrama2019}
\BIBentryALTinterwordspacing
S.~Abeywickrama, R.~Zhang, and C.~Yuen, ``Intelligent reflecting surface:
  Practical phase shift model and beamforming optimization,'' \emph{arXiv},
  Feb. 2020. [Online]. Available: \url{https://arxiv.org/abs/1907.06002.}
\BIBentrySTDinterwordspacing

\bibitem{Yxu2020resource}
\BIBentryALTinterwordspacing
Y.~Xu, Z.~Qin, Y.~Zhao, G.~Li, G.~Gui, and H.~Sari, ``Resource allocation for
  intelligent reflecting surface enabled heterogeneous networks,''
  \emph{arXiv}, Jan. 2020. [Online]. Available:
  \url{https://arxiv.org/abs/2001.11729.}
\BIBentrySTDinterwordspacing

\bibitem{xie2019max}
\BIBentryALTinterwordspacing
H.~Xie, J.~Xu, and Y.-F. Liu, ``{Max-Min} fairness in {IRS}-aided multi-cell
  {MISO} systems via joint transmit and reflective beamforming,'' \emph{arXiv},
  Dec. 2019. [Online]. Available: \url{https://arxiv.org/abs/1912.12827.}
\BIBentrySTDinterwordspacing

\bibitem{gao2020reconfigurable}
\BIBentryALTinterwordspacing
Y.~Gao, C.~Yong, Z.~Xiong, D.~Niyato, Y.~Xiao, and J.~Zhao, ``Reconfigurable
  intelligent surface for {MISO} systems with proportional rate constraints,''
  Jan. 2020. [Online]. Available: \url{https://arxiv.org/abs/2001.10845.}
\BIBentrySTDinterwordspacing

\bibitem{Pan2019b}
\BIBentryALTinterwordspacing
C.~Pan, H.~Ren, K.~Wang, W.~Xu, M.~Elkashlan, A.~Nallanathan, and L.~Hanzo,
  ``Multicell {MIMO} communications relying on intelligent reflecting
  surface,'' \emph{arXiv}, Dec. 2019. [Online]. Available:
  \url{https://arxiv.org/abs/1907.10864.}
\BIBentrySTDinterwordspacing

\bibitem{Guo2019}
H.~Guo, Y.-C. Liang, J.~Chen, and E.~G. Larsson, ``Weighted sum-rate
  optimization for intelligent reflecting surface enhanced wireless networks,''
  \emph{IEEE Global. Conf. Communications}, pp. 1--6, Dec. 2019.

\bibitem{Zhou2019b}
\BIBentryALTinterwordspacing
G.~Zhou, C.~Pan, H.~Ren, K.~Wang, and A.~Nallanathan, ``Intelligent reflecting
  surface aided multigroup multicast {MISO} communication systems,''
  \emph{arXiv}, Sep. 2019. [Online]. Available:
  \url{https://arxiv.org/abs/1909.04606.}
\BIBentrySTDinterwordspacing

\bibitem{Yang2019}
Y.~Yang, B.~Zheng, S.~Zhang, and R.~Zhang, ``Intelligent reflecting surface
  meets {OFDM}: Protocol design and rate maximization,'' \emph{IEEE Trans.
  Commun.}, IEEE Early Access, 2020.

\bibitem{Li2019c}
\BIBentryALTinterwordspacing
H.~Li, R.~Liu, M.~Li, and Q.~Liu, ``{IRS}-enhanced wideband {MU-MISO-OFDM}
  communication systems,'' \emph{arXiv}, Sep. 2019. [Online]. Available:
  \url{https://arxiv.org/abs/1909.11314.}
\BIBentrySTDinterwordspacing

\bibitem{Yang2019c}
Y.~Yang, S.~Zhang, and R.~Zhang, ``{IRS}-enhanced {OFDMA}: Joint resource
  allocation and passive beamforming optimization,'' \emph{IEEE Wireless
  Commun. Lett.}, IEEE Early Access, 2020.

\bibitem{Han2019}
\BIBentryALTinterwordspacing
H.~Han, J.~Zhao, D.~Niyato, M.~{Di~Renzo}, and Q.-V. Pham, ``Intelligent
  reflecting surface aided network: Power control for physical-layer
  broadcasting,'' \emph{arXiv}, Oct. 2019. [Online]. Available:
  \url{https://arxiv.org/abs/1910.14383.}
\BIBentrySTDinterwordspacing

\bibitem{Wu2019b}
Q.~Wu and R.~Zhang, ``Intelligent reflecting surface enhanced wireless network
  via joint active and passive beamforming,'' \emph{IEEE Trans. Wireless
  Commun.}, vol.~18, no.~11, pp. 5394--5409, Aug. 2019.

\bibitem{zhou2020framework}
\BIBentryALTinterwordspacing
G.~Zhou, C.~Pan, H.~Ren, K.~Wang, and A.~Nallanathan, ``A framework of robust
  transmission design for {IRS}-aided {MISO} communications with imperfect
  cascaded channels,'' \emph{arXiv}, Jan. 2020. [Online]. Available:
  \url{https://arxiv.org/abs/2001.07054.}
\BIBentrySTDinterwordspacing

\bibitem{shao2020minimum}
\BIBentryALTinterwordspacing
M.~Shao, Q.~Li, and W.-K. Ma, ``Minimum symbol-error probability symbol-level
  precoding with intelligent reflecting surface,'' \emph{arXiv}, Jan. 2020.
  [Online]. Available: \url{https://arxiv.org/abs/2001.06840.}
\BIBentrySTDinterwordspacing

\bibitem{liu2019joint}
\BIBentryALTinterwordspacing
R.~Liu, M.~Li, Q.~Liu, and A.~L. Swindlehurst, ``Joint symbol-level precoding
  and reflecting designs for {RIS}-enhanced {MU-MISO} systems,'' \emph{arXiv},
  Dec. 2019. [Online]. Available: \url{https://arxiv.org/abs/1912.11767.}
\BIBentrySTDinterwordspacing

\bibitem{jiang2019over}
\BIBentryALTinterwordspacing
T.~Jiang and Y.~Shi, ``Over-the-air computation via intelligent reflecting
  surfaces,'' \emph{arXiv}, Apr. 2019. [Online]. Available:
  \url{https://arxiv.org/abs/1904.12475.}
\BIBentrySTDinterwordspacing

\bibitem{Zhou2019}
\BIBentryALTinterwordspacing
G.~Zhou, C.~Pan, H.~Ren, K.~Wang, M.~{Di~Renzo}, and A.~Nallanathan, ``Robust
  beamforming design for intelligent reflecting surface aided {MISO}
  communication systems,'' \emph{arXiv}, Nov. 2019. [Online]. Available:
  \url{https://arxiv.org/abs/1911.06237.}
\BIBentrySTDinterwordspacing

\bibitem{guo2019model}
\BIBentryALTinterwordspacing
H.~Guo, Y.-C. Liang, and S.~Xiao, ``Model-free optimization for reconfigurable
  intelligent surface with statistical {CSI},'' \emph{arXiv}, Dec. 2019.
  [Online]. Available: \url{https://arxiv.org/abs/1912.10913.}
\BIBentrySTDinterwordspacing

\bibitem{li2020weighted}
\BIBentryALTinterwordspacing
Z.~Li, M.~Hua, Q.~Wang, and Q.~Song, ``Weighted sum-rate maximization for
  multi-{IRS} aided cooperative transmission,'' \emph{arXiv}, Feb. 2020.
  [Online]. Available: \url{https://arxiv.org/abs/2002.04900.}
\BIBentrySTDinterwordspacing

\bibitem{Yu2019}
X.~Yu, D.~Xu, and R.~Schober, ``{MISO} wireless communication systems via
  intelligent reflecting surfaces,'' \emph{IEEE Int. Conf. Communications in
  China}, pp. 735--740, Aug. 2019.

\bibitem{Li2019b}
\BIBentryALTinterwordspacing
X.~Li, J.~Fang, F.~Gao, and H.~Li, ``Joint active and passive beamforming for
  intelligent reflecting surface-assisted massive {MIMO} systems,''
  \emph{arXiv}, Dec. 2019. [Online]. Available:
  \url{https://arxiv.org/abs/1912.00728.}
\BIBentrySTDinterwordspacing

\bibitem{jia2020analysis}
\BIBentryALTinterwordspacing
Y.~Jia, C.~Ye, and Y.~Cui, ``Analysis and optimization of an intelligent
  reflecting surface-assisted system with interference,'' \emph{arXiv}, Feb.
  2020. [Online]. Available: \url{https://arxiv.org/abs/2002.00168.}
\BIBentrySTDinterwordspacing

\bibitem{razavizadeh20203d}
\BIBentryALTinterwordspacing
S.~M. Razavizadeh and T.~Svensson, ``{3D} beamforming in reconfigurable
  intelligent surfaces-assisted wireless communication networks,''
  \emph{arXiv}, Jan. 2020. [Online]. Available:
  \url{https://arxiv.org/abs/2001.06653.}
\BIBentrySTDinterwordspacing

\bibitem{yan2019passive}
\BIBentryALTinterwordspacing
W.~Yan, Z.-Q. He, and X.~Kuai, ``{IRS}-aided large-scale {MIMO} systems with
  passive constant envelope precoding,'' \emph{arXiv}, Dec. 2019. [Online].
  Available: \url{https://arxiv.org/abs/1912.10209.}
\BIBentrySTDinterwordspacing

\bibitem{yu2020optimal}
\BIBentryALTinterwordspacing
X.~Yu, D.~Xu, and R.~Schober, ``Optimal beamforming for {MISO} communications
  via intelligent reflecting surfaces,'' \emph{arXiv}, Jan. 2020. [Online].
  Available: \url{https://arxiv.org/abs/2001.11429.}
\BIBentrySTDinterwordspacing

\bibitem{gao2020reflection}
\BIBentryALTinterwordspacing
Y.~Gao, C.~Yong, Z.~Xiong, J.~Zhao, Y.~Xiao, and D.~Niyato, ``Reflection
  resource management for intelligent reflecting surface aided wireless
  networks,'' \emph{arXiv}, Feb. 2020. [Online]. Available:
  \url{https://arxiv.org/abs/2002.00331.}
\BIBentrySTDinterwordspacing

\bibitem{zhang2020joint}
\BIBentryALTinterwordspacing
Z.~Zhang and L.~Dai, ``A joint precoding framework for wideband reconfigurable
  intelligent surface-aided cell-free network,'' \emph{arXiv}, Feb. 2020.
  [Online]. Available: \url{https://arxiv.org/abs/2002.03744.}
\BIBentrySTDinterwordspacing

\bibitem{zhou2019ris}
\BIBentryALTinterwordspacing
Z.~Zhou, N.~Ge, W.~Liu, and Z.~Wang, ``{RIS}-aided offshore communications with
  adaptive beamforming and service time allocation,'' \emph{arXiv}, Nov. 2019.
  [Online]. Available: \url{https://arxiv.org/abs/1911.03240.}
\BIBentrySTDinterwordspacing

\bibitem{he2020coordinated}
\BIBentryALTinterwordspacing
J.~He, K.~Yu, and Y.~Shi, ``Coordinated passive beamforming for distributed
  intelligent reflecting surfaces network,'' \emph{arXiv}, Feb. 2020. [Online].
  Available: \url{https://arxiv.org/abs/2002.05915.}
\BIBentrySTDinterwordspacing

\bibitem{xu2020resource}
\BIBentryALTinterwordspacing
D.~Xu, X.~Yu, and R.~Schober, ``Resource allocation for intelligent reflecting
  surface-assisted cognitive radio networks,'' \emph{arXiv}, Jan. 2020.
  [Online]. Available: \url{https://arxiv.org/abs/2001.11729.}
\BIBentrySTDinterwordspacing

\bibitem{atapattu2020reconfigurable}
\BIBentryALTinterwordspacing
S.~Atapattu, R.~Fan, P.~Dharmawansa, G.~Wang, J.~Evans, and T.~A. Tsiftsis,
  ``Reconfigurable intelligent surface assisted two-way communications:
  Performance analysis and optimization,'' \emph{arXiv}, Jan. 2020. [Online].
  Available: \url{https://arxiv.org/abs/2001.07907.}
\BIBentrySTDinterwordspacing

\bibitem{cao2020sum}
\BIBentryALTinterwordspacing
Y.~Cao and T.~Lv, ``Sum rate maximization for reconfigurable intelligent
  surface assisted device-to-device communications,'' \emph{arXiv}, Jan. 2020.
  [Online]. Available: \url{https://arxiv.org/abs/2001.03344.}
\BIBentrySTDinterwordspacing

\bibitem{du2019multiple}
\BIBentryALTinterwordspacing
L.~Du, J.~Ma, Q.~Liang, and Y.~Tang, ``Multiple antenna multicast transmission
  assisted by reconfigurable intelligent surfaces,'' \emph{arXiv}, Dec. 2019.
  [Online]. Available: \url{https://arxiv.org/abs/1912.07960.}
\BIBentrySTDinterwordspacing

\bibitem{yuan2019intelligent}
\BIBentryALTinterwordspacing
J.~Yuan, Y.-C. Liang, J.~Joung, G.~Feng, and E.~G. Larsson, ``Intelligent
  reflecting surface-assisted cognitive radio system,'' \emph{arXiv}, Dec.
  2019. [Online]. Available: \url{https://arxiv.org/abs/1912.10678.}
\BIBentrySTDinterwordspacing

\bibitem{ye2019joint}
\BIBentryALTinterwordspacing
J.~Ye, S.~Guo, and M.-S. Alouini, ``Joint reflecting and precoding designs for
  ser minimization in reconfigurable intelligent surfaces assisted mimo
  systems,'' \emph{arXiv}, Jun. 2019. [Online]. Available:
  \url{https://arxiv.org/abs/1906.11466.}
\BIBentrySTDinterwordspacing

\bibitem{Ning2019b}
\BIBentryALTinterwordspacing
B.~Ning, Z.~Chen, W.~Chen, and J.~Fang, ``Intelligent reflecting surface design
  for {MIMO} system by maximizing sum-path-gains,'' \emph{arXiv}, Sep. 2019.
  [Online]. Available: \url{https://arxiv.org/abs/1909.07282.}
\BIBentrySTDinterwordspacing

\bibitem{sanchez2019iterative}
\BIBentryALTinterwordspacing
J.~R. Sanchez, F.~Rusek, O.~Edfors, and L.~Liu, ``An iterative interference
  cancellation algorithm for large intelligent surfaces,'' \emph{arXiv}, Nov.
  2019. [Online]. Available: \url{https://arxiv.org/abs/1911.10804.}
\BIBentrySTDinterwordspacing

\bibitem{Hu2018}
S.~Hu, K.~Chitti, F.~Rusek, and O.~Edfors, ``User assignment with distributed
  large intelligent surface {(LIS)} systems,'' \emph{IEEE Int. Symp. Personal,
  Indoor and Mobile Radio Communications}, pp. 1--6, Sep. 2018.

\bibitem{wu2019beamforming}
Q.~Wu and R.~Zhang, ``Beamforming optimization for wireless network aided by
  intelligent reflecting surface with discrete phase shifts,'' \emph{IEEE
  Trans. Commun.}, vol.~68, no.~11, pp. 1838--1851, Dec. 2019.

\bibitem{omid2020irs}
\BIBentryALTinterwordspacing
Y.~Omid, S.~M. Shahabi, C.~Pan, Y.~Deng, and A.~Nallanathan, ``{IRS}-aided
  large-scale {MIMO} systems with passive constant envelope precoding,''
  \emph{arXiv}, Feb. 2020. [Online]. Available:
  \url{https://arxiv.org/abs/2002.10965.}
\BIBentrySTDinterwordspacing

\bibitem{han2019large}
Y.~Han, W.~Tang, S.~Jin, C.-K. Wen, and X.~Ma, ``Large intelligent
  surface-assisted wireless communication exploiting statistical {CSI},''
  \emph{IEEE Trans. Veh. Technol.}, vol.~68, no.~8, pp. 8238--8242, Jun. 2019.

\bibitem{li2019sum}
\BIBentryALTinterwordspacing
Q.~Li, X.~Cui, S.~X. Wu, and J.~Lin, ``Sum rate maximization for multiuser
  {MISO} downlink with intelligent reflecting surface,'' \emph{arXiv}, Dec.
  2019. [Online]. Available: \url{https://arxiv.org/abs/1912.09315.}
\BIBentrySTDinterwordspacing

\bibitem{zhao2019intelligent}
\BIBentryALTinterwordspacing
M.-M. Zhao, Q.~Wu, M.-J. Zhao, and R.~Zhang, ``Intelligent reflecting surface
  enhanced wireless network: Two-timescale beamforming optimization,''
  \emph{arXiv}, Dec. 2019. [Online]. Available:
  \url{https://arxiv.org/abs/1912.01818.}
\BIBentrySTDinterwordspacing

\bibitem{BoyaDi20}
B.~Di, H.~Zhang, L.~Li, L.~Song, Y.~Li, and Z.~Han, ``Practical hybrid
  beamforming with limited-resolution phase shifters for reconfigurable
  intelligent surface based multi-user communications,'' \emph{IEEE Trans. Veh.
  Technol.}, IEEE Early Access, 2020.

\bibitem{Liu2019}
R.~Liu, H.~Li, M.~Li, and Q.~Liu, ``Symbol-level precoding design for
  intelligent reflecting surface assisted multi-user {MIMO} systems,''
  \emph{IEEE Int. Conf. Wireless Communications and Signal Processing}, pp.
  1--6, Oct. 2019.

\bibitem{Wu2019c}
Q.~Wu and R.~Zhang, ``Beamforming optimization for intelligent reflecting
  surface with discrete phase shifts,'' \emph{IEEE Int. Conf. Acoustics, Speech
  and Signal Processing}, pp. 7830--7833, May. 2019.

\bibitem{Di2019}
\BIBentryALTinterwordspacing
B.~Di, H.~Zhang, L.~Song, Y.~Li, Z.~Han, and H.~V. Poor, ``Hybrid beamforming
  for reconfigurable intelligent surface based multi-user communications:
  Achievable rates with limited discrete phase shifts,'' \emph{arXiv}, Oct.
  2019. [Online]. Available: \url{https://arxiv.org/abs/1910.14328.}
\BIBentrySTDinterwordspacing

\bibitem{mu2020capacity}
\BIBentryALTinterwordspacing
X.~Mu, Y.~Liu, L.~Guo, J.~Lin, and N.~Al-Dhahir, ``Capacity and optimal
  resource allocation for {IRS}-assisted multi-user communication systems,''
  \emph{arXiv}, Jan. 2020. [Online]. Available:
  \url{https://arxiv.org/abs/2001.03913.}
\BIBentrySTDinterwordspacing

\bibitem{Chu2019}
Z.~{Chu}, W.~{Hao}, P.~{Xiao}, and J.~{Shi}, ``Intelligent reflecting surface
  aided multi-antenna secure transmission,'' \emph{IEEE Wireless Commun.
  Lett.}, vol.~9, no.~1, pp. 108--112, Sep. 2019.

\bibitem{Cui2019}
M.~Cui, G.~Zhang, and R.~Zhang, ``Secure wireless communication via intelligent
  reflecting surface,'' \emph{IEEE Wireless Commun. Lett.}, vol.~8, pp.
  1410--1414, May. 2019.

\bibitem{Feng2019b}
\BIBentryALTinterwordspacing
K.~Feng and X.~Li, ``Physical layer security enhancement exploiting intelligent
  reflecting surface,'' \emph{arXiv}, Nov. 2019. [Online]. Available:
  \url{https://arxiv.org/abs/1911.02766.}
\BIBentrySTDinterwordspacing

\bibitem{Shen2019}
H.~Shen, W.~Xu, S.~Gong, Z.~He, and C.~Zhao, ``Secrecy rate maximization for
  intelligent reflecting surface assisted multi-antenna communications,''
  \emph{IEEE Commun. Lett.}, vol.~23, pp. 1488--1492, Jun. 2019.

\bibitem{Feng2019}
B.~Feng, Y.~Wu, and M.~Zheng, ``Secure transmission strategy for intelligent
  reflecting surface enhanced wireless system,'' \emph{IEEE Int. Conf. Wireless
  Communications and Signal Processing}, pp. 1--6, Oct. 2019.

\bibitem{Yu2019c}
\BIBentryALTinterwordspacing
X.~Yu, D.~Xu, and R.~Schober, ``Enabling secure wireless communications via
  intelligent reflecting surfaces,'' \emph{arXiv}, Apr. 2019. [Online].
  Available: \url{http://arxiv.org/abs/1904.09573.}
\BIBentrySTDinterwordspacing

\bibitem{Guan2019}
\BIBentryALTinterwordspacing
X.~Guan, Q.~Wu, and R.~Zhang, ``Intelligent reflecting surface assisted secrecy
  communication via joint beamforming and jamming,'' \emph{arXiv}, Jul. 2019.
  [Online]. Available: \url{https://arxiv.org/abs/1907.12839.}
\BIBentrySTDinterwordspacing

\bibitem{Yu2019b}
\BIBentryALTinterwordspacing
X.~Yu, D.~Xu, Y.~Sun, D.~W.~K. Ng, and R.~Schober, ``Robust and secure wireless
  communications via intelligent reflecting surfaces,'' \emph{arXiv}, Dec.
  2019. [Online]. Available: \url{https://arxiv.org/abs/1912.01497.}
\BIBentrySTDinterwordspacing

\bibitem{Lu2019}
\BIBentryALTinterwordspacing
X.~Lu, E.~Hossain, T.~Shafique, S.~Feng, H.~Jiang, and D.~Niyato, ``Intelligent
  reflecting surface ({IRS})-enabled covert communications in wireless
  networks,'' \emph{arXiv}, Nov. 2019. [Online]. Available:
  \url{https://arxiv.org/abs/1911.00986.}
\BIBentrySTDinterwordspacing

\bibitem{Chen2019}
J.~Chen, Y.-C. Liang, Y.~Pei, and H.~Guo, ``Intelligent reflecting surface: A
  programmable wireless environment for physical layer security,'' \emph{IEEE
  Access}, vol.~7, pp. 82\,599--82\,612, Jun. 2019.

\bibitem{hong2020artificial}
\BIBentryALTinterwordspacing
S.~Hong, C.~Pan, H.~Ren, K.~Wang, and A.~Nallanathan, ``Artificial-noise-aided
  secure {MIMO} wireless communications via intelligent reflecting surface,''
  Feb. 2020. [Online]. Available: \url{https://arxiv.org/abs/2002.07063.}
\BIBentrySTDinterwordspacing

\bibitem{dong2020secure}
L.~Dong and H.-M. Wang, ``Secure {MIMO} transmission via intelligent reflecting
  surface,'' \emph{IEEE Wireless Commun. Lett.}, IEEE Early Access, 2020.

\bibitem{jiang2020intelligent}
\BIBentryALTinterwordspacing
W.~Jiang, Y.~Zhang, J.~Wu, W.~Feng, and Y.~Jin, ``Intelligent reflecting
  surface assisted secure wireless communications with multiple-transmit and
  multiple-receive antennas,'' \emph{arXiv}, Jan. 2020. [Online]. Available:
  \url{https://arxiv.org/abs/2001.08963.}
\BIBentrySTDinterwordspacing

\bibitem{wang2020angle}
\BIBentryALTinterwordspacing
S.~Wang, M.~Wen, M.~Xia, R.~Wang, Q.~Hao, and Y.-C. Wu, ``Angle aware user
  cooperation for secure massive {MIMO} in rician fading channel,''
  \emph{arXiv}, Feb. 2020. [Online]. Available:
  \url{https://arxiv.org/abs/2002.10327.}
\BIBentrySTDinterwordspacing

\bibitem{xu2019resource}
\BIBentryALTinterwordspacing
D.~Xu, X.~Yu, Y.~Sun, D.~W.~K. Ng, and R.~Schober, ``Resource allocation for
  secure {IRS}-assisted multiuser {MISO} systems,'' \emph{arXiv}, Jul. 2019.
  [Online]. Available: \url{https://arxiv.org/abs/1907.03085.}
\BIBentrySTDinterwordspacing

\bibitem{lyu2020irs}
\BIBentryALTinterwordspacing
B.~Lyu, D.~T. Hoang, S.~Gong, D.~Niyato, and D.~I. Kim, ``{IRS}-based wireless
  jamming attacks: When jammers can attack without power,'' \emph{arXiv}, Jan.
  2020. [Online]. Available: \url{https://arxiv.org/abs/2001.01887.}
\BIBentrySTDinterwordspacing

\bibitem{makarfi2019physical}
\BIBentryALTinterwordspacing
A.~U. Makarfi, K.~M. Rabie, O.~Kaiwartya, X.~Li, and R.~Kharel, ``Physical
  layer security in vehicular networks with reconfigurable intelligent
  surfaces,'' \emph{arXiv}, Dec. 2019. [Online]. Available:
  \url{https://arxiv.org/abs/1912.12183.}
\BIBentrySTDinterwordspacing

\bibitem{Fu2019}
\BIBentryALTinterwordspacing
M.~Fu, Y.~Zhou, and Y.~Shi, ``Reconfigurable intelligent surface empowered
  downlink non-orthogonal multiple access,'' \emph{arXiv}, Oct. 2019. [Online].
  Available: \url{https://arxiv.org/abs/1910.07361.}
\BIBentrySTDinterwordspacing

\bibitem{Li2019}
\BIBentryALTinterwordspacing
Y.~Li, M.~Jiang, Q.~Z., and J.~Qin, ``Joint beamforming design in multi-cluster
  {MISO} {NOMA} intelligent reflecting surface-aided downlink communication
  networks,'' \emph{arXiv}, Sep. 2019. [Online]. Available:
  \url{https://arxiv.org/abs/1909.06972.}
\BIBentrySTDinterwordspacing

\bibitem{Mu2019}
\BIBentryALTinterwordspacing
X.~Mu, Y.~Liu, L.~Guo, J.~Lin, and N.~Al-Dhahir, ``Exploiting intelligent
  reflecting surfaces in multi-antenna aided {NOMA} systems,'' \emph{arXiv},
  Oct. 2019. [Online]. Available: \url{https://arxiv.org/abs/1910.13636.}
\BIBentrySTDinterwordspacing

\bibitem{Yang2019b}
\BIBentryALTinterwordspacing
G.~Yang, X.~Xu, and Y.-C. Liang, ``Intelligent reflecting surface assisted
  non-orthogonal multiple access,'' \emph{arXiv}, Jul. 2019. [Online].
  Available: \url{http://arxiv.org/abs/1907.03133.}
\BIBentrySTDinterwordspacing

\bibitem{zheng2020intelligent}
B.~Zheng, Q.~Wu, and R.~Zhang, ``Intelligent reflecting surface-assisted
  multiple access with user pairing: {NOMA or OMA}?'' \emph{IEEE Commun.
  Lett.}, vol.~24, no.~4, pp. 753--757, Jan. 2020.

\bibitem{ding2019simple}
\BIBentryALTinterwordspacing
Z.~Ding and H.~V. Poor, ``A simple design of {IRS-NOMA} transmission,''
  \emph{arXiv}, Jul. 2019. [Online]. Available:
  \url{https://arxiv.org/abs/1907.09918.}
\BIBentrySTDinterwordspacing

\bibitem{ding2020impact}
\BIBentryALTinterwordspacing
Z.~Ding, R.~Schober, and H.~V. Poor, ``On the impact of phase shifting designs
  on {IRS-NOMA},'' \emph{arXiv}, Jan. 2020. [Online]. Available:
  \url{https://arxiv.org/abs/2001.10909.}
\BIBentrySTDinterwordspacing

\bibitem{fu2019intelligent}
\BIBentryALTinterwordspacing
M.~Fu, Y.~Zhou, and Y.~Shi, ``Intelligent reflecting surface for downlink
  non-orthogonal multiple access networks,'' \emph{arXiv}, Jun. 2019. [Online].
  Available: \url{https://arxiv.org/abs/1906.09434.}
\BIBentrySTDinterwordspacing

\bibitem{yue2020performance}
\BIBentryALTinterwordspacing
X.~Yue and Y.~Liu, ``Performance analysis of intelligent reflecting surface
  assisted {NOMA} networks,'' \emph{arXiv}, Feb. 2020. [Online]. Available:
  \url{https://arxiv.org/abs/2001.09907.}
\BIBentrySTDinterwordspacing

\bibitem{zuo2020resource}
\BIBentryALTinterwordspacing
J.~Zuo, Y.~Liu, Z.~Qin, and N.~Al-Dhahir, ``ower efficient {IRS-assisted
  NOMA},'' \emph{arXiv}, Feb. 2020. [Online]. Available:
  \url{https://arxiv.org/abs/2002.01765.}
\BIBentrySTDinterwordspacing

\bibitem{zhu2019power}
\BIBentryALTinterwordspacing
J.~Zhu, Y.~Huang, J.~Wang, K.~Navaie, and Z.~Ding, ``Power efficient
  {IRS-assisted NOMA},'' \emph{arXiv}, Dec. 2019. [Online]. Available:
  \url{https://arxiv.org/abs/1912.11768.}
\BIBentrySTDinterwordspacing

\bibitem{hou2019reconfigurable}
\BIBentryALTinterwordspacing
T.~Hou, Y.~Liu, Z.~Song, X.~Sun, Y.~Chen, and L.~Hanzo, ``Reconfigurable
  intelligent surface aided {NOMA} networks,'' Dec. 2019. [Online]. Available:
  \url{https://arxiv.org/abs/1912.10044.}
\BIBentrySTDinterwordspacing

\bibitem{park2020intelligent}
S.~Y. Park and D.~I. Kim, ``Intelligent reflecting surface-aided phase-shift
  backscatter communication,'' \emph{IEEE Int. Conf. Ubiquitous Information
  Management and Communication}, pp. 1--5, Jan. 2020.

\bibitem{zhang2020large}
\BIBentryALTinterwordspacing
Q.~Zhang, Y.-C. Liang, and H.~V. Poor, ``Large intelligent surface/antennas
  ({LISA}) assisted symbiotic radio for {IoT} communications,'' \emph{arXiv},
  Feb. 2020. [Online]. Available: \url{https://arxiv.org/abs/2002.00340.}
\BIBentrySTDinterwordspacing

\bibitem{ZhaoPerformanceAnalysis20}
W.~Zhao, G.~Wang, S.~Atapattu, T.~A. Tsiftsis, and X.~Ma, ``Performance
  analysis of large intelligent surface aided backscatter communication
  systems,'' \emph{IEEE Wireless Commun. Lett.}, IEEE, Early Access, 2020.

\bibitem{makarfi2019reconfigurable}
\BIBentryALTinterwordspacing
A.~U. Makarfi, K.~M. Rabie, O.~Kaiwartya, O.~S. Badarneh, X.~Li, and R.~Kharel,
  ``Reconfigurable intelligent surface enabled {IoT} networks in generalized
  fading channels,'' \emph{arXiv}, Dec. 2019. [Online]. Available:
  \url{https://arxiv.org/abs/1912.06250.}
\BIBentrySTDinterwordspacing

\bibitem{Li2019d}
S.~Li, B.~Duo, X.~Yuan, Y.-C. Liang, and M.~{Di~Renzo}, ``Reconfigurable
  intelligent surface assisted {UAV} communication: Joint trajectory design and
  passive beamforming,'' \emph{IEEE Wireless Commun. Lett.}, IEEE Early Access,
  2020.

\bibitem{Ma2019}
\BIBentryALTinterwordspacing
D.~Ma, M.~Ding, and M.~Hassan, ``Enhancing cellular communications for {UAVs}
  via intelligent reflective surface,'' \emph{arXiv}, Nov. 2019. [Online].
  Available: \url{https://arxiv.org/abs/1911.07631.}
\BIBentrySTDinterwordspacing

\bibitem{lu2020enabling}
\BIBentryALTinterwordspacing
H.~Lu, Y.~Zeng, S.~Jin, and R.~Zhang, ``Enabling panoramic full-angle
  reflection via aerial intelligent reflecting surface,'' \emph{arXiv}, Jan.
  2020. [Online]. Available: \url{https://arxiv.org/abs/2001.07339.}
\BIBentrySTDinterwordspacing

\bibitem{Zhang2019d}
\BIBentryALTinterwordspacing
Q.~Zhang, W.~Saad, and M.~Bennis, ``Reflections in the sky: Millimeter wave
  communication with {UAV}-carried intelligent reflectors,'' \emph{arXiv}, Aug.
  2019. [Online]. Available: \url{https://arxiv.org/abs/1908.03271.}
\BIBentrySTDinterwordspacing

\bibitem{ZhengIRS20}
Y.~Zheng, S.~Bi, Y.~J. Zhang, Z.~Quan, and H.~Wang, ``Intelligent reflecting
  surface enhanced user cooperation in wireless powered communication
  networks,'' \emph{IEEE Wireless Commun. Lett.}, Feb. 2020.

\bibitem{Pan2019}
\BIBentryALTinterwordspacing
C.~Pan, H.~Ren, K.~Wang, M.~Elkashlan, A.~Nallanathan, J.~Wang, and L.~Hanzo,
  ``Intelligent reflecting surface aided {MIMO} broadcasting for simultaneous
  wireless information and power transfer,'' \emph{arXiv}, Aug. 2019. [Online].
  Available: \url{https://arxiv.org/abs/1908.04863.}
\BIBentrySTDinterwordspacing

\bibitem{Tang2019}
\BIBentryALTinterwordspacing
Y.~Tang, G.~Ma, H.~Xie, J.~Xu, and X.~Han, ``Joint transmit and reflective
  beamforming design for {IRS}-assisted multiuser {MISO} swipt systems,''
  \emph{arXiv}, Oct. 2019. [Online]. Available:
  \url{https://arxiv.org/abs/1910.07156.}
\BIBentrySTDinterwordspacing

\bibitem{Wu2019d}
\BIBentryALTinterwordspacing
Q.~Wu and R.~Zhang, ``Joint active and passive beamforming optimization for
  intelligent reflecting surface assisted {SWIPT} under {QoS} constraints,''
  \emph{arXiv}, Oct. 2019. [Online]. Available:
  \url{https://arxiv.org/abs/1910.06220.}
\BIBentrySTDinterwordspacing

\bibitem{shi2019enhanced}
\BIBentryALTinterwordspacing
W.~Shi, X.~Zhou, L.~Jia, Y.~Wu, F.~Shu, and J.~Wang, ``Enhanced secure wireless
  information and power transfer via intelligent reflecting surface,''
  \emph{arXiv}, Nov. 2019. [Online]. Available:
  \url{https://arxiv.org/abs/1911.01001.}
\BIBentrySTDinterwordspacing

\bibitem{Bai2019}
\BIBentryALTinterwordspacing
T.~Bai, C.~Pan, Y.~Deng, M.~Elkashlan, and A.~Nallanathan, ``Latency
  minimization for intelligent reflecting surface aided mobile edge
  computing,'' \emph{arXiv}, Oct. 2019. [Online]. Available:
  \url{https://arxiv.org/abs/1910.07990.}
\BIBentrySTDinterwordspacing

\bibitem{Hua2019}
\BIBentryALTinterwordspacing
S.~Hua, Y.~Zhou, K.~Yang, and Y.~Shi, ``Reconfigurable intelligent surface for
  green edge inference,'' \emph{arXiv}, Dec. 2019. [Online]. Available:
  \url{https://arxiv.org/abs/1912.00820.}
\BIBentrySTDinterwordspacing

\bibitem{cao2019intelligent}
\BIBentryALTinterwordspacing
Y.~Cao and T.~Lv, ``Intelligent reflecting surface enhanced resilient design
  for {MEC} offloading over millimeter wave links,'' \emph{arXiv}, Dec. 2019.
  [Online]. Available: \url{https://arxiv.org/abs/1912.06361.}
\BIBentrySTDinterwordspacing

\bibitem{liu2020intelligent}
\BIBentryALTinterwordspacing
Y.~Liu, J.~Zhao, Z.~Xiong, D.~Niyato, Y.~Chau, C.~Pan, and B.~Huang,
  ``Intelligent reflecting surface meets mobile edge computing: Enhancing
  wireless communications for computation offloading,'' \emph{arXiv}, Jan.
  2020. [Online]. Available: \url{https://arxiv.org/abs/2001.07449.}
\BIBentrySTDinterwordspacing

\bibitem{Cao2019}
\BIBentryALTinterwordspacing
Y.~Cao and T.~Lv, ``Intelligent reflecting surface aided multi-user
  millimeter-wave communications for coverage enhancement,'' \emph{arXiv}, Oct.
  2019. [Online]. Available: \url{https://arxiv.org/abs/1908.10734.}
\BIBentrySTDinterwordspacing

\bibitem{chaccour2020risk}
\BIBentryALTinterwordspacing
C.~Chaccour, M.~N. Soorki, W.~Saad, M.~Bennis, and P.~Popovski, ``Risk-based
  optimization of virtual reality over terahertz reconfigurable intelligent
  surfaces,'' \emph{arXiv}, Feb. 2020. [Online]. Available:
  \url{https://arxiv.org/abs/2002.09052.}
\BIBentrySTDinterwordspacing

\bibitem{ying2020gmd}
\BIBentryALTinterwordspacing
K.~Ying, Z.~Gao, S.~Lyu, Y.~Wu, H.~Wang, and M.-S. Alouini, ``{GMD}-based
  hybrid beamforming for large reconfigurable intelligent surface assisted
  millimeter-wave massive {MIMO},'' \emph{arXiv}, Jan. 2020. [Online].
  Available: \url{https://arxiv.org/abs/2001.05763.}
\BIBentrySTDinterwordspacing

\bibitem{cao2019delay}
\BIBentryALTinterwordspacing
Y.~Cao and T.~Lv, ``Delay-constrained joint power control, user detection and
  passive beamforming in intelligent reflecting surface assisted uplink mmwave
  system,'' \emph{arXiv}, Dec. 2019. [Online]. Available:
  \url{https://arxiv.org/abs/1912.10030.}
\BIBentrySTDinterwordspacing

\bibitem{xiu2020irs}
\BIBentryALTinterwordspacing
Y.~Xiu, Y.~Zhao, Y.~Liu, J.~Zhao, O.~Yagan, and N.~Wei, ``{IRS}-assisted
  millimeter wave communications: Joint power allocation and beamforming
  design,'' \emph{arXiv}, Jan. 2020. [Online]. Available:
  \url{https://arxiv.org/abs/2001.07467.}
\BIBentrySTDinterwordspacing

\bibitem{chen2019sum}
W.~Chen, X.~Ma, Z.~Li, and N.~Kuang, ``Sum-rate maximization for intelligent
  reflecting surface based terahertz communication systems,'' \emph{IEEE Int.
  Conf. Communications in China}, pp. 153--157, Aug. 2019.

\bibitem{jamali2019intelligent}
\BIBentryALTinterwordspacing
V.~Jamali, A.~Tulino, G.~Fischer, R.~M{\"u}ller, and R.~Schober, ``Intelligent
  reflecting and transmitting surface aided millimeter wave massive {MIMO},''
  \emph{arXiv}, Feb. 2019. [Online]. Available:
  \url{https://arxiv.org/abs/1902.07670.}
\BIBentrySTDinterwordspacing

\bibitem{najafi2019intelligent}
\BIBentryALTinterwordspacing
M.~Najafi and R.~Schober, ``Intelligent reflecting surfaces for free space
  optical communications,'' \emph{arXiv}, May. 2019. [Online]. Available:
  \url{https://arxiv.org/abs/1905.01094.}
\BIBentrySTDinterwordspacing

\bibitem{Perovi2019}
\BIBentryALTinterwordspacing
N.~S. Perovi, M.~{Di~Renzo}, and M.~F. Flanagan, ``Channel capacity
  optimization using reconfigurable intelligent surfaces in indoor mmwave
  environments,'' \emph{arXiv}, Oct. 2019. [Online]. Available:
  \url{https://arxiv.org/abs/1910.14310.}
\BIBentrySTDinterwordspacing

\bibitem{Pradhan2019}
C.~Pradhan, A.~Li, L.~Song, B.~Vucetic, and Y.~Li, ``Hybrid precoding design
  for reconfigurable intelligent surface aided mm{W}ave communication
  systems,'' \emph{IEEE Wireless Commun. Lett.}, IEEE Early Access, 2020.

\bibitem{Wang2019b}
\BIBentryALTinterwordspacing
P.~Wang, J.~Fang, X.~Yuan, Z.~D. Chen, H.~Duan, and H.~Li, ``Intelligent
  reflecting surface-assisted millimeter wave communications: Joint active and
  passive precoding design,'' \emph{arXiv}, Aug. 2019. [Online]. Available:
  \url{https://arxiv.org/abs/1908.10734.}
\BIBentrySTDinterwordspacing

\bibitem{Wang2019}
\BIBentryALTinterwordspacing
P.~Wang, J.~Fang, and H.~Li, ``Joint beamforming for intelligent reflecting
  surface-assisted millimeter wave communications,'' \emph{arXiv}, Oct. 2019.
  [Online]. Available: \url{https://arxiv.org/abs/1910.08541.}
\BIBentrySTDinterwordspacing

\bibitem{wang2020performance}
\BIBentryALTinterwordspacing
H.~Wang, Z.~Zhang, B.~Zhu, J.~Dang, L.~Wu, L.~Wang, K.~Zhang, and Y.~Zhang,
  ``Performance of wireless optical communication with reconfigurable
  intelligent surfaces and random obstacles,'' \emph{arXiv}, Jan. 2020.
  [Online]. Available: \url{https://arxiv.org/abs/2001.05715.}
\BIBentrySTDinterwordspacing

\bibitem{Jamieson_2017}
A.~Welkie, L.~Shangguan, J.~Gummeson, W.~Hu, and K.~Jamieson, ``Programmable
  radio environments for smart spaces,'' \emph{ACM Work. Hot Topics in
  Networks}, pp. 36--42, Nov. 2017.

\bibitem{Jamieson_2019}
Z.~Li, Y.~Xie, L.~Shangguan, R.~I. Zelaya, J.~Gummeson, W.~Hu, and K.~Jamieson,
  ``Towards programming the radio environment with large arrays of inexpensive
  antennas,'' \emph{USENIX Symp. Networked Systems Design and Implementation},
  pp. 285--299, Feb. 2019.

\bibitem{Bharadia_2020}
M.~Dunna, C.~Zhang, D.~Sievenpiper, and D.~Bharadia, ``{ScatterMIMO}: Enabling
  virtual {MIMO} with smart surfaces,'' \emph{ACM Annual Int. Conf. Mobile
  Computing and Networking}, 14 pages, Sep. 2020.

\bibitem{Hu2017}
S.~{Hu}, F.~{Rusek}, and O.~{Edfors}, ``Beyond massive-{MIMO}: The potential of
  data-transmission with large intelligent surfaces,'' \emph{IEEE Trans. Signal
  Process}, vol.~66, no.~10, pp. 2746--2758, Apr. 2018.

\bibitem{Bjornson2019}
E.~Bj{\"o}rnson, {\"O}.~{\"O}zdogan, and E.~G. Larsson, ``Intelligent
  reflecting surface vs. decode-and-forward: How large surfaces are needed to
  beat relaying?'' \emph{IEEE Wireless Commun. Lett.}, vol.~9, no.~2, pp.
  244--248, Feb. 2020.

\bibitem{bjornson2020power}
\BIBentryALTinterwordspacing
E.~Bj{\"o}rnson and L.~Sanguinetti, ``Power scaling laws and near-field
  behaviors of massive {MIMO} and intelligent reflecting surfaces,''
  \emph{arXiv}, Feb. 2020. [Online]. Available:
  \url{https://arxiv.org/abs/2002.04960.}
\BIBentrySTDinterwordspacing

\bibitem{dardari2019communicating}
\BIBentryALTinterwordspacing
D.~Dardari, ``Communicating with large intelligent surfaces: Fundamental limits
  and models,'' \emph{arXiv}, Dec. 2019. [Online]. Available:
  \url{https://arxiv.org/abs/1912.01719.}
\BIBentrySTDinterwordspacing

\bibitem{VisorSurf__6}
S.~Abadal, C.~Liaskos, A.~Tsioliaridou, S.~Ioannidis, A.~Pitsillides,
  J.~Sole-Pareta, E.~Alarcan, and A.~Cabellos-Aparicio, ``Computing and
  communications for the software-defined metamaterial paradigm: A context
  analysis,'' \emph{IEEE Access}, vol.~5, pp. 6225--6235, Apr. 2017.

\bibitem{VisorSurf__7}
S.~Abadal, A.~Mestres, J.~Torrellas, E.~Alarcon, and A.~Cabellos-Aparicio,
  ``Medium access control in wireless network-on-chip: A context analysis,''
  \emph{IEEE Commun. Mag.}, vol.~56, no.~6, pp. 172--178, Jun. 2018.

\bibitem{VisorSurf__4}
C.~Liaskos, A.~Tsioliaridou, S.~Nie, A.~Pitsillides, S.~Ioannidis, and I.~F.
  Akyildiz, ``On the network-layer modeling and configuration of programmable
  wireless environments,'' \emph{IEEE/ACM Trans. Netw.}, vol.~27, no.~4, pp.
  1696--1713, Apr. 2019.

\bibitem{VisorSurf__5}
C.~Liaskos, S.~Nie, A.~Tsioliaridou, A.~Pitsillides, S.~Ioannidis, and I.~F.
  Akyildiz, ``A novel communication paradigm for high capacity and security via
  programmable indoor wireless environments in next generation wireless
  systems,'' \emph{Elsevier Ad Hoc Networks}, vol.~87, pp. 1--16, May 2019.

\bibitem{VisorSurf__3}
A.~C. Tasolamprou, A.~Pitilakis, S.~Abadal, O.~Tsilipakos, X.~Timoneda,
  H.~Taghvaee, M.~S. Mirmoosa, F.~Liu, C.~Liaskos, A.~Tsioliaridou,
  S.~Ioannidis, N.~V. Kantartzis, D.~Manessis, J.~Georgiou,
  A.~Cabellos-Aparicio, E.~Alarcon, I.~F. Akyildiz, S.~A. Tretyakov, E.~N.
  Economou, M.~Kafesaki, and C.~M. Soukoulis, ``Exploration of intercell
  wireless millimeter-wave communication in the landscape of intelligent
  metasurfaces,'' \emph{IEEE Access}, vol.~7, pp. 122\,931--122\,948, Aug.
  2019.

\bibitem{VisorSurf__1}
H.~Taghvaee, A.~Cabellos-Aparicio, J.~Georgiou, and S.~Abadal, ``Error analysis
  of programmable metasurfaces for beam steering,'' \emph{IEEE Trans. Emerg.
  Sel. Topics Circuits Syst}, vol.~10, no.~1, pp. 62--74, Mar. 2020.

\bibitem{VisorSurf__2}
D.~Kouzapas, C.~Skitsas, L.~Petrou, M.~Lestas, A.~Philippou, C.~Liaskos,
  V.~Soteriou, J.~Georgiou, S.~Abadal, T.~Saeed, and A.~Pitsillides, ``Towards
  fault adaptive routing in metasurface controller networks,'' \emph{Elsevier
  J. Systems Architecture}, vol. 106, no. 101703, Jun. 2020.

\bibitem{liaskos2019interpretable}
C.~Liaskos, A.~Tsioliaridou, S.~Nie, A.~Pitsillides, S.~Ioannidis, and
  I.~Akyildiz, ``An interpretable neural network for configuring programmable
  wireless environments,'' \emph{IEEE Int. Work. Signal Processing Advances in
  Wireless Communications}, pp. 1--5, Jul. 2019.

\bibitem{Huang2019}
C.~{Huang}, G.~C. {Alexandropoulos}, C.~{Yuen}, and M.~{Debbah}, ``Indoor
  signal focusing with deep learning designed reconfigurable intelligent
  surfaces,'' \emph{IEEE Int. Workshop on Signal Processing Advances in
  Wireless Communications}, pp. 1--5, Jul. 2019.

\bibitem{gao2020unsupervised}
\BIBentryALTinterwordspacing
J.~Gao, C.~Zhong, X.~Chen, H.~Lin, and Z.~Zhang, ``Unsupervised learning for
  passive beamforming,'' \emph{arXiv}, Jan. 2020. [Online]. Available:
  \url{https://arxiv.org/abs/2001.02348.}
\BIBentrySTDinterwordspacing

\bibitem{huang2020reconfigurable}
\BIBentryALTinterwordspacing
C.~Huang, R.~Mo, and C.~Yuen, ``Reconfigurable intelligent surface assisted
  multiuser {MISO} systems exploiting deep reinforcement learning,'' Feb. 2020.
  [Online]. Available: \url{https://arxiv.org/abs/2002.10072.}
\BIBentrySTDinterwordspacing

\bibitem{taha2020deep}
\BIBentryALTinterwordspacing
A.~Taha, Y.~Zhang, F.~B. Mismar, and A.~Alkhateeb, ``Deep reinforcement
  learning for intelligent reflecting surfaces: Towards standalone operation,''
  \emph{arXiv}, Feb. 2020. [Online]. Available:
  \url{https://arxiv.org/abs/2002.11101.}
\BIBentrySTDinterwordspacing

\bibitem{feng2020deep}
K.~Feng, Q.~Wang, X.~Li, and C.-K. Wen, ``Deep reinforcement learning based
  intelligent reflecting surface optimization for {MISO} communication
  systems,'' \emph{IEEE Wireless Commun. Lett.}, IEEE Early Access, 2020.

\bibitem{liu2020ris}
\BIBentryALTinterwordspacing
X.~Liu, Y.~Liu, Y.~Chen, and H.~V. Poor, ``{RIS} enhanced massive
  non-orthogonal multiple access networks: Deployment and passive beamforming
  design,'' \emph{arXiv}, Jan. 2020. [Online]. Available:
  \url{https://arxiv.org/abs/2001.10363.}
\BIBentrySTDinterwordspacing

\bibitem{Khan2019}
\BIBentryALTinterwordspacing
S.~Khan and S.~Y. Shin, ``Deep-learning-aided detection for reconfigurable
  intelligent surfaces,'' \emph{arXiv}, Oct. 2019. [Online]. Available:
  \url{https://arxiv.org/abs/1910.09136.}
\BIBentrySTDinterwordspacing

\bibitem{elbir2020deep}
\BIBentryALTinterwordspacing
A.~M. Elbir, A.~Papazafeiropoulos, P.~Kourtessis, and S.~Chatzinotas, ``Deep
  channel learning for large intelligent surfaces aided mm-wave massive {MIMO}
  systems,'' \emph{arXiv}, Jan. 2020. [Online]. Available:
  \url{https://arxiv.org/abs/2001.11085.}
\BIBentrySTDinterwordspacing

\bibitem{taha2019enabling}
\BIBentryALTinterwordspacing
A.~Taha, M.~Alrabeiah, and A.~Alkhateeb, ``Enabling large intelligent surfaces
  with compressive sensing and deep learning,'' \emph{arXiv}, Apr. 2019.
  [Online]. Available: \url{https://arxiv.org/abs/1904.10136.}
\BIBentrySTDinterwordspacing

\bibitem{yang2020deep}
\BIBentryALTinterwordspacing
H.~Yang, Z.~Xiong, J.~Zhao, D.~Niyato, and L.~Xiao, ``Deep reinforcement
  learning based intelligent reflecting surface for secure wireless
  communications,'' \emph{arXiv}, Feb. 2020. [Online]. Available:
  \url{https://arxiv.org/abs/2002.12271.}
\BIBentrySTDinterwordspacing

\bibitem{Hu2018c}
S.~{Hu}, F.~{Rusek}, and O.~{Edfors}, ``Beyond massive {MIMO}: The potential of
  positioning with large intelligent surfaces,'' \emph{IEEE Trans. Signal
  Process}, vol.~66, no.~7, pp. 1761--1774, Apr. 2018.

\bibitem{AlegriaASILOMAR}
J.~V. Alegria and F.~Rusek, ``Cram{\'e}r-{R}ao lower bounds for positioning
  with large intelligent surfaces using quantized amplitude and phase,'' in
  \emph{IEEE Asilomar Conf. Signals, Systems, and Computers}, Nov. 2019.

\bibitem{he2019large}
\BIBentryALTinterwordspacing
J.~He, H.~Wymeersch, L.~Kong, O.~Silv{\'e}n, and M.~Juntti, ``Large intelligent
  surface for positioning in millimeter wave {MIMO} systems,'' \emph{arXiv},
  Sep. 2019. [Online]. Available: \url{https://arxiv.org/abs/1910.00060.}
\BIBentrySTDinterwordspacing

\bibitem{He2019}
\BIBentryALTinterwordspacing
J.~He, H.~Wymeersch, T.~Sanguanpuak, O.~Silv\'{e}n, and M.~Juntti, ``Adaptive
  beamforming design for mmwave ris-aided joint localization and
  communication,'' \emph{arXiv}, Nov. 2019. [Online]. Available:
  \url{https://arxiv.org/abs/1911.02813.}
\BIBentrySTDinterwordspacing

\bibitem{wymeersch2019radio}
\BIBentryALTinterwordspacing
H.~Wymeersch, J.~He, B.~Denis, A.~Clemente, and M.~Juntti, ``Radio localization
  and mapping with reconfigurable intelligent surfaces,'' \emph{arXiv}, Dec.
  2019. [Online]. Available: \url{https://arxiv.org/abs/1912.09401.}
\BIBentrySTDinterwordspacing

\bibitem{DocomoMetaWave}
\BIBentryALTinterwordspacing
{NTT~DOCOMO}, ``{DOCOMO} and {M}etawave announce successful demonstration of 28
  {GHz} band,'' Dec. 2018. [Online]. Available:
  \url{https://www.businesswire.com/news/home/20181204005253/en/NTT-DOCOMO-Metawave-Announce-Successful-Demonstration-28GHz-Band.}
\BIBentrySTDinterwordspacing

\bibitem{Tan_ICC2016}
X.~Tan, Z.~Sun, J.~M. Jornet, and D.~Pados, ``Increasing indoor spectrum
  sharing capacity using smart reflect-array,'' \emph{IEEE Int. Conf.
  Communications}, pp. 1--6, May. 2016.

\bibitem{Tan2018}
X.~Tan, Z.~Sun, D.~Koutsonikolas, and J.~Jornet, ``Enabling indoor mobile
  millimeter-wave networks based on smart reflect-arrays,'' \emph{IEEE Int.
  Conf. Computer Communications}, pp. 270--278, Apr. 2018.

\bibitem{Buffalo2019}
S.~K. Saha, Y.~Ghasempour, M.~K. Haider, T.~Siddiqui, P.~De~Melo, N.~Somanchi,
  L.~Zakrajsek, A.~Singh, O.~Torres, O.~Uvaydov, J.~M. Jornet, E.~Knightly,
  D.~Koutsonikolas, D.~Pados, and Z.~Sun, ``{X60}: A programmable testbed for
  wideband 60 {GHz} {WLANs} with phased arrays,'' \emph{Elsevier Computer
  Commun.}, vol. 133, pp. 77--88, Jan. 2019.

\bibitem{dai2019reconfigurable}
L.~{Dai}, B.~{Wang}, M.~{Wang}, X.~{Yang}, J.~{Tan}, S.~{Bi}, S.~{Xu},
  F.~{Yang}, Z.~{Chen}, M.~{Di Renzo}, C.~{Chae}, and L.~{Hanzo},
  ``Reconfigurable intelligent surface-based wireless communication: Antenna
  design, prototyping and experimental results,'' \emph{IEEE Access}, vol.~8,
  pp. 45\,913--45\,923, Mar. 2020.

\bibitem{hu2019reconfigurable}
\BIBentryALTinterwordspacing
J.~Hu, H.~Zhang, B.~Di, L.~Li, L.~Song, Y.~Li, Z.~Han, and H.~V. Poor,
  ``Reconfigurable intelligent surfaces based {RF} sensing: Design,
  optimization, and implementation,'' \emph{arXiv}, Dec. 2019. [Online].
  Available: \url{https://arxiv.org/abs/1912.09198.}
\BIBentrySTDinterwordspacing

\bibitem{Maci_Huawei}
S.~Maci, ``Present and future trends in metasurface antennas,'' in \emph{Huawei
  Technology Summit}, Munich, Germany, Oct. 2019.

\end{thebibliography}

\end{document}